\documentclass[preprint,11pt, texlive=2016]{elsarticle}
\usepackage[T1]{fontenc}
\usepackage[utf8]{inputenc}
\usepackage{charter}
\usepackage{wrapfig}
\usepackage{arydshln}

\usepackage[colorlinks,citecolor=blue,linktoc=all,linkcolor=cyan]{hyperref}
\usepackage{amssymb,amsbsy,amsmath,amsfonts}
\usepackage{mathrsfs} 
\numberwithin{equation}{section}
\numberwithin{table}{section}
\numberwithin{figure}{section}
\usepackage{nicefrac}
\usepackage{esint}
\usepackage{upgreek}
\usepackage{slashed}
\usepackage{multirow}
\usepackage{caption}
\usepackage{graphicx}
\usepackage{setspace} 
\usepackage{fancyhdr}
\usepackage{titlesec} 
\usepackage{helvet}
\usepackage{tocloft} 
\usepackage{blindtext} 
\usepackage{xcolor}
\usepackage{colortbl}
\usepackage{tabularx}
\usepackage{hyperref}
\usepackage[left]{lineno}
\usepackage{hyperref}
\usepackage{fancyhdr}
\usepackage{tabularx}
\usepackage{hyperref}
\usepackage{wrapfig}

\DeclareMathAlphabet{\mathantt}{OML}{antt}{l}{it}
\DeclareMathAlphabet{\mathpzc}{OT1}{pzc}{m}{n}

\include{preamble}

\journal{Progress of Particle and Nuclear Physics}

\newcommand{\qqbar} {\ensuremath{q\overline q}\xspace}

\newcommand{\ccbar} {\ensuremath{c\overline c}\xspace}

\newcommand{\bbbar} {\ensuremath{b\overline b}\xspace}

\newcommand{\msbar}{\ensuremath{\overline{\small{\rm MS}}}\xspace}
\newcommand{\mW}{\ensuremath{M_{W}}\xspace}
\newcommand{\MW}{\ensuremath{M_{W}}\xspace}
\newcommand{\mZ}{\ensuremath{M_{Z}}\xspace}
\newcommand{\MZ}{\ensuremath{M_{Z}}\xspace}
\newcommand{\mH}{\ensuremath{M_{H}}\xspace}
\newcommand{\MH}{\ensuremath{M_{H}}\xspace}
\newcommand{\GammaW}{\ensuremath{\Gamma_{W}}\xspace}
\newcommand{\GammaZ}{\ensuremath{\Gamma_{Z}}\xspace}
\newcommand{\mt}{\ensuremath{m_{t}}\xspace}
\mathchardef\Upsilon="7107
\def\Y#1S{\ensuremath{\Upsilon{(#1S)}}\xspace}

\newcommand{\as}{\ensuremath{\alpha_{\scriptscriptstyle s}}\xspace}

\newcommand{\Blue}{\textsc{Blue}}
\newcommand{\Pythia}{\textsc{Pythia}}
\newcommand{\Powheg}{\textsc{PowhegBox}}
\newcommand{\Resbos}{\textsc{Resbos}}

\newcommand{\DZero}{D\O\xspace}
\newcommand{\Aleph}{ALEPH\xspace}
\newcommand{\Opal}{OPAL\xspace}
\newcommand{\LThree}{L3\xspace}
\newcommand{\Delphi}{DELPHI\xspace}

\newcommand{\sintheta}{\ensuremath{\sin\!^2\theta_{W}}\xspace}
 
\newcommand{\seffsf}[1]{\sin\!^2\theta^{#1}_{{\rm eff}}}
\newcommand{\seffsffour}[1]{\sin\!^4\theta^{#1}_{{\rm eff}}}

\newcommand{\sinleff}{\ensuremath{\seffsf{l}}\xspace}
\newcommand{\sinfeff}{\ensuremath{\seffsf{f}}\xspace}

\newcommand{\dahadZf}{\ensuremath{\Delta\alpha_{\rm had}^{(5)}(M_Z^2)}\xspace}

\newcommand{\Kbar    }{\kern 0.2em\overline{\kern -0.2em K}{}\xspace}

\newcommand{\Kz      }{\ensuremath{K^0}\xspace}
\newcommand{\Kzb     }{\ensuremath{\Kbar^0}\xspace}
\newcommand{\KzKzb   }{\ensuremath{\Kz \kern -0.16em \Kzb}\xspace}
\newcommand{\Kp      }{\ensuremath{K^+}\xspace}
\newcommand{\Km      }{\ensuremath{K^-}\xspace}

\newcommand{\KpKm    }{\ensuremath{\Kp \kern -0.16em \Km}\xspace}

\newcommand\Dbar    {\kern 0.18em\overline{\kern -0.18em D}{}\xspace}

\newcommand\Bbar    {\kern 0.18em\overline{\kern -0.18em B}{}\xspace}

\newcommand\Bz      {\ensuremath{B^0}\xspace}

\newcommand\Bzb     {\ensuremath{\Bbar^0}\xspace}
\newcommand\Bu      {\ensuremath{B^+}\xspace}
\newcommand\Bub     {\ensuremath{B^-}\xspace}

\newcommand\BpBm    {\ensuremath{\Bu {\kern -0.16em \Bub}}\xspace}
\newcommand\Bs      {\ensuremath{B^0_{s}}\xspace}
\newcommand\Bsb     {\ensuremath{\Bbar^0_{s}}\xspace}

\newcommand\BzBzb   {\ensuremath{\Bz {\kern -0.16em \Bzb}}\xspace}
\newcommand\BszBszb {\ensuremath{\Bs {\kern -0.16em \Bsb}}\xspace}

\newcommand{\tev}{\ensuremath{\mathrm{Te\kern -0.1em V}}\xspace}
\newcommand{\gev}{\ensuremath{\mathrm{Ge\kern -0.1em V}}\xspace}
\newcommand{\GeV}{\ensuremath{\mathrm{Ge\kern -0.1em V}}\xspace}
\newcommand{\mev}{\ensuremath{\mathrm{Me\kern -0.1em V}}\xspace}
\newcommand{\MeV}{\ensuremath{\mathrm{Me\kern -0.1em V}}\xspace}
\newcommand{\kev}{\ensuremath{\mathrm{ke\kern -0.1em V}}\xspace}
\newcommand{\ev}{\ensuremath{\mathrm{e\kern -0.1em V}}\xspace}
\newcommand{\gevc}{\ensuremath{{\mathrm{Ge\kern -0.1em V\!/}c}}\xspace}
\newcommand{\mevc}{\ensuremath{{\mathrm{Me\kern -0.1em V\!/}c}}\xspace}
\newcommand{\gevcc}{\ensuremath{{\mathrm{Ge\kern -0.1em V\!/}c^2}}\xspace}
\newcommand{\mevcc}{\ensuremath{{\mathrm{Me\kern -0.1em V\!/}c^2}}\xspace}

\newcommand{\bei}{\begin{itemize}}
\newcommand{\eei}{\end{itemize}}

\newcommand\ie{{\it i.e.}\xspace} 
\newcommand\eg{{\it e.g.}\xspace}

\begin{document}

\begin{frontmatter}
\title{Electroweak Precision Tests of the Standard Model after the Discovery of the Higgs Boson}

\author[L1,L2]{Jens Erler}
\author[L1,L3]{Matthias Schott}

\address[L1]{PRISMA Cluster of Excellence, Institute of Physics, Johannes Gutenberg-University, 55099 Mainz, Germany}
\address[L2]{Departamento de F\'isica Te\'orica, Instituto de F\'isica, Universidad Nacional Aut\'onoma de M\'exico, 04510 CDMX, M\'exico}
\address[L3]{European Center for Nuclear Research, CERN, Geneva, Switzerland}

\begin{abstract}
The global fit of the Standard Model predictions to electroweak precision data, which has been routinely performed in the past decades by several groups, 
led to the prediction of the top quark and the Higgs boson masses before their respective discoveries. 
With the measurement of the Higgs boson mass at the Large Hadron Collider (LHC) in 2012 by the ATLAS and CMS collaborations, 
the last free parameter of the Standard Model of particle physics has been fixed, and the global electroweak fit can be used 
to test the full internal consistency of the electroweak sector of the Standard Model and constrain models beyond. 
In this article, we review the current state-of-the-art theoretical calculations, as well as the precision measurements performed at the LHC,
and interpret them within the context of the global electroweak fit. Special focus is drawn in the impact of the Higgs boson mass on the fit.
\end{abstract}

\end{frontmatter}

\thispagestyle{empty}
\tableofcontents


\section{Introduction\label{sec:Introduction}}

The birth of the leptonic sector of the Standard Model (SM) of particle physics can be dated back to 1967, 
when Steven Weinberg~\cite{Weinberg:1967tq} and Abdus Salam~\cite{Salam:1968rm} applied the Higgs mechanism~\cite{Higgs:1964pj,Englert:1964et,Guralnik:1964eu} 
to the electroweak unified theory of Sheldon Lee Glashow~\cite{Glashow:1961tr}.
The electroweak part of the SM combines two of the four known elementary forces of nature at a unification energy of $v = 246$~\GeV 
and is described by four gauge bosons:
the massless photon as the gauge boson of the electromagnetic interaction and the massive $W^+$, $W^-$ and $Z$ bosons as force carriers of the weak interaction. 
The Higgs mechanism is needed to consistently attribute masses to the $W^\pm$ and $Z$ bosons, 
which allows to construct a renormalizable theory as was first proved by Gerardus 't Hooft and Martinus Veltman~\cite{tHooft:1972tcz} in 1972.

The bosonic sector of the SM is determined by two parameters in the Higgs potential, 
\begin{equation}
\label{eqn:VH}
V_H = \mu^2 \phi^\dagger \phi + \frac{\lambda^2}{2} \left( \phi^\dagger \phi \right)^2,
\end{equation}
as well as the three gauge couplings $g_S$, $g$ and $g'$ associated with the $SU(3)_C \times SU(2)_L \times U(1)_Y$ gauge factors acting on the strong color (QCD), 
weak isospin and hypercharge quantum numbers, respectively.
$SU(2)_L$ acts on left-handed quark and lepton doublets, while the right-handed fermions transform trivially.
The hypercharge assignment is chiral, as well.
In this way, the $V-A$ structure of the charged current weak interaction is incorporated into the SM
and both vector and axial-vector couplings are predicted to appear in the $Z$ boson mediated neutral current.
It is convenient to utilize the alternative parameter quintet of very precisely measured quantities comprised of the electromagnetic and strong couplings,
\begin{equation}
\alpha \equiv \frac{e^2}{4 \pi} = \frac{g^2 g'^2}{4\pi(g^2 + g'^2)} \ , \qquad\qquad\qquad
\as \equiv \frac{g_S^2}{4 \pi}\ ,
\end{equation}
as well as the Fermi constant $G_F$ (see Section~\ref{sec:theofermi}) and the masses of the $Z$ and Higgs bosons, which to lowest order are given, respectively, by
\begin{equation}
\label{higgsvev}
\sqrt{2} G_F \equiv v^{-2} = \frac{\lambda^2}{2 |\mu^2|}\ , \qquad\qquad\qquad
\mZ = \frac{1}{2} \sqrt{g^2 + g'^2}\ v\ , \qquad\qquad\qquad
\mH = \lambda v\ .
\end{equation}

Other quantities can be expressed in terms of these inputs\footnote{In global analyses such as those described in later sections,
the distinction between input and derived observables is inessential and purely illustrative.}.
For example, the weak mixing angle $\theta_W$ obeys the tree level relations,
\begin{equation}
\label{eqn:sintreelevel}
\sintheta \equiv \frac{g'^2}{g^2 + g'^2} = 1 - \frac{g^2}{g^2 + g'^2} = 1 - \frac{\MW^2}{\MZ^2}\ ,
\end{equation}
where the $W$ boson mass is in turn related to $G_F$ and the fine-structure constant $\alpha$ {\em via\/}
\begin{equation}
\label{eqn:mwtreelevel}
\mW = \frac{1}{2} g v = \sqrt{\frac{\alpha \pi}{\sqrt{2} G_F \sintheta}} = \frac{\mZ}{\sqrt{2}} \sqrt{1+\sqrt{1-\frac{\sqrt{8} \pi \alpha}{G_F \MZ^2}}}\ .
\end{equation}
Equation~(\ref{eqn:sintreelevel}) has been used in the last step and subsequently solved for $\mW$.

\begin{figure*}[tb]
\centering
\resizebox{0.9\textwidth}{!}{\includegraphics{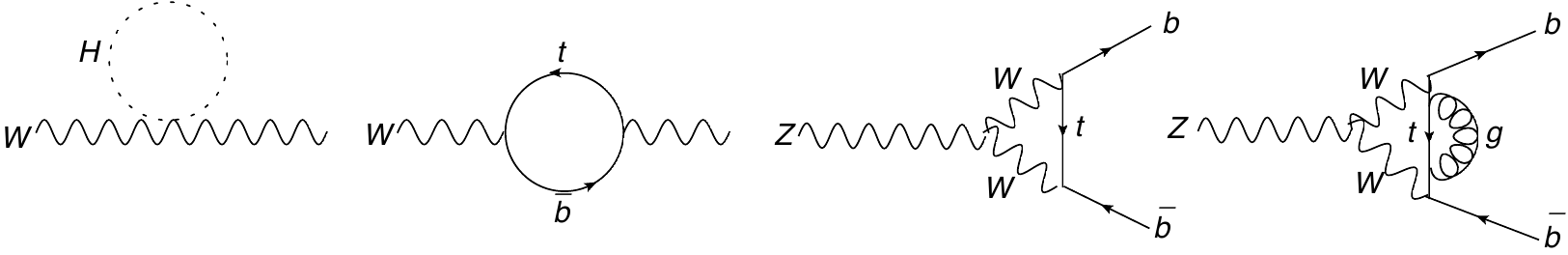}}
\caption{\label{fig:LoopCorrections} Feynman diagrams exemplifying loop corrections to the W boson propagator and the $Z\rightarrow b\bar b$ vertex.}
\end{figure*}

Inserting the measured values of \MZ, $G_F$ and $\alpha$ into Equation~(\ref{eqn:mwtreelevel}), 
a value of $\MW \approx 80.94$~\GeV is predicted. 
Comparison with the current experimental world average, $\MW \approx 80.38$~\GeV, reveals a significant discrepancy,
which is due to higher order electroweak and QCD corrections to the tree level relations. 
As examples of higher-loop Feynman diagrams we show loop corrections to the $W$ boson self-energy
and similar correction to the $Z \to b\bar b$ vertex in Figure~\ref{fig:LoopCorrections}.

The electroweak corrections can be absorbed into the quantity $\Delta r$~\cite{Sirlin:1980nh} 
describing the electroweak radiative corrections~\cite{Passarino:1978jh} to $\mu$ decay~\cite{Aoki:1982ed},
as well as into form factors~\cite{Hollik:1988ii} $\rho_Z^f$, modifying the vector and axial-vector couplings of fermion $f$ to the $Z$ boson, and $\kappa_Z^f$, 
modifying the additional corrections to the vector coupling,
\begin{eqnarray}
\label{Eqn:CorrectedObs}
\MW^2			&=& 	\frac{\MZ^2}{2} \left( 1 + \sqrt{1 - \frac{\sqrt{8} \pi \alpha (1 + \Delta r)}{G_F \MZ^2}} \right), \\
\sinfeff		 	&=& 	\kappa_Z^f \sintheta\ , \label{eq:sineff} \\
g_V^f 			&=&	 \sqrt{\rho_Z^f} \left( I_3^f - 2 Q^f \sinfeff \right), \label{eq:gveff}\\
g_A^f 			&=&	 \sqrt{\rho_Z^f} I_3^f\ ,\label{eq:gaeff}
\end{eqnarray}
where $Q^f$ and $I_3^f$ denote the electric charge and the third component of isospin, respectively. 
In general, the form factors are momentum dependent quantities and except at the $Z$ resonance they are gauge dependent. 
For the following discussion of the effective mixing angle, the $Z$ boson mass scale, \MZ, is chosen. 
If not stated otherwise, $\sintheta$ will from now on refer to the on-shell definition of the weak mixing angle,
{\em i.e.\/} the last form in Equation~(\ref{eqn:sintreelevel}) is {\em defined\/} to hold to all orders in perturbation theory. 

The one-loop radiative corrections depend logarithmic on $\MH$, 
and their dependence on the quark masses is dominated by quadratic terms in the mass of the heaviest SM particle, {\em i.e.}, the top quark mass $\mt$.
Hence, precise measurements of all observables of the electroweak sector plus the heavy quark masses and $\as$, allow for tests of the consistency of the SM,
and are therefore best analyzed in a global context.

\begin{figure*}[tb]
\centering
\resizebox{0.2455\textwidth}{!}{\includegraphics{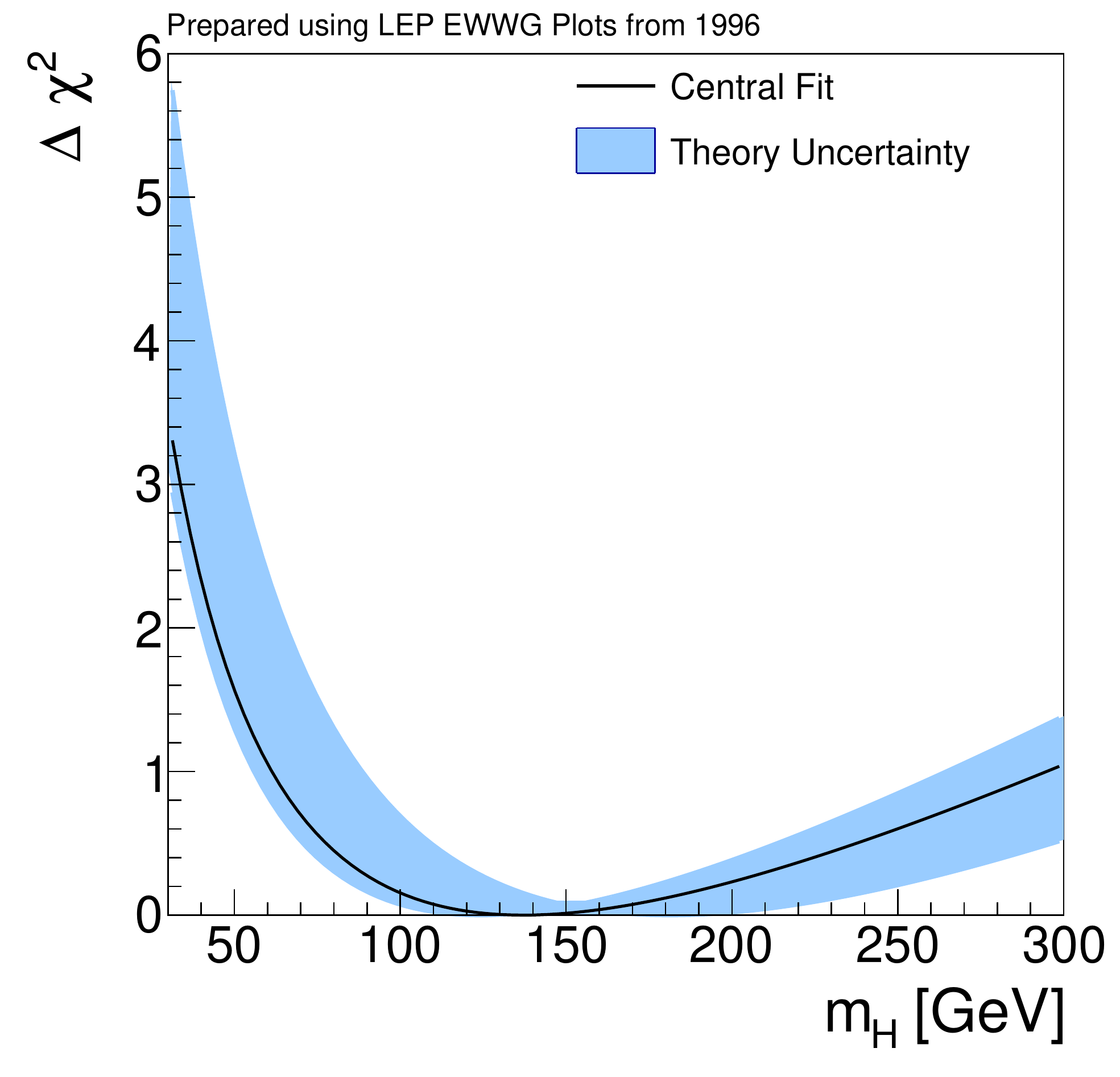}}
\resizebox{0.2455\textwidth}{!}{\includegraphics{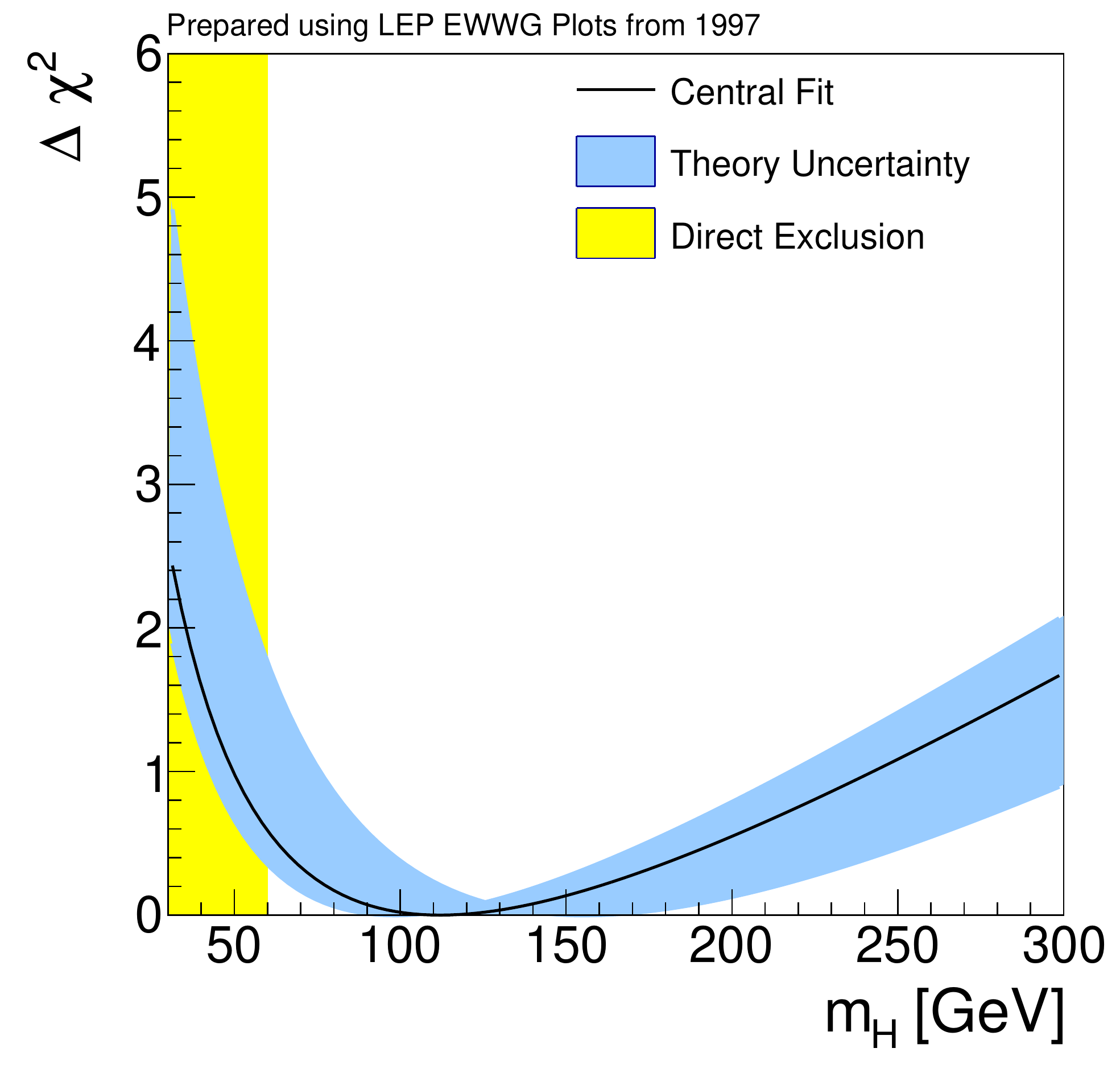}}
\resizebox{0.2455\textwidth}{!}{\includegraphics{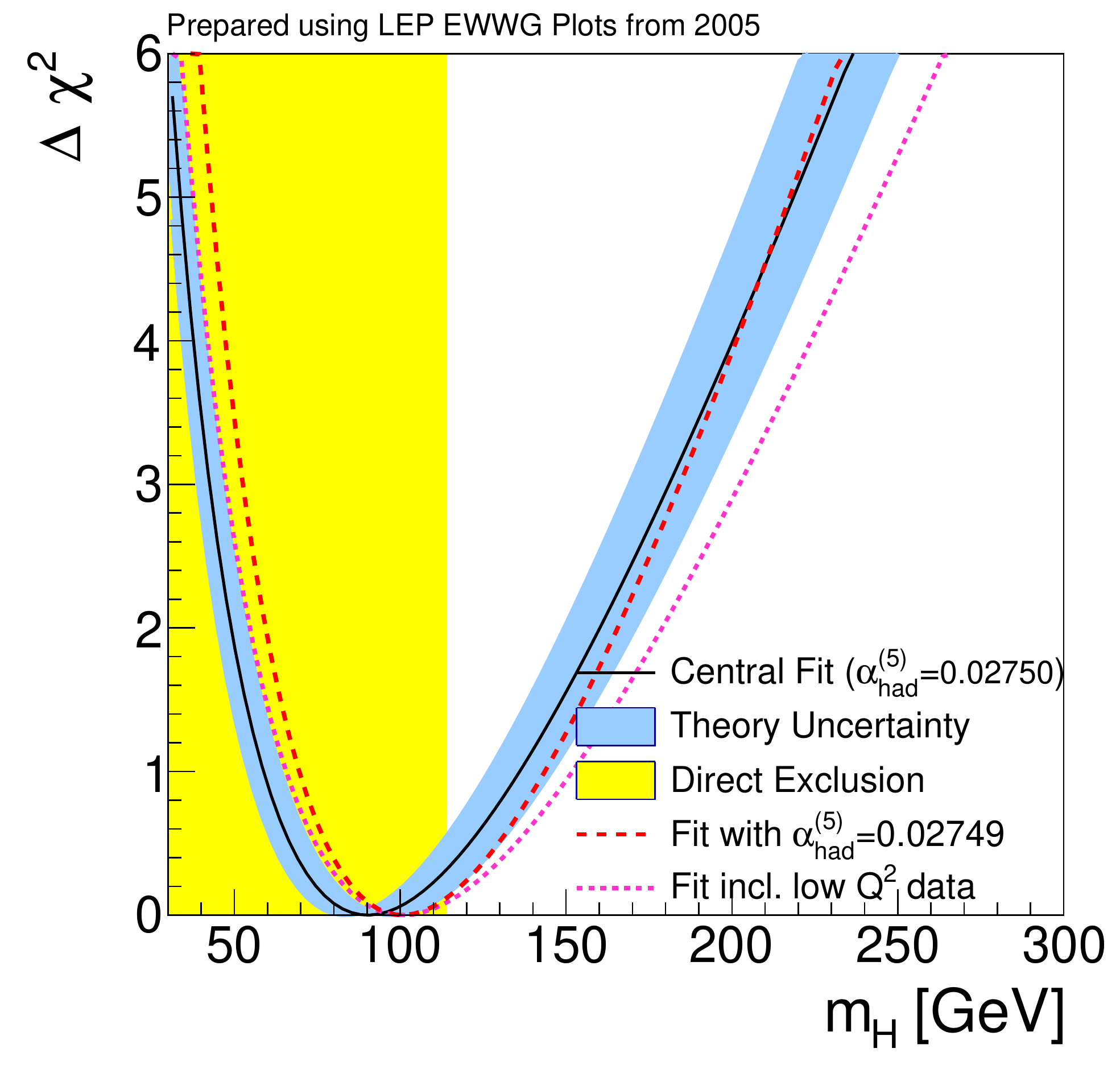}}
\resizebox{0.2455\textwidth}{!}{\includegraphics{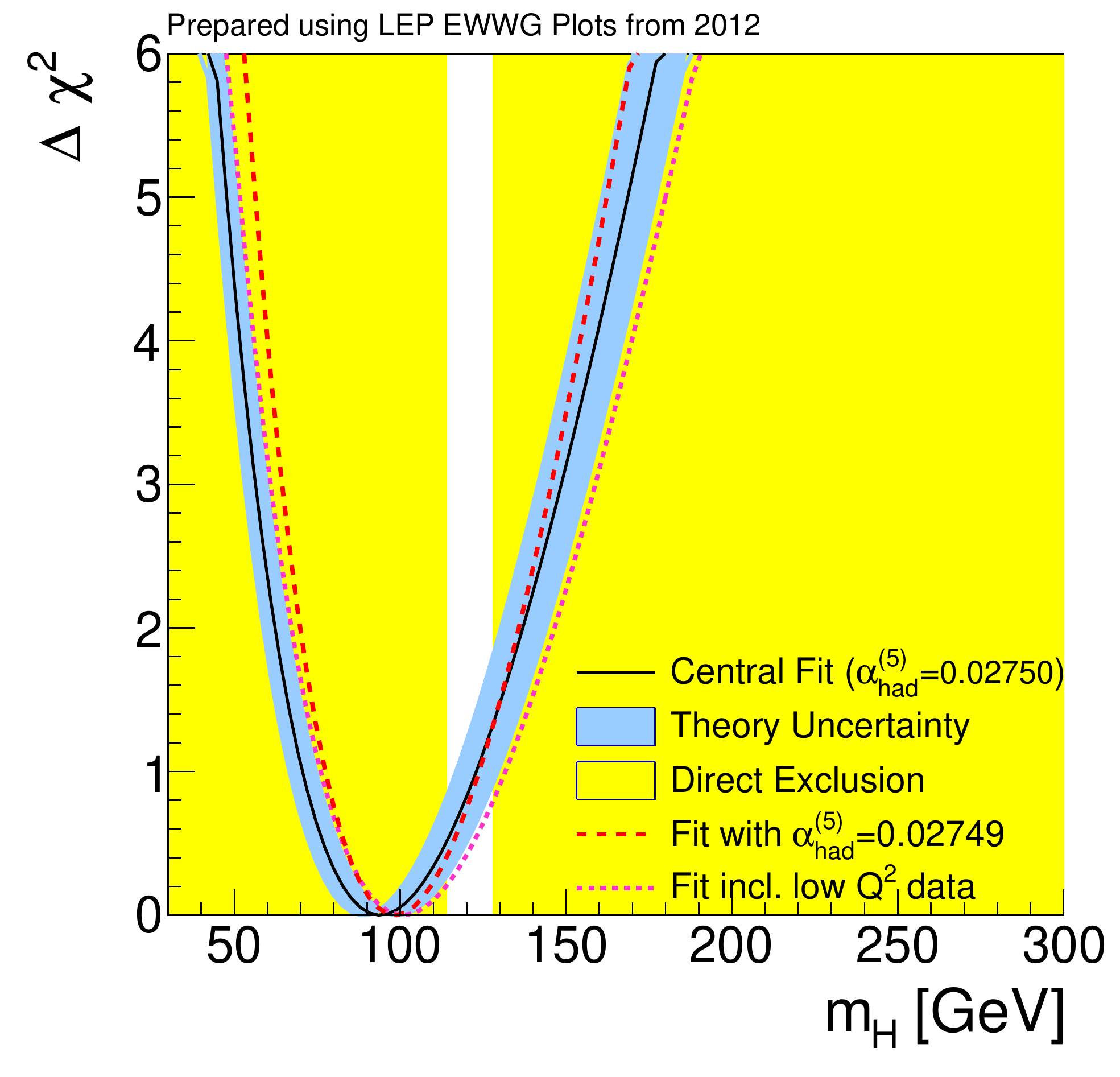}}
\caption{Historic development of the 'blue-band' plots~\cite{ALEPH:2005ab}, illustrating the estimate of the Higgs Boson mass in the Standard Model from electroweak precision measurements. 
The excluded area by direct searches is shown in yellow, while the blue band illustrate the $\chi^2$ distribution of the global electroweak fit. The original plots have been remade for this article.}
\label{fig:BlueBandHistory}
\end{figure*}

Global electroweak analyses and fits have a long history in particle physics and were pioneered by Paul Langacker and collaborators~\cite{Kim:1980sa,Amaldi:1987fu}, as well as other group~\cite{Costa:1987qp},
starting already before the discovery of the $W$~\cite{Arnison:1983rp,Banner:1983jy} and $Z$~\cite{Arnison:1983mk,Bagnaia:1983zx} bosons by the UA1 and UA2 experiments.
The $Z$~boson factories, LEP and SLC~\cite{ALEPH:2005ab}, produced high-precision measurements of $\mZ$ and $\sinleff$ and of many other observables.
Including these, the global analyses at the time~\cite{ALEPH:2005ab,Kennedy:1990ib,Langacker:1991an,Erler:1994fz} successfully predicted $\mt$ 
in the range between 140 and 190~\GeV before the top quark discovery by the CDF~\cite{Abe:1995hr} and D\O~\cite{D0:1995jca} detectors at the Tevatron Collider in 1995.
There were even first hints~\cite{Erler:1994fz} --- albeit statistically weak --- 
at a relatively light Higgs with a mass of ${\cal O}(100~\mbox{\GeV})$ rather than ${\cal O}(1~\mbox{TeV})$.
LEP~2 running (near and above the $W^+ W^-$ production threshold)~\cite{Schael:2013ita} and the increasingly precise results from the Tevatron, especially on the top quark and $W$~boson masses, constrained electroweak physics even further~\cite{Schael:2013ita,Erler:1998df,Erler:2004cx,Flacher:2008zq},
revealing that any new physics beyond the SM can at most represent a small perturbation of the SM and 
culminating in successful predictions of $\mH$~\cite{Schael:2013ita,Erler:2010wa,Erler:2012uu,Baak:2011ze} before the Higgs boson was observed~\cite{Aad:2012tfa,Chatrchyan:2012xdj} in 2012. In fact, the probably most famous and influential result of the global electroweak fit is the indirect determination of the Higgs boson mass, illustrated by the \textit{'blue-band'} plots, which show the $\chi^2$ distribution of the fit to all available electroweak observable in dependence of $\MH$. The historic development of the blue-band plots is shown in Figure \ref{fig:BlueBandHistory}.

The most recent fits were performed after the Higgs boson discovery~\cite{Erler:2013xha,Ciuchini:2013pca,Baak:2014ora,Erler:2017ozu, Haller:2018nnx}.
The steady progress that has been achieved in global electroweak fits over the years --- from both the experimental and theoretical sides ---
can best be seen from the biennially updated section on the {\em Electroweak Model and Constraints on New Physics}
in the {\sl Review of Particle Properties} produced by the {\sl Particle Data Group} (PDG)~\cite{Tanabashi:2018}.

It should be emphasized, that the results of many of the groups mentioned in the previous paragraph, which are based on mostly independent electroweak libraries,
are generally in very good agreement with each other.  
Residual variations can usually be traced to differences in experimental inputs or lack of some higher-order correction in one work relative to another. 
For example, the ZFITTER program~\cite{Bardin:1999yd,Arbuzov:2005ma,Akhundov:2013ons} is a FORTRAN package based on the on-shell renormalization scheme,
and was tailored to the physics program at LEP. Gfitter~1.0~\cite{Flacher:2008zq} and later versions ostensibly consist of independent object-oriented {\bf\small C\hspace{-1pt}+\hspace{-1pt}+} code.
GAPP (Global Analysis of Particle Properties)~\cite{Erler:1999ug} is a FORTRAN library for the evaluations of pseudo-observables
which are implemented in the $\overline{\small{\rm MS}}$ renormalization scheme~\cite{Bardeen:1978yd}, exploiting its better convergence properties.
It offers unique capabilities to constrain physics beyond the SM, such as extra neutral gauge bosons~\cite{Erler:2009jh} or a fourth fermion generation~\cite{Erler:2010sk}.
Most numerical results presented here been obtained with GAPP and Gfitter. 

Now that all free parameters defining the SM are fixed, the focus of global electroweak fits has shifted to tests of the full internal consistency of the theory 
and to the search for possible hints of theories beyond the SM (BSM). 
Since new, so far unobserved particles could also appear in the loop diagrams of the types shown in Figure~\ref{fig:LoopCorrections}, 
specific BSM scenarios could alter the radiative corrections and thus the relations~(\ref{Eqn:CorrectedObs}) in characteristic ways.
Possible deviations between the predicted and the measured values could indicate the presence of BSM physics effects. 

In this review article we will discuss the status of the electroweak precision tests of the SM after the discovery of the Higgs boson. 
The state of the art of the theoretical calculations in radiative corrections to precision observables will be discussed in Section~\ref{sec:ewpotheory}, 
where we will focus on the most recent developments. 
The experimental status of all relevant measurements will be summarized in Section~\ref{sec:ewpo}. 
After the current knowledge of all electroweak observables and their relations has been reviewed, 
we discuss the internal consistency of the SM, as well as the constraints on BSM scenarios (Section~\ref{sec:SMfit}) by performing up-to-date global electroweak fits.
The impact of future electroweak precision measurements at upcoming or planned colliders will also be briefly reviewed. 
The article concludes with a summary in Section~\ref{sec:summary}.

For recent discussions focussed on specific aspects of electroweak precision physics~\cite{Langacker:2010zza}, we refer to the dedicated reviews 
on low energy tests of the weak interaction~\cite{Erler:2004cx},
the weak neutral current~\cite{Erler:2013xha},
low energy measurements of the weak mixing angle~\cite{Kumar:2013yoa}, 
and weak polarized electron scattering~\cite{Erler:2014fqa}.

\section{Status of Theoretical Calculations in the Precision Electroweak Sector\label{sec:ewpotheory}}

Significant progress has been made in the calculation of electroweak radiative corrections in recent years, leading to reduced 
theoretical uncertainties in observables, and thus to a higher sensitivity of the global electroweak fit to possible contributions of 
new physics. 
Here, we briefly review the status of the theoretical calculations of $\Delta r$, $\kappa_Z^f$ and $\rho_Z^f$, which absorb 
the radiative corrections for the relations between \MW, $\sin ^2\theta_W$, $g_{V,f}$ and $g_{A,f}$. 
Moreover, we summarize the theoretical description of the scale dependences of the electromagnetic and strong coupling constants, 
as well as the precision calculations which are required for the extraction of the Fermi constant $G_F$.

\subsection{Radiative corrections to the $W$ boson mass}

One of the most important observables in the global electroweak fit is the mass of the $W$ boson.
The parameter $\Delta r$~\cite{Sirlin:1980nh} in Equation~(\ref{Eqn:CorrectedObs}) absorbs the remaining radiative corrections to 
$\mu$ decay after the factorizing QED corrections already present and calculable 
in the Fermi $V-A$ theory~\cite{Berman:1958ti,Kinoshita:1958ru,vanRitbergen:1998yd} have been removed~\cite{Ferroglia:1999tg}.
It has dominantly quadratic dependence on $\mt$ which enters through the correction $\Delta\rho \equiv \rho - 1$, 
where the electroweak $\rho$ parameter describes the ratio of neutral-current to charged-current interaction strengths.
One can write,
\begin{equation}
\Delta r = \Delta\alpha - \cot^2\theta_W \Delta\rho + \Delta_r^{\rm rem}\ ,
\end{equation}
where $\Delta\alpha$ accounts for the scale dependence of $\alpha$ (QED running) and is numerically very important due to 
logarithmic singularities regulated by the fermion masses.
$\Delta_r^{\rm rem}$ collects the remaining radiative corrections.
The $M_H$ dependence is much milder than the dependence on $\mt$ and at one-loop order it is only logarithmic for asymptotically large values.
Since the radiative corrections depend themselves on $\MW$ and $\sin^2\theta_W$, an iterative procedure is necessary 
to solve Equation~(\ref{Eqn:CorrectedObs}).

\begin{figure*}
\begin{center}
\begin{minipage}{0.48\textwidth}
\resizebox{1.0\textwidth}{!}{\includegraphics{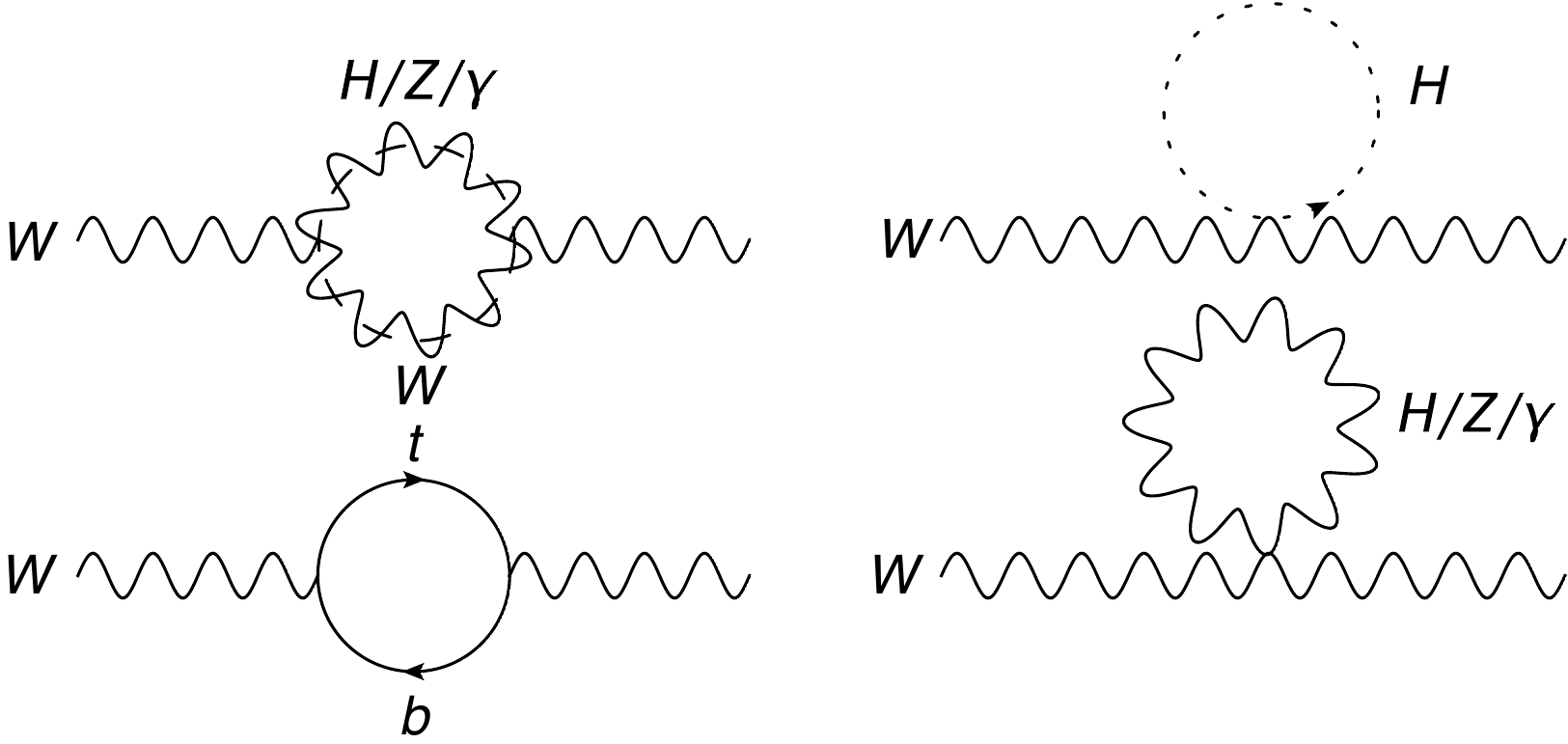}}
\caption{\label{fig:WMassPropagator}Feynman diagrams of loop corrections to the $W$ boson propagator.}
\end{minipage}
\hspace{0.3cm}
\begin{minipage}{0.48\textwidth}
\centering
\resizebox{0.8\textwidth}{!}{\includegraphics{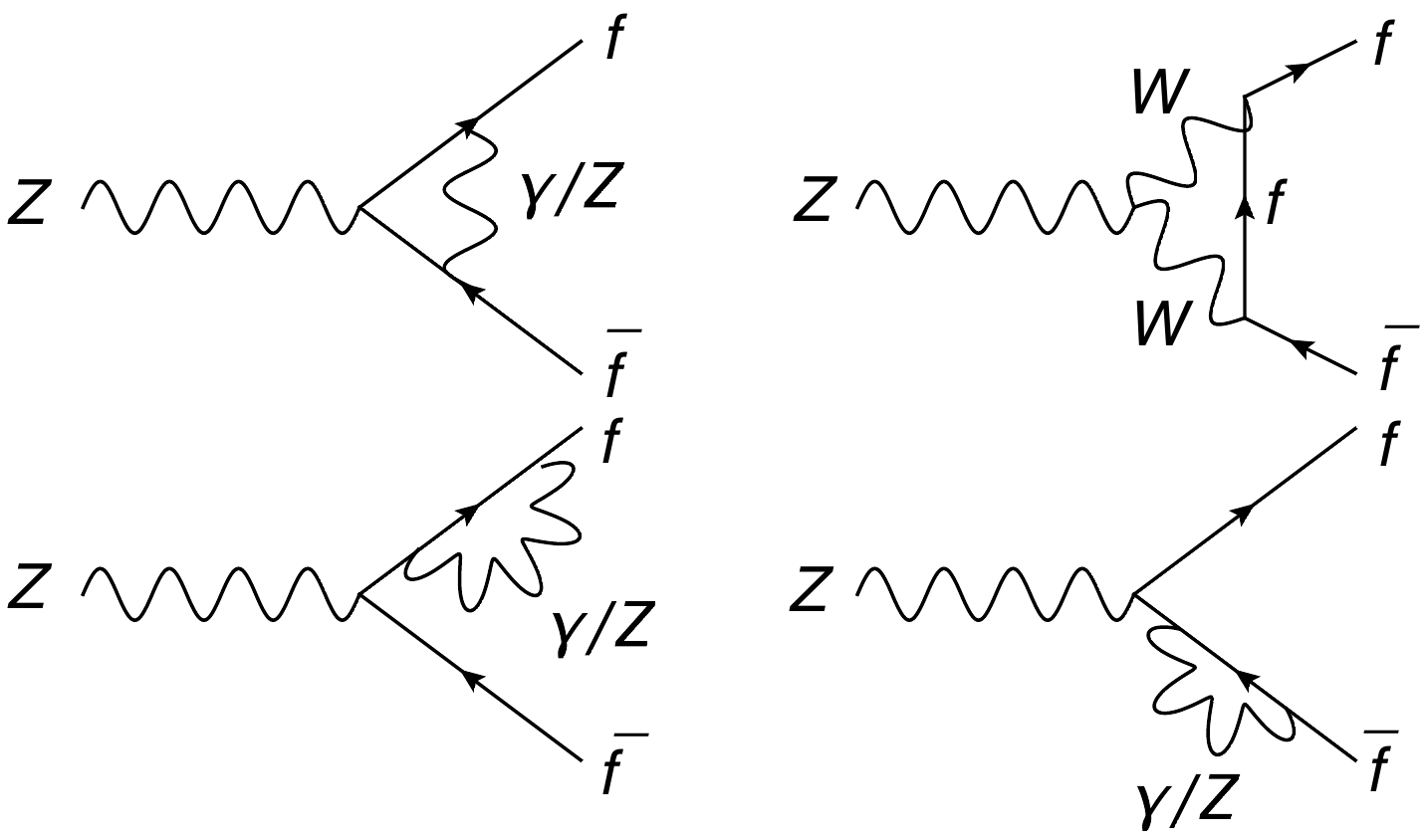}}
\caption{\label{fig:SinThetaCorrections}Feynman diagrams of loop corrections to the decay of the $Z$ boson and to $\sinleff$.}
\end{minipage}
\end{center}
\end{figure*}

The Feynman diagrams of the most important radiative corrections to the $W$ boson propagator are illustrated in 
Figure~\ref{fig:WMassPropagator}. 
The correction term of ${\cal O}(\alpha m_t^2)$~\cite{Veltman:1977kh} is dominant because it is enhanced by the large value of 
the top quark mass.
The full one-loop calculation was completed in the 1980s~\cite{Sirlin:1980nh,Hollik:1988ii,Marciano:1980pb}, and strategies were developed to re-sum 
certain reducible higher-order terms, both in the on-shell~\cite{Sirlin:1983ys,Consoli:1989fg} and \msbar~\cite{Degrassi:1990tu} 
renormalization schemes.
For example, writing $\Delta r$ in the form~\cite{Consoli:1989fg} 
\begin{equation}
1 + \Delta r = \frac{1}{(1 - \Delta\alpha) \left( 1 + \cot^2\theta_W \Delta\rho \right)} + \Delta_r^{\rm rem}\ ,
\end{equation}
correctly anticipates the terms of the form $(\Delta\alpha)^n$, $(\Delta\rho)^2$, $\Delta\alpha\Delta\rho$ and 
$\Delta\alpha\Delta_{\rm rem}$.

Mixed electroweak-QCD two-loop corrections also arrived at around that time, 
again starting with the $m_t^2$ enhanced term of ${\cal O}(\alpha \as m_t^2)$~\cite{Djouadi:1987gn,Djouadi:1987di},
followed by the residual correction of ${\cal O}(\alpha \as)$~\cite{Kniehl:1989yc,Halzen:1990je} 
with the full quark mass dependence~\cite{Djouadi:1993ss} soon thereafter.
As for the purely electroweak two-loop corrections, the leading ${\cal O}(\alpha^2 m_t^4)$ correction was first obtained for 
$M_H \ll m_t$~\cite{vanderBij:1986hy},
and then for arbitrary $M_H$~\cite{Barbieri:1992nz,Fleischer:1993ub}.
The next terms in the inverse $m_t^2$ expansion of ${\cal O}(\alpha^2 m_t^2)$~\cite{Degrassi:1996mg} took a rather complex form 
and turned out to be surprisingly sizable. 
The full ${\cal O}(\alpha^2)$ result, first from diagrams containing at least one fermion loop~\cite{Freitas:2000gg,Freitas:2002ja,Awramik:2003ee} 
and then from the purely bosonic diagrams~\cite{Awramik:2002wn,Onishchenko:2002ve}, is complete by now without further approximation
in the on-shell scheme, as well as in the hybrid~\cite{Degrassi:2014sxa} and pure~\cite{Martin:2015lxa} $\msbar$ scheme.

Enhanced three-loop contributions are also important.  
The ${\cal O}(\alpha \alpha^2_s m_t^2)$ correction reduces $M_W$ by about 10~MeV~\cite{Avdeev:1994db,Chetyrkin:1995ix}.
However, this shift is almost entirely due to the use of the pole mass definition, and amounts to less than 3~MeV if the definition
based on the $\msbar$ scheme is employed instead.
In this case, the correction is dominated by the axial anomaly that enters through double-triangle (singlet) diagrams 
which had been obtained first~\cite{Anselm:1993uq}.
Thus, use of the pole mass necessitates the inclusion even of the four-loop ${\cal O}(\alpha \alpha^3_s m_t^2)$ singlet~\cite{Schroder:2005db} 
and non-singlet~\cite{Chetyrkin:2006bj,Boughezal:2006xk} terms which are, however, completely negligible in the $\msbar$ scheme.
Residual effects of ${\cal O}(\alpha \alpha^2_s)$ from the third~\cite{Chetyrkin:1995js} 
and the first two~\cite{Chetyrkin:1996cf} generations are also known.
The corrections of ${\cal O}(\alpha^2 \alpha_s m_t^4)$ and ${\cal O}(\alpha^3 m_t^6)$ were obtained first for $M_H = 0$~\cite{vanderBij:2000cg}
and then in the general case~\cite{Faisst:2003px}.

There is a numerical approximation~\cite{Awramik:2003rn} expressing $M_W$ in terms of other input parameters.
Updating this expression to correspond closer to the currently favored values and fixing $M_H = 125$~GeV we find,
\begin{equation}
\label{eq:mw}
M_W = \left[ M_W^0 + c_t \Delta_t + c_t^\prime \Delta_t^2 + c_Z \Delta_Z + c_\alpha \Delta_\alpha + c_{\as} \Delta_{\as} \right]~{\rm MeV},
\end{equation}
with the definitions,
\begin{equation}
\label{eq:param}
\Delta_t \equiv \left( \frac{\mt}{173~{\rm GeV}} \right)^2 - 1, \qquad
\Delta_Z \equiv \frac{\MZ}{91.1876~{\rm GeV}} - 1, \qquad
\Delta_\alpha \equiv \frac{\Delta\alpha_{\rm had}^{(5)}(\MZ^2)}{0.0276} - 1, \qquad
\Delta_{\as} \equiv \frac{\as(\MZ^2)}{0.119} - 1,
\end{equation}
and the numerical values,
\begin{equation}
M_W^0 = 80359.5 \qquad
c_t = 520.5 \qquad
c_t^\prime = - 67.7 \qquad
c_Z = 115000. \qquad
c_\alpha = - 503. \qquad
c_{\alpha_s} = - 71.6
\end{equation}
where $m_t$ is the top quark pole mass and $\Delta\alpha_{\rm had}^{(5)}(\MZ^2)$ is the hadronic contribution
due to the five light quarks to the running of $\alpha$ from the Thomson limit to the scale $M_Z$.
Equation~(\ref{eq:mw}) reproduces the full theoretical calculations to the level of 0.1~MeV precision.
One can see that the prediction for \MW correlates positively with $\mt$ and $M_Z$
and negatively with $\alpha(M_Z)$ and $\alpha_s(M_Z)$ .

\subsection{Radiative corrections to the weak mixing angle\label{sec:theoSin}}
The radiative corrections collected in $\kappa_Z^f$ relate the on-shell weak mixing angle {\em via\/} ~(\ref{eq:sineff}) 
to effective mixing angles $\sinfeff$~\cite{Arbuzov:2005ma}.
The observable values therefore depend on the fermion type $f$ at the $Z$ vertex as different radiative corrections contribute. 
Thus one can write,
\begin{equation}
\sinfeff = \frac{1}{4 |Q^f|} \left( 1 - {\rm Re} \frac{g_V^f}{g_A^f} \right),
\end{equation}
where $g_V^f$ and $g_A^f$ are the effective vector and axial-vector couplings of fermions to the $Z$ boson as defined in Equations~(\ref{eq:gveff}) 
and (\ref{eq:gaeff}).
Most $Z$ pole observables with leptons either in the initial or in the final state are directly sensitive to the leptonic definition with $f = \ell$.
This includes the leptonic forward-backward asymmetries at the Tevatron and the LHC, as well.

The leading corrections are illustrated in Figure~\ref{fig:SinThetaCorrections}. 
The most important radiative corrections are related to those in $M_W$, entering through $\Delta\alpha$ and $\Delta\rho$,
where the latter appears as an artifact of using the on-shell definition of the weak mixing angle in Equation~(\ref{eq:sineff}).
Similar to $M_W$, the two-loop ${\cal O}(\alpha^2)$ fermionic~\cite{Awramik:2004ge, Hollik:2005va} 
and bosonic~\cite{Awramik:2006ar,Hollik:2006ma} corrections are fully known, marking an important improvement over the next-term expansion in the top-quark mass \cite{Degrassi:1996ps}.

Fixing $M_H = 125$~GeV, we write the corresponding formula for \sinleff~\cite{Awramik:2006uz} in terms of the parameters in Equations~(\ref{eq:param}) as,
\begin{equation}
\label{eq:s2tw}
\sinleff  = \sin^2\theta_{\rm eff}^{\ell,0} - (d_t \Delta_t + d_t^\prime \Delta_t^2 + d_Z \Delta_Z + d_\alpha \Delta_\alpha + d_{\as} \Delta_{\as})\times 10^{-4} ,
\end{equation}
where,
\begin{equation}
\sin^2\theta_{\rm eff}^{\ell,0} = 0.231533 \qquad
d_t = 27.14 \qquad
d_t^\prime = - 1.62 \qquad
d_Z = 6550. \qquad
d_\alpha = - 96.7 \qquad
d_{\alpha_s} = - 4.05
\end{equation}
This approximation reproduces the full calculation to better than $5\times 10^{-6}$.
The sign configuration in the $d_i$ coefficients is identical to the one in the $c_i$ coefficients, reconfirming that the effects from $\Delta\alpha$ and $\Delta\rho$ dominate.

The predictions for the effective weak mixing angles $\sinleff$ of the four light quarks ($f =u, d, s$ and $c$) differ slightly from the prediction for charged leptons. 
For example, there are flavor dependent corrections of ${\cal O}(\alpha\alpha_s)$ that do not factorize in the total $Z$ width 
and need to be included~\cite{Czarnecki:1996ei,Harlander:1997zb}.
For bottom quarks additional ${\cal O}(\alpha m_t^2)$~\cite{Akhundov:1985fc,Bernabeu:1987me,Beenakker:1988pv,Degrassi:1990ec} and 
${\cal O}(\alpha^2 m_t^4)$~\cite{Barbieri:1992nz,Fleischer:1992fq} enhanced effects enter the $Zb\bar b$-vertex,
resulting in a qualitatively different dependence on the input parameters.
The leading two-loop corrections of ${\cal O}(\alpha\alpha_s m_t^2)$ were obtained in Ref.~\cite{Fleischer:1992fq,Degrassi:1993ij}.
The full two-loop electroweak fermionic~\cite{Awramik:2008gi} and bosonic~\cite{Dubovyk:2016aqv} corrections have been completed more recently.

\subsection{Radiative corrections to gauge boson decay rates}
The flavor-dependent normalization factors $\rho _f^Z$ defined through Equations~(\ref{eq:gveff}) and~(\ref{eq:gaeff}) absorb the remaining electroweak radiative corrections 
to the vector and axial-vector couplings $g^f_V$ and $g^f_A$, and thus to the $W$ and $Z$ boson partial and total decay widths.
They have been computed alongside the $\kappa _f^Z$, and we refer to the previous subsection for the corresponding references.

\begin{figure*}
\begin{center}
\begin{minipage}{0.99\textwidth}
\resizebox{1.0\textwidth}{!}{\includegraphics{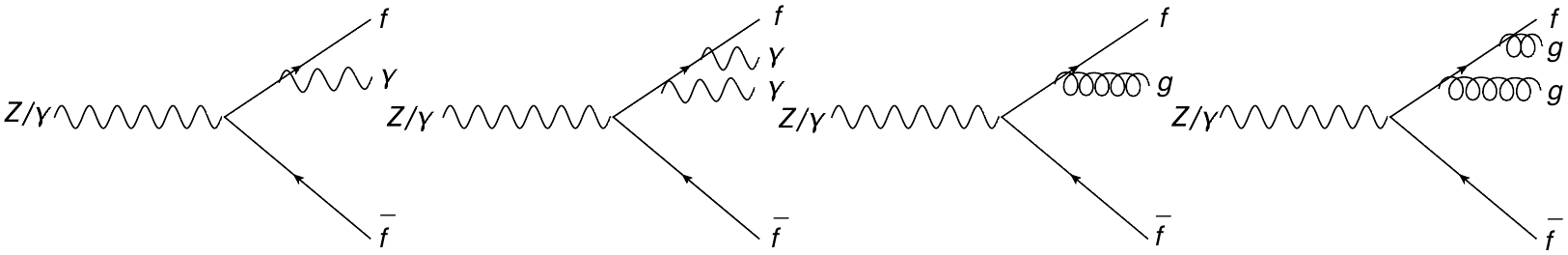}}
\caption{\label{fig:radiatorcor}Feynman diagrams generating the QED (left) and QCD (right) radiator functions.}
\end{minipage}
\end{center}
\end{figure*}

The partial width of the $Z$ boson to decay into an $f\bar f$ pair plus any number of photons and gluon jets is given by,
\begin{equation}
\label{eqn:PartialZWidth}
\Gamma^{f\bar f}_Z = \frac{\sqrt{2} G_F \MZ^3}{12 \pi} N_c^f \left[ |g^f_V|^2 R_V^f (M_Z) + |g^f_A|^2 R_A^f (M_Z) \right].
\end{equation}
$N_c^f$ denotes the number of colors, so that $N_c^f = 1$ for leptons and $N_c^f = 3$ for quarks. 
It should be noted that scale for the effective couplings is \MZ and that the effective coupling $g^f_V$ and $g^f_A$ are in general complex-valued,
a fact that starts to be relevant starting at two-loop precision.
The vector and axial-vector radiator functions $R_V^f(\MZ)$ and $R_A^f(\MZ)$ describe QED and QCD corrections~\cite{Chetyrkin:1994js} 
to the final state particles and are illustrated in Figure~\ref{fig:radiatorcor}. 
For example, for massless quarks they are available up to four-loop order in QCD and take the form,
\begin{equation}
\label{eqn:radqed}
R_V^q = R_A^q = 1 + \frac{\as(\MZ)}{\pi} + 1.409 \frac{\as^2}{\pi^2} - 12.77 \frac{\as^3}{\pi^3} - 80.0 \frac{\as^4}{\pi^4} + 
Q_q^2 \left[ \frac{3}{4} - \frac{\as}{4\pi} - \left( 1.106 + \frac{3}{32} Q_q^2 \right) \frac{\alpha}{\pi} \right] \frac{\alpha(\MZ)}{\pi}\ .
\end{equation}
The one-loop correction was known~\cite{Appelquist:1973uz,Zee:1973sr} already before the discovery of the charm quark.
The non-Abelian character of QCD became fully explicit with the advent of the two-loop result~\cite{Chetyrkin:1979bj,Dine:1979qh,Celmaster:1979xr},
which also radically reduced the scale setting ambiguity in $\alpha_s(\mu)$.
The three-loop~\cite{Surguladze:1990tg,Gorishnii:1990vf} and four-loop~\cite{Baikov:2008jh} calculations brought the uncertainty 
in the massless series to a currently negligible level. Fermion mass effects~\cite{Poggio:1975af,Barnett:1980sm}, other than from $m_t$, lead to $R_V^f(\MZ) \neq R_A^f(\MZ)$ 
and are small at the electroweak scale, provided one uses the $\msbar$ quark mass definitions evaluated at the $Z$ mass scale.
The last term gives the QED and mixed QED/QCD corrections~\cite{Kataev:1992dg}. Expressions for finite quark masses can also be found in \cite{Chetyrkin:1994js}.

The radiator functions account for the so-called non-singlet diagrams where both gauge bosons in the two-point correlation function couple 
to the same fermion.
The non-singlet QCD corrections to $W$ and $Z$ decays are therefore identical in the massless limit 
(except for the scale at which $\alpha_s$ is evaluated).
On the other hand, for $Z$ decays there are also singlet contributions with purely gluonic intermediate states.
As a result of Furry's theorem, they cannot occur for vector currents before three-loop order
where they have been obtained~\cite{Surguladze:1990tg} including tiny top quark decoupling terms~\cite{Larin:1994va}.
These effects turned out to be much smaller than the corresponding non-singlet effects, 
a fact which is also true for the very recently completed four-loop result (for $M_Z \ll m_t$)~\cite{Baikov:2012er}.

The axial-vector current, however, is very different.  
Furry's theorem does not apply here, and singlet effects appear already at two-loop order in QCD~\cite{Kniehl:1989bb}.
The contributions basically cancel within degenerate families, but large non-decoupling effects arise due to the large mass splitting of the third family.
Including the corresponding contributions at three-loop~\cite{Larin:1993ju} and four-loop~\cite{Baikov:2012er} order,
we write the effective QCD expansion for hadronic $Z$ decays for $\hat m_t (M_Z) = 171.4$~GeV (in the $\msbar$ scheme) 
and $m_q = 0$ for the other quarks,
\begin{equation}
\label{eqn:radqcd}
\Gamma_Z^{\rm had} \propto \rho\left( 1 + \frac{\as(\MZ)}{\pi} + 0.79 \frac{\as^2}{\pi^2} - 15.52 \frac{\as^3}{\pi^3} - 69.3 \frac{\as^4}{\pi^4}\right).
\end{equation}

One can address individual sources of uncertainty by assuming a geometric growth of the higher order terms.
For example, one can estimate the unknown terms of order $\alpha_s^{n+4}$ with $n \geq 1$ as
\begin{equation}
{\cal O}(\alpha_s^{n+4}) \sim \frac{{\cal O}(\alpha_s^4)^{n+1}}{{\cal O}(\alpha_s^3)^n}\ ,
\end{equation}
and then sum them up either in quadrature (assuming that the signs are random) or linearly (which is most conservative).
The latter applied to Equation~(\ref{eqn:radqed}) gives an error from perturbative QCD (PQCD) of
\begin{equation}
\label{eq:PQCD}
\delta_{\rm PQCD}\approx \pm \frac{\as^5}{\pi^4}\frac{(80.0)^2 }{12.77\pi - 80.0\as} = \pm 5.1 \times 10^{-5}.
\end{equation}
It affects the $\alpha_s$ extractions from the lineshape parameters $\Gamma_Z$, $R_\ell$ and $\sigma^0_{\rm had}$ in identical ways,
but it is nevertheless negligible compared to the experimental error in $\as$ of $\pm 0.0028$ (see Sec.~\ref{sec:SMfit}). 
By contrast, uncertainties from higher-order electroweak corrections~\cite{Freitas:2014hra,Freitas:2013dpa,Degrassi:1999jd} to the vector and axial-vector couplings 
can affect the lineshape observables in different, albeit correlated ways, so there is complementary information when analyzed in a global fit.
They are discussed in Sec.~\ref{sec:theoerrors} and shift $\alpha_s$ at levels that are also negligible at present\footnote{There 
are further theory uncertainties that enter when the lineshape observables
are derived from the underlying cross section measurements, but these are conventionally included in the experimental errors.}.

Radiative corrections to the decay width of the $W$ boson have been calculated 
at ${\cal O}(\alpha)$~\cite{Denner:1990tx,Bardin:1986fi,Rosner:1993rj} and at ${\cal O}(\alpha\as)$~\cite{Kara:2013dua}.
The theoretical uncertainty due to missing higher order corrections is significantly smaller than the current experimental precision, 
making the calculation of further corrections not necessary for now.

\subsection{Theoretical uncertainties due to unknown higher-order electroweak corrections\label{sec:theoerrors}}
The theoretical uncertainties in quantities such as $M_W$, $\sinleff$, and $\Gamma_Z$, due to unknown higher-order electroweak corrections,
arise from those in the $W$ and $Z$ boson self-energies, in the vertex and box corrections, and in further non-factorizable corrections,
{\em i.e.\/}, those that are not captured by the improved Born approximation.

The first type can be described by uncertainties in the so-called oblique parameters $S$, $T$, and $U$~\cite{Peskin:1991sw}, 
which have been originally introduced to parameterize potential new physics contributions to electroweak radiative corrections. 
There are many variants of oblique parameters in the literature, they may describe SM or new physics contributions or both, and are scheme dependent.
However, we are only interested in uncertainties parameterized by them, so that we do not specify such details here.
By using these oblique parameters, one can account for some of the theoretical correlations they induce in the suite of observables analyzed in global fits. 
Specifically, we use $T$ to parameterize the uncertainty in weak isospin breaking.
Other types of uncertainty might cancel in $T$, and instead appear in the energy-dependence of the $W$ ($Z$) boson vacuum polarization function
called $S_W = S + U$ ($S_Z = S$)~\cite{Marciano:1990dp}.
These uncertainties can be estimated by considering the expansion parameters involved.
Including the SM fermion content of three full generations as an enhancement factor, these are
\begin{equation}
\label{eq:enhance1}
\frac{3 \alpha_s}{\pi} \approx 0.116\ , \qquad\qquad 
\frac{8 \alpha}{\pi} \approx 0.020\ ,
\end{equation}
for QED and QCD, and
\begin{equation}
\label{eq:enhance2}
\frac{3 \alpha}{\pi \sin^2\theta_W} \approx 0.032\ , \qquad\qquad
\frac{3 - 6 \sin^2\theta_W + 8 \sin^4\theta_W}{\pi \sin^2\theta_W \cos^2\theta_W}\ \alpha \approx 0.029\ ,
\end{equation}
for the charged and neutral current interactions, respectively.
For the numerical estimates we evaluated the couplings in the \msbar scheme at the $W$ mass scale, where we have
$\alpha^{-1} \approx 128$, $\alpha_s \approx 0.121$, and $\sin^2\theta_W \approx 0.2311$.
Other possible enhancements can arise through the eigenvalue of the quadratic Casimir operator in the adjoint representation in QCD, $C_A = 3$, 
and through $m_t^2/M_W^2 \approx 4$ (for the $\msbar$ top quark mass) effects.
Coincidentally, these amount to the same factors as in the first of Equations~(\ref{eq:enhance1}) and Equations~(\ref{eq:enhance2}), respectively.

For example, the theoretical uncertainty in $\alpha S_W/4\sin^2\theta_W$ from the unknown electroweak corrections of ${\cal O}(\alpha^3)$ are then given by
\begin{equation}
\label{error1}
\left( \frac{3 \alpha}{\pi\sin^2\theta_W} \right)^3 \approx 3.4 \times 10^{-5},
\end{equation}
where we used $\alpha^{-1}(M_W) \approx 128$.
Likewise, we estimate the mixed electroweak/QCD corrections of ${\cal O}(\alpha^2\alpha_s^n)$ and ${\cal O}(\alpha\alpha_s^{n+2})$ with $n \geq 1$ as
\begin{equation}
\label{error2}
\frac{C_F \alpha_s}{\pi - C_A \alpha_s} \left( \frac{9 \alpha}{4\pi\sin^2\theta_W} \right)^2 \approx 3.4 \times 10^{-5}, \qquad\qquad
\frac{C_F C_A^2}{\pi - C_A \alpha_s} \left( \frac{\alpha_s}{\pi} \right)^3 \frac{9 \alpha}{4\sin^2\theta_W} \approx 1.9 \times 10^{-5},
\end{equation}
respectively, where $C_F = 4/3$ refers to the Casimir eigenvalue in the fundamental (quark) representation, 
where we took $\alpha_s = \alpha_s(M_W) \approx 0.121$, and where we performed a summation similar to the one in Equation~(\ref{eq:PQCD}).
Adding these in quadrature, we find a theoretical uncertainty $\Delta S_W = \pm 0.0061$.

The uncertainty in $\alpha S_Z/\sin^2 2\theta_W$ can be estimated in a similar way, amounting to $\Delta S_Z = \pm 0.0034$.
It is convenient and perhaps more realistic to correlate these by assuming that $\Delta S_W$ is the quadratic sum of $\Delta S_Z$ and $\Delta U = \pm 0.0051$
and to add $\alpha \Delta S_Z/\sin^2 2\theta_W = \pm 3.7 \times 10^{-5}$ to the uncertainty in $\Delta\alpha_{\rm had}$ (see next subsection) which enters in the same way.

In addition to those uncertainties that are analogous to $S_W$, there are additional ones contributing to $T$ associated from the top-bottom doublet.
These are enhanced by powers of $m_t$ and arise from the unknown electroweak three-loop orders ${\cal O}(\alpha^3 m_t^{2m+2})$ with $m \geq 1$,
summing up to
\begin{equation}
\label{error3}
\frac{m_t^6}{M_W^4 (m_t^2 - M_W^2)} \left( \frac{3 \alpha}{4\pi\sin^2\theta_W} \right)^3 \approx
\frac{64}{3} \left( \frac{3 \alpha}{4\pi\sin^2\theta_W} \right)^3 \approx 
9 \left( \frac{\alpha}{\pi\sin^2\theta_W} \right)^3 \approx 1.1 \times 10^{-5}.
\end{equation}
Similarly, the missing mixed ${\cal O}(\alpha^2 \alpha_s^n m_t^{2m})$ and ${\cal O}(\alpha^2 \alpha_s^{n+1} m_t^{2m+2})$ contributions induce errors of
\begin{equation}
\label{error4}
\frac{4\alpha_s}{\pi - 3 \alpha_s} \left( \frac{\alpha}{\pi\sin^2\theta_W} \right)^2 \approx 2.0 \times 10^{-5}, \qquad\qquad
\frac{48\pi}{\pi - 3 \alpha_s} \left( \frac{\alpha\alpha_s}{\pi^2\sin^2\theta_W} \right)^2 \approx 0.9 \times 10^{-5},
\end{equation}
respectively.
Adding the estimates (\ref{error1})--(\ref{error4}) in quadrature yields $\Delta T \approx 0.0073$.
As a reality check, the error estimate of the ${\cal O}(\alpha^2 \alpha_s m_t^4)$ term which one would find from this method, $\pm 5.4 \times 10^{-5}$, 
results in about twice the known result~\cite{Faisst:2003px} of $2.8 \times 10^{-5}$.
Similarly, the comparison of the ${\cal O}(\alpha^3 m_t^6)$ estimate, $\pm 3.4 \times 10^{-5}$, with the calculated term~\cite{Faisst:2003px} 
shows an overestimate by more than an order of magnitude.

The second (non-oblique) type of uncertainty is from non-universal corrections to specific observables, such as to the partial $Z$ decay widths or $\sigma_{\rm had}^0$.
In particular, there are unknown corrections of ${\cal O}(\alpha\alpha_s^{n+1})$ and ${\cal O}(\alpha^2 \alpha_s^n)$ that affect the hadronic partial $Z$ width
at estimated levels of $1.1 \times 10^{-4}$ and $1.8 \times 10^{-5}$, respectively.
These include flavor-dependent, non-factorizable singlet effects of ${\cal O}(\alpha \as^2)$ with two gluons and a $Z$ boson in the intermediate state
which have no counterpart in lower orders.
Together with the error from pure PQCD in Equation~(\ref{eq:PQCD}) we arrive at a fractional uncertainty in the hadronic partial width of $\delta_{\rm PQCD/EW} = \pm 0.00013$.
Starting at two-loop order, there are also non-resonant corrections to the Breit-Wigner lineshape (BW) of the $Z$ boson~\cite{Grassi:2000dz,Freitas:2014hra},
which are dominated by uncertainties of ${\cal O}(\alpha^2\alpha_s^n)$ and ${\cal O}(\alpha^3)$,
and which according to our method result in a fractional error of $3 \times 10^{-5}$ in $\sigma_{\rm had}^0$.

The effects of these uncertainties are illustrated in Table~\ref{tab:errors}.
One can implement them in a global fit by adding $\Delta S$ to $\Delta\alpha_{\rm had}$ as already mentioned, 
and by allowing $T$ and $U$ to float subject to the indicated constraints. 
The uncertainty due to $\delta_{\rm PQCD/EW}$ is relevant for the extraction of $\alpha_s$ 
and leads to an error of $\delta_{\rm theory}\alpha_s = \pm 0.0004$, which, however, 
is virtually negligible compared to the experimental error of $\pm 0.0028$ resulting from the lineshape extraction of~$\alpha_s$.
Finally, the error from the Breit-Wigner shape correction could be added to the much larger $\pm 37$~pb experimental error in $\sigma_{\rm had}^0$,
but can also be neglected.

\begin{table}[t]
\begin{center}
\begin{tabular}{c|c|c|c|c|c|c}
\hline
& $\Delta T = \pm 0.0073$ & $\Delta S = \pm 0.0034$ & $\Delta U = \pm 0.0051$ & $\delta_{\rm PQCD/EW} $ & ${\rm BW}$ & ${\rm total}$ \\ \hline
$M_W$ & $\pm 3.3~{\rm MeV}$ & $\mp 0.6~{\rm MeV}$ & $\pm 1.8~{\rm MeV}$ & --- & --- & $3.8~{\rm MeV}$ \\
$\sinleff$ & $\mp 1.9 \times 10^{-5}$ & $\pm 1.2 \times 10^{-5}$ & $0$ & --- & --- & $2.2 \times 10^{-5}$ \\
$\hat\rho$ & $\pm 5.9 \times 10^{-5}$ & $0$ & $\pm 4.4 \times 10^{-5}$ & --- & --- & $7.4 \times 10^{-5}$ \\ \hline
$\Gamma_Z$ & $\pm 0.19~{\rm MeV}$ & $\mp 0.03~{\rm MeV}$ & $0$ & $\pm 0.22~{\rm MeV}$ & --- & $0.29~{\rm MeV}$ \\
$R_\ell$ & $\pm 0.3 \times 10^{-3}$ & $\mp 0.2 \times 10^{-3}$ & $0$ & $\pm 2.6 \times 10^{-3}$ & --- & $2.6 \times 10^{-3}$ \\
$\sigma_{\rm had}^0$ & $\mp 0.1~{\rm pb}$ & $\pm 0.1~{\rm pb}$ & $0$ & $\mp 2.1~{\rm pb}$ & $\pm 1.2~{\rm pb}$ & $2.4~{\rm pb}$ \\ \hline
\end{tabular}
\end{center}
\caption{Uncertainties from missing higher-order electroweak corrections to precision observables.
The parameter $\hat\rho$ is a high-energy variant~\cite{Fanchiotti:1989wv} of the parameter $\rho = 1 + \alpha T$.
The uncertainties within each column are fully correlated, while those between columns are treated as independent and uncorrelated.}
\label{tab:errors}
\end{table}

Our estimate for the error in $M_W$ agrees very well with the result of Ref.~\cite{Awramik:2003rn}, which used a method
based on the assumption of a geometric growth of the perturbation series leading to a theory uncertainty of $\lesssim 4$~\MeV in $M_W$.
The error of $\pm 0.37$~MeV~\cite{Dubovyk:2018rlg} based on the same method is also in reasonable agreement with Table~\ref{tab:errors}.
On the other hand, the theory errors in $\sinleff$ of $\pm 4.7\times 10^{-5}$ as obtained in Ref.~\cite{Awramik:2006uz},
and those in $R_\ell$ ($6\times 10^{-3}$~\cite{Dubovyk:2018rlg}) and $\sigma_{\rm had}^0$ ($5.6~{\rm pb}^{-1}$~\cite{Freitas:2014hra}) are more than twice as large.
We note, however, that these larger uncertainties may be a consequence of using the on-shell renormalization scheme.
This is best illustrated by the partial decay rate $\Gamma^{\nu\bar\nu}_Z$.  
The one-loop correction amounts to almost 4\% of the tree-level value, and the two-loop electroweak corrections to 0.4\%.
By contrast, in the $\msbar$ scheme~\cite{Degrassi:1990ec} the tree-level expression approximates the fully known result to better than per mille precision.  
There is even a 0.3\% ${\cal O}(\alpha\as)$ correction which can, of course, not be a genuine (irreducible) effect in a leptonic $Z$ decay observable.
Indeed, normalizing the total $Z$ width in terms of $G_F$ mitigates this problem, with the result 
that the ${\cal O}(\alpha^2)$ and ${\cal O}(\alpha \as)$ two-loop corrections amount to only $-0.85$~MeV and $-1.21$~MeV, respectively~\cite{Dubovyk:2018rlg},
which is an order of magnitude less than normalizing in terms of $\alpha$ and the on-shell weak mixing angle.
In view of this, and also in order to have an independent check, an {\em ab initio\/} two-loop calculation within the $\msbar$ scheme
of all relevant precision observables would be more than welcome.

Here it is interesting to note that Ref.~\cite{Baak:2014ora} finds a theory uncertainty in $\alpha_s$ of about $\pm 0.0009$.
This reference implements the error estimates of Ref.~\cite{Dubovyk:2018rlg}, however, deviates from the geometric growth philosophy
for $\delta_{\rm PQCD}$ where the quoted theory uncertainty exceeds the last term in Equation~(\ref{eqn:radqcd}).

In any case, the theoretical uncertainties are currently well under control. 
In the future, however, with the possible advent of an ultra-precise lepton collider such as the Circular Electron Positron Collider (CEPC) 
or the Future Circular Collider in $e^\pm$ mode (FCC-ee), 
the calculations of radiative corrections will need to be pushed by at least another loop-order~\cite{Blondel:2018mad}.

\subsection{Running of the electromagnetic and strong coupling constants\label{sec:constants}}
\begin{figure*}[t]
\begin{center}
\begin{minipage}{0.70\textwidth}
\begin{minipage}{0.90\textwidth}
\resizebox{1.0\textwidth}{!}{\includegraphics{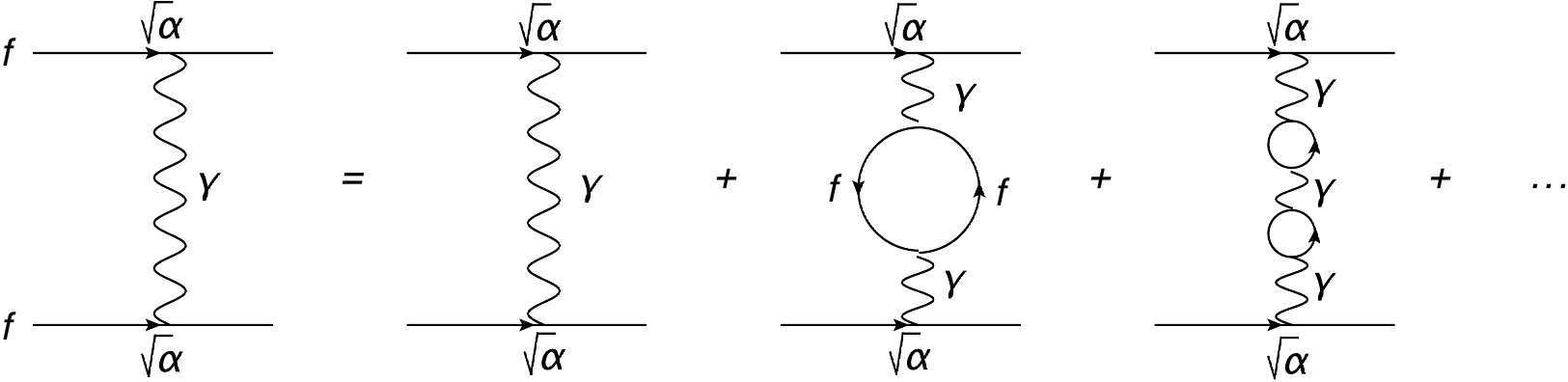}}
\vspace{0.5cm}
\caption{\label{fig:alphaEMLoop}Feynman diagrams of loop corrections to the fine structure constant $\alpha$.\vspace{1.2cm}}
\end{minipage}
\begin{minipage}{0.90\textwidth}
\resizebox{1.0\textwidth}{!}{\includegraphics{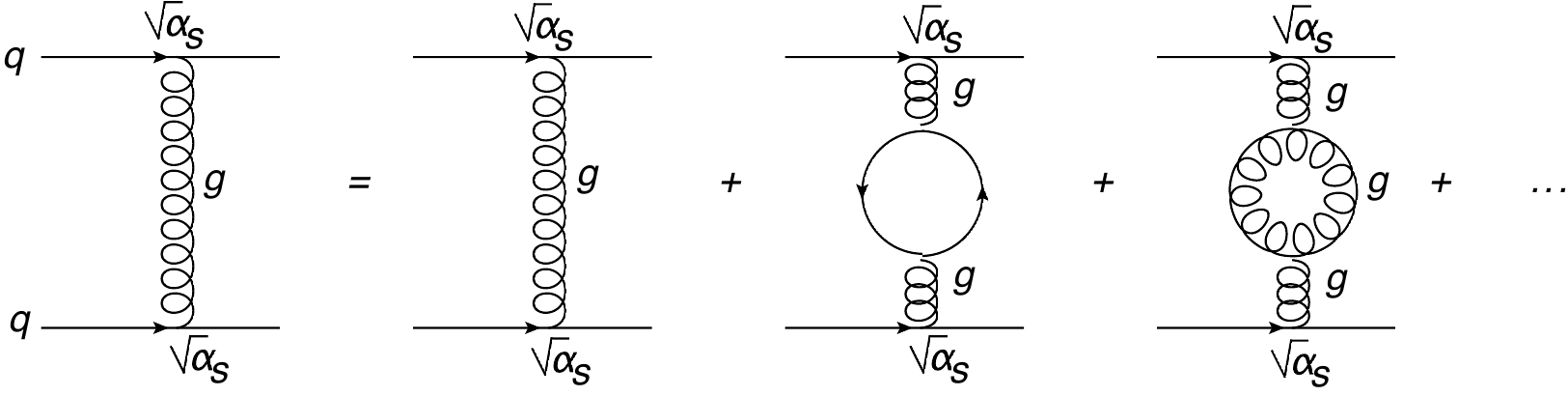}}
\vspace{0.9cm}
\caption{\label{fig:alphaSLoop}Feynman diagrams of loop corrections to the strong coupling constant $\alpha_s$.\vspace{0.7cm}}
\end{minipage}
\end{minipage}
\begin{minipage}{0.29\textwidth}
\centering
\resizebox{0.70\textwidth}{!}{\includegraphics{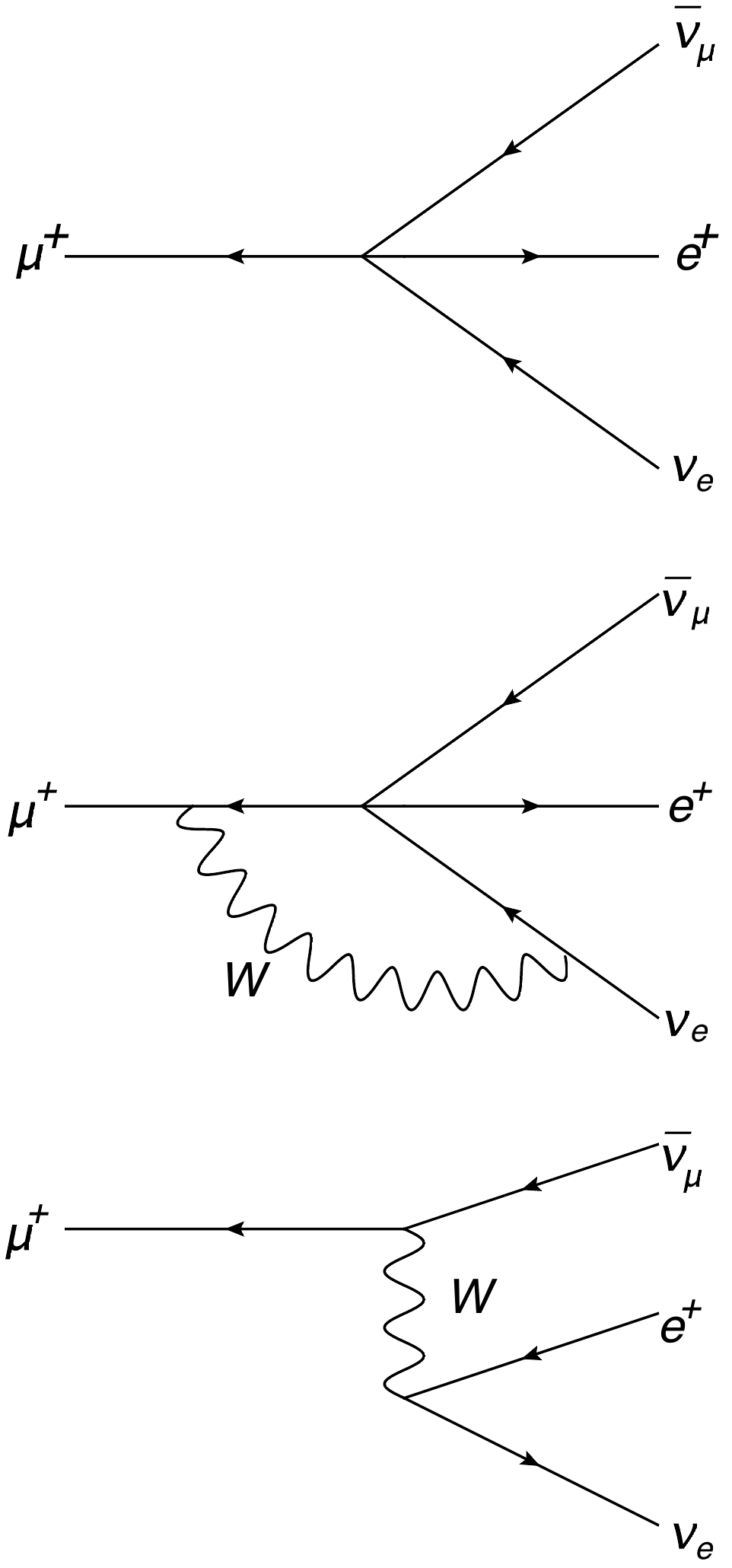}}
\caption{\label{fig:GFTheory}Feynman diagram for muon decay in the Fermi theory (upper), a one loop correction to the Fermi theory (middle) and the diagram at tree level in the Standard Model (lower).}
\end{minipage}
\end{center}
\end{figure*}

A certain class of loop corrections can be absorbed into the electromagnetic coupling constant rendering it energy scale dependent.
One example is the QED radiator function (\ref{eqn:radqed}) which requires the value of the electromagnetic coupling constant 
at the $Z$ boson mass scale with a precision at the level of $10^{-4}$ in order to be a subdominant uncertainty within the full global electroweak fit. 
The energy dependence (running) of the fine structure constant can be parameterized as
\begin{equation}
\alpha(s) = \frac{\alpha}{1  - \Delta \alpha(s)}\ ,
\end{equation}
where $\alpha^{-1} = 137.035 999 139(31)$~\cite{Tanabashi:2018} is the fine structure constant in the Thomson limit and $\Delta \alpha (s)$ 
includes all corrections to an all-order resummation of vacuum polarization diagrams such as those in Figure~\ref{fig:alphaEMLoop}. 
The term $\Delta \alpha (s)$ can be decomposed as
\begin{equation}
\label{eqn:delalp}
\Delta \alpha(s) = \Delta \alpha_{\rm lep}(M_Z^2) + \Delta \alpha^{(5)}_{\rm had}(M_Z^2) + \Delta \alpha_{\rm top}(M_Z^2)\ ,
\end{equation}
where the leptonic contribution, $\Delta \alpha_{\rm lep}(M_Z^2) = 314.97 \times 10^{-4}$, is included up to three-loop order in 
the domain where $s \gg m^2_l$~\cite{Steinhauser:1998rq}, 
but the three-loop contribution (where quarks can appear in internal loops) is only of order $10^{-6}$ and negligible.
The corresponding four-loop result is also known~\cite{Sturm:2013uka}.
The contribution of top-quark loops at \MZ is known to second order in $\as$ and yields a correction of 
$-0.72 \times 10^{-4}$~\cite{Davier:2017zfy} with a negligible uncertainty.
$\Delta \alpha^{(5)}_{\rm had}$ collects the contributions from the five lighter quarks.
Equation~(\ref{eqn:delalp}) is used in the on-shell scheme, and ignores the bosonic contribution.
Conversely, in the $\msbar$ scheme~\cite{Erler:1998sy} one includes the $W^\pm$ contribution, but may decouple the top quark~\cite{Fanchiotti:1992tu}.
Electroweak two-loop renormalization in the $\msbar$ scheme has been completed in Ref.~\cite{Degrassi:2003rw}.

The hadronic loop contributions of $\Delta \alpha^{(5)}_{\rm had}(M_Z^2)$ can be predicted perturbatively for energy scales sufficiently larger than
the strong interaction scale $\Lambda _{\rm QCD}$, but experimental data has to be used to constrain the contributions from hadronic scales. 
There are four very recent evaluations of $\Delta \alpha^{(5)}_{\rm had}(M_Z^2)$, 
which differ not only in the precise sets of experimental input data, but also in the statistical methods, theoretical refinements, and the renormalization schemes.
Two evaluations are mostly data driven, using hadron production data in $e^+ e^-$ annihilation, $R(s)$ (normalized to the muon point cross section),
and the theoretical prediction~\cite{Chetyrkin:1996cf,Baikov:2008jh} for $R(s)$ in the perturbative regime away from the heavy quark resonance and threshold regions.  
These yield values of $\Delta\alpha^{(5)}_{\rm had}(M_Z^2) = (276.00 \pm 0.95) \times 10^{-4}$~\cite{Davier:2017zfy} and 
$\Delta\alpha^{(5)}_{\rm had}(M_Z^2) = (276.11 \pm 1.11) \times 10^{-4}$~\cite{Keshavarzi:2018mgv}.
To obtain his most precise result, the author of Ref.~\cite{Jegerlehner:2017zsb} works in the space-like region (in what he calls the Adler function approach) 
computing $\Delta\alpha^{(5)}_{\rm had}(-M_Z^2)$ from his data-extracted $\Delta\alpha^{(5)}_{\rm had}(-4~{\rm GeV}^2)$,
and then converts to $\Delta\alpha^{(5)}_{\rm had}(M_Z^2) = (275.23 \pm 1.19) \times 10^{-4}$.
Finally, Ref.~\cite{Erler:2017knj} first computes $\Delta\hat\alpha^{(3)}_{\rm had}(4~{\rm GeV}^2)$ in the $\msbar$ scheme 
and then solves the renormalization group equation to obtain 
$\Delta\hat\alpha^{(5)}_{\rm had}(M_Z^2) = 127.959 \pm 0.010$,
which corresponds to the value $\Delta\alpha^{(5)}_{\rm had}(M_Z^2) = (276.1 \pm 0.8) \times 10^{-4}$.
Thus, good agreement is seen, except that Ref.~\cite{Jegerlehner:2017zsb} finds a somewhat lower central value\footnote{The result
of an alternative evaluation in the time-like region~\cite{Jegerlehner:2017zsb} yields the higher value
$\alpha^{(5)}_{\rm had}(M_Z^2) = (277.38 \pm 1.58) \times 10^{-4}$.}.
Subject to these kinds of constraints, parameters such as $\Delta\alpha^{(5)}_{\rm had}(M_Z^2)$ or $\Delta\hat\alpha^{(3)}_{\rm had}(4~{\rm GeV}^2)$ 
are usually allowed to float in electroweak fits.
A summary of the current experimental knowledge of $\Delta \alpha^{(5)}_{\rm had}$ is given in Section~\ref{sec:dalpha}.

Just as in QED, the strong coupling constant has to be evaluated at the $Z$ boson mass scale in the QCD radiator function \ref{eqn:radqcd}. 
Diagrams of loop corrections, responsible for the running of $\as$ are illustrated schematically in Figure~\ref{fig:alphaSLoop}. 
The renormalisation group equation, describing the running of $\as$ between two different energy scales $\mu$, is known to five-loop 
order, and is given for the case of five massless quarks as,
\begin{equation}
\frac{d}{d\ln\mu^2} \left( \frac{\as}{\pi} \right) = - \left[ \frac{23}{12} \frac{\as^2}{\pi^2} + \frac{29}{12} \frac{\as^3}{\pi^3} + 2.827 \frac{\as^4}{\pi^4} 
+ 18.85 \frac{\as^5}{\pi^5} + 15.1 \frac{\as^6}{\pi^6} + {\cal O} \left( \frac{\as^7}{\pi^7} \right) \right],
\end{equation}
where the first~\cite{Gross:1973id,Politzer:1973fx} and second~\cite{Caswell:1974gg,Jones:1974mm} term are scheme-independent and 
the third~\cite{Tarasov:1980au,vanRitbergen:1997va}, the fourth~\cite{vanRitbergen:1997va,Czakon:2004bu}, and the fifth~\cite{Baikov:2016tgj,Herzog:2017ohr} 
order term are given in the $\msbar$ scheme.
Interestingly, no factorial growth of the coefficients is seen in the known orders.

Consistency requires that five-loop running be accompanied by four-loop matching~\cite{Schroder:2005hy} (decoupling) at the heavy quark thresholds.
For a pedagogical review on the role of $\as$ in the Standard Model, we refer to Ref.~\cite{Dissertori:2015tfa}. 

Similar to the energy dependence of $\as$, also running masses, especially those of the $b$ and $c$ quarks, need to be considered,
as they enter, for example, into phase space expressions (using pole masses in those would introduce very large, but spurious logarithms).
The corresponding anomalous dimensions are also known to five-loop precision~\cite{Baikov:2014qja}, 
and so are the corresponding four-loop matching conditions~\cite{Liu:2015fxa}.

\subsection{The Fermi constant and the muon lifetime\label{sec:theofermi}}
The Fermi constant $G_F$ governs the coupling strength of the $V-A$ current-current Fermi interaction in an effective theory for the description of charged-current weak interactions. 
Even though the Fermi theory is not renormalizable, it is fully sufficient to describe weak interactions in the low-energy limit. 
From Equation~(\ref{Eqn:CorrectedObs}) one can see that $G_F$ can be used as a replacement of the $SU(2)_L$ gauge coupling $g$. 
Moreover, the Fermi constant can be directly related to the Higgs vacuum expectation value, $v = \left( \sqrt{2} G_F \right)^{-1/2} = 246.22$~\GeV. 

$G_F$ can be extracted {\em via\/} the muon-lifetime $\tau_\mu$~\cite{Tishchenko:2012ie}, using the relation
\begin{equation}
\label{eqn:taumuon}
\frac{1}{\tau_\mu} = \frac{G_F^2 m_\mu^5}{192\pi^3} (1+\Delta q),
\end{equation}
where $\Delta q$ accounts for higher-order QED corrections. 
The leading order and next-to-leading order Feynman diagrams for the decay of a positively charged muon into positrons in the Fermi Theory 
and the Standard Model are illustrated in Figure \ref{fig:GFTheory}.

The theoretical accuracy in Equation~(\ref{eqn:taumuon}) was limited until 1999 by the missing two-loop QED corrections to the muon decay rate. 
Ref.~\cite{vanRitbergen:1999fi} provided these corrections, leading to a negligible theoretical uncertainty from $G_F$ in current electroweak fits.

\section{Experimental Aspects of Electroweak Precision Observables\label{sec:ewpo}}

An overview of the electroweak precision observables used in the global electroweak fit is shown in Table~\ref{tab:Overview}. We will discuss the experimental status of all observables in the following. While the final experimental values of measurements, performed at the lepton colliders LEP and SLC, have been published since more than ten years ago, significant progress has been made in particular in hadron collider measurements. Hence the focus will be on the latter. 

\begin{table}[tb]
\footnotesize
\begin{center}
\begin{tabular}{l c |l c} 

\hline
parameter				\hspace{3.5cm}	& experimental value		& parameter				\hspace{3.5cm} 	& experimental value \\	
\hline
\multicolumn{4}{c}{measurements from hadron colliders}\\
\hline

$\MH$ 					& $125.09 \pm 0.15\,\GeV$	&							&						\\
$\MW$ 					& $80.380 \pm 0.013\,\GeV$	&	$\Gamma_W$				& $2.085 \pm 0.042\,\GeV$	\\
$\sinleff (\rm hadron~colliders)$& $0.23140 \pm 0.00023$		&	$m_t$		& $172.90 \pm 0.47\,\GeV$	\\
\hline
\multicolumn{4}{c}{measurements from $e^+ e^-$ colliders}\\
\hline 
$\MZ$					& $91.1875 \pm 0.0021\,\GeV$	&	$\Gamma_Z$				& $2.4952 \pm 0.0023\,\GeV$	\\
$\sigma^0_{had}$ 			& $41.540 \pm 0.037$\,nb		&	$R_l$					& $20.767 \pm 0.025$		\\
$R_c$					& $0.1721 \pm 0.0030$		&	$R_b$					& $0.21629 \pm 0.00066$		\\
\hline 
$A_c$					& $0.670 \pm 0.027$			&	$A_l(\rm LEP)$				& $0.1465 \pm 0.0033$		\\
$A_b$					& $0.923 \pm 0.020$			&	$A_l(\rm SLD)$				& $0.1513 \pm 0.0021$		\\
$A_{FB}^{l}$				& $0.0171 \pm 0.0010$		&	$\sinleff (\rm LEP)$				& $0.2324 \pm 0.0012$		\\
$A_{FB}^{c}$				& $0.0707 \pm 0.0035$		&	$A_{FB}^{b}$				& $0.0992 \pm 0.0016$		\\
\hline 
\multicolumn{4}{c}{further observables}\\
\hline 
$\Delta \alpha^{(5)}_{had}(\MZ^2)\,[\times 10^{-5}]$	& $2758 \pm 10$		&	$\alpha_s(\MZ^2)$			& 						\\
$G_F$						& $1.1663787(6)\times10^{-5}\,\GeV^{-2}$ &			&			\\
\hline
\end{tabular}
\caption{Overview of observables, their values und current uncertainties which are used or determined within the global electroweak fit. See text for additional details. \label{tab:Overview}}
\end{center}
\end{table}

\subsection{The $H$ boson\label{sec:higgs}}

\begin{figure}[t!]
\begin{center}
\begin{minipage}{0.37\textwidth}
\resizebox{0.9\textwidth}{!}{\includegraphics{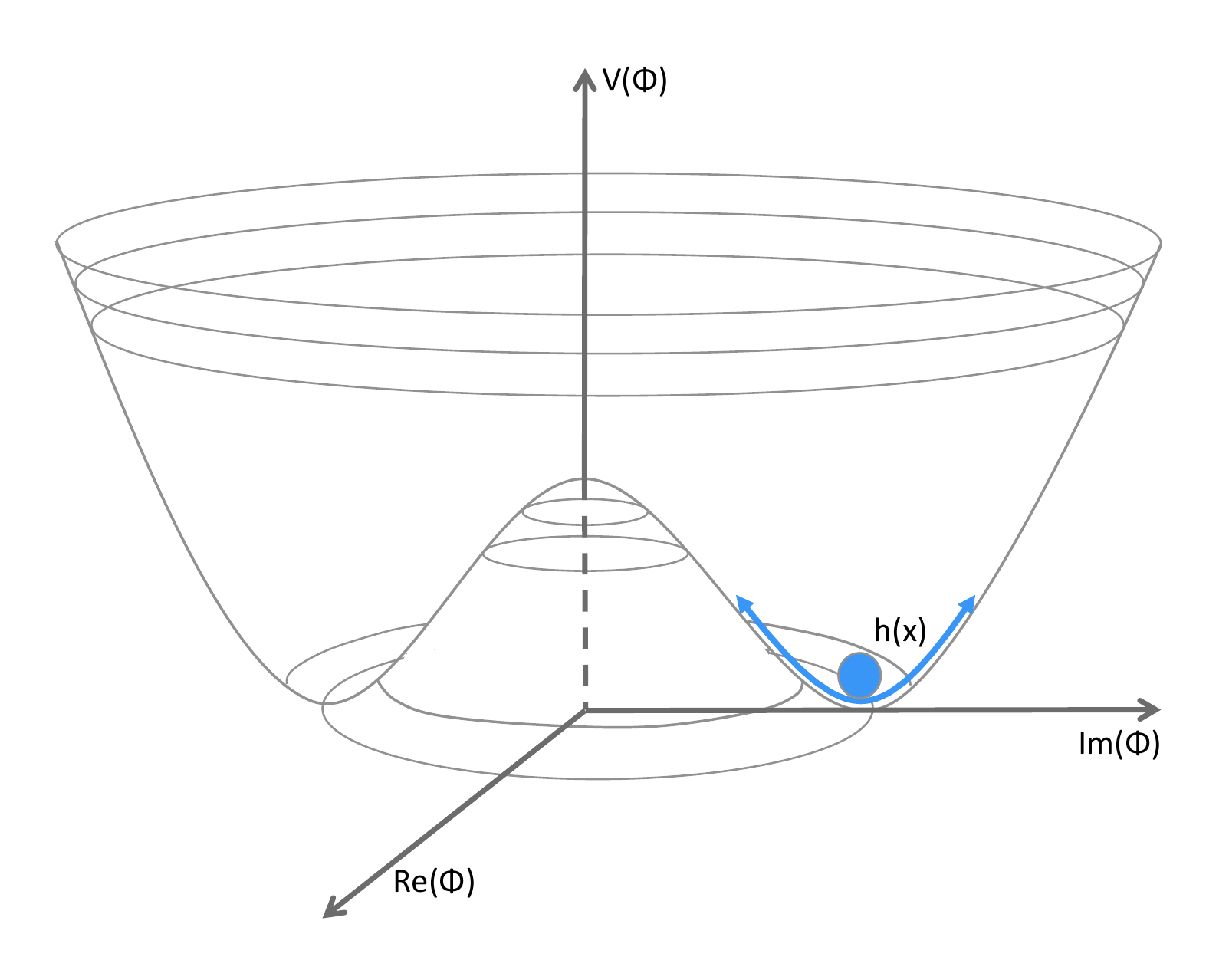}}
\caption{\label{fig:HiggsPotential}Schematic illustration of the Higgs potential.\vspace{0.3cm}}
\end{minipage}
\hspace{0.3cm}
\begin{minipage}{0.6\textwidth}
\resizebox{1.0\textwidth}{!}{\includegraphics{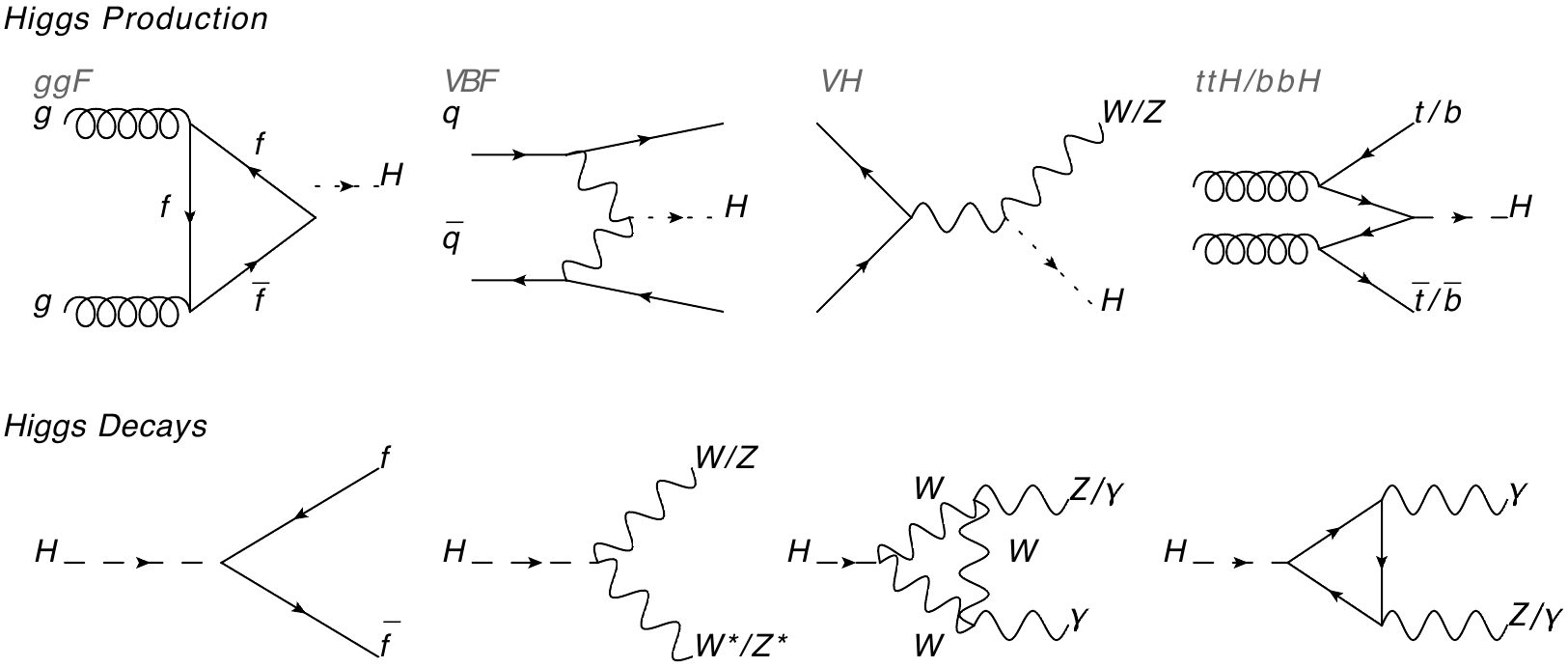}}
\caption{\label{fig:HiggsProduction} Feynman diagrams for dominant Higgs production (upper row) and decay processes (lower row).}
\end{minipage}
\end{center}
\end{figure}

The Brout-Englert-Higgs electroweak symmetry breaking mechanism~\cite{Higgs:1964pj,Englert:1964et} in the Standard Model is described by two parameters, 
$\mu$ and $\lambda$, which define the shape of the underlying scalar field (Higgs field) potential. 
According to Equation~(\ref{higgsvev}), for $\mu^2 < 0$ (one needs $\lambda^2 > 0$ to bound $V_H$ in Equation~(\ref{eqn:VH}) below), the Higgs potential has a minimum at
\begin{equation}
v =\frac{\sqrt{2}|\mu|}{\lambda} = \frac{2 \mW}{g} = 246.022~{\rm GeV},
\end{equation}
known as the vacuum expectation value, which sets the electroweak scale. 
The Higgs field itself is a weak isospin doublet with four components which fluctuates around the minimum $v$ spontaneously breaking the rotational symmetry 
(in field space) of the Higgs field. 
The physical Higgs boson is therefore a scalar field $h(x)$, expanded around $v$. 
Expanding the Higgs potential (see Figure~\ref{fig:HiggsPotential}) to second order in $h(x)$,
\begin{equation}
V_H = V_0 + \frac{\mu^2}{2} \left[ 2v h(x) + h(x)^2 \right] + \frac{\lambda^2}{8} \left [4 v^3 h(x) + 6 v^2 h(x)^2 \right] = V_0 + \frac{\lambda^2 v^2}{2} h(x)^2,
\end{equation}
introduces directly the Higgs boson mass term, $\mH = \lambda v$. 
After $v$ is fixed experimentally by the Fermi constant (from $\mu$ decay)
the parameter $\lambda$ (and thus $\mH$) is the only remaining free parameter in the Higgs potential. 
When gravity is included, $V_0$ corresponds to a contribution of order $v^4$ to the cosmological constant which is some 60 orders of magnitude too large compared
to the value deduced from the acceleration rate of the universe (the infamous and unsolved cosmological constant problem).

The couplings of the Higgs field to the electroweak bosons are encoded by their masses in Equations~(\ref{higgsvev}) and (\ref{eqn:mwtreelevel}).
Likewise, the coupling of the Higgs field to fermions can be introduced {\em via\/} the Lagrangian (after spontaneous symmetry breaking),
\begin{equation}
L_Y = g_f (\bar f_L f_R + \bar f_R f_L) (v + h),
\end{equation}
where the term proportional to $v$ can be interpreted as a fermion mass term, $m_f = g_f v$, and the other term represents the (Yukawa) coupling to the physical Higgs field $h$. 
The coupling constant $g_f$ of the Higgs field to the fermion field $f$ is therefore proportional to the fermion mass $m_f$,
and the measurement of fermion masses is equivalent to the measurement of their couplings to the Higgs field. 
For very large values of $\mH \gg \nu$, the total Higgs width scales like $\sim G_F \mH^3$, so that the Higgs boson ceases to represent a particle resonance.
This can be traced to the Higgs self-coupling $\lambda$ which couples to the longitudinal components of the $W$ and $Z$ bosons.
Ultimately, tree-level unitarity of the partial S-wave amplitude of elastic Goldstone boson scattering implies the bound~\cite{PhysRevD.16.1519, PhysRevD.40.1725},
$$ \frac{M_H^2}{v^2} < \frac{16\pi}{5}\ , \qquad\qquad\qquad\qquad\qquad M_H < 781~{\rm GeV}. $$

Once the Higgs boson mass is known, all parameters in the electroweak sector are fixed and all other relations, 
such as the Higgs boson width or the couplings to the electroweak vector bosons can be derived within the framework of the Standard Model. 
New physics models can lead to differences in the Higgs boson couplings as well as to a change in the relation of the Higgs boson mass to other electroweak parameters. 
We will focus in this review article on the electroweak fit within the Standard Model itself, and we will only discuss the measurement of $\mH$ in more detail, 
but not the studies of the Higgs boson couplings. 
The underlying assumption is that the new particle observed at the LHC in 2012~\cite{Aad:2012tfa, Chatrchyan:2012xdj}, with a mass of 125~GeV, is indeed the SM Higgs boson. 
The most recent studies of Higgs properties are therefore summarized in Section~\ref{sec:HiggsLHC} 
together with a brief discussion of Higgs boson production and decay.

\subsubsection{The Higgs boson at the LHC\label{sec:HiggsLHC}}

The leading order diagrams for the most relevant Standard Model Higgs boson production processes in proton-proton collisions,
as well as its relevant decay channels are illustrated in Figure~\ref{fig:HiggsProduction}. 
A Higgs boson with a mass of 125~GeV is expected to be produced via gluon fusion ($ggH$) with a probability of 87\%. 
The vector boson fusion Higgs production (VBF-H) has a contribution of 6.8\%, followed by the Higgs boson radiation process (VH) with 4\%, 
and the associated production with top ($ttH$) or bottom quarks ($bbH$) with 0.9\%. 
The different production processes can be distinguished experimentally to some extent. 
The VBF-H process is typically accompanied by two (forward) jets with a large rapidity gap, since no significant color connection is expected in the initial state. 
The VH process implies an additional vector boson in the final state, which can be reconstructed in both leptonic and hadronic decay channels, 
depending on the decay channel of the Higgs. 
While the $bbH$ production is experimentally difficult to distinguish from the ggH process, the $ttH$ process has two top-quarks in the final state which can be detected. 
Apart from the $bbH$ production, all other Higgs production processes have been  experimentally confirmed~\cite{Khachatryan:2016vau,Aaboud:2017jvq,Sirunyan:2018hoz}. 

\begin{figure}[t!]
\begin{center}
\begin{minipage}{0.481\textwidth}
\resizebox{1.0\textwidth}{!}{\includegraphics{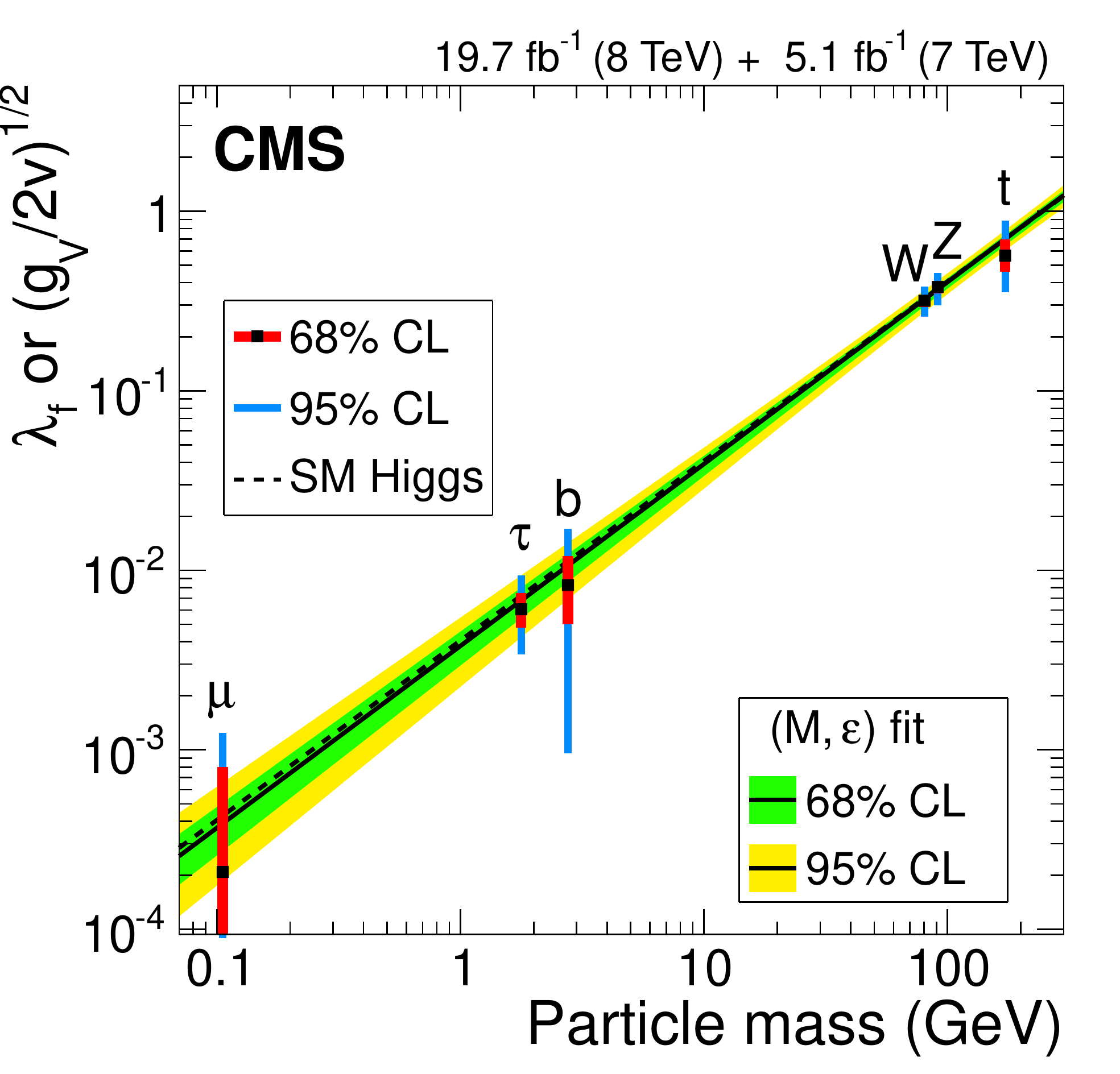}}
\caption{\label{fig:HiggsCoupling}CMS Collaboration~\cite{Khachatryan:2014jba}: 
summary of the fits to Higgs couplings expressed as a function of the particle mass. 
For the fermions, the values of the fitted Yukawa couplings (here called $y_f$) are shown, 
and for vector bosons the square root of the $h_{VV}$ coupling divided by twice the vacuum expectation value of the Higgs boson field. 
Particle masses for leptons and weak boson, and the vacuum expectation value of the Higgs boson are taken from the PDG.\vspace{0.3cm}}
\end{minipage}
\hspace{0.3cm}
\begin{minipage}{0.45\textwidth}
\resizebox{1.0\textwidth}{!}{\includegraphics{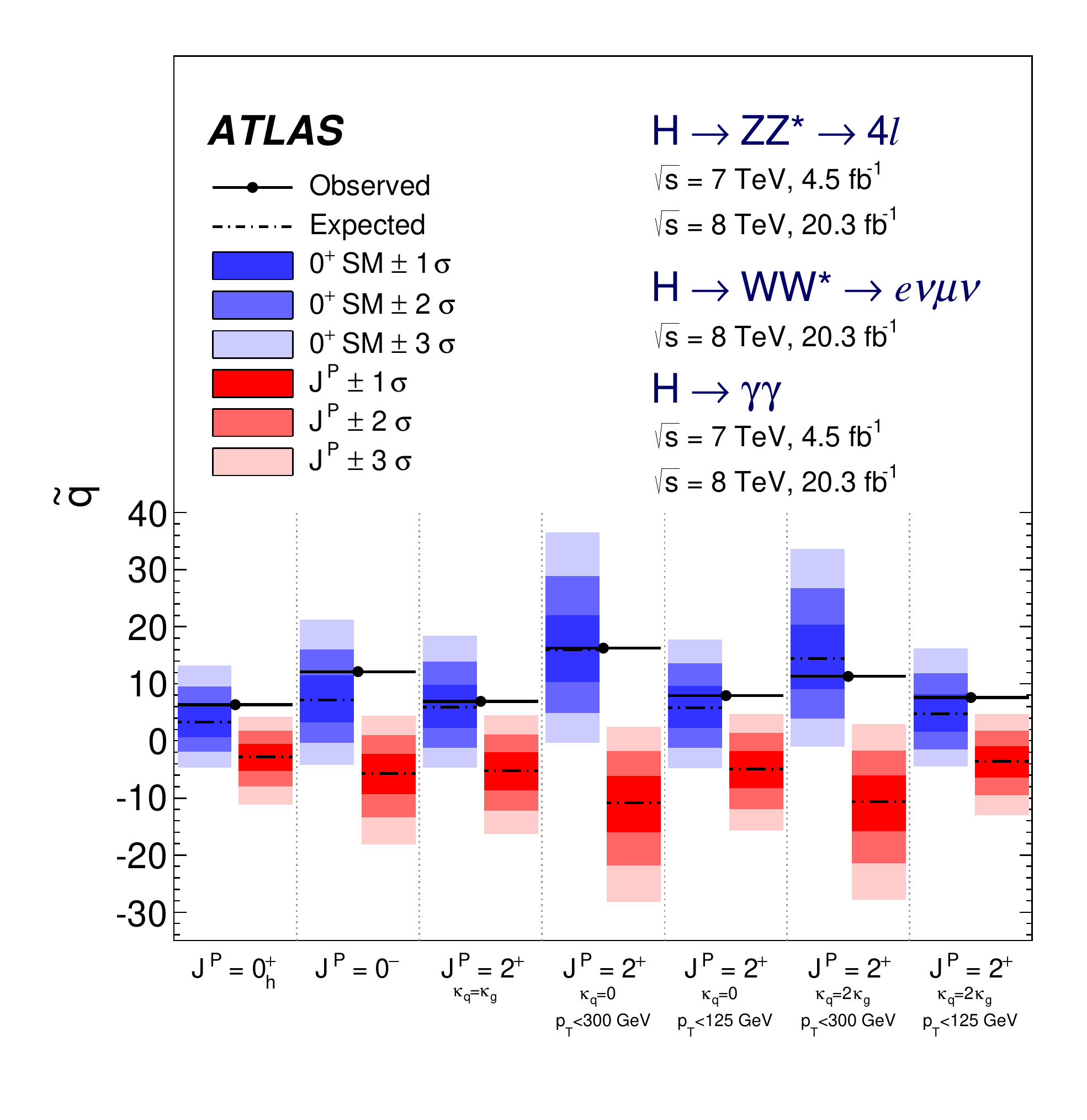}}
\caption{\label{fig:HiggsSpin} ATLAS Collaboration~\cite{Aad:2015mxa}: 
distributions of the test statistic $\tilde q$ for the SM Higgs boson and for the $J^P$ alternative hypotheses. 
They are obtained by combining the $H\rightarrow ZZ\rightarrow 4l$, $H\rightarrow WW\rightarrow e\nu\mu\nu$ and $H\rightarrow\gamma\gamma$ decay channels. 
The expected median (black dashed line) and the $\pm1$, $\pm2$ and $\pm3$ $\sigma$ regions for the SM Higgs boson (blue) 
and for the alternative $J^P$ hypotheses (red) are shown for the signal strength fitted to data. 
The observed $\tilde q$ values are indicated by the black points.}
\end{minipage}
\end{center}
\end{figure}

The branching ratios of the Higgs boson can be classified similiarly to the different production processes. 
The dominant decay of a 125~GeV Higgs boson is to bottom-quarks ($H\rightarrow b\bar b$) with a branching ratio of 0.57. 
A significant branching ratio for the decay into other fermions exists for the $\tau^+ \tau^-$ and the $c \bar c$ final states with values of 0.063 and 0.029, respectively. 
Due to their small mass the decay into muons is highly suppressed yielding $BR(H\rightarrow\mu^+\mu^-)=0.02\%$. 
The Higgs boson decay into $WW$ or $ZZ$ implies that one of the vector bosons must be highly off-shell, leading to branching ratios of 0.22 and 0.028, respectively. 
In addition to the direct decay channels also loop induced decays, illustrated in Figure~\ref{fig:HiggsProduction}, have to be considered. 
The two gluon decay has a branching ratio of 0.082, but it is experimentally not accessible due to the overwhelming multi-jet background in proton-proton collisions. 
The situation is significantly different for the loop decay into two photons with a branching ratio of 0.23\%. 
The $H\rightarrow Z\gamma$ decay has only a slightly smaller branching ratio of 0.15\% but requires an additional leptonic decay of the $Z$ boson to be identified in the data. 
Higgs decays into $WW$, $ZZ$, $\gamma\gamma$, $\tau^+ \tau^-$ and $b\bar b$ 
have been observed~\cite{Khachatryan:2016vau,Aaboud:2017jvq,Sirunyan:2018hoz,Aaboud:2018urx}, since its discovery in 2012. 

\begin{figure}[t!]
\begin{center}
\begin{minipage}{0.45\textwidth}
\resizebox{1.\textwidth}{!}{\includegraphics{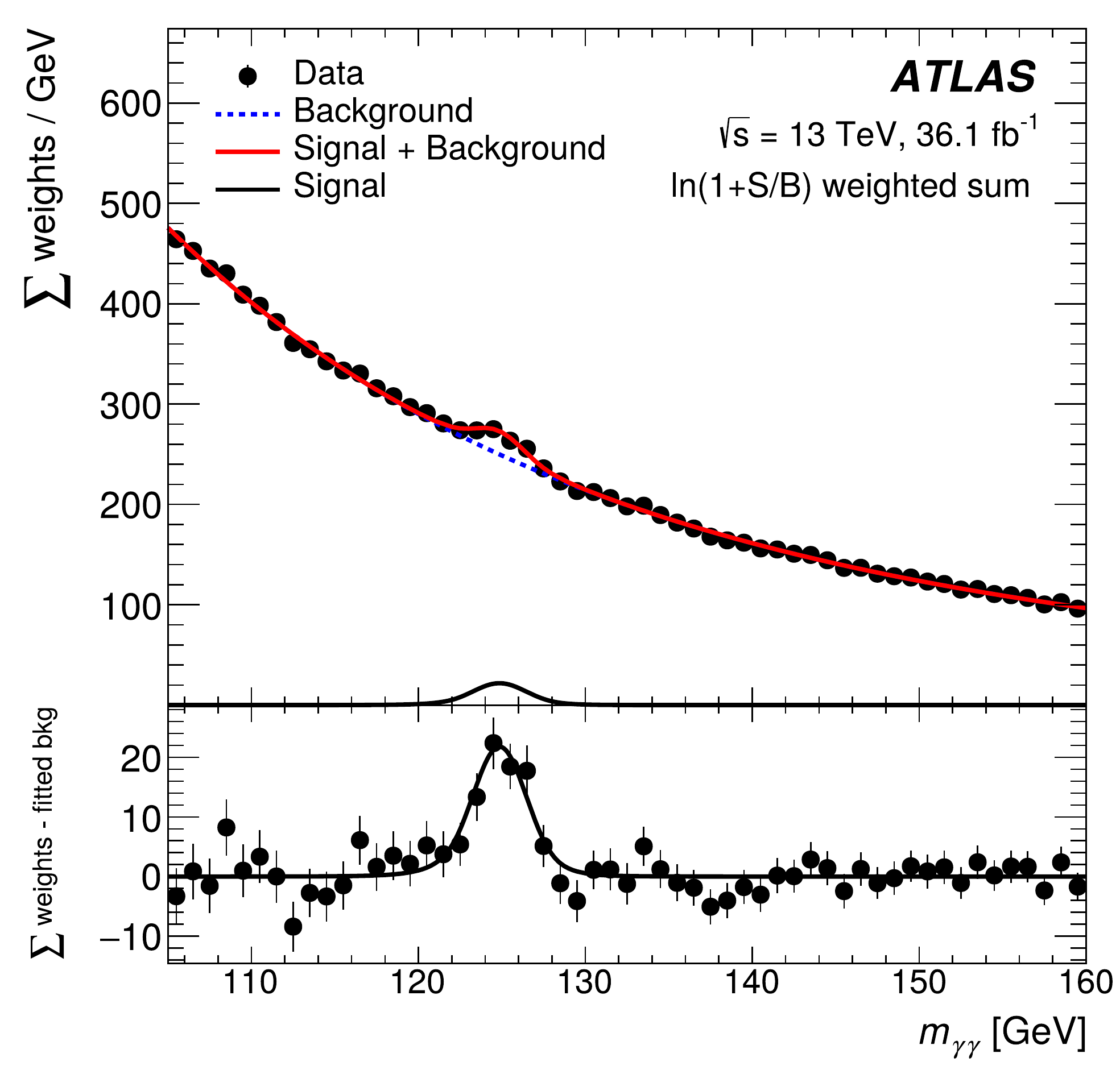}}
\caption{\label{fig:HiggsHGG}ATLAS Collaboration~\cite{Aaboud:2018wps}: 
diphoton invariant mass distribution of all selected data events, overlaid with the result of the fit (solid red line). 
Both for data and for the fit, each category is weighted by a factor $\ln(1 + S/B)$, where $S$ and $B$ are the fitted signal and background yields 
in an $m_{ \gamma \gamma}$ interval containing 90\% of the expected signal. 
The dotted line describes the background component of the model.}
\end{minipage}
\hspace{0.3cm}
\begin{minipage}{0.45\textwidth}
\resizebox{1.0\textwidth}{!}{\includegraphics{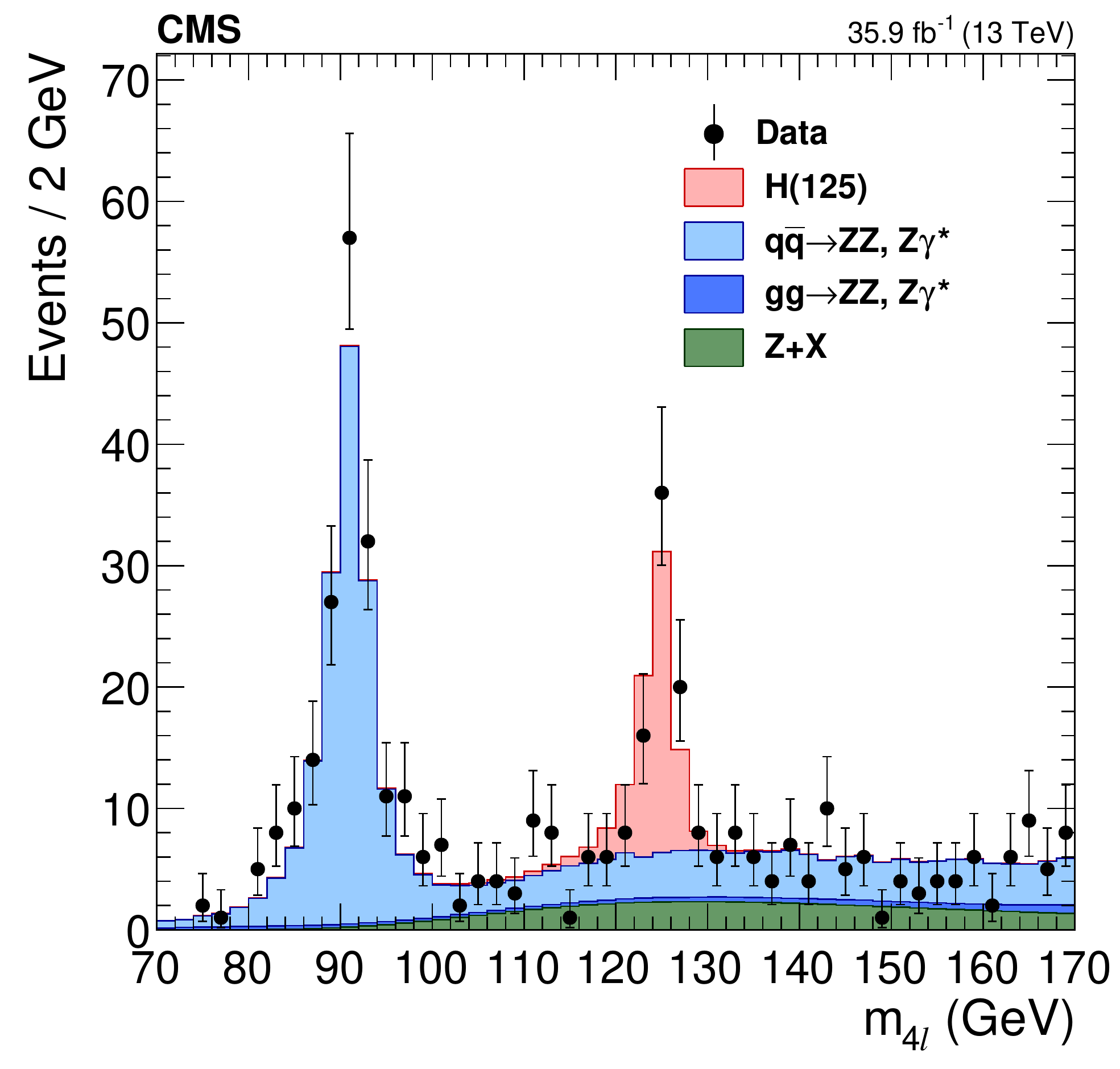}}
\caption{\label{fig:HiggsZZ} CMS Collaboration~\cite{Sirunyan:2017exp}: 
distribution of the reconstructed four-lepton invariant mass $m_{4l}$. 
Points with error bars represent the data and stacked histograms represent expected signal and background distributions. 
The SM Higgs boson signal with $\mH=125\,\GeV$, denoted as $H(125)$, and the $ZZ$ backgrounds are normalized to the SM expectation, 
whilst the $Z+X$ background is normalized to the estimation from data.}
\end{minipage}
\end{center}
\end{figure}

The experimental identification of the production processes of the Higgs boson, 
as well as of the different decay channels allows for measurements of several (differential) cross-sections in mutually exclusive regions of phase space. 
A combined fit of these cross sections or their ratios allows then to constrain the Higgs boson couplings~\cite{Khachatryan:2016vau}. 
It is crucial to note that only ratios of Higgs couplings can be measured in the most generic approach, since the total width of the Higgs boson 
is not accessible with sufficient precision at the LHC. 
Up to now, no deviations from Standard Model expectations have been observed in the available measurements of Higgs coupling strength parameters. 
These findings are schematically summarized in Figure~\ref{fig:HiggsCoupling}, which shows the measured coupling strengths of the Higgs boson to several SM particles 
in dependence of their mass. As predicted by the SM, a linear relation can be seen.

The Standard Model Higgs boson is predicted to have even parity and is the only scalar, \ie, spin-0 particle, in the SM. 
The study of angular distributions of the decay products in $H\rightarrow ZZ\rightarrow 4l$, $H\rightarrow WW\rightarrow e\nu\mu\nu$ and $H\rightarrow\gamma\gamma$ 
allow to test this prediction as well as other hypotheses of the underlying $J^{P}$ structure. 
All tested alternative models are disfavored against the SM Higgs boson hypothesis at more than 99.9\% CL~\cite{Aad:2015mxa, Khachatryan:2014kca}. 
This is illustrated in Figure~\ref{fig:HiggsSpin}, where the distributions of the test statistic $\tilde q$ for the SM Higgs boson and for the $J^P$ alternative hypotheses are shown. 

\subsubsection{Higgs boson mass measurements}

For a precise measurement of the Higgs boson mass, the full final state of the Higgs boson decay has to be reconstructed 
with an excellent energy and momentum resolution of the final state particles. 
Hence, the mass measurements of the Higgs boson are performed in the $H\rightarrow Z Z^*$ and $H\rightarrow \gamma \gamma$ decay channels, 
where the fully reconstructed invariant mass of the final state system leads to a narrow peak over smooth background. 
The mass value can therefore be extracted from the peak position in a model independent way, \ie, without assumptions concerning the Higgs boson production and decay yields. 
Since the SM expectation of the Higgs boson width is only 4~MeV, the width of the observed signal is purely an artifact of the detection resolution 
and the mass peak shift due to the interference between the SM background and Higgs signal can be neglected. 
Both collaborations reported their final Higgs mass measurements of Run~I and published a combination~\cite{Aad:2015zhl}.

The Higgs boson decay into photons, through a loop of heavy particles, 
has a small branching ratio but a very large event yield can be obtained with a narrow peak on top of a smoothly falling background. 
The most important irreducible background process in the $H\rightarrow \gamma \gamma$ channel is di-photon production ($q\bar{q} \rightarrow \gamma \gamma$), 
but large contributions from reducible background processes such as $\gamma+\rm jet$ or $\rm jet+jet$ production are also expected, where one or two jets are mis-identified as photons. 
In order to suppress fake photons, typically tight isolation and identification criteria are applied. 
The events need to clear a di-photon trigger and are required to pass a minimal transverse energy requirement 
on the order of 25-35~GeV within the geometric acceptance of the tracking detectors of ATLAS and CMS, \ie, with a maximal pseudo-rapidity $|\eta|\approx2.5$. 
The Higgs signal in the invariant mass distribution can be modeled by a sum of two Gaussian distributions and a Crystal Ball function (ATLAS) 
or a sum of three to five Gaussian functions (CMS). 
The background is modeled by both collaborations by smooth function, \eg, an exponential function, 
which can be fitted and tested directly at the invariant mass distribution outside the signal region, \ie, {\em via\/} a side-band approach. 
The bias due to the choice of the background function can be either studied by simulated samples or by the comparison of different functional choices. 
The mass determination is then performed by maximizing a profile likelihood function, depending on \mH and further nuisance parameters. The likelihood function is constructed using probability density functions based on expected signal and background distributions, also parameterizing detector response effects\footnote{In fact, a ratio of likelihood-functions is used, however, we refer to Ref.~\cite{Aad:2015zhl} for a more detailed discussion}.

In order to gain sensitivity, the $H\rightarrow \gamma \gamma$ events are split into disjoint categories, 
which have different di-photon mass resolutions, systematic uncertainties and signal-to-background ratios. 
The invariant di-photon mass distribution including the fitted signal and background of the recorded data sets in 2015/16 by the ATLAS detector, 
corresponding to an integrated luminosity of 36.1~fb$^{-1}$ at a center of mass energy of 13~TeV is shown as an example in Figure~\ref{fig:HiggsHGG}. 
Here, a signal to background ratio of roughly 0.03 is observed within a mass range of 120 to 130~GeV. 
An overview of the $\mH$ measurements in the $H\rightarrow \gamma \gamma$ channels is given in Table~\ref{tab:HiggsMass} for both collaborations. 
The dominant systematic uncertainties are due to non-linearity effects of the electromagnetic calorimeter response, 
uncertainties in the material in front of the calorimeter and shower-shape uncertainties which play a role in the photon identification. 
The differences in the statistical uncertainties between both experiments can be partly explained by the treatment of the bias due to the choice of the background fit function, 
which can be interpreted as a statistical or as systematic uncertainty. 

\begin{table}[t]
\footnotesize
\begin{tabular}{l | l ccc | l | l ccc}
\hline
\multicolumn{5}{c|}{$H\rightarrow\gamma\gamma$}	& \multicolumn{5}{c}{$H\rightarrow ZZ^* \rightarrow 4l$ } \\
\hline
experiment & data set & $\mH$ & stat.\ unc. & syst.\ unc. & experiment & data set & $\mH$ & stat.\ unc. & syst.\ unc. 	\\
\hline
ATLAS~\cite{Aad:2015zhl} & LHC Run 1 & 126.02 & 0.43 & 0.27 & ATLAS~\cite{Aad:2015zhl} & LHC Run 1 & 124.51 & 0.52 & 0.04 \\
CMS	\cite{Aad:2015zhl} & LHC Run 1 & 124.70	 & 0.31 & 0.15 & CMS~\cite{Aad:2015zhl}	& LHC Run 1 & 125.59 & 0.42 & 0.16 \\
\hline
ATLAS~\cite{Aaboud:2018wps}	 & LHC 2015/16 & 124.93	& 0.21 & 0.34 & ATLAS~\cite{Aaboud:2018wps} & LHC 2015/16	& 124.79	& 0.36 & 0.05 \\
& &  &  & & CMS~\cite{Sirunyan:2017exp} & LHC 2015/16 & 125.26 & 0.20 & 0.08 \\
\hline
\end{tabular}
\centering
\caption{\label{tab:HiggsMass}Overview of the most precise determinations of the Higgs boson mass in the two-photon and four-lepton final state by the ATLAS and CMS collaborations.
All values are given in GeV.}
\end{table}

The Higgs boson decay to two $Z$ bosons has a very good signal over background ratio, when both $Z$ bosons decay further into leptons, i.e. $H\rightarrow ZZ^* \rightarrow 4l$, 
where only electrons and muons ($l=e,\mu$) are considered for the mass measurements. 
Since one $Z$ boson has to be off-shell, \ie, with a mass around $|\mH-\mZ|\approx 34\,\GeV$, 
the energy and momentum of its decay leptons are significantly lower compared to the on-shell $Z$ boson decay. 
The minimal requirement on the transverse energy and momentum of the electrons and muons is therefore reduced to the lowest possible values 
which still allow a clean reconstruction and identification in the detector of approximately 5~GeV within the geometrical acceptance of the tracking systems. 
Events with four isolated reconstructed leptons are selected, of which two opposite charged, same flavor lepton pairs have to be formed. 
Two of the leptons are required to have an invariant mass close to the Z boson mass. 
The dominant background in the $H\rightarrow ZZ^* \rightarrow 4l$ channel is the non-resonant $ZZ$ di-boson production $qq\rightarrow ZZ^*$, 
which is typically estimated and constrained in control and side-band regions. 
Additional background sources are top quark pair production and $Z$ boson production in association with jets. 
They are estimated in the signal region using data-driven techniques according to the flavor of the sub-leading lepton pairs. 
The expected signal to background ratio for $H\rightarrow ZZ^* \rightarrow 4l$ events ranges between 1.5 and 2, 
\ie, it is significantly larger than in the $H\rightarrow \gamma \gamma$ channel. 
For example, the CMS collaboration selected 176~candidate events with an invariant mass between 120 and 130~GeV in the 2015/2016 data set of the LHC, 
taken at a center of mass energy of 13~TeV and corresponding to an integrated luminosity of 35.9~fb$^{-1}$ (Figure~\ref{fig:HiggsZZ}). 
The expected number of signal and background events has been approximately 110 and 64, respectively. 

The events can be categorized in four classes depending on their final states, \ie, $4e, 2e2\mu, 4\mu$. 
Moreover, further categories can be built, \eg, utilizing production channel characteristics, to enhance the statistical sensitivity further. 
The $\mH$ mass determination is then performed either via a per-event approach or a template based method, 
employing a simultaneous profile likelihood fit to all measurement categories. 
The free parameters of the fit are the Higgs boson mass, $\mH$, as well as the nuisance parameters associated with systematic uncertainties. 
The resulting values of $\mH$ including their statistical and systematic uncertainties of all currently available measurements of the ATLAS and CMS collaborations 
are also summarized in Table~\ref{tab:HiggsMass}. 
The dominant systematic uncertainties are due to the limited knowledge in the lepton energy and resolution, but the individual measurements are still statistically limited.

\begin{figure}[tb]
\begin{center}
\begin{minipage}{0.49\textwidth}
\resizebox{1.0\textwidth}{!}{\includegraphics{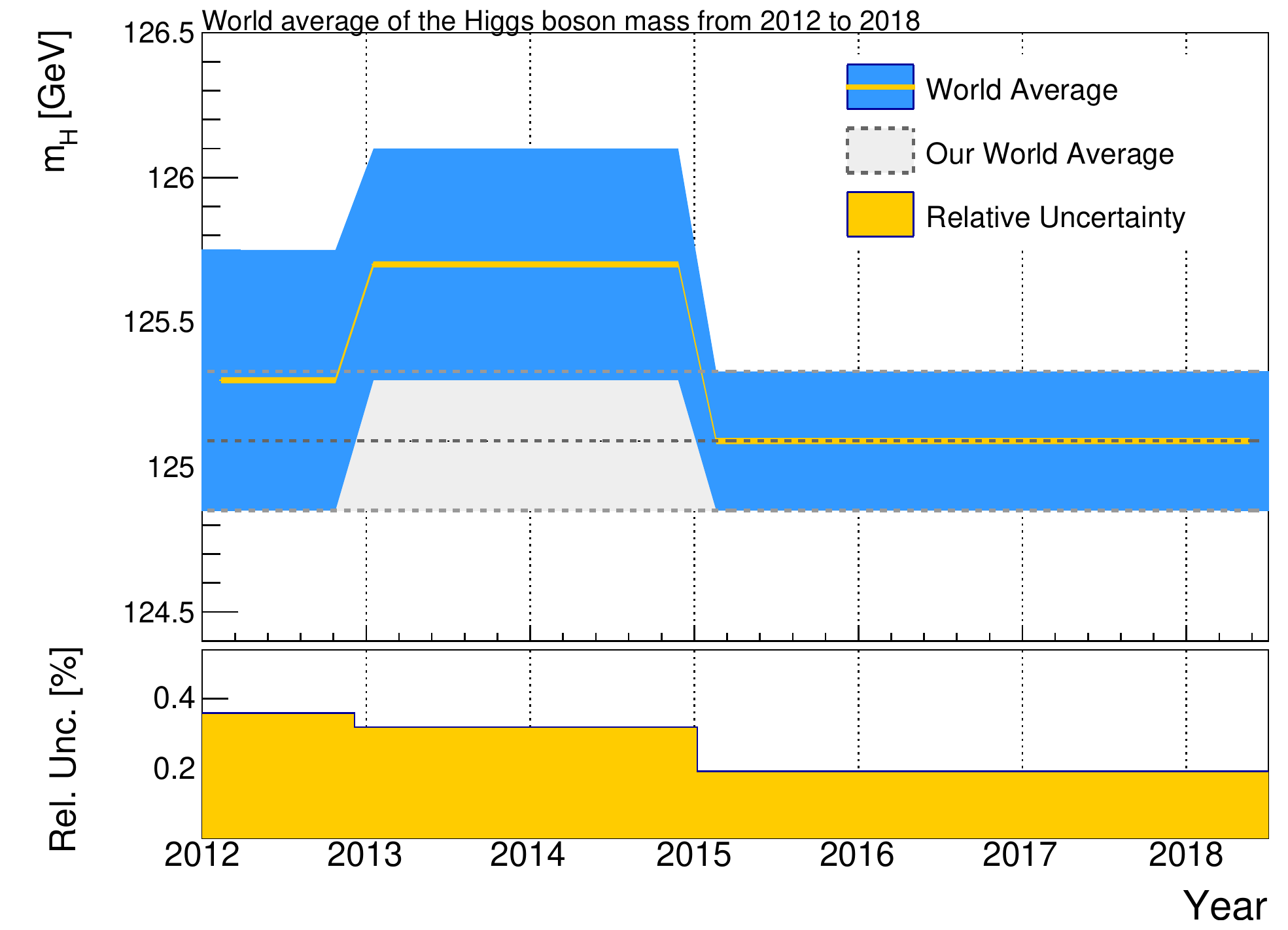}}
\caption{\label{fig:HMassEvolution}Evolution of the world average of the Higgs boson mass and its uncertainties {\em vs.\/}\ time. 
Values are taken from previous editions of the PDG review~\cite{Olive:2016xmw}.}
\end{minipage}
\hspace{0.1cm}
\begin{minipage}{0.49\textwidth}
\resizebox{1.0\textwidth}{!}{\includegraphics{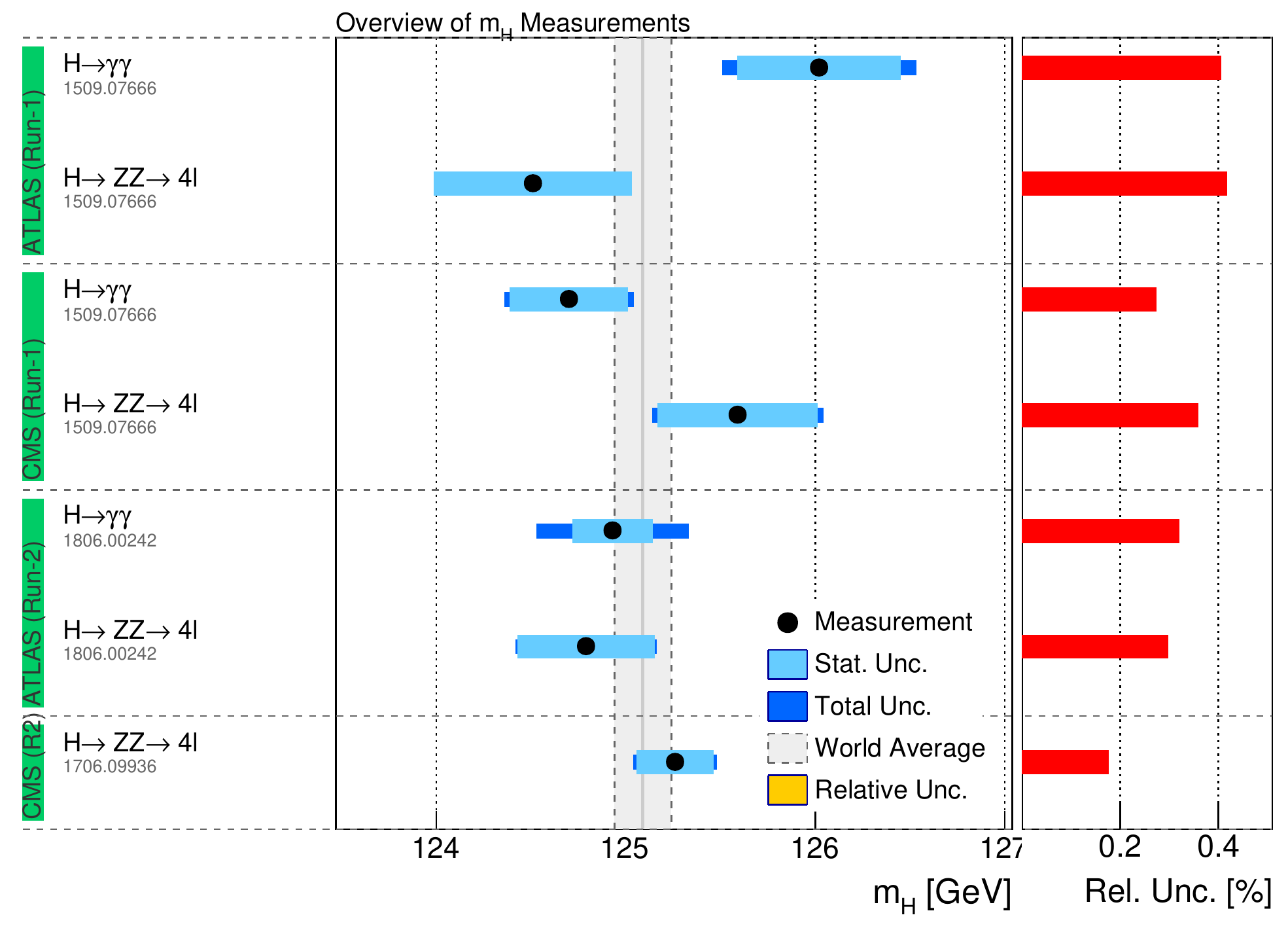}}
\caption{\label{fig:HMassOverview}Overview of the measurements of the Higgs boson mass by ATLAS and CMS, as well as their combination.\vspace{0.4cm}}
\end{minipage}
\end{center}
\end{figure}

The evolution of the world average of the Higgs boson mass is shown in Figure~\ref{fig:HMassEvolution}, 
while an overview of the relevant $\mH$ measurements from the ATLAS and CMS collaborations is given in Figure~\ref{fig:HMassOverview}. 
So far, only the measurements of the first LHC run at a center of mass energy of 7 and 8~TeV have been combined by the experimental collaborations. 
Here, three signal strength parameters have been introduced which scale, respectively, the production rates {\em via\/} fermions 
and vector boson initial states for the di-photon decay channel, and the total production rate for the $ZZ^*$ channel. 
The production rates are assumed to be the same for both experiments. 
Further theory uncertainties are negligible. 
Since no official combination of all measurements including the most recent ones, given in Table~\ref{tab:HiggsMass}, is currently available, 
we perform here a simplified combination using the \Blue-method~\cite{Lyons:1988rp, Nisius:2014wua}. 
The assumed correlations between the individual measurements have been validated by reproducing the official available combinations and range between 0 and 0.3. 
We find a new world average value,
\[\mH = 125.10 \pm 0.14~{\rm GeV}, \]
with a $\chi^2/{\rm n.d.f.} = 8.9/6$, corresponding to a likelihood 
value\footnote{When assuming no correlations between all available measurements, we find $\mH = 125.12 \pm 0.14$~GeV with a p-value of 0.22.} of $\rm p=0.18$. 
This value is also illustrated in Figure~\ref{fig:HMassOverview} together with previous measurements. 
The total 0.14~GeV uncertainty has approximately a 0.12~GeV statistical and a 0.07~GeV systematic component.
Hence, the overall combined uncertainty is still dominated by the limited statistics which will be come negligible after future runs of the LHC.

\subsection{The $W$ boson\label{sec:wboson}}

\subsubsection{Principle and challenges of the precision $W$ boson mass measurement\label{sec:wmassdis}}

\begin{figure*}[tb]
\begin{center}
\begin{minipage}{0.48\textwidth}
\resizebox{1.0\textwidth}{!}{\includegraphics{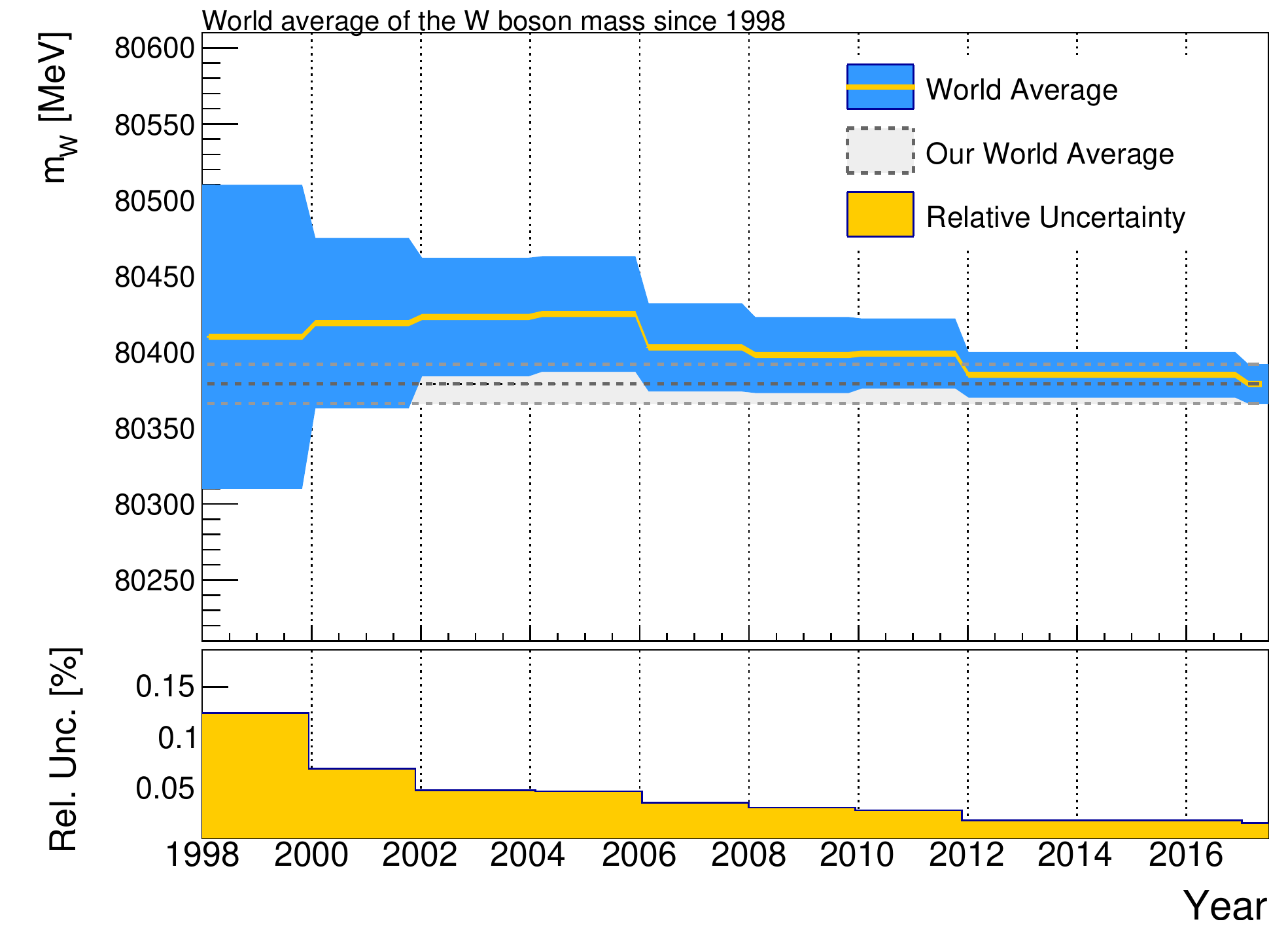}}
\caption{\label{fig:WMassEvolution}Evolution of the world average of \MW and its uncertainties {\em vs.\/}\ time. 
Values are taken from previous editions of the PDG review~\cite{Olive:2016xmw}.}
\end{minipage}
\hspace{0.3cm}
\begin{minipage}{0.48\textwidth}
\resizebox{1.0\textwidth}{!}{\includegraphics{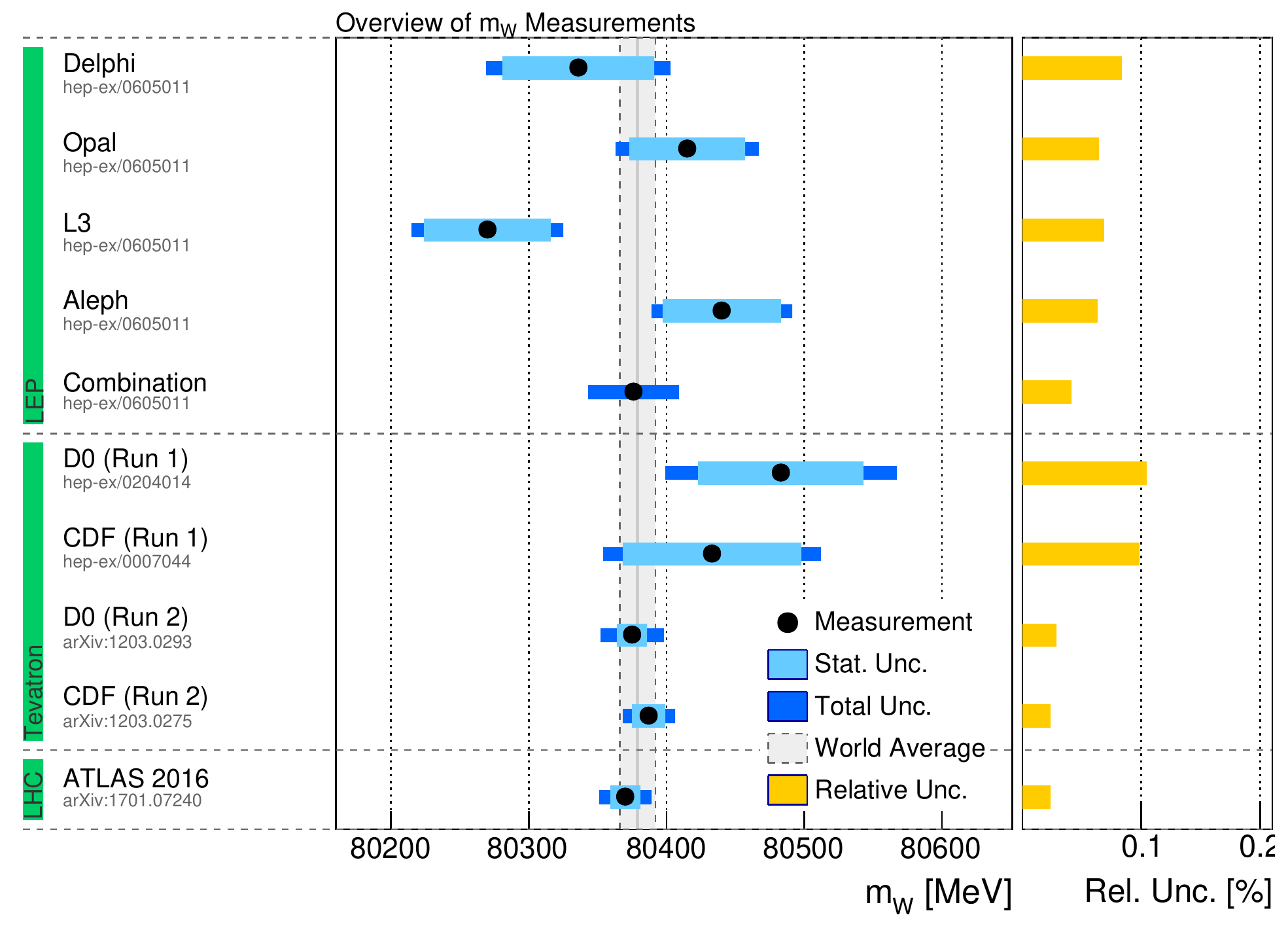}}
\caption{\label{fig:WMassOverview}Overview of selected measurements of \MW, 
including the most precise measurements from LEP~\cite{Schael:2013ita}, Tevatron~\cite{Aaltonen:2012bp,Abazov:2012bv}, and the LHC~\cite{Aaboud:2017svj}.}
\end{minipage}
\end{center}
\end{figure*}

Soon after the discovery of the $W$ boson at the UA1 and UA2 experiments~\cite{Arnison:1983rp,Banner:1983jy}, its mass was known with a precision of 5~GeV, 
based on the measurement of its decay product kinematics. 
A first precision measurement of \MW was possible during the LEP 2 runs, when the energy threshold for $W^+W^-$ production was reached. 
These measurements have been performed by the four LEP experiments, \Aleph, \Delphi, \LThree and \Opal~\cite{Schael:2013ita}, 
once extracting it making use of the dependence of the $WW$ cross section close to the production threshold 
but also {\em via\/} the direct reconstruction of the kinematics of the full hadronic decay channel (four quarks), semi-hadronic decay channel 
(two quarks, one charged lepton, one neutrino) and partially also the full leptonic decay channel with two charged leptons and two neutrinos in the final state. 
The latter measurements contribute mostly to the final LEP combined value of
\[\MW^{\rm LEP} = 80412 \pm 29 \pm 3~{\rm MeV}.\]
The statistical and systematic uncertainties are of similar order and 
the dominant systematic contribution is due to uncertainties in the description of fragmentation and hadronisation processes~\cite{Schael:2013ita}. 
The evolution of the world average of the $W$ boson mass and its associated uncertainty from 2000 to 2017 are illustrated in Figure~\ref{fig:WMassEvolution}, 
based on values of the PDG as well as on the combination within this work for the year 2018. 
The LEP measurements have been contributing mostly to the world average until 2007, 
when the first precision measurement of \MW at the Tevatron collider by the CDF collaboration was published, reaching a similar level of precision. 
We will therefore review the principle and the challenges of the $W$ boson mass measurement at hadron colliders in more detail, 
\ie, the measurements by the CDF and \DZero Collaborations at the Tevatron, and by the ATLAS Collaboration at the LHC. 

Similar to the LEP experiments, the determination of the $W$ boson mass at hadron colliders exploits the kinematic distributions of its decay products. 
Technically this is realized by a template fit approach. 
The expected final state distributions, \ie templates, are predicted by Monte Carlo (MC) simulations for varying $W$ boson masses 
and compared in a second step to the observed kinematic distributions. 
The minimal residual difference between the simulated \MW hypotheses and the measured distribution provides a handle on \MW. 

\begin{figure*}[t]
\centering
\resizebox{0.49\textwidth}{!}{\includegraphics{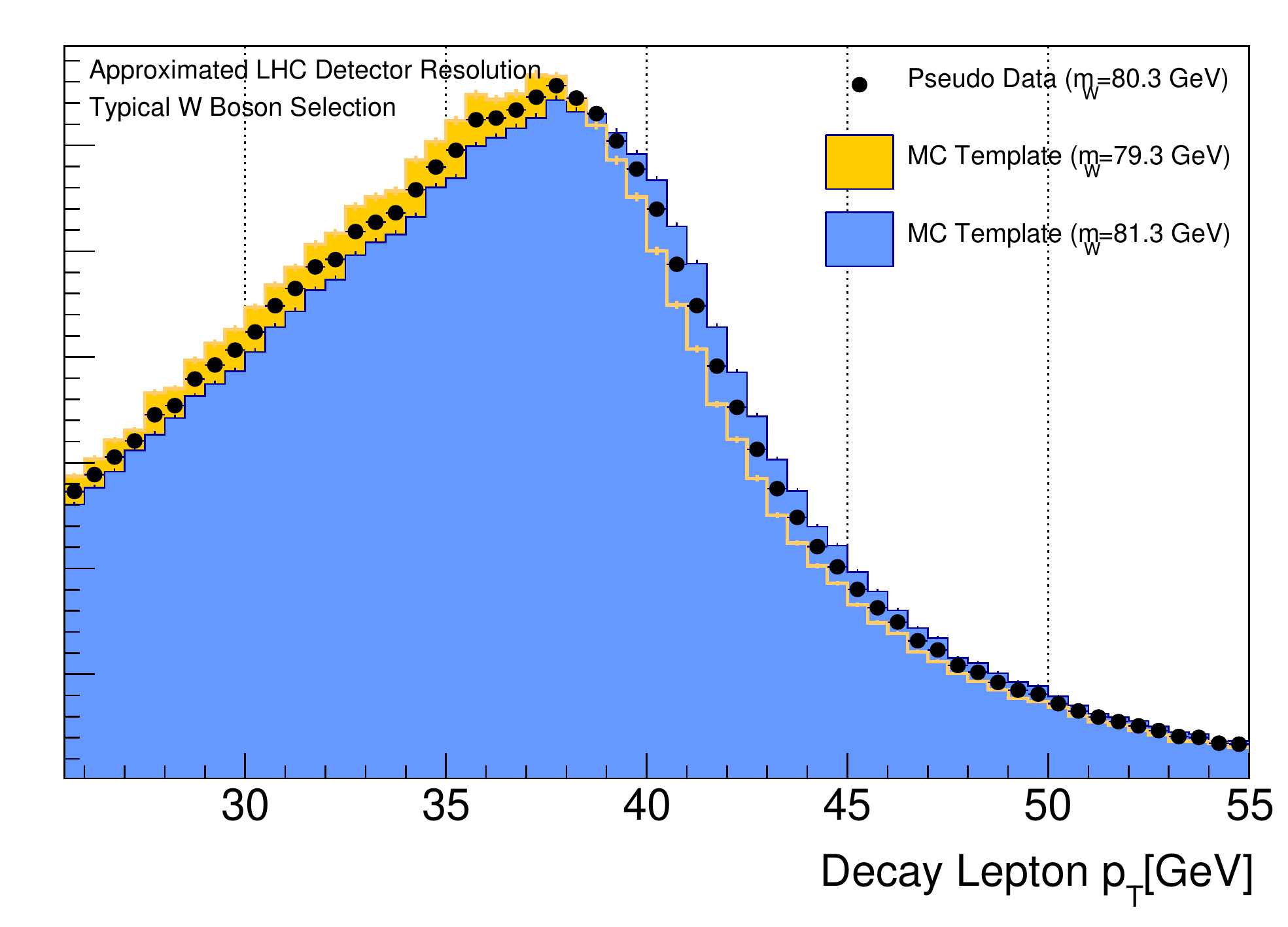}}
\resizebox{0.49\textwidth}{!}{\includegraphics{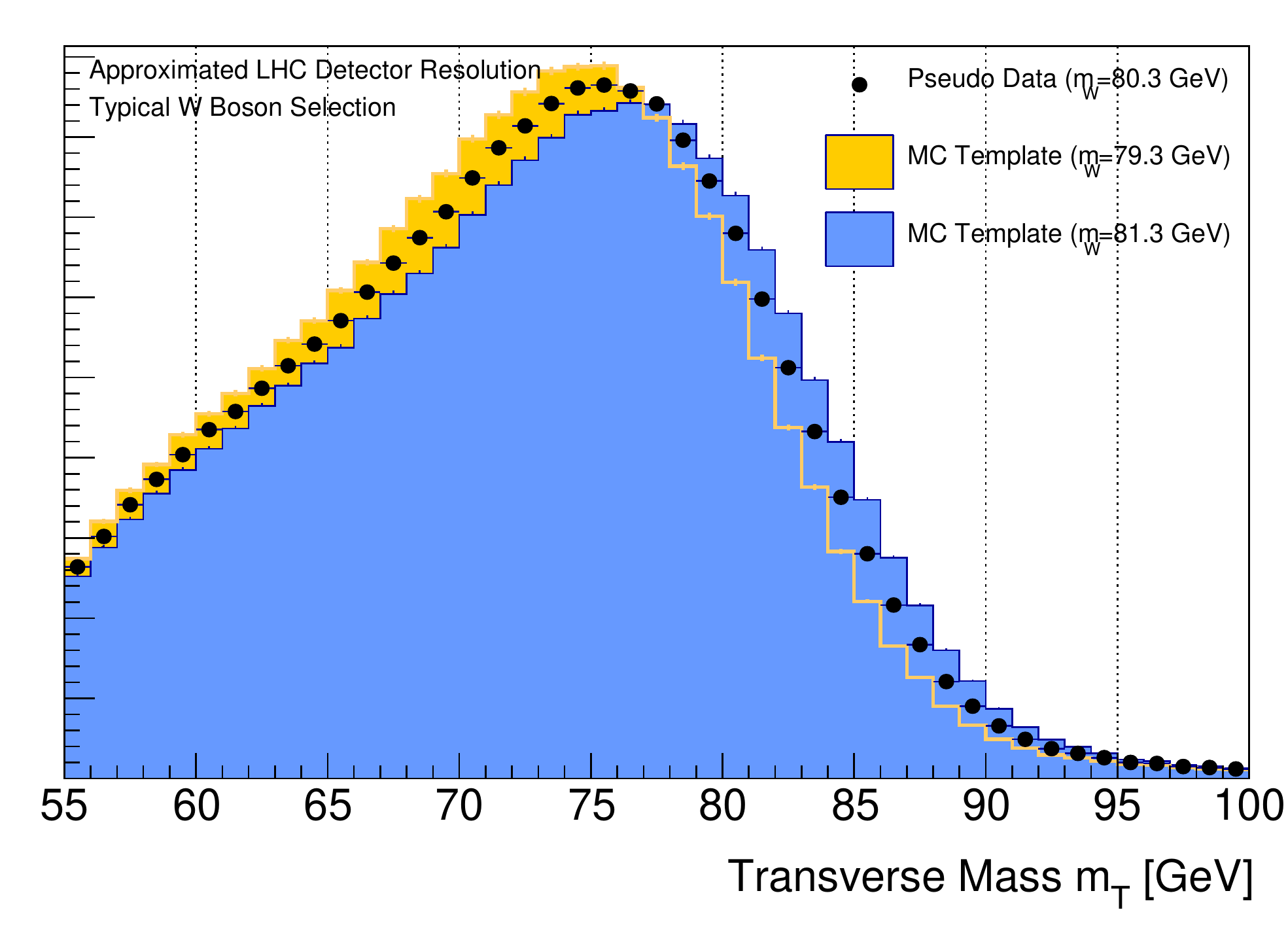}}
\caption{Templates of the $p_T$ (left) and $m_T$ (right) distributions for three different assumed \MW masses using a typical LHC detector response, based on \textsc{Pythia8} \cite{Sjostrand:2007gs} and  \textsc{Delphes} \cite{deFavereau:2013fsa}.}
\label{fig:WMassTemplates}
\end{figure*}

In contrast to the LEP experiments, the kinematics of the $W$ boson decay can only be fully reconstructed in the transverse plane w.r.t.\ the beam axis in hadron collisions, 
since the initial collision energy of the interacting partons in beam direction is unknown. 
Hence, only the momentum conservation in the transverse plane can be used to obtain relevant information on the decay kinematics. 
The measurement of \MW is performed in the electron and muon decay channels due to the overwhelming multi-jet background in case of the hadronic decay channels. 
The relevant observables sensitive to \MW are the transverse momentum spectrum of the charged decay lepton, ${\vec p\, }_T^l$, 
the missing transverse energy distribution, $\vec E_T^{miss} = - {\vec p\, }_T^l + \vec p_T(W)$, 
and the transverse mass distribution, 
$$m_T = \sqrt{2 p_T^l E_T^{\rm miss} [ 1 - \cos \Delta \Phi ({\vec p\, }_T^l, \vec E_T^{\rm miss})]}.$$ 
Here, $\Delta \Phi$ denotes the angle between the lepton and $\vec E_T^{\rm miss}$ in the transverse plane w.r.t.\ the beam axis, 
while $p_T(W)$ denotes the transverse momentum of the $W$ boson in that plane. 
Experimentally it is measured by the vectorial sum of reconstructed energy clusters in the calorimeters of the detector, known as hadronic recoil $\vec u_T = \sum \vec E^{\rm calo}_T$. 
The hadronic recoil is only a proxy for $p_T(W)$ as it is significantly affected by the number of simultaneous hadron collisions during one recorded event. 
The transverse mass $m_T$ corresponds to the invariant mass of the dilepton system when the $W$ boson decays fully in the transverse plane. 
Examples of $p_T$ and $m_T$ templates for three different assumed \MW masses are shown in Figure~\ref{fig:WMassTemplates}. 

Generally, the $p_T$ and $E_T^{\rm miss}$ spectra peak at $\approx \MW/2$, but the transverse mass spectrum peaks near $\MW$. 
These distributions depend on different aspects of the detector response, as well as the underlying physics modeling, 
and therefore allow for partially uncorrelated measurements of \MW. 
Experimentally the $p_T$ distribution is limited by the knowledge of the momentum and energy scale of the tracking system and the electromagnetic calorimeter, respectively, 
while the $E_T^{\rm miss}$ distribution is also affected by the calorimeter response parameters, leading to a significantly reduced sensitivity to \MW. 

Two aspects are important for a precision measurement of the $W$ boson mass {\em via\/} a template fit approach at hadron colliders: 
Firstly, the detector response for the decay products of the $W$ boson have to be modeled to the highest precision. 
Secondly, the modeling of the $W$ boson production and its decay has to be described sufficiently well. 
Both aspects will be briefly discussed.

The calibration of the detector response is typically performed using well known resonances, such as the $Z$ boson or the $J/\psi$, 
in their leptonic decay channels as their masses are known with a relative precision of 0.002\% and 0.0002\%, respectively. 
The invariant mass distribution of the $Z$ boson is precisely known from the LEP experiments and has the advantage that it is kinematically very close to the decay of $W$ bosons. 
In particular, corrections of the momentum and energy scales of leptons, their identification and reconstruction efficiencies, 
as well as the modeling of the hadronic recoil measurements can be derived. 
Any dependencies on the lepton $p_T$ and the transverse mass are of special importance, 
as they directly affect the template shapes and therefore introduce a bias in the \MW measurement. 
Since the kinematics of $Z$ and $W$ boson decays are similar but not identical, 
several methodologies have to be developed to transfer the corrections from the $Z$ to the $W$ boson events. 
The consistency of the detector calibration can be tested by measuring \MW in two decay channels, \ie, in the electron and the muon final states. 
Most experimental uncertainties, such as lepton scales, lepton identification efficiencies and background estimations are uncorrelated between these two channels. 
Hence, consistent results for fitted \MW values in both channels provide an important consistency check of various experimental calibrations. 
The related systematic uncertainties are typically limited by the available statistics of the $Z$ boson calibration sample.

In order to discuss the physics modeling in more detail, it is useful to decompose the fully differential leptonic Drell-Yan cross section in four terms,
\begin{equation} 
\label{eqn:decay}
\frac{d\sigma}{dp_1 dp_2} = \left[ \frac{d\sigma(m_{ll})}{dm_{ll}} \right] \left[ \frac{d\sigma(y_{ll})}{dy_{ll}} \right] \left[ \frac{d\sigma(p_T,y_{ll})}{dp_T}\frac{1}{\sigma(y_{ll})} \right] 
\left[ (1 + \cos^2\theta) + \sum _{i=0}^7 A_i (p_T,y_{ll}) P_i(\cos\theta, \phi) \right],
\end{equation}
where $p_1$ and $p_2$ are the four-momenta of the decay leptons. 
This decomposition is valid in the limit of massless leptons in a 2-body phase space and helicity conservation in the decay. 
The kinematics of the dilepton system are described by its invariant mass $m_{ll}$, its transverse momentum $p_T(ll)$ and its rapidity $y_{ll}$. 
The angles $\theta$ and $\phi$ describe the polar and the azimuthal angle of one lepton in the rest frame of the dilepton system. 
The impact of the helicity and polarization effects on the decay kinematics can be described by eight spherical harmonics $P_i$ of the order zero, one and two, 
weighted by the eight numerical coefficients $A_i$, which will be discussed in more detail in Section~\ref{sec:sinhad}.

The model uncertainties in these terms impact the final measurement uncertainty in \MW in several ways. 
The corrections due to electroweak effects \cite{Calame_2006} can be mostly described by final state radiation of the photons from the decay leptons, while EW loop corrections
are described via the mass-dependent term in Equation~(\ref{eqn:decay}). 
The dominant physics modeling uncertainties occur due to the limited knowledge of parton distribution functions (PDFs), 
the description of the transverse momentum distribution of the $W$ boson, and the prediction of QCD angular coefficients. 
The latter have been measured in $Z$ boson events and compared to the NNLO prediction, showing good agreement. 
It is therefore expected that their uncertainty is small compared to the other effects. 

The kinematic distributions of the decay leptons are broadened by the intrinsic transverse momentum distribution of the $W$ boson, $p_T(W)$. 
The $p_T(W)$ spectrum is caused by multiple gluon emissions of the initial state partons. 
Uncertainties in the $p_T(W)$ modeling imply changes in the decay lepton $p_T$ distribution and hence directly affect the \MW measurement. 
The associated model uncertainties originate from various approximations and free parameters in perturbative and non-perturbative calculations. 
However, they can be precisely constrained by a direct measurement of the corresponding $Z$ boson $p_T(Z)$ distribution.

The limited knowledge of proton PDFs induces the largest model uncertainty in \MW and impact several terms in Equation~(\ref{eqn:decay}). 
The most important aspects are uncertainties caused by the average polarization of $W$ bosons and acceptance effects. 
The polarization influences significantly the $p_T$ spectrum of the decay leptons and therefore impacts the \MW measurement. 
In order to illustrate the relation between the $W$ boson polarization and PDFs, it is instructive to discuss a simplified example at leading order. 
Here, the differential cross sections for $W$ bosons in the two helicity states $\lambda = \pm 1$ can be decomposed as,
\[ \sigma_{W^{+}} (y) \sim u(x_1) \bar d(x_2) + \bar d(x_1) u(x_2), \qquad\qquad\qquad\qquad\qquad
\sigma_{W^{-}} (y) \sim d(x_1) \bar u(x_2) + \bar u(x_1) d(x_2), \]
where $u$ and $d$ are the PDFs for the respective quarks. 
The parameters $x_1$ and $x_2$ denote the corresponding Bjorken parameters. 
This implies unpolarized $W$ bosons at central rapidity $y_W=0$, but polarized $W$ bosons for $|y_W|>0$, 
since both terms contribute differently to the average net polarization. 
Therefore, the polarization depends on the relative contributions of the $u, \bar u, d$ and $\bar d$ quarks in the initial state. 
The gluon induced production of the $W$ boson becomes relevant at NLO calculations and permits also helicity states with $\lambda = 0$. 
The uncertainty from quark PDFs in the proton is significantly reduced at the Tevatron, as the incoming direction of the interacting anti-quarks is known to higher confidence.

\begin{table}[tb]
\footnotesize
\begin{tabular}{l |clcclcclcc}
\hline
experiment				&	\multicolumn{2}{c|}{\DZero}	&	\multicolumn{2}{c|}{CDF}	&	\multicolumn{2}{c}{ATLAS}	\\
\hline
observable				&	\multicolumn{1}{c|}{$p_T^{\rm lep}$[\MeV]}	&	\multicolumn{1}{c|}{$m_T$ [\MeV]}	&	\multicolumn{1}{c|}{$p_T^{\rm lep}$[\MeV]}	&	\multicolumn{1}{c|}{$m_T$ [\MeV]}	&	\multicolumn{1}{c|}{$p_T^{\rm lep}$[\MeV]}	&	\multicolumn{1}{c}{$m_T$ [\MeV]}	\\
\hline
$M_W$					&	\multicolumn{2}{c|}{80367}	&	\multicolumn{1}{c|}{80390}	&	\multicolumn{1}{c|}{80366}	&	\multicolumn{1}{c|}{80376}	&	\multicolumn{1}{c}{80370	}\\
\hline
stat.\ unc.					&	13	&	14	&	12	&	14	&	10	&	7	\\
syst.\ unc. 					&	18	&	20	&	12	&	11	&	20	&	11	\\
model unc. 				&	13	&	14	&	11	&	13	&	14	&	13	\\
\hline
total unc.   				&	26	&	28	&	20	&	22	&	25	&	19	\\
\hline
lepton calibration unc.			&	17	&	18	&	7	&	7	&	10	&	9	\\
hadronic calibration unc.			&	5	&	6	&	9	&	8	&	15	&	3	\\
other exp.\ unc.			&	1	&	2	&	3	&	3	&	8	&	5	\\
PDF						&	11	&	11	&	10	&	9	&	10	&	8	\\
QED effects				&	7	&	7	&	4	&	4	&	3	&	6	\\
$p_T(W)$ modelling			&	2	&	5	&	3	&	9	&	10	&	9	\\
\hline
reference				&	\multicolumn{2}{c}{\cite{Abazov:2012bv}}	&	\multicolumn{2}{c}{\cite{Aaltonen:2012bp}}		&	\multicolumn{2}{c}{\cite{Aaboud:2017svj}}	\\
\hline
final result of collaboration	&	\multicolumn{2}{c}{$80375\pm23$}			&	\multicolumn{2}{c}{$80387\pm19$}	&	\multicolumn{2}{c}{$80370\pm19$}		\\
(stat., exp.\ syst., model unc.)	&	\multicolumn{2}{c}{$80375\pm11\pm15\pm13$}			&	\multicolumn{2}{c}{$80387\pm12\pm10\pm12$}	&	\multicolumn{2}{c}{$80370\pm7\pm11\pm14$}		\\
\hline
\end{tabular}
\centering
\caption{\label{tab:WMass}Overview of the most precise determinations of the $W$ boson mass at hadron colliders using the lepton $p_T$ and the transverse mass ($m_T$) distributions. 
The experimental systematic uncertainties are broken down to effects due to the lepton response calibration, the calibration of the hadronic recoil reconstruction 
and further experimental uncertainties such as backgrounds. 
Contributions to modeling uncertainties from PDFs, QED effects, as well as the modeling of $p_T(W)$ are shown separately.}
\end{table}

An additional uncertainty at the LHC arises from the uncertainty of the charm quark PDFs in protons. 
The mass of a $c$-quark is roughly 1~GeV~and affects the kinematics of the $W$ boson production in the $c\bar d$ or $d \bar c$ initial states, leading to a harder $p_T(W)$ spectrum. 
This, in turn, affects the resulting decay lepton $p_T$ distribution and thus the \MW measurements.

\subsubsection{Discussion and prospects of $W$ boson mass measurements at hadron colliders}

The most recent $W$ boson mass measurements with leading precision have been published in 2012 by 
the CDF and \DZero Collaborations at the Tevatron~\cite{Aaltonen:2012bp,Abazov:2012bv}, using 20\% and 50\% of the available data of the Tevatron Run~2 
in proton anti-proton collisions at a center of mass energy of 1.96 TeV, respectively, 
as well as in 2017 by the ATLAS Collaboration~\cite{Aaboud:2017svj} based on the full dataset of proton-proton collisions collected at a center of mass energy of 7~TeV.

While the ATLAS and \DZero collaborations use only $Z$ boson events for the detector response calibration, 
the CDF measurements also makes use of $J/\psi$ events for the calibration of the lepton momentum scale. 
All measurements have been performed in the electron decay channel; the CDF and ATLAS collaboration use the muon decay channel in addition. 
An overview of the experimental uncertainties for these three measurements is given in Table~\ref{tab:WMass}. 
The experimental uncertainties in the lepton calibration are of similar size. 
However, it should be noted that the measurements using the $p_T$ and $m_T$ distributions have a very different impact on the combined measurement. 
While the Tevatron measurements are mainly driven by the $m_T$ distribution, the measurement of ATLAS is driven by the $p_T$ distribution. 
This is due to the significantly smaller pile-up contribution\footnote{Pile-up refers to the number of simultaneous proton-proton collisions within one recorded event, \ie, bunch-crossing.} 
at the Tevatron, which leads to a better resolution of the hadronic recoil, and therefore a higher sensitivity of $m_T$. 
Future improvements in the experimental systematics are expected mainly due to an increase of the calibration samples. 

An overview of all associated physics modeling uncertainties is also given in Table~\ref{tab:WMass}. 
The PDF related uncertainties are dominant for all measurements performed at hadron colliders, but they arise from different origins. 
In particular, the PDF uncertainties could be reduced significantly for the Tevatron experiments, when including also forward leptons in the analyses. 
Currently leptons are restricted to the central detector with a pseudo-rapidity of $|\eta|<1.0$. 
Another significant difference is the treatment of the uncertainties associated to the modeling of $p_T(W)$. 
Here, the \DZero and CDF measurements rely on a prediction based on re-summed calculations provided by the \textsc{\Resbos} event generator~\cite{Ladinsky:1993zn, Balazs:1997xd} 
at next-to-next-to leading order, which was tuned to the measured transverse momentum of the $Z$ boson at the Tevatron collider. 
This approach was not adapted by the ATLAS Collaboration, as the currently available event generators based on re-summation techniques 
or next-to-leading order generators with parton shower approaches fail to describe the observed ratio of $p_T(W)$ over $p_T(Z)$ at the LHC. 
One explanation could be an insufficient description of heavy quark mass effects, which are important at the LHC but play only a minor role at the Tevatron. 
Hence, ATLAS chose to use a pure leading-order parton shower modeling based on \textsc{\Pythia8}~\cite{Sjostrand:2007gs}, which incorporates heavy quark mass effects. 
The corresponding parton shower model was tuned to $p_T(Z)$ data and transferred within assigned systematic uncertainties to the $W$ boson production.

When combining the most recent measurements at hadron colliders, 
it is fair to assume that there are no correlations between statistical and experimental systematic uncertainties between the three detectors. 
The situation is more complicated for the physics modeling uncertainties. 
Since different theoretical descriptions of the transverse momentum distribution of $W$ and $Z$ bosons were chosen at the LHC and the Tevatron, 
and also the underlying measurement of the $p_T(Z)$ distribution was different, we assume additionally no correlations for the $p_T(W)$ uncertainties. 
The chosen baseline PDF sets are different at both colliders and also the source of PDF uncertainties on \MW is different as previously discussed. 
Hence, neither full nor zero correlation for the PDF uncertainties is expected. 
A full study of the correlations is beyond the scope of this review article. 
We therefore assume a 0.5~correlation coefficient for the PDF-related uncertainties. 
The final combination is performed using the \Blue\ method~\cite{Lyons:1988rp, Nisius:2014wua} and includes the combined LEP measurement, 
which is assumed to be completely uncorrelated with the measurements at hadron colliders. 
The input measurements as well as their contribution to the final fit are summarized in Table~\ref{tab:WMass}. 
We find a new world average value of
\[\MW = 80380 \pm 13~{\rm MeV}, \]
with a likelihood value\footnote{Assuming no correlations, the central value does not change, but the uncertainty reduces to 12~\MeV.} of ${\rm p} = 0.74$. 
This value is shown in Figure~\ref{fig:WMassOverview}, together with previous measurements.

The ultimate precision in the measurement of the $W$ boson mass at the LHC was previously estimated to be $\Delta M_W = 7$~\MeV~\cite{Besson:2008zs}. 
The data sets collected in the years 2012--2019 will provide sufficient statistics to reduce the detector response related uncertainties to a minimum. 
Several developments of the modeling of the vector boson production, in particular the treatment of heavy flavors 
and the interplay between high-order corrections and re-summation approaches have been triggered by the first measurement of \MW at the LHC. 
Moreover, new precision measurements of vector boson production became available or are foreseen in the near future. 
Those will reduce the corresponding PDF uncertainties. 
The study of a dedicated low pile-up run at the LHC in 2017 might even allow for a direct measurement of $p_T(W)$ and hence reduce the corresponding modeling uncertainties. 
In view of the upcoming measurements of \MW from CMS and potentially also the LHCb collaboration, a final uncertainty in the world average of 7~MeV does not seem unrealistic. 

\subsubsection{The $W$ boson width}

Within the Standard Model, the total decay width of the $W$ boson is predicted to be equal to the sum of the partial widths over three generations of lepton doublets 
and two generations of quark doublets. 
Its partial widths are expressed as,
\begin{equation}
\label{eqnwidth}
\Gamma_{W\rightarrow f\bar f'} = \frac{|M_{f\bar f'}|^2 N_C G_F \MW^3}{6\pi\sqrt{2}} \left[ 1+\delta_f^{\textrm{rad}}(\mt, \MH, \dots)\right],
\end{equation} 
where $N_C = 3 [1 + \as(\MW)/\pi + \cdots ]$ is the color and QCD correction factor \cite{Chang:1981qq} and $M_{f\bar f'}$ corresponds to the CKM matrix elements for quark decay modes, 
while $M_{f\bar f'} = N_C = 1$ for leptonic decays. 
Corrections for the non-vanishing final state fermion masses need to be included, as well.
Electroweak radiative corrections~\cite{Denner:1990tx}, including $\alpha_s$ corrections~\cite{Kara:2013dua}, are given by $\delta_\ell^{\textrm{rad}}\approx -0.34\%$ for leptons 
and $\delta_q^{\textrm{rad}}\approx -0.40\%$ for quarks~\cite{Renton:2008ub}, 
which are small in the Standard Model since a large part of the corrections is absorbed in the measured values of $G_F$ and \MW. 
New particle candidates that couple to the $W$ boson and are lighter than \MW, would therefore open a new decay channel and alter \GammaW. 
One very prominent example are supersymmetric models in which the $W$ boson can decay to the lightest super-partner of the charged gauge bosons 
and the lightest super-partner of the neutral gauge bosons. 
In contrast to the $Z$ boson width, the $W$ boson width is known only to one electroweak loop. 

The total width of the $W$ boson can be measured directly by kinematic fits to the measured decay lepton spectra, 
such as the transverse momentum of the charged lepton decay $p_T$ or the high-mass tail of the transverse mass $m_T$ as was performed 
at CDF and \DZero~\cite{Aaltonen:2007ai,Abazov:2009vs,Aaltonen:2013iut}. 
Alternatively, fits to the invariant mass distributions in the $qqqq$ and $qql\nu$ final states at the LEP experiments~\cite{Schael:2013ita} have been studied. 
A combination of these direct results, based on kinematic measurements, leads to $\GammaW = 2085 \pm 42$~MeV, 
which is currently used as the world average value~\cite{Tanabashi:2018} and did not change since the discovery of the Higgs boson.

\begin{figure*}[tb]
\begin{center}
\begin{minipage}{0.48\textwidth}
\resizebox{1.0\textwidth}{!}{\includegraphics{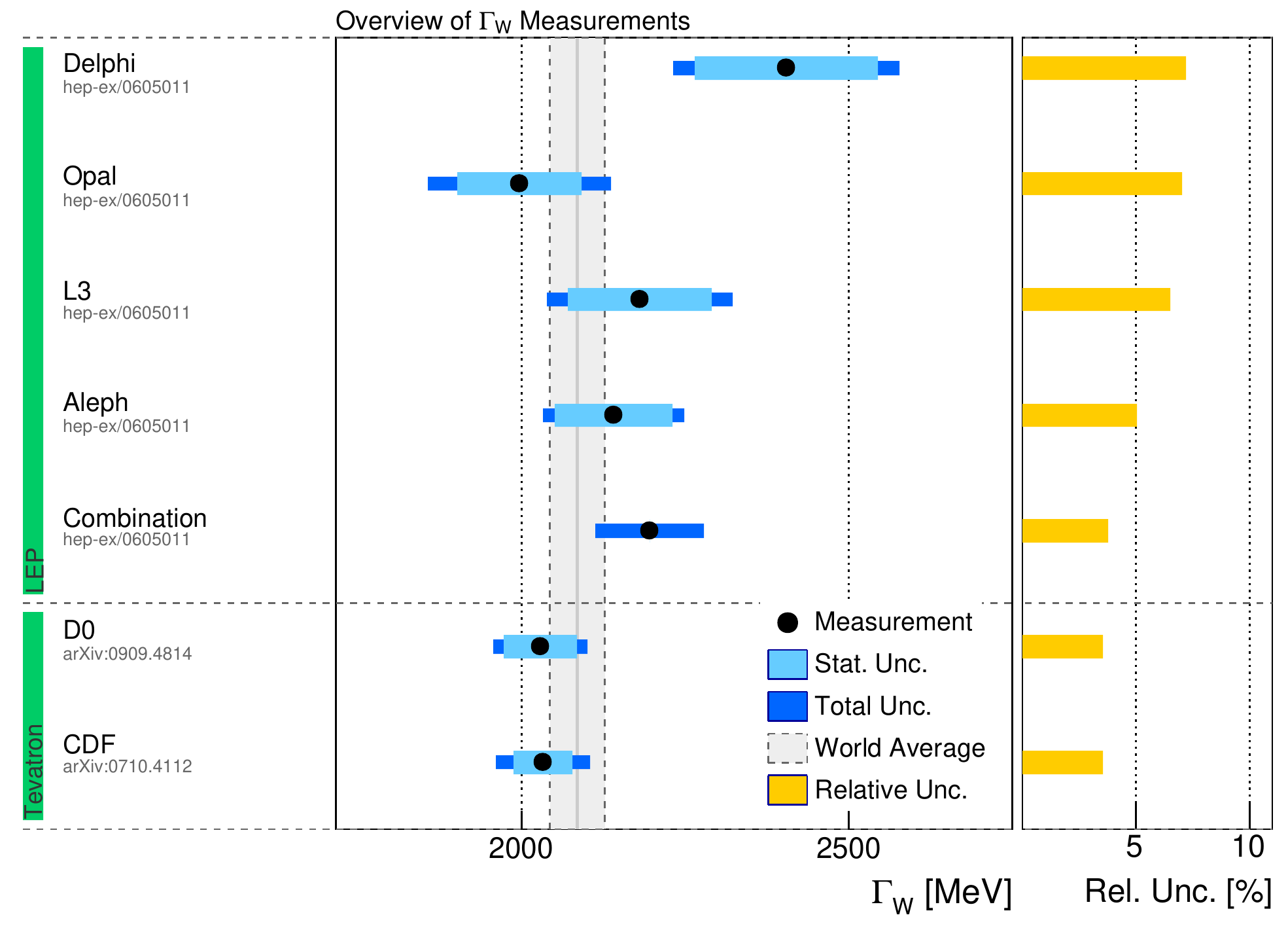}}
\caption{\label{fig:WWidthOverview}Overview of selected measurements of \GammaW, including the most precise measurements from LEP~\cite{Schael:2013ita} 
and the Tevatron~\cite{Aaltonen:2007ai, Abazov:2009vs, Aaltonen:2013iut}.}
\end{minipage}
\hspace{0.3cm}
\begin{minipage}{0.48\textwidth}
\resizebox{1.0\textwidth}{!}{\includegraphics{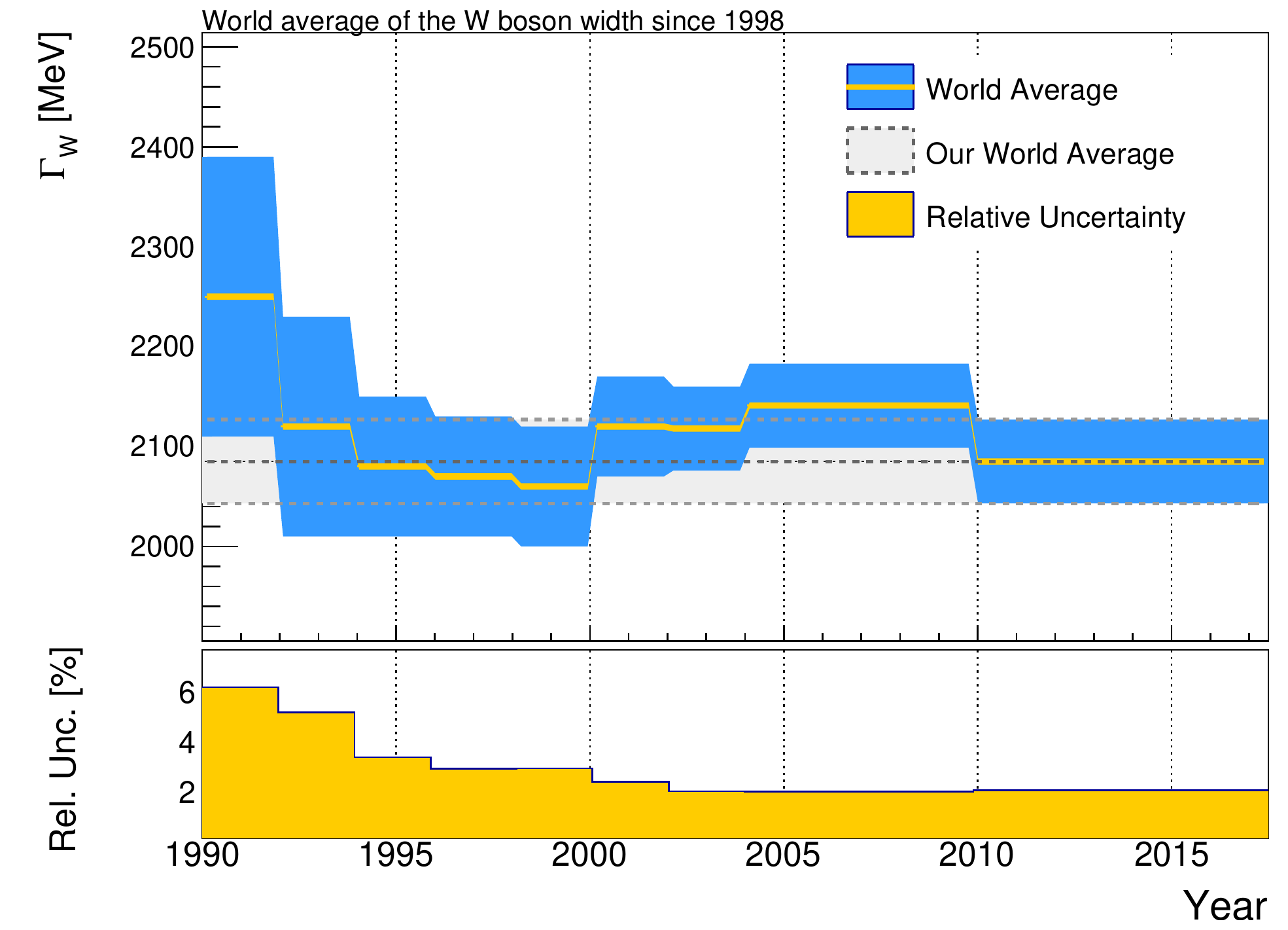}}
\caption{\label{fig:WWidthEvolution}Evolution of the World Average of \GammaW and its uncertainties {\em vs.\/}\ time. 
Values are taken from previous editions of the PDG review~\cite{Olive:2016xmw}.}
\end{minipage}
\end{center}
\end{figure*}

An independent determination of the $W$ boson width is based on the measurement of the ratio of cross sections of $W$ and $Z$ boson production in hadron collisions, \ie,
\begin{equation}
R = \frac{\sigma(pp' \rightarrow W^\pm + X) \textrm{BR}(W\rightarrow \ell\nu)}{ \sigma(pp' \rightarrow Z^0 + X) \textrm{BR}(Z\rightarrow \ell^+\ell^-)}\ ,
\end{equation}
where $\textrm{BR}(V\rightarrow \ell\ell') = \Gamma_{V\rightarrow\ell\ell'} / \Gamma_V$ denotes the leptonic branching ratio of the vector boson ($V=W,Z$) decays. 
The ratio $R$ can be written as
\begin{equation}
\label{eqn:r}
R = \frac{\sigma_W}{\sigma_Z} \frac{\Gamma_{W\rightarrow \ell\nu}}{\GammaW} \frac{\GammaZ}{\Gamma_{Z\rightarrow \ell\ell}}\ ,
\end{equation}
where the total production cross section ratio $\sigma_W/\sigma_Z$ is known theoretically to high accuracy~\cite{Catani:2009sm}. 
The ratio $\sigma_{Z\rightarrow \ell\ell} / \sigma_{Z}$ was measured precisely by the LEP experiments and thus the leptonic branching ratio of the $W$ boson, 
$ \textrm{BR}(W^\pm \rightarrow \ell^\pm \nu)$, can be inferred from the measurement of $R$. 
The advantage of extracting $\textrm{BR}(W^\pm\rightarrow \ell^\pm \nu)$ from the cross section ratio $R$ lies in the fact 
that many experimental uncertainties approximately cancel in the ratio, such as the uncertainty on the integrated luminosity. 
The leptonic width of the $W$ boson in the SM can be predicted by Equation~(\ref{eqnwidth})
and is given by $\Gamma(W\rightarrow \ell\nu) = 226.32 \pm 0.04$~MeV~\cite{Tanabashi:2018} using updated values of \MW, $G_F$ and $\as(\MW)$. 
The dominant uncertainty is due to the accuracy of \MW. 
Using this value, the total width of the $W$ boson can be extracted by a measurement of the leptonic branching ratio. 
This approach for the determination of the $W$-boson width has already been pursued by several experiments, in particular CDF~\cite{CDFWZ}, \DZero~\cite{D0WZ}, 
and CMS~\cite{CMSWZ2}, leading to measurements of \GammaW with an accuracy comparable to the current world average. 
A combination of these indirect measurements was presented in Ref.~\cite{Camarda:2016twt} and yields a value of $\GammaW = 2113 \pm31$~MeV. 
It should be noted that the indirect determination of \GammaW\ {\em via\/} Equation~(\ref{eqn:r}) assumes the Standard Model branching ratio. 
However, possible loop corrections arising from contributions of new physics to the $W$ boson width 
could alter the term $\delta^{\textrm{rad}}$ in Equation~(\ref{eqnwidth}) independently of the decay channel. 
Hence, the branching ratio is insensitive to effects that could appear in the corresponding loop correction terms and the value of \GammaW
resulting from the cross section measurements should not be used in the context of the global electroweak fit, as it is based on assuming the Standard Model relations.

The indirect determination {\em via\/} the electroweak fit yields the value $\GammaW = 2089.5 \pm 0.6$~\MeV~\cite{Tanabashi:2018}, 
which is in good agreement with the world average. 
Equation~(\ref{eqnwidth}) shows that \GammaW depends, among other SM parameters, on \MW, $\as$, and $\MH$. 
However, the small uncertainties in the determination of $\GammaW$ indicate that the sensitivity of \GammaW to these parameters of the SM is rather weak. 
An overview of the most precise measurements of $\Gamma_W$ 
and the evolution of the world average is shown in Figures~\ref{fig:WWidthOverview} and~\ref{fig:WWidthEvolution}, respectively.

\subsection{The $Z$ Boson and its lineshape\label{sec:zboson}}

\subsubsection{Lineshape measurements at LEP and SLC}

Even though the precision measurements of the $Z$ boson line shape has not changed after the discovery of the Higgs boson, 
we will briefly introduce the basic measurement principles and limitations due to their enormous importance for electroweak precision tests of the Standard Model. 
Detailed reviews can be found for example in~\cite{ALEPH:2005ab} and~\cite{Tanabashi:2018}. 

\begin{table}[tp]
\footnotesize
\begin{center}
\begin{tabular}{l | c| c | cc | cc}
\hline
data-taking period 	& $\sqrt{s}\,[\GeV]$	& number of $Z/\gamma^*\rightarrow q\bar q$ events & number of $Z/\gamma^*\rightarrow l\bar l$ events		\\
\hline
1990--1991		& 88.2--94.2 (7 points) & $1.660\times 10^{6}$		& $1.86\times 10^{5}$	\\
1992    			& 91.3			& $2.741\times 10^{6}$		& $2.94\times 10^{5}$	\\
1993				& 89.4, 91.2, 93.0	& $2.607\times 10^{6}$		& $2.96\times 10^{5}$	\\
1994				& 91.2			& $5.910\times 10^{6}$		& $6.57\times 10^{5}$	\\
1995				& 89.4, 91.3, 93.0	& $2.579\times 10^{6}$		& $2.91\times 10^{5}$	\\
\hline
\end{tabular}
\caption{Overview of selected $Z$ boson events at LEP at different energies. \label{tab:LEPZBosonSelected}}
\end{center}
\end{table}

The measurements of the $Z$ boson line shape have been performed by all LEP experiments, \Aleph, \Delphi, \LThree and \Opal, 
as well as at the Standford Linear Collider (SLC) by the SLD collaboration. 
We will focus on the former for the discussion of most observables connected to the $Z$ boson line shape, since those yield significantly smaller uncertainties.

The cross section for the process $e^+ e^- \rightarrow Z \rightarrow f \bar f$ can be written in lowest order as,
\begin{equation}
\label{eqn:Zlineshape}
\sigma(e^+ e^- \rightarrow Z \rightarrow f \bar f) = \frac{12\pi}{\MZ^2} \frac{s \Gamma_{ee} \Gamma_{f\bar{f}}}{(s-\MZ^2)^2 + s \Gamma _Z^2\MZ^{-2}}\ ,
\end{equation}
where $\sqrt{s}$ is the center of mass energy of the colliding $e^+e^-$ pairs, \MZ is the mass of the $Z$ boson, 
$\Gamma _Z$ its total decay width and $\Gamma_{f\bar{f}}$ the corresponding partial decay width for fermions in the final state of type $f$. 
The maximum is reached near $\sqrt{s}=\MZ$, and the peak cross section is defined as
\begin{equation}
\sigma_{f\bar f}^0 = \frac{12\pi}{\MZ^2} \frac{\Gamma_{ee}\Gamma_{ff}}{\GammaZ^2}\ .
\end{equation}
To lowest order, the total decay width \GammaZ can be expressed as the sum of partial widths,
\begin{equation}
\Gamma _Z = \Gamma_{ee} + \Gamma_{\mu\mu} + \Gamma_{\tau\tau}  + \sum_{q=u,d,s,c,b} \Gamma_{q\bar q} + \sum_{i=1,2,3} \Gamma_{\nu_{i}\bar \nu_{i}}\ .
\end{equation}
Therefore, a measurement of the peak cross section for a given final state provides constraints of both the total decay width and the corresponding partial width.

\begin{figure*}[tb]
\begin{center}
\begin{minipage}{0.25\textwidth}
\resizebox{1.0\textwidth}{!}{\includegraphics{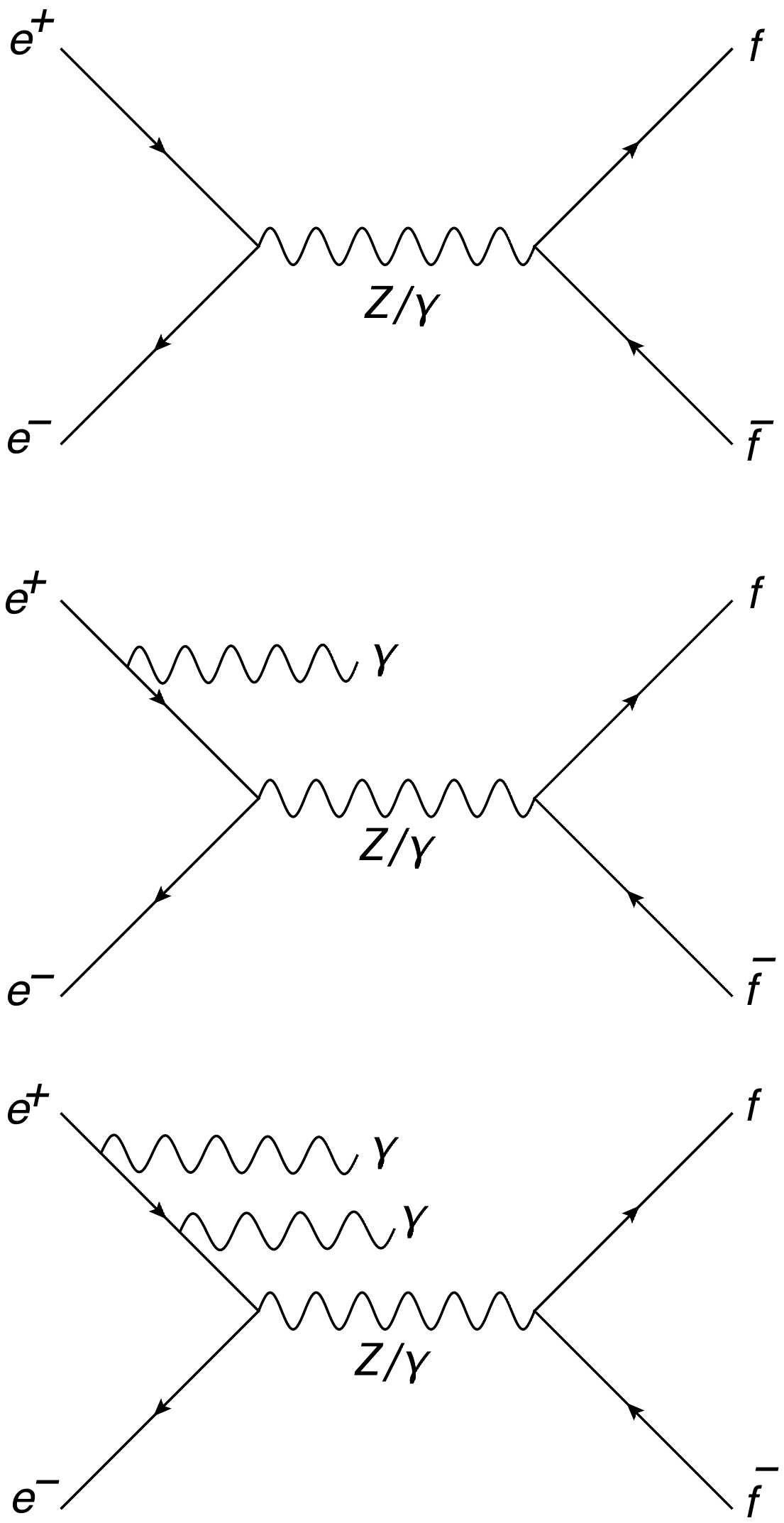}}
\caption{\label{fig:LEPISR}Sample Feynman diagrams of initial state radiation in the $e^+e^-\rightarrow \gamma^*/Z \rightarrow f\bar f$ process, 
that lead to a shift of the available center of mass energy.}
\end{minipage}
\hspace{0.3cm}
\begin{minipage}{0.7\textwidth}
\centering
\resizebox{0.9\textwidth}{!}{\includegraphics{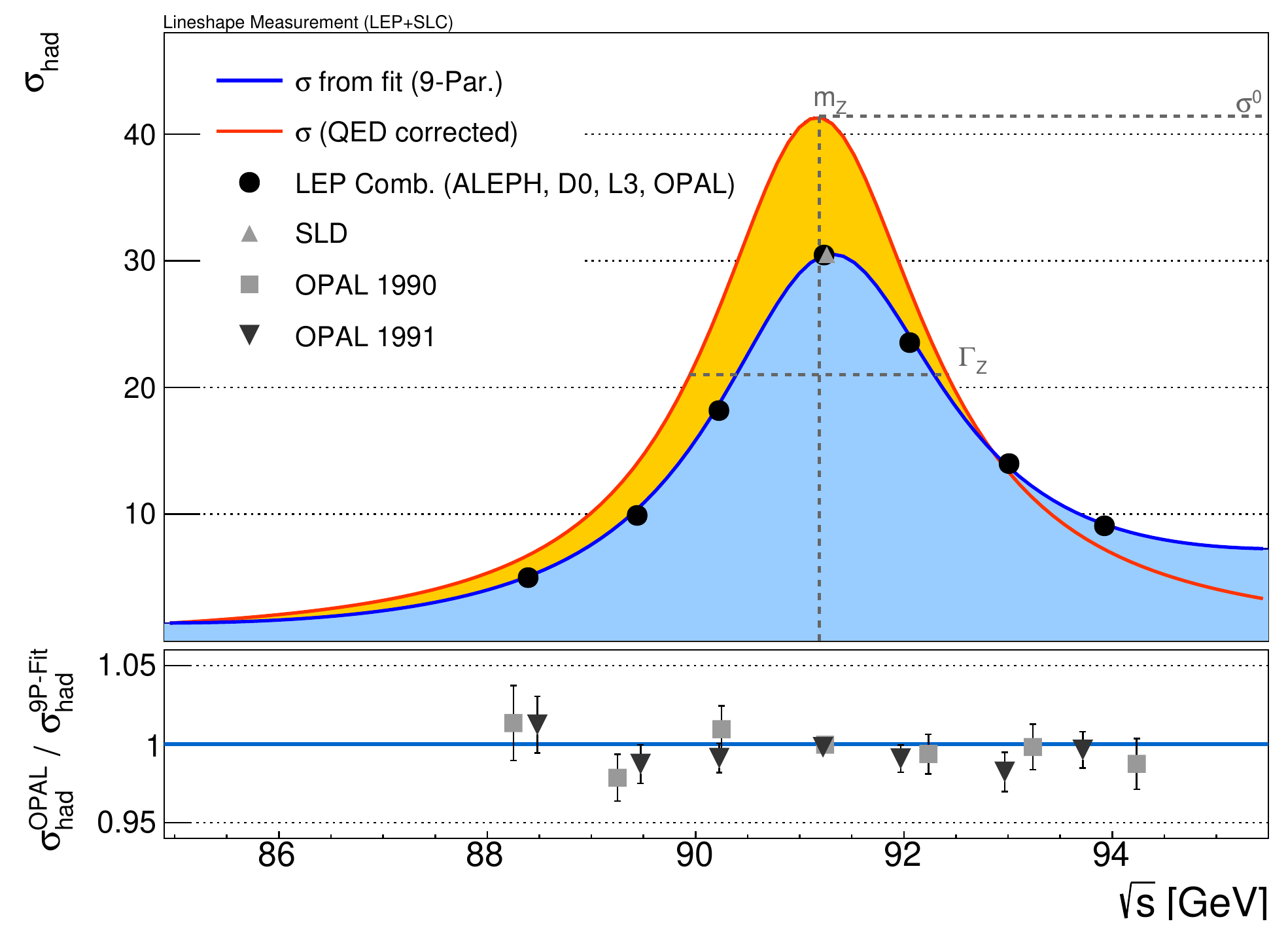}}\vspace*{0.6cm}
\caption{\label{fig:ZLineshape}Measured hadronic cross sections from the combination of the LEP experiments, 
as well as the fitted $Z$ boson lineshape using nine free parameters before and after QED corrections. 
The original version of this plot can be found in Ref.~\cite{ALEPH:2005ab}. 
The measured hadronic cross sections of the \Opal collaboration of the years 1990 and 1991 are shown compared to the fitted line shape. 
The center of mass energy of the SLD measurements is also indicated.}
\end{minipage}
\end{center}
\end{figure*}

The free parameters in Equation~(\ref{eqn:Zlineshape}) can be deduced by fits to relativistic Breit-Wigner distributions with $s$-dependent widths
--- as in Equation~(\ref{eqn:Zlineshape}) --- of the measured cross sections at different center of mass energies around the $Z$ boson mass. 
Experimentally, the cross section measurements at a given $\sqrt{s}$ reduce to counting experiments of different reconstructed final state objects, determined by 
\begin{equation}
\sigma(e^+ e^- \rightarrow Z \rightarrow f \bar f) = \frac{N_{\rm sig} }{C \int L dt}\ ,
\end{equation}
where $N_{\rm sig}$ is the number of selected signal candidates in a specific final state, $\int L dt$ is the integrated luminosity of the analyzed data set, and $C$ accounts for detector and acceptance efficiencies. 
These measurements have been performed in the electron, muon, and tau final states, as well as final states involving particle jets. 
The latter can be experimentally distinguished from jets stemming from light quarks, as well as from $b$- and $c$-quarks. 
All experimental uncertainties, such as reconstruction efficiencies or energy scales of leptons are absorbed into uncertainties in the proportionality factor $C$. 
The integrated luminosity is determined by measurements of small angle Bhabha scattering with associated uncertainties of about $0.06\%$ affecting all experiments. 
It is important to note that the actual energy scale calibration of final state objects has only a negligible impact on the fitting procedure of the line shape.

The four LEP experiments measured cross sections in the different decay modes in three data-taking periods from 1990 to 1992, from 1993 to 1994, 
and in 1995 at different center of mass energies, which are summarized in Table~\ref{tab:LEPZBosonSelected}, along with the selected number of signal events. 
The measured cross sections have to be corrected for QED effects~\cite{Bardin:1990fu} such as initial state radiation (Figure~\ref{fig:LEPISR}), 
which reduce the effective center of mass energy at the $e^+e^-Z$ vertex. 
Moreover, the pure $Z$ boson amplitude interferes with the pure electromagnetic amplitude $e^+ e^- \rightarrow \gamma \rightarrow f \bar f$. 
Since close to the $Z$ boson mass the contribution of this purely electromagnetic amplitude is expected to be at the percent level, 
it was assumed by the LEP collaborations that this interference term follows the prediction of the Standard Model, \ie, it was fixed during the fit, leading to reduced uncertainties. 
More general fits were also performed as consistency checks.
The combination of measured cross sections for the hadronic final states of all LEP experiments,
as well as the resulting fit before and after the QED corrections, is shown in Figure~\ref{fig:ZLineshape}.

\begin{table}[tb]
\footnotesize
\begin{tabular}{l | cc| cc| cc| cc}
\hline
experiment		&	\multicolumn{2}{c|}{ALEPH}	& \multicolumn{2}{c|}{OPAL}& \multicolumn{2}{c|}{DELPHI}		&	\multicolumn{2}{c}{L3}	\\
\hline
observable		&	value	&	unc.		&	value	&	unc.		&	value	&	unc.		&	value	&	unc.	\\
\hline
$M_Z$ [GeV]		&	91.1891	&	0.0031	&	91.1858	&	0.003	&	91.1864	&	0.0028	&	91.1897	&	0.003	\\
$\Gamma_Z$ [GeV]	&	2.4959	&	0.0043	&	2.4948	&	0.0041	&	2.4876	&	0.0041	&	2.5025	&	0.0041	\\
$\sigma_0$ [nb]	&	41.558	&	0.057	&	41.501	&	0.055	&	41.578	&	0.069	&	41.535	&	0.054	\\
\hline
$R_e$			&	20.69	&	0.075	&	20.901	&	0.084	&	20.88	&	0.12		&	20.815	&	0.089	\\
$R_\mu$			&	20.801	&	0.056	&	20.811	&	0.058	&	20.65	&	0.076	&	20.861	&	0.097	\\
$R_\tau$			&	20.708	&	0.062	&	20.832	&	0.091	&	20.84	&	0.13		&	20.79	&	0.13	\\
\hline
$A_{\rm FB}^{e}$		&	0.0184	&	0.0034	&	0.0089	&	0.0045	&	0.0171	&	0.0049	&	0.0107	&	0.0058	\\
$A_{\rm FB}^{\mu}$	&	0.0172	&	0.0024	&	0.0159	&	0.0023	&	0.0165	&	0.0025	&	0.0188	&	0.0033	\\
$A_{\rm FB}^{\tau}$	&	0.017	&	0.0028	&	0.0145	&	0.003	&	0.0241	&	0.0037	&	0.026	&	0.0047	\\
\hline
$R_l$			&	20.729	&	0.039	&	20.823	&	0.044	&	20.73	&	0.06		&	20.809	&	0.06	\\
$A_{\rm FB}^{l}$	&	0.0173	&	0.0016	&	0.0146	&	0.0017	&	0.0187	&	0.0019	&	0.0192	&	0.0024	\\
\hline
$R_b$			&	0.2159	&	0.2178	&	0.2178	&	0.0016	&	0.2163	&	0.0009	&	0.2174	&	0.0027	\\
\hline
\end{tabular}
\centering
\caption{\label{tab:ZExp}Overview of the results of the $Z$ boson line shape measurements and some related observables at LEP~\cite{ALEPH:2005ab}. 
Lepton universality has only been assumed for $R_l$ and $A_{FB}^{l}$. }
\end{table}

Each experiment performed a combined fit of all measured cross sections. 
Using the partial and total widths directly as free fitting parameters would lead to a highly correlated parameter set, 
since the cross sections themselves depend on products of these parameters. 
Moreover, the peak cross sections $\sigma_0$ for each fermion species would have the same statistical and systematic uncertainties of the luminosity determination. 
It was therefore chosen to use instead the ratios of peak cross sections for the leptonic final states relative to that of hadrons,
\begin{equation}
R_l = \frac{\sigma^0_{\rm had}}{\sigma^0_{l^{+}l^{-}}} = \frac{\Gamma_{\rm had}}{\Gamma_{l^{+}l^{-}}}\ ,
\end{equation}
where $l^\pm = e^\pm, \mu^\pm$ or $\tau^-$.
Historically, all jet final states have been combined into a common hadronic width parameter $\Gamma_{\rm had}$. 
Once final states originating from $b$ and $c$ quarks could be experimentally distinguished, two new associated ratios have been introduced, namely
\begin{equation}
\label{Rbc}
R_b =  \frac{\Gamma_{b\bar{b}}}{\Gamma_{\rm had}}\ , \qquad \qquad \qquad \qquad \qquad R_c = \frac {\Gamma_{c\bar{c}}} {\Gamma_{\rm had}}\ ,
\end{equation}
which were treated as independent parameters during the fitting procedure\footnote{In fact, the LEP and SLC experiments did not measure cross-sections with flavour tagging, instead measured the b- and c-rates in a fiducial reason normalised to the hadronic rate on the peak.}. While SLC could not significantly contribute to the lineshape measurement itself, it reached competitive measurements of $R_b$ and $R_c$,
as well as of the asymmetry parameters (see Section~\ref{sec:sineff}). due to its small beam pipe and the low bunch crossing frequency.

The most general fit performed by the LEP collaborations to the measured cross sections, depends (before quark flavor tagging) on nine parameters, given by 
\MZ, $\Gamma _Z$, $\sigma^0_{\rm had}$, $R_e$, $R_\mu$ and $R_\tau$, as well as the three forward-backward asymmetries, 
$A_{FB}^e$, $A_{FB}^\mu$, and $A_{FB}^\tau$, which are discussed in detail in Section~\ref{sec:sineff}. 
Assuming lepton universality, the fit can be reduced to five free parameters including the quantities $R_l$ and $A_{FB}^l$,
provided a correction for the final state $\tau$ mass phase space effect of $\delta_\tau = -0.23\%$ in $\Gamma_{\tau^+\tau^-}$ has been taken into account.
Since the $R_l$ are determined by the fermionic couplings to the $Z$ boson and are therefore sensitive to the electroweak mixing angle, 
we will discuss their experimental status together with further $\sintheta$ sensitive observables in Section~\ref{sec:sineff},
and focus in the following only on \MZ, $\Gamma _Z$ and $\sigma^0_{\rm had}$.

\subsubsection{$Z$ boson mass, width and pole cross section}

The $Z$ boson mass, width and pole cross section are directly determined by the combined fit to the measured cross sections, as illustrated in Figure~\ref{fig:ZLineshape}. 
The uncertainties in the fit parameters depend on the cross section uncertainties themselves as well as in the beam energy uncertainties. 
The luminosity uncertainties in the cross sections range from 0.03\% for the \Opal measurements to 0.09\% for \Delphi. 
The further experimental uncertainties depend largely on the final state, where the highest relative precision of $0.04\%$ is reached by \LThree for the hadronic final state 
and the largest uncertainty of $0.6\%$ is seen in the $\tau$ decay channel by \Delphi. 

However, the dominant uncertainties stem from the limited knowledge of the center of mass energy, hence the energy scale at which the measurements have been performed. 
An absolute uncertainty in the energy scale impacts the position of the measurements on the x-axis in Figure~\ref{fig:ZLineshape}, \ie, translates directly to the fitted value of \MZ, 
while relative differences in the energy scales between measurement points affect $\Gamma _Z$. 

The precise calibration of the LEP energy has been performed {\em via\/} the resonant depolarization technique constantly outside normal data-taking~\cite{Arnaudon:1994zq}.
Sources of systematic errors related to the LEP energy measurement are the non-linear response of the magnets to currents, 
interdependencies of the dipole fields and the beam energy, as well as further unknown effects such as tidal effects or temperatures. 
While the LEP energy uncertainties are largely uncorrelated between different data-taking periods, 
they are nearly fully correlated between the experiments and dominate the combined uncertainty in \MZ. 
Further uncertainties that are common between the experiments are theoretical uncertainties in the luminosity determination {\em via\/} Bhabha scattering 
or in the calculation of QED radiative effects. 

\begin{figure*}[t!]
\begin{center}
\begin{minipage}{0.48\textwidth}
\resizebox{1.0\textwidth}{!}{\includegraphics{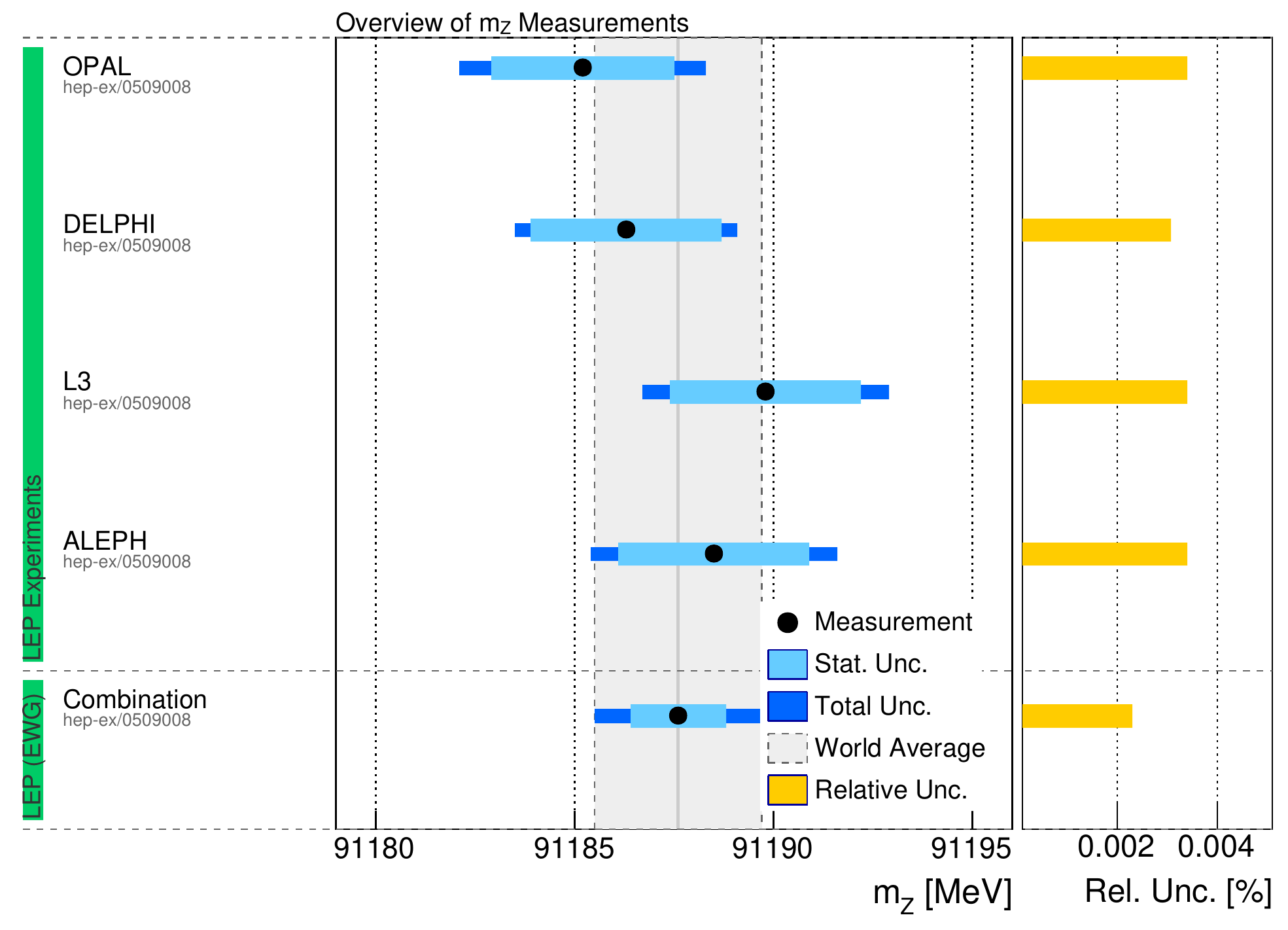}}
\caption{\label{fig:ZMassOverview}Overview of the measurements of \MZ by the LEP collaborations (\Aleph, \LThree, \Delphi, \Opal) as well as their combination.}
\end{minipage}
\hspace{0.3cm}
\begin{minipage}{0.48\textwidth}
\resizebox{1.0\textwidth}{!}{\includegraphics{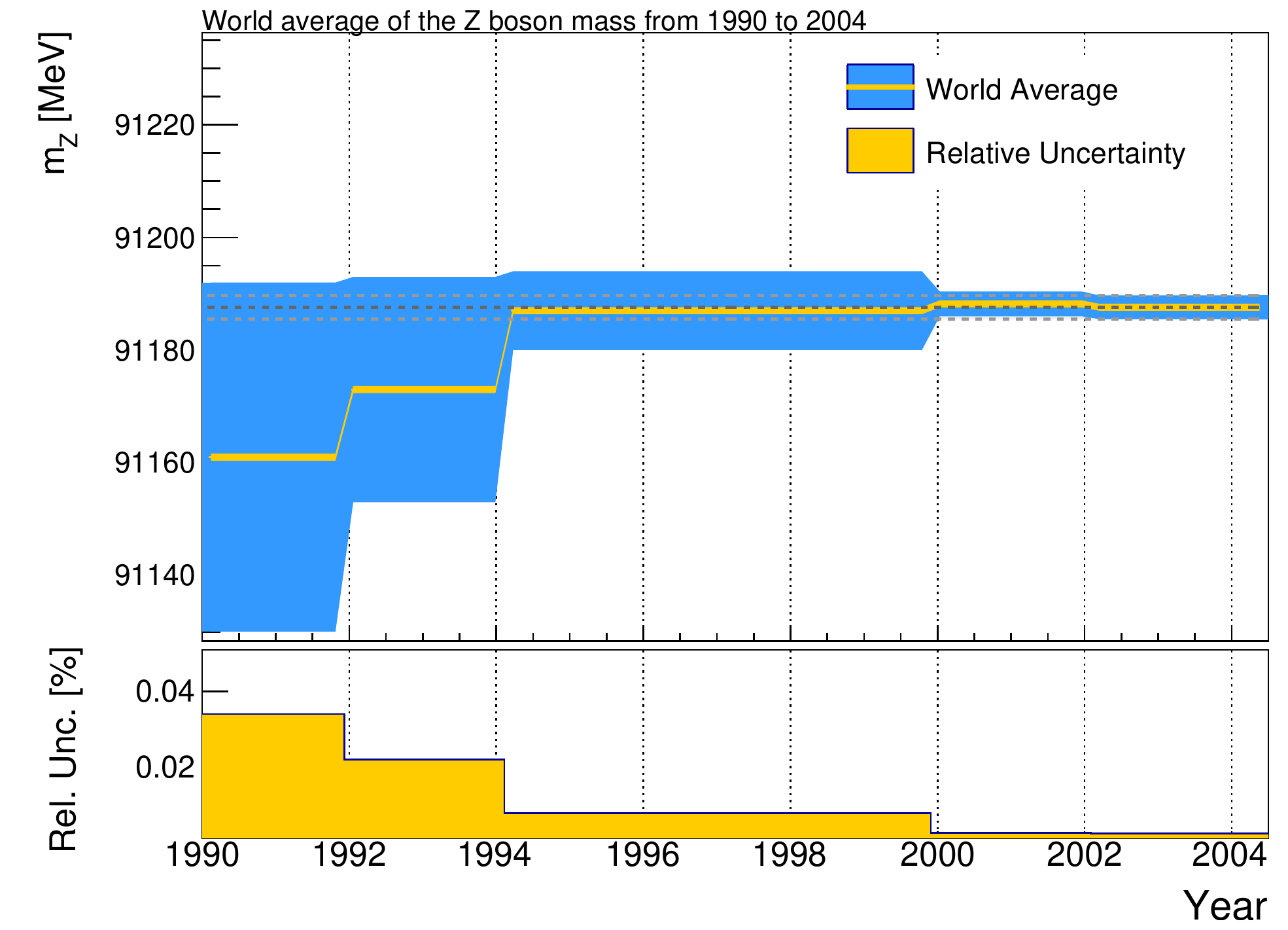}}
\caption{\label{fig:ZMassEvolution}Evolution of the world average of \MZ and its uncertainties {\em vs.\/}\ time. 
Values are taken from previous editions of the PDG review~\cite{Olive:2016xmw}.}
\end{minipage}
\end{center}
\end{figure*}

\begin{figure*}[t]
\begin{center}
\begin{minipage}{0.48\textwidth}
\resizebox{1.0\textwidth}{!}{\includegraphics{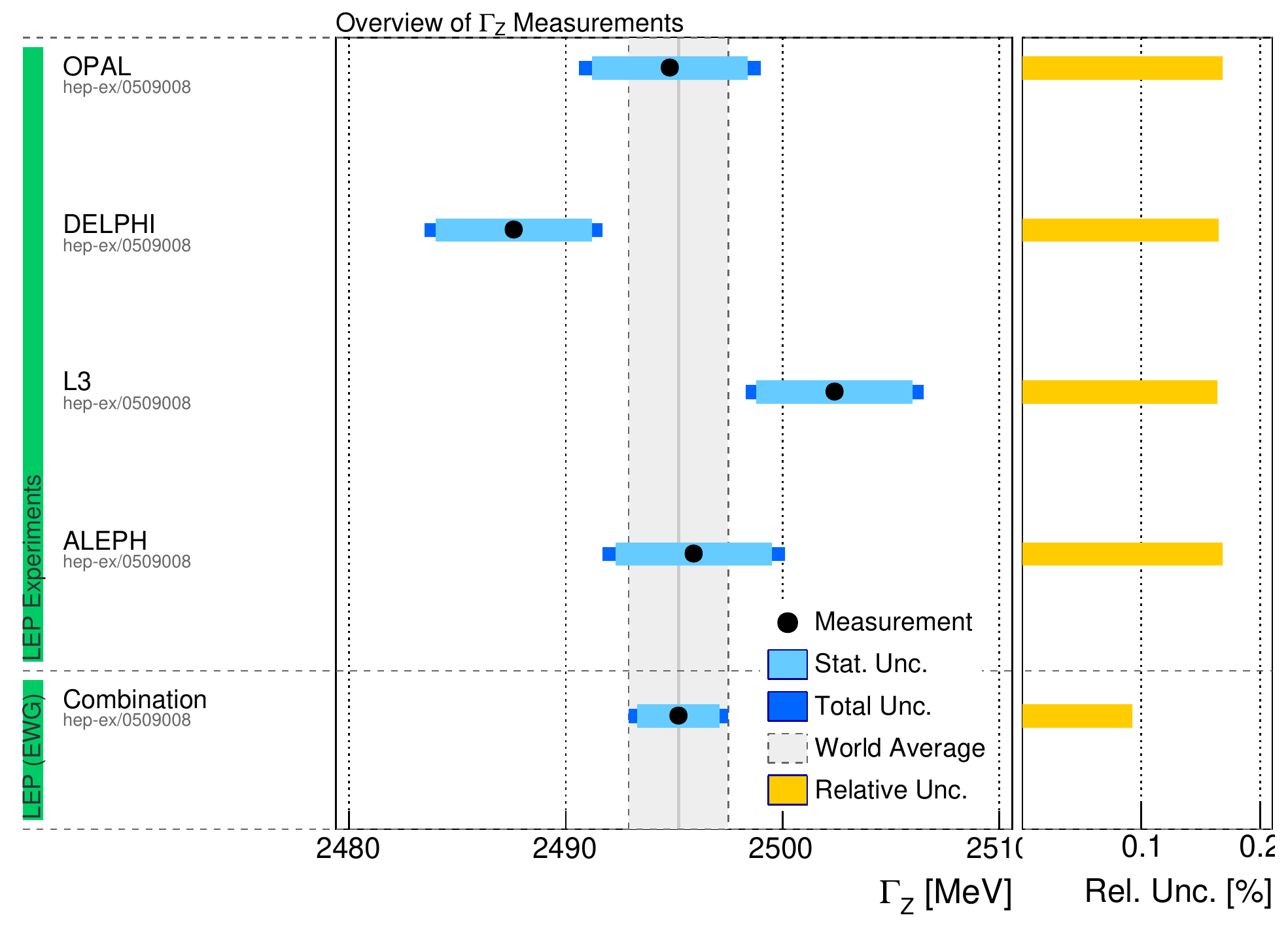}}
\caption{\label{fig:ZWidthOverview}Overview of the measurements of \GammaZ by the LEP collaborations (\Aleph, \LThree, \Delphi, \Opal) as well as their combination.}
\end{minipage}
\hspace{0.3cm}
\begin{minipage}{0.48\textwidth}
\resizebox{1.0\textwidth}{!}{\includegraphics{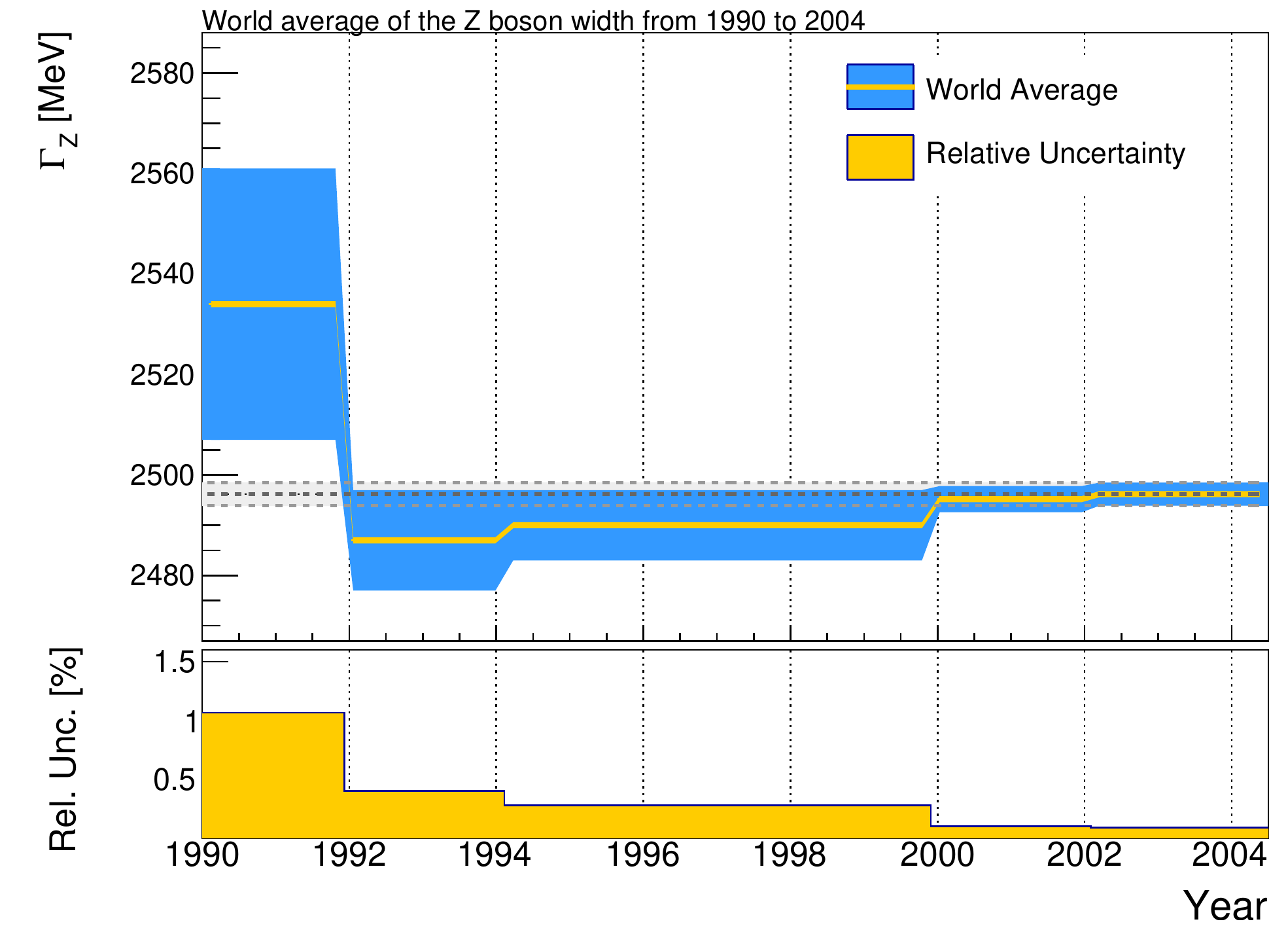}}
\caption{\label{fig:ZWidthEvolution}Evolution of the world average of \GammaZ and its uncertainties {\em vs.\/}\. time. 
Values are taken from previous editions of the PDG review~\cite{Olive:2016xmw}.}
\end{minipage}
\end{center}
\end{figure*}

The individual measurements of \MZ, \GammaZ and $\sigma_{0}^{\rm had}$ and their combined values are summarized 
in Tables~\ref{tab:ZExp} and~\ref{tab:ZLEP}, respectively, and illustrated for \MZ and \GammaZ in Figures~\ref{fig:ZMassOverview} and~\ref{fig:ZWidthOverview}. 
Statistical uncertainties are relevant for the individual measurements, but are subdominant for the combined values. 
Systematic experimental uncertainties except of the LEP energy scale calibration are of minor importance for the measurement of \MZ and $\Gamma _Z$. While the results of \MZ and $\Gamma _Z$ are largely uncorrelated, there is a significant correlation between $\Gamma _Z$ and $\sigma^0_{\rm had}$. 

The evolution of the experimental precision of \MZ and \GammaZ over the years is shown in Figures~\ref{fig:ZMassEvolution} and~\ref{fig:ZWidthEvolution}, respectively. 
With the analysis of the data collected during the energy scans in 1993 and 1995, the final precision was reached 
and did not significantly change after the final publication of the combined fit of the LEP Electroweak Working Group in 2006~\cite{ALEPH:2005ab}. 

\begin{table}[tb]
\footnotesize
\begin{tabular}{l | cc| cc cc| cc}
\hline
observable		&	value	&	total	unc.	&	stat.\ unc.		&	beam unc. & further syst.\ unc. & theo.\ unc. & \multicolumn{2}{c}{SM expectation ({\em via.\/}\ EW fit)}	\\
\hline
$M_Z$ [GeV]		&	91.1876	&	0.0021	&	0.0012	&	0.0018	&	0.0000	&	0.0004	&	\multicolumn{2}{c}{$91.1874\pm0.0004$}	\\
$\Gamma_Z$ [GeV]	&	2.4952	&	0.0023	&	0.0019	&	0.0013	&	0.0002	&	0.0004	&	\multicolumn{2}{c}{$2.4956\pm0.019$}	\\
$\sigma_0$ [nb]	&	41.541	&	0.037	&	0.014	&	0.011	&	0.031	&	0.009	&	\multicolumn{2}{c}{$41.476\pm0.015$}	\\
\hline
$R_e$			&	20.804	&	0.05		&	0.03		&	0.01		&	0.03		&	0.02		&	\multicolumn{2}{c}{$20.744\pm0.019$}	\\
$R_\mu$			& 	20.785	&	0.033	&	0.026	&	0.003	&	0.020	&	0.004	&	\multicolumn{2}{c}{$20.745\pm0.019$}	\\
$R_\tau$			& 	20.764	&	0.045	&	0.027	&	0.003	&	0.036	&	0.004	&	\multicolumn{2}{c}{$20.792\pm0.019$}	\\
\hline
$A_{FB}^{e}$		&	0.0145	&	0.0025	&	0.0017	&	0.0003	&	0.0016	&	0.0007	&	\multicolumn{2}{c}{$0.01627\pm0.0003$}	\\
$A_{FB}^{\mu}$	&	0.0169	&	0.0013	&	0.0011	&	0.0003	&	0.0006	&	0.0001	&	\multicolumn{2}{c}{$0.01627\pm0.0003$}	\\
$A_{FB}^{\tau}$	&	0.0188	&	0.0017	&	0.0013	&	0.00025	&	0.0011	&	0.0001	&	\multicolumn{2}{c}{$0.01627\pm0.0003$}	\\
\hline
\end{tabular}
\centering
\caption{\label{tab:ZLEP}Combination by the LEP electroweak working group of different $Z$ boson line shape and related observables~\cite{ALEPH:2005ab},
including a breakdown of statistical und systematic uncertainties. 
The breakdown was not published in Ref.~\cite{ALEPH:2005ab} but derived within this work using results 
of the ALEPH~\cite{Barate:1999ce} and OPAL~\cite{Akers:1993is} collaborations. 
Therefore these values should be considered as estimates only.}
\end{table}

\subsection{The weak mixing angle\label{sec:sineff}}

As discussed in Section~\ref{sec:theoSin}, the effective weak mixing angle can be determined by observables 
which are sensitive to the ratio of the fermionic vector and axial-vector couplings to the $Z$ boson. 
We turn our attention here to the differential cross section of the $e^+ e^- \rightarrow f \bar f$ annihilation process.
The case $f = e$ has an additional $t$-channel contribution and correspondingly larger uncertainties, but these cross sections have been measured, as well.
The unpolarized cross section at lowest order is given by,
\begin{eqnarray}
\frac{d\sigma}{d\cos\theta} = \frac{N^f_C G_F^2 \MZ^4}{16\pi} \frac{s}{(s-\MZ^2)^2 + s^2 \GammaZ^2 \MZ^{-2}}  
\left[ (g_v^{e2} + g_a^{e2}) (g_v^{f2} + g_a^{f2}) (1+\cos^2 \theta) + 2 g_v^e g_a^e g_v^f g_a^f \cos\theta \right],
\end{eqnarray}
and can be schematically written as
\begin{equation} 
\frac{d\sigma}{d\cos\theta} = \kappa [A(1+\cos^2 \theta) + B \cos\theta],
\end{equation}
\noindent where $\cos\theta$ is the angle between the incoming and outgoing fermions (or between incoming and outgoing anti-fermions).  
Several things should be noted. 
First of all, the $(1+\cos^2\theta)$ term would also appear in a pure $\gamma$ exchange diagram. 
However, the vector- and axial-vector couplings of the $Z$ boson boson introduce an additional $\cos\theta$ dependence. 
Secondly, the coefficients $A$ and $B$ depend on the electroweak vector and axial-vector couplings. 
Thus, for a given center-of-mass energy the tree level differential cross section depends only on the weak mixing angle $\sintheta$, 
after fixing the electric charges, the weak hypercharges, \MZ, and \GammaZ. 

By defining forward and background cross sections by,
\[\sigma_F = \int_0^1 \frac{d\sigma}{d\cos\theta}\ d\cos\theta, \hspace{0.8cm} \sigma_B = \int_{-1}^0 \frac{d\sigma}{d\cos\theta}\ d\cos\theta,\]
the forward-backward (FB) asymmetry can be defined as,
\begin{equation}
\label{eqn:AFB}
A_{\rm FB}^f \equiv \frac{\sigma_F-\sigma_B}{\sigma_F + \sigma_B}\ ,
\end{equation}
At tree level and ignoring photon exchange and $Z/\gamma^*$ interference, $A^f_{\rm FB}$ is given by the idealized expression,
\begin{equation}
\label{AFB}
A^f_{\rm FB} = 3 \frac{(1-4|Q_e|\seffsf{f})}{1+(1-4|Q_e|\seffsf{f})^2} \frac{(1-4|Q_f|\seffsf{f})}{1+(1-4|Q_f|\seffsf{f})^2}\ .
\end{equation}
Experimentalists remove the photon exchange amplitude and radiative corrections. 
Due to the observed value of $\sintheta \sim 1/4$, it happens that $A_{\rm FB}$ in Equation~(\ref{AFB}) is rather small, and $Z/\gamma^*$ interference effects are relatively enhanced.
This even causes $A_{\rm FB}^f$ to change sign not far from $s = M_Z^2$ in the cases $f = e, \mu, \tau$ where both factors in Equation~(\ref{AFB}) are suppressed.
The sensitivity of $A_{\rm FB}^f$ to $\sintheta$ is then largest at the $Z$ pole, while the interference terms dominate for energies sufficiently smaller or larger than \MZ. 

It is convenient to introduce asymmetry parameters defined by,
\begin{equation}
\label{eqn:asymlep}
A_f \equiv \frac{2 g_v^f g_a^f}{g_a^{f2} + g_v^{f2}} = \frac{1 - 4 |Q_f | \seffsf{f} }{1-4 |Q_f | \seffsf{f}+8 |Q_f|^2 \seffsffour{f}}\ ,
\end{equation}
in terms of which the forward-backward asymmetry takes the simple form,
\begin{equation}
\label{eqn:AFBpol2}
A^f_{\rm FB} = \frac{3}{4} A_e A_f\ .
\end{equation}
$A_e$ is the asymmetry parameter corresponding to the initial $e^+ e^- \rightarrow Z$ vertex and $A_f$ corresponds to the final state $Z \rightarrow f \bar f$ vertex. 
The electroweak mixing angle can then be extracted using Equation~(\ref{eqn:asymlep}). 
It should be noted that $A_b$ is close to one, due to the $Q_b=-1/3$ charge of the bottom quark and consequently $A^{0,b}_{\rm FB}$ is only weakly sensitive to $\seffsf{b}$.

Measurements of $A_{\rm FB}$ close to the $Z$ pole have been performed at lepton and at hadron colliders, yielding relative precisions below 0.1\% and 0.2\% on $\sinleff$, respectively. 
The underlying process at hadron colliders is Drell-Yan production of $Z$ bosons, where the initial $e^+e^-$ state at LEP and SLC is replaced by $u\bar u$ and $d\bar d$.
Less precise measurements of $\sinleff$ have also been performed using measurements of atomic parity violation, 
as well as neutrino and polarized electron scattering experiments on fixed targets. 
While these measurements do not contribute significantly to the world average of the electroweak mixing angle, they can test its energy dependence. 
Therefore, we also briefly summarize these measurements.

Interestingly, the measurement of $A_{\rm FB}$ can also be used to search for new physics. 
While $A_{\rm FB}$ at the $Z$ boson mass is used for the determination of $\sinleff$, large invariant masses are governed by virtual $\gamma^*$ and $Z^*$ amplitudes
with comparable magnitudes. 
A direct search for a new resonance in the electroweak sector {\em via\/} the study of the invariant mass spectra of di-lepton events 
might not show an excess if the new resonance has a large width. 
However, such a new resonance would also interfere with the Standard Model amplitudes and hence introduce a structure in the measured asymmetries $A_{\rm FB}$ near its mass
leading to a measurable effect.

\subsubsection{Low energy observables\label{sec:lowEnergy}}
At lowest energies, the measurements of parity violation in atoms yield a value of the electroweak mixing angle. 
The sensitivity of these experiments lies in the vector couplings of the proton and the axial-vector couplings of the electrons in the atomic shell~\cite{Erler:2013xha}, \ie
\begin{equation}
g_{AV}^{ep} = 2  g_{AV}^{eu} + g_{AV}^{ed} \approx -\frac{1}{2} + 2  \sintheta .
\end{equation}
This directly translates to the weak charge of the atomic nucleus~\cite{Bouchiat:1974zz},
\begin{equation}
Q_W^{Z,N} \approx -2 \left( Z g_{AV}^{ep} + N g_{AV}^{en} \right) \left( 1-\frac{\alpha}{2\pi} \right),
\end{equation}
where $Z$ and $N$ are the numbers of protons and neutrons of the atom, respectively. 
The nuclear weak charge $Q_W^{Z,N}$ impacts the ratio of the parity violating amplitude and the Stark vector polarizability which can be measured. 
These measurements have been performed for several atoms such as cesium~\cite{Wood:1997zq,Guena:2004sq} 
and yield uncertainties in $\sintheta$ at the 0.4\% level for momentum transfers of order a few MeV. 
However, the interpretation of these measurements needs sophisticated atomic structure calculations which introduce additional theory errors of similar size~\cite{Ginges:2003qt}.
On the other hand, most of these uncertainties cancel when determining atomic parity violation in isotope ratios. 
Very recently, the first series of measurements of this type was achieved in a chain of ytterbium isotopes~\cite{Budker:2018}.

A second approach to measure $\sintheta$ at low energies is based on the scattering of a left- and right-handed polarized electron beam on a deuteron target, 
\ie, the reaction $e_{L,R} N \rightarrow e X$. 
Here, the measurement of the left-right asymmetry of the cross section,
\begin{equation}
\label{eqn:APV}
A_{\rm LR} = \frac{\sigma_L - \sigma_R}{\sigma_L + \sigma_R}\ ,
\end{equation}
provides an observable that is sensitive to the electroweak mixing angle, since the dominant QED cross section is parity conserving and drops out from the numerator.
These measurements have been performed by several groups and laboratories at different center of mass energies~\cite{Beise:2004py}.
This includes deep inelastic scattering (DIS) at SLAC~\cite{Prescott:1979dh} and at the 6~GeV CEBAF at JLab~\cite{Wang:2014bba},
where the latter reached about 5\% precision in $A_{\rm LR}$.
A new detector, SoLID, to operate at the upgraded CEBAF to12~GeV is projected to reduce the uncertainty to the 0.5\% level~\cite{Souder:2016xcn}.
Most recently, the Qweak collaboration published a measurement using a beam of longitudinally polarized electrons accelerated to 1.16~GeV 
on an unpolarized liquid hydrogen target~\cite{Androic:2018kni}, resulting in a precision of 5\% at a momentum transfer of 157~MeV, \ie, in the elastic regime.
$A_{\rm LR}$ in Equation~(\ref{eqn:APV}) has also been measured in the purely leptonic reaction $e^- e^- \rightarrow e^- e^-$ using the electron arm of the SLC
aimed at liquid hydrogen~\cite{Anthony:2005pm}. 
This measurement was performed by the SLAC--E158 experiment, yielding a value of $\sinleff = 0.2397 \pm 0.0013$ at $Q^2 = 0.026$~GeV$^2$~\cite{Anthony:2005pm}.

Several future experiments are planned to significantly improve the electroweak mixing angle from low energy observables 
and therefore to test the running of the electroweak mixing angle to higher precision than was previously possible. 
Notably, the future measurement of the parity violating asymmetry in the elastic electron proton scattering by the P2 experiment
at the new MESA (Mainz Energy-Recovery Superconducting Accelerator) accelerator, currently built at the University of Mainz, 
is expected to reach a relative precision of $0.13\%$ in the electroweak mixing angle~\cite{Becker:2018ggl}, which is comparable to the $Z$ pole measurements at LEP and the SLC. 
The MESA accelerator provides a high intensity beam with an energy of 155~\MeV with high degree of longitudinal polarization. 
The extraction of \sintheta is performed at very low momentum transfer of $Q^2=0.005$~GeV$^2$, 
leading to a measurement complementary to previous and upcoming measurements around the $Z$ pole, including those at the LHC. 
Similarly, the asymmetry in M\o ller scattering may be determined with a fivefold greater precision compared to SLAC--E158 
by the MOLLER collaboration at JLab~\cite{Benesch:2014bas}.

\begin{figure*}[t]
\begin{center}
\begin{minipage}{0.48\textwidth}
\resizebox{1.0\textwidth}{!}{\includegraphics{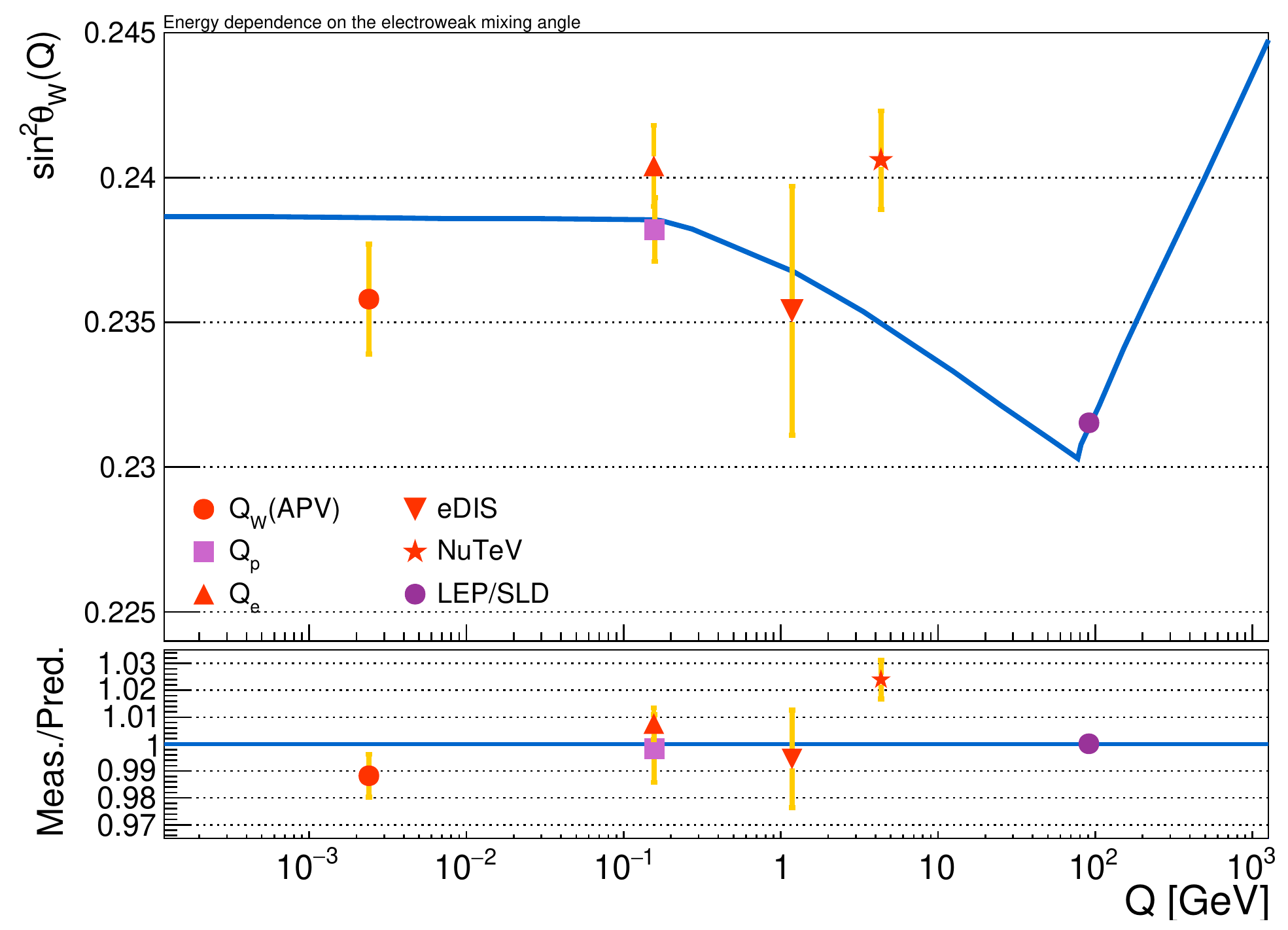}}
\caption{\label{fig:EnergySin2Theta}Running of $\sintheta$ in the $\msbar$ renormalization scheme~\cite{Erler:2017knj,Erler:2004in}, 
together with measurements from low energy experiments including atomic parity violation~\cite{Wood:1997zq,Guena:2004sq}, 
parity violating electron scattering~\cite{Anthony:2005pm,Androic:2018kni}, e-DIS~\cite{Prescott:1979dh,Wang:2014bba}, 
and $\nu$-DIS~\cite{Zeller:2001hh}. All measurements have been translated to the $\msbar$ scheme using the relations from
Refs.~\cite{Degrassi:1996ps,Gambino:1993dd}. Numerically, the difference between the $\msbar$ definition and $\sinleff$
is small and only mildly $m_t$ dependent. \vspace{-0.65cm}}
\end{minipage}
\hspace{0.3cm}
\begin{minipage}{0.48\textwidth}
\resizebox{1.0\textwidth}{!}{\includegraphics{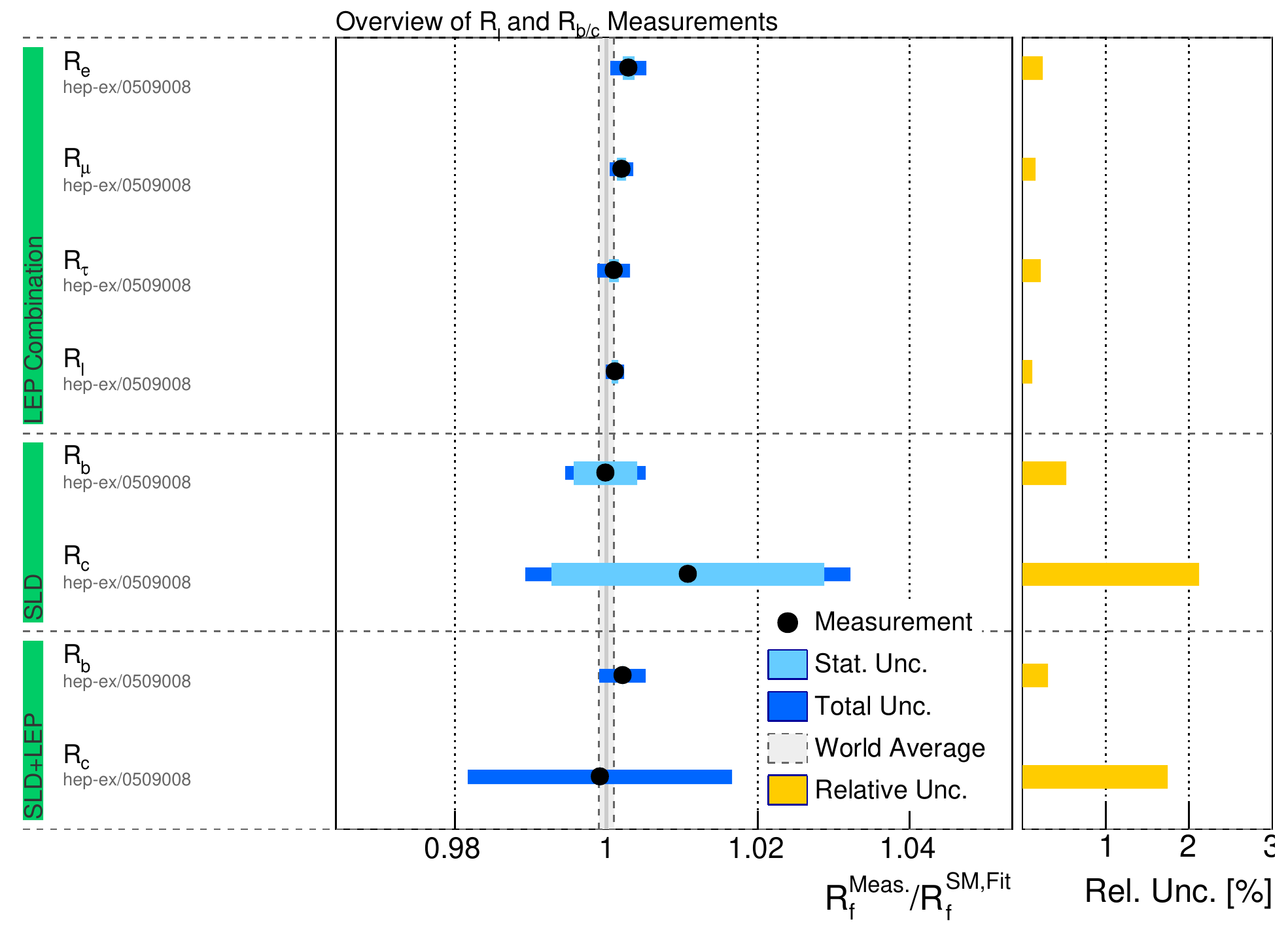}}
\caption{\label{fig:RatiosPull}Ratio of measured values of $R_f$ at LEP and SLD~\cite{ALEPH:2005ab} and their Standard Model expectations.\vspace{1.75cm}}
\end{minipage}
\end{center}
\end{figure*}

The measurement of the couplings $g_V$ and $g_A$ in neutrino scattering experiments is typically performed 
using the processes $\nu_\mu e \rightarrow \nu_\mu e$ and $\bar \nu_\mu e \rightarrow \bar \nu_\mu e$. 
The cross sections for these reactions in the limit of large neutrino energies compared to the electron mass, is given by
\begin{equation}
\sigma_{\nu e} = \frac{G_F^2 m_e E_\nu}{2\pi} \left[ (g_V^{\nu e}\pm g_A^{\nu e})^2 + \frac{(g_V^{\nu e}\mp g_A^{\nu e})^2}{3} \right],
\end{equation}
where the upper (lower) sign refers to (anti)-neutrino scattering.
The most precise experimental observable is given by the ratio of the cross sections of neutrino and anti-neutrino scattering,
\begin{equation}
R = \frac{\sigma_{\nu_{\mu} e} }{\sigma_{\bar \nu_{\mu} e}}\ .
\end{equation}
Similar to neutrino scattering on leptons, the cross sections for DIS events of neutrinos on nuclear targets can also be studied. 
A direct sensitivity on $\sintheta$ is achieved by measuring the ratio of neutral current (NC) to charged current (CC) scattering events, \ie,
\begin{equation}
R_\nu = \frac{\sigma^{\rm NC}_{\nu N}}{\sigma^{\rm CC}_{\nu N}}\ ,
\end{equation}
where $N$ denotes an isoscalar nucleon target. 
Such measurements have been performed in the 1980s and 1990s by the CDHS~\cite{Blondel:1989ev}, CHARM~\cite{Allaby:1987vr} and CCFR~\cite{McFarland:1997wx} collaborations 
with an experimental precision on the order of 1\%. 
Several QCD related theoretical uncertainties cancel in this ratio. 
The remaining dominant theoretical uncertainty is associated with the mass threshold of the $c$-quark which affects mainly $\sigma^{\rm CC}_{\nu N}$. 
Since this uncertainty is of similar size for neutrinos and anti-neutrinos, the definition of the Paschos-Wolfenstein ratio~\cite{Paschos:1972kj},
\begin{equation}
R^- = \frac{\sigma^{\rm NC}_{\nu N} - \sigma^{\rm NC}_{\bar \nu N}}{\sigma^{\rm CC}_{\nu N} - \sigma^{\rm CC}_{\bar \nu N}}\ ,
\end{equation}
provides an experimental observable which is still sensitive to $\sintheta$, but with a significantly reduced theoretical uncertainty. 
This quantity was measured in 2002 by the NuTeV collaboration~\cite{Zeller:2001hh} with sub-percent precision, leading to a value for the on-shell mixing angle of $\sintheta = 0.2277\pm 0.0016$. 
This is $3\sigma$ larger than the expectation of the SM and hence raised significant attention. Several effects which might lead to significant changes in the interpretation of the originally published value, such as isospin violation in the parton distributions, 
the strange sea quark content of the nucleons, as well as QED splitting and nuclear effects, have been discussed since. 
However, a final conclusion has not yet been reached. 
A full review of this topic can be found in~Refs.~\cite{Erler:2013xha,Agashe:2014kda}.
The energy dependence of the weak mixing angle, together with the discussed experimental results of low energy measurements is illustrated in Figure~\ref{fig:EnergySin2Theta}.

\subsubsection{Partial decay widths and fermionic couplings of the $Z$ boson\label{sec:sinratio}}

The leptonic partial widths of the $Z$ boson are precisely determined alongside the lineshape fit measurements performed at LEP (see Section~\ref{sec:zboson}) 
{\em via\/} the ratios of leptonic to hadronic cross sections, $R_l$. 
The corresponding ratios for $c$- and $b$-quark final states, $R_b$ and $R_c$ defined in Equation~(\ref{Rbc}), have also been measured by the LEP experiments, 
as well as by SLD. 
The latter measurements reached a competitive precision due to a very small beam size and a precise CCD pixel detector at the SLC, 
allowing for a clean identification of $b$- and $c$-quark induced particle jets. 
A summary of all measured ratios including dominant uncertainties, as well as their combination, is given in Table~\ref{tab:ZRatios}. 
All measurements are statistically limited. 
Figure~\ref{fig:RatiosPull} shows pull distributions for all precision measurements of $R_f$ relative to their Standard Model predictions. 
In general, good agreement is observed. 

\begin{table}[tb]
\footnotesize
\begin{tabular}{l | c | cc| cc | c}
\hline
obs.	& collider					&	value	&	total unc.	&	stat.\ unc.	&	sys.\ unc. 	&	SM expectation (EW fit)	\\
\hline
$R_l$		& LEP			&	20.767	&	0.025	&	0.017	&	0.018	&	$20.722\pm0.026$	\\
\hline
$R_b$		& LEP (ALEPH)	&	0.2159	&	0.0013	&	0.0009	&	0.0009	&	$0.2158\pm0.0004$	\\
$R_b$		& LEP (DELPHI)	&	0.2163	&	0.0009	&	0.0007	&	0.0005	&	$0.2158\pm0.0004$	\\
$R_b$		& LEP (OPAL)		&	0.2174	&	0.0027	&	0.0011	&	0.0012	&	$0.2158\pm0.0004$	\\
$R_b$		& LEP (L3)		&	0.2173	&	0.0015	&	0.0015	&	0.0023	&	$0.2158\pm0.0004$	\\
$R_b$		& SLD			&	0.2158	&	0.0011	&	0.0009	&	0.0007	&	$0.2158\pm0.0004$	\\
$R_c$		& SLD			&	0.1741	&	0.0037	&	0.0031	&	0.002	&	$0.1722\pm0.0001$	\\
\hline
$R_b$		& SLD+LEP		&	0.21629	&	0.0007	&	 --- 		&	 --- 		&	$0.2158\pm0.0004$	\\
$R_c$		& SLD+LEP		&	0.1721	&	0.003	&	 --- 		&	 --- 		&	$0.1722\pm0.0001$	\\
\hline
\end{tabular}
\centering
\caption{\label{tab:ZRatios}Overview of values of $R_f$ measured at from the LEP~\cite{ALEPH:2005ab, Barate:1999ce, Akers:1993is} and SLD~\cite{Abe:1996nj,Abe:2005nqa} collaborations.}
\end{table}

\subsubsection{Asymmetry measurements at LEP and SLC\label{sec:asymLEP}}

The measurements of \sinleff at LEP reduce to counting experiments by measuring the forward-backward asymmetries in Equation~(\ref{eqn:AFB}).
Effective weak mixing angle for various flavors can be extracted using Equations~(\ref{eqn:asymlep}) and~(\ref{eqn:AFBpol2}),
but most asymmetries are sensitive to the angle for charged leptons, \sinleff. 
This observable is therefore used as reference in the following.

The forward backward asymmetries $A^e_{\rm FB}$, $A^\mu_{\rm FB}$, and $A^\tau_{\rm FB}$, have all been measured at the $Z$ pole at LEP.
Since most systematic detector effects cancel in the ratio, the asymmetry measurements are statistically limited. 
The largest uncertainties are observed in the electron decay channel, due to the $t$-channel contributions and associated uncertainties. 
Common systematic uncertainties between all experiments, like the beam energy calibration, are small. 
An overview of the combined LEP measurements of the leptonic forward-backward asymmetries along with the SM expectation is given in Table~\ref{tab:ZAsymmetry}.

The measurements of hadronic asymmetries are more complicated, since the final state quarks are only reconstructed as particle jets. 
The charge of the outgoing quark has to be determined in order to distinguish particles from anti-particles, so as to allow for a correct definition of the angle $\theta$. 
The charge tagging of reconstructed particle jets at the LEP experiments is based on the reconstruction of lepton charges, $D$-mesons, kaons and a jet-charge observable, 
which is defined as the sum of individual charges of reconstructed particle tracks, weighted by their momenta. 
This approach allowed for a determination of \sinleff\ {\em via\/} the measurement of the forward-backward asymmetry inclusively in hadronic final states. 

The inclusive measurements of hadronic final states can be split up in measurements of individual quark asymmetry parameters $A^q_{\rm FB}$, when applying flavor tagging techniques, 
similar to the measurements of $R_c$ and $R_b$, discussed in Section~\ref{sec:sinratio}. 
Those rely on the lifetime tagging of $B$-mesons, lepton tagging or $D$-meson tagging in reconstructed particle jets. 
Again several associated uncertainties cancel in the forward-backward asymmetry ratio, allowing for determinations of 
$A^b_{\rm FB}$, $A^c_{\rm FB}$, and also $A^s_{\rm FB}$, the latter, however, with a limited precision. 
An overview of the combined LEP quark asymmetry measurements along with the SM expectation is also given in 
Table~\ref{tab:ZAsymmetry}.
Notice, that the value for \sinleff\ derived from $A^b_{\rm FB}$ is somewhat larger than the SM expectation.
However, an update of the two-loop QCD correction to the asymmetry~\cite{Bernreuther:2016ccf} 
with massive $b$ quarks reduces the deviation by about $1/4~\sigma$.
Previous ${\cal O}(\alpha_s^2)$ calculations were obtained in the $m_b = 0$ 
limit~\cite{Altarelli:1992fs,Ravindran:1998jw,Catani:1999nf,Weinzierl:2006yt}
(for the one-loop calculations see Refs.~\cite{Jersak:1979uv,Djouadi:1989uk}).

\begin{table}[tb]
\footnotesize
\begin{tabular}{l | c|cc|c|r|c}
\hline
observable		&collider	&	value	&	total	unc.	&	SM expectation		&	pull 	&	corresponding	$\sin^2\theta_{\rm eff}^{l}$	\\
\hline
$A_e$			&	LEP	&	0.1498	&	0.0049	&	$0.1473	\pm	0.0012$	&	0.5	&	$0.23117	\pm	0.00062$*	\\
$A_\tau$			&	LEP 	&	0.1439	&	0.0043	&	$0.1473	\pm	0.0012$	&	$-0.8	$&	$0.23192	\pm	0.00055$	\\
$A^{0,e}_{\rm FB}$		&	LEP 	&	0.0145	&	0.0025	&	$0.01627	\pm	0.00027$	&	$-0.7$	&	$0.23254	\pm	0.0015$*	\\
$A^{0,\mu}_{\rm FB}$	&	LEP	&	0.0169	&	0.0013	&	$0.01627	\pm	0.00027$	&	0.5	&	$0.23113	\pm	0.0007$*	\\
$A^{0,\tau}_{\rm FB}$	&	LEP	&	0.0188	&	0.0017	&	$0.01627	\pm	0.00027$	&	1.5	&	$0.23000	\pm	0.0009$*	\\
$A^{0,l}_{\rm FB}$		&	LEP	&	0.0171	&	0.001	&	$0.01627	\pm	0.00027$	&	0.8	&	$0.23099	\pm	0.00053$	\\
$A^{0,c}_{\rm FB}$		&	LEP	&	0.0699	&	0.0036	&	$0.07378	\pm	0.00068$	&	$-1.1	$&	$0.23220	\pm	0.00081$	\\
$A^{0,b}_{\rm FB}$		&	LEP	&	0.0992	&	0.0017	&	$0.10324	\pm	0.00088$	&	$-2.4$	&	$0.23221	\pm	0.00029$	\\
\hline
$A_e$			&	SLD	&	0.1516	&	0.0021	&	$0.1473	\pm	0.0012$	&	2.0	&	$0.23094	\pm	0.00027$*	\\
$A_\mu$			&	SLD	&	0.142	&	0.015	&	$0.1473	\pm	0.0012$	&	$-0.4$	&	$0.23216	\pm	0.002$*	\\
$A_\tau$			&	SLD	&	0.136	&	0.015	&	$0.1473	\pm	0.0012$	&	$-0.8$	&	$0.23259	\pm	0.002$*	\\
$A_l$			&	SLD	&	0.1513	&	0.0021	&	$0.1473	\pm	0.0012$	&	1.9	&	$0.23098	\pm	0.00026$	\\
$A_c$			&	SLD	&	0.67		&	0.027	&	$0.66798	\pm	0.00055$	&	0.1	&	$0.231	\pm	0.008$*	\\
$A_b$			&	SLD	&	0.923	&	0.02		&	$0.93462	\pm	0.00018$	&	$-0.6$	&	$0.25	\pm	0.03$*	\\
\hline
\end{tabular}
\centering
\caption{\label{tab:ZAsymmetry}Overview of the measured asymmetries at the $Z$ pole from the LEP and SLD experiments~\cite{ALEPH:2005ab}. 
The values are compared to the SM prediction and a pull value for each observable, $({\cal O}_{\rm measured} - {\cal O}_{\rm predicted})/\Delta {\cal O}$, is calculated. 
In addition, the corresponding effective weak mixing angle $\sin^2\theta_{\rm eff}^{l}$ is given. 
The values indicated with an asterisk have been derived within this work.}
\end{table}

During LEP~1 direct determinations of $A_e$ and $A_\tau$ have also been performed by measuring the polarization of $\tau^-$ leptons in $Z\rightarrow\tau^+\tau^-$ events. 
The $\tau$ polarization is given by
\begin{equation}
P_\tau (\cos\theta_{\tau}) = \frac{\sigma^+(\theta_\tau) - \sigma^-(\theta_\tau) }{\sigma^+(\theta_\tau) + \sigma^-(\theta_\tau)} 
= - \frac{A_\tau (1 + \cos^2\theta_\tau) + 2 A_e \cos\theta_\tau}{(1+\cos^2\theta_\tau) + \frac{8}{3} A^\tau_{\rm FB} \cos\theta_\tau}\ ,
\end{equation}
where $\sigma^\pm$ denotes the cross section for the production of $\tau^-$ leptons with positive and negative helicity, 
and $\theta_\tau$ is the angle between the incoming $e^-$ beam and the outgoing $\tau^-$. 
The measurement of the differential $\tau^-$ polarization therefore allows the simultaneous extraction of $A_\tau$ and $A_e$. 
The corresponding LEP results are also summarized in Table~\ref{tab:ZAsymmetry}.

The high polarization of the initial electron beam at the SLC allowed for the measurement of the left-right asymmetry $A_{\rm LR}$, as in Equation~(\ref{eqn:APV}),
leading to competitive precision in \sinleff.
By virtue of the relation $A_{LR} = A_e P_e$. this asymmetry translates directly to $A_e$ provided the average polarization of the initial electron beam, $P_e$, is known. 
Similarly, the forward-backward asymmetry can be measured for specified final states, except that Equation~(\ref{eqn:AFBpol2}) has to be modified to
\begin{equation}
\label{eqn:AFBpol}
A^f_{\rm FB} = \frac{3}{4} A_f \frac{A_e + P_e}{1+ P_e A_e} \ .
\end{equation}
Once the polarization of the initial state electron is measured, also the left-right forward-backward cross section ratio
\begin{equation}
A^f_{FB,LR} = \frac{\sigma^f_{LF} + \sigma^f_{RB} -\sigma^f_{LB} - \sigma^f_{RF} }{\sigma^f_{LF} + \sigma^f_{RB} +\sigma^f_{LB} + \sigma^f_{RF}} = \frac{3}{4} A_f
\end{equation}
can be extracted, where $f$ is the fermion flavor in the final state, the subscripts $L,R$ denote the left- or right-hand initial state electrons 
to be detected in the forward ($F$) or backward ($B$) hemisphere of the detector. 
Thus, $A_f$ can be measured directly, again with the advantage of being a counting experiments of events in different hemispheres of the detector 
where several systematic and theoretical uncertainties cancel in the ratio. 
This approach was followed by SLD, providing precision measurements of $A_s$, $A_c$, $A_b$, $A_\mu$ and $A_\tau$. 
The corresponding SLD results are also summarized in Table~\ref{tab:ZAsymmetry}.

\begin{figure*}[t]
\begin{center}
\begin{minipage}{0.47\textwidth}
\resizebox{1.0\textwidth}{!}{\includegraphics{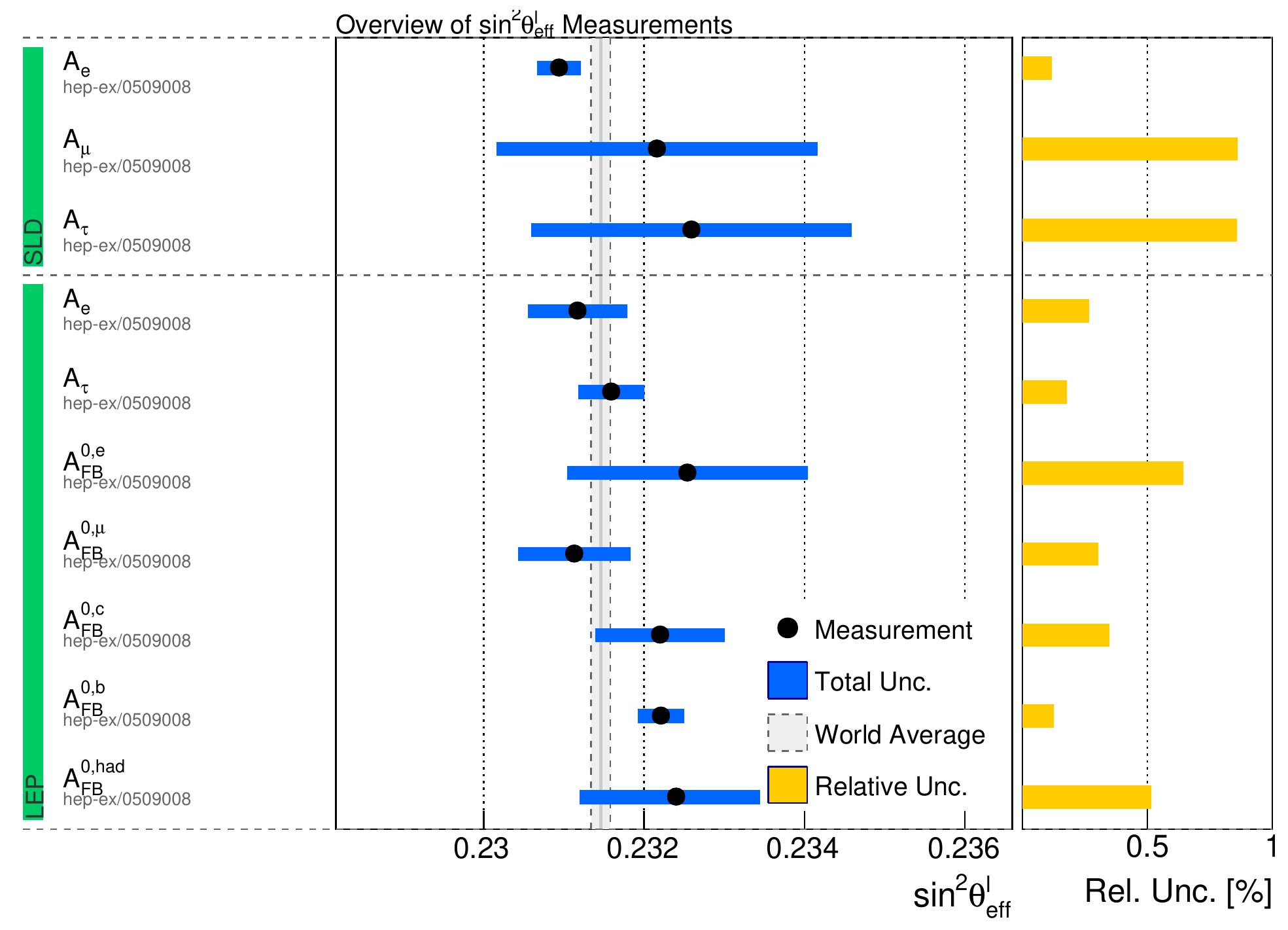}}
\caption{\label{fig:SinThetaLEP}Overview of the extracted \sinleff values from asymmetry measurements at LEP and SLD, 
as well as from forward-backward asymmetry measurements at LEP~\cite{ALEPH:2005ab}.\vspace{0.15cm}}
\end{minipage}
\hspace{0.3cm}
\begin{minipage}{0.5\textwidth}
\resizebox{1.0\textwidth}{!}{\includegraphics{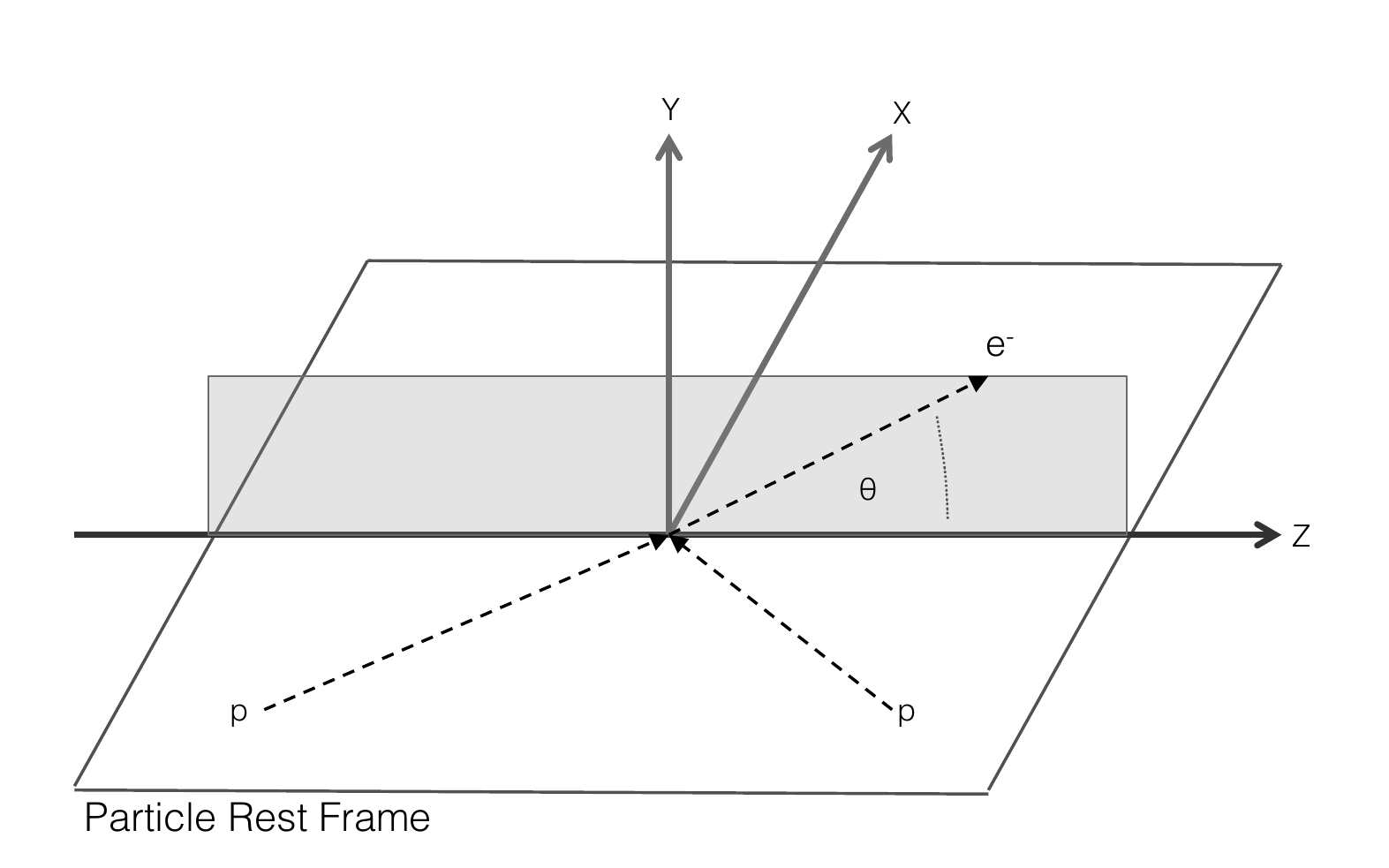}}
\caption{\label{fig:frames}Illustration of the Collins-Soper frame axes in the lepton pair rest frame, relative to the $z$-axis of the laboratory frame.}
\end{minipage}
\end{center}
\end{figure*}

All measured asymmetries can be translated to values of \sinleff by using Equations~(\ref{eqn:asymlep}) and~(\ref{eqn:AFBpol}). 
The results are shown in Table~\ref{tab:sin2theta} and are illustrated in Figure~\ref{fig:SinThetaLEP} 
together with the world average and the Standard Model expectation. 
In general, good agreement between all measurements can be observed, with the exception of the two most precise measurements, 
namely $A_e$ from SLD and $A_{\rm FB}^{b}$ at LEP, which differ by more than $3\sigma$. Both measurements are statistically limited.

\begin{table}[tb]
\footnotesize
\begin{tabular}{l | c|l|l|l|l|l|c}
\hline
$\sin^2\theta_{\rm eff}^{l}$	&	value	&	stat.\ unc.	&	syst.\ unc.	&	PDF unc.	&	model unc.	&	total unc.	&	reference	\\
\hline
\DZero				&	0.23095	&	0.00035	&	0.00007	&	0.00019	&	0.00008		&	0.00047	&	\cite{Abazov:2017gpw}	\\
CDF					&	0.23221	&	0.00043	&	0.00003	&	0.00016	&	0.00006		&	0.00046	&	\cite{Aaltonen:2016nuy}	\\
Tevatron (combined) 	&	0.23148	&	0.00027	&	0.00005	&	0.00018	&	0.00006		&	0.00033	&	\cite{Aaltonen:2018dxj}	\\
\hline
CMS 				&	0.23101	&	0.00036	&	0.00018	&	0.00030	&	0.00016		&	0.00053	&	\cite{Sirunyan:2018swq}\\
ATLAS (central)		&	0.23119	& 	0.00031	& 	0.00018	& 	0.00033	& 	0.00006 		&	0.00049	&	\cite{ATLAS:2018gqq}	\\
ATLAS (forward)		&	0.23166	& 	0.00029	& 	0.00021	& 	0.00022	& 	0.00010 		&	0.00043	&	\cite{ATLAS:2018gqq}	\\
ATLAS (combined)		&	0.23140	& 	0.00021	& 	0.00014	& 	0.00024	& 	0.00007 		&	0.00036	&	\cite{ATLAS:2018gqq}	\\
LHCb 				&	0.23142	&	0.00073	&	0.00052	&	0.00043$^*$	&	0.00036$^*$		&	0.00106	&	\cite{Aaij:2015lka}	\\
\hline
$A^{\rm had}_{FB}$ (LEP)	&	0.23240	&	0.00070		&	0.00100		&	 --- 		&	 --- 			&	0.00120	&	\cite{ALEPH:2005ab}	\\
$A_{l}$ (LEP)	  		&	0.23099	&	0.00042$^*$	&	0.00032$^*$	&	 --- 		&	 --- 			&	0.00053	&	\cite{ALEPH:2005ab}	\\
$A_\tau+A_e$ (LEP)		&	0.23159	&	0.00037$^*$	&	0.00018$^*$	&	 --- 		&	 --- 			&	0.00041	&	\cite{ALEPH:2005ab}	\\
$A^{b}_{\rm FB}$ (LEP)	&	0.23221	&	0.00023$^*$	&	0.00017$^*$	&	 --- 		&	 --- 			&	0.00029	&	\cite{ALEPH:2005ab}	\\
\hline
$A_l$ (SLD)			&	0.23098	&	0.00024		&	0.00013		&	 --- 		&	 --- 			&	0.00026	&	\cite{ALEPH:2005ab}	\\
\hline
\end{tabular}
\centering
\caption{\label{tab:sin2theta}Overview of selected measurements at LEP, SLD, Tevatron and the LHC of the effective leptonic electroweak mixing angle $\sin^2\theta_{\rm eff}^{l}$ 
using different observables including a breakdown of different sources of uncertainties. 
Values which are indicated with an asterisk have not been published and hence only estimated within this work.}
\end{table}

\subsubsection{\label{sec:sinhad}Principle and challenges of $\sin^{2} \theta^l_{\rm eff}$ measurements at hadron colliders}

At electron-positron colliders, the incoming direction of fermions and anti-fermions is known explicitly, 
allowing for a natural definition of an angle $\theta$ between the incoming $e^-$ and the outgoing final state particle. 
At hadron colliders, the initial state of the $Z$ boson production consists of one quark and one anti-quark, but it is not possible to determine the flavor of these quarks. 
Hence a similar forward-backward asymmetry measurement at hadron colliders always has to involve the convolution of parton density functions in the initial state, 
as well as an averaging over all initial state flavors. 
Similar to $e^+e^-$ colliders, a unique angle $\theta$ has to be defined. A natural choice would be the direction of the incoming quark. 
However, this can not be done for two reasons:

Firstly, the direction of the incoming quark is not precisely known in hadron collisions. 
However, the beam direction of protons and anti-protons is known at the Tevatron and the majority of $Z$ bosons is produced in a valence quark annihilation process. 
The situation is even more challenging at the LHC, since the anti-quark in the initial state is necessarily a sea quark and it cannot be determined 
on an event-by-event basis by which of the two protons it was provided.
Secondly, the incoming partons are subject to initial state radiation, which leads to a non-negligible transverse momentum, $p_T$, of the vector boson
arising from the annihilation. 
Therefore, the directions of the incoming partons are not co-linear.

To overcome these problems, the rest frame of the vector boson is typically chosen as reference frame in which the angular distributions of the decay leptons are measured. 
The definition of the axes in this rest frame is still ambiguous. 
To minimize the effect arising from the lack of information about the kinematics and the incoming partons one proceeds as follows.
The hadron plane is defined in the rest frame of the vector boson such that it contains both incoming protons.
The $x$-axis within the hadron plane is defined to bisect the angle between the protons.
The $y$-axis is defined as the normal vector to the hadron plane, 
and the $z$-axis is chosen such that a right-handed Cartesian coordinate system is defined (Figure~\ref{fig:frames}). 
The resulting reference frame is called Collins-Soper (CS) frame~\cite{Collins:1977iv}. 
The angle $\theta^*_{CS}$ in the CS frame can be expressed as
\begin{equation}
\label{eqn:CSFrame}
\cos\theta^*_{CS} = \frac{p_z(ll)}{|p_z(ll)|} \frac{2(p_1^+ p_2^- - p_1^- p_2^+)}{m_{ll}\sqrt{m_{ll}^2 + p_T(ll)^2}}\, ,
\end{equation}
with $p_i^\pm = (E_i \pm p_{z,i})/\sqrt{2} $, where $E_i$ and $p_{z,i}$ are the energy and longitudinal momentum of lepton $i$. 
The first factor in this equation defines the sign and hence the direction of the incoming quark. 
The second factor corrects for the measured boost due to the hadronic activity in the initial state of the event 
and defines an average angle between the decay leptons and the quarks.  

\begin{figure}[t]
\resizebox{0.495\textwidth}{!}{\includegraphics{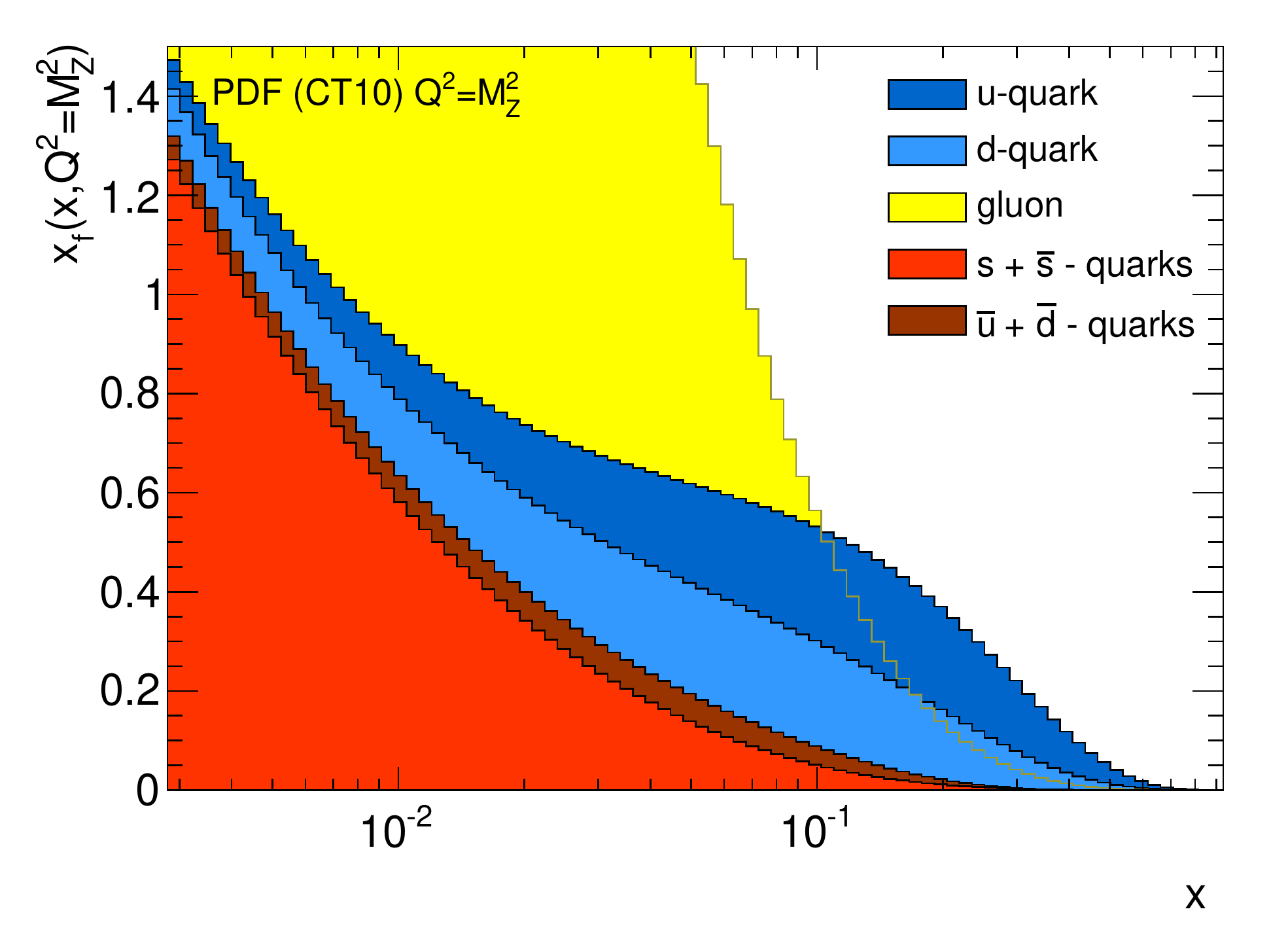}}
\resizebox{0.495\textwidth}{!}{\includegraphics{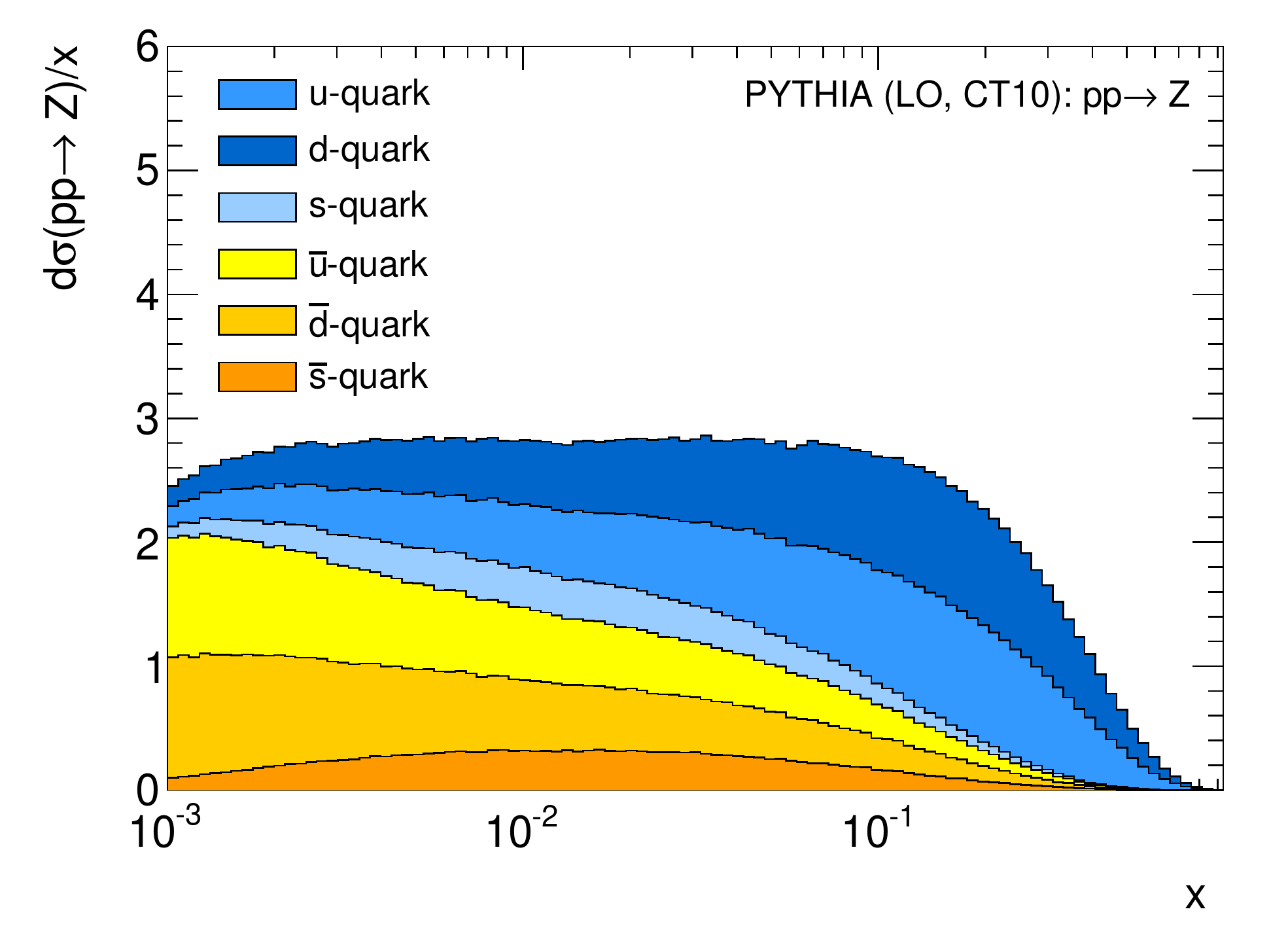}}
\caption{\label{fig:CTEQPDFX}PDF distribution of the CT10 PDF set at $Q^2=\MZ^2$ (left) and distribution of Bjorken $x$-values of partons 
that are involved in the leading-order production of $Z$ bosons at 7~TeV $pp$ collisions as a stacked histogram (right)~\cite{Schott:2014sea}. 
It should be noted that each participating valence quark has to be matched with a corresponding sea quark.}
\end{figure}

The correct assignment of the quark direction in Equation~(\ref{eqn:CSFrame}) can be addressed on a statistical basis. 
Vector bosons with a longitudinal momentum $p_z(V)$ have been produced by partons which have significantly different Bjorken-$x$ values. 
Figure~\ref{fig:CTEQPDFX} illustrates that large $x$-values enhance the probability for having valence quarks in the interaction, 
and that the corresponding anti-quark can be associated with the smaller $x$-values. 
Hence, the measurement of $p_z(V)$ allows to assign the longitudinal quark and anti-quark directions on a statistical basis. 
It should be noted that large $p_z(Z)$ values also imply large rapidities and hence the statistical precision for the correct quark/anti-quark assignment 
is enhanced for $Z$ bosons in the forward region of the detectors.

Thus, one measures the forward-backward asymmetry $A_{\rm FB}$ in Equation~(\ref{eqn:AFB}) at hadron colliders with respect to $\theta^*_{CS}$. 
The knowledge of the parton density functions of the proton plays a crucial role here: 
firstly, to determine the contributions of the different quark flavors in the initial state which contribute with different $A_{\rm FB}$ values, 
and secondly --- and more importantly at the LHC --- to estimate the dilution effect due to the wrong assignment of the quark direction in Equation~(\ref{eqn:CSFrame}). 
The actual extraction of the weak mixing angle can then be achieved by a template fit approach for the measured $A_{\rm FB}$ spectrum, 
based on the variation of \sinleff in the underlying MC generator program. 

In order to understand the problematic aspect of the parton density functions in the context of $A_{\rm FB}$ in more detail, 
it is helpful to further decompose the general form of the differential cross section of the Drell-Yan process,
$pp\rightarrow Z(W) + X\rightarrow l^+ l^- (l \nu) + X$~\cite{Mirkes:1994eb,Mirkes:1994dp} following Equation~(\ref{eqn:decay}) as 
\begin{equation}
\label{EQN:DECOMP}
\frac{d\sigma}{dp_T^2 dy d\cos\theta d\phi} =  \frac{d\sigma_{\rm unpol}}{dp_T^2 dy} 
\left[ (1+\cos^2\theta)  + \frac{A_0}{2} (1-3 \cos^2\theta) + A_1 \sin2\theta \cos\phi + \frac{A_2}{2} \sin^2\theta \cos2\phi \right.
\end{equation}
$$  \left. + A_3 \sin\theta\cos\phi + A_4 \cos\theta + A_5 \sin^2\theta\sin2\phi + A_6 \sin2\theta\sin\phi + A_7 \sin\theta\sin\phi \right], $$
where $\theta$ and $\phi$ are the polar and azimuthal angles in the CS frame defined before. 
The $\cos\theta$ and $\phi$ dependence of the differential Drell Yan cross-section is therefore completely described analytically, 
while the dependence on $p_T$, rapidity and invariant mass is entirely contained in the $A_i$ coefficients. 
These nine terms describe the polarization states of the boson. 
Hence the QCD dynamics of the $Z$ boson production mechanism can be factorized from the decay kinematics in the $Z$ boson rest frame. 
The measurement of the angular coefficients is therefore independent from QCD and QED effects related to the $Z$ boson production 
and decay\footnote{QED and EW corrections between the initial state and final state particles have only a negligible impact at the $Z$ boson mass.}. 

The coefficients $A_i$ depend on boson kinematics and, with the exception of $A_4$, vanish for small transverse momenta of the vector boson. 
All coefficients are subject to higher-order perturbative and non-perturbative corrections, structure functions, and renormalisation and factorization scale uncertainties. 
Since the PDFs of the proton impact the vector boson kinematics, the $A_i$ also depend indirectly on the PDFs themselves. 
The longitudinal and transverse states of polarization are described by the coefficients $A_0$ and $A_2$, while the interference between both polarization states is described by $A_1$. 
A special role is played by the $A_4$ coefficient, as the $\cos\theta$ term is odd under parity, and hence does not vanish at tree-level for $Z$ exchange.
Thus, it is directly connected to  the forward-backward asymmetry and the electroweak mixing angle. 
The coefficients $A_5, A_6$ and $A_7$ appear in NLO calculations in $\as$. 
The first measurement of the angular coefficients in hadron colliders was done at CDF~\cite{Aaltonen:2011nr}.
All angular coefficients have also been measured to high precision in Drell-Yan production 
at the LHC~\cite{Khachatryan:2015paa,Aad:2016izn} and agree with NNLO predictions. 
The uncertainty in their functional dependence played an important role for the $W$ boson mass measurement at the LHC, 
as it directly affects the $p_T$ spectrum of the vector boson decay leptons.

\begin{figure}[t]
\resizebox{0.495\textwidth}{!}{\includegraphics{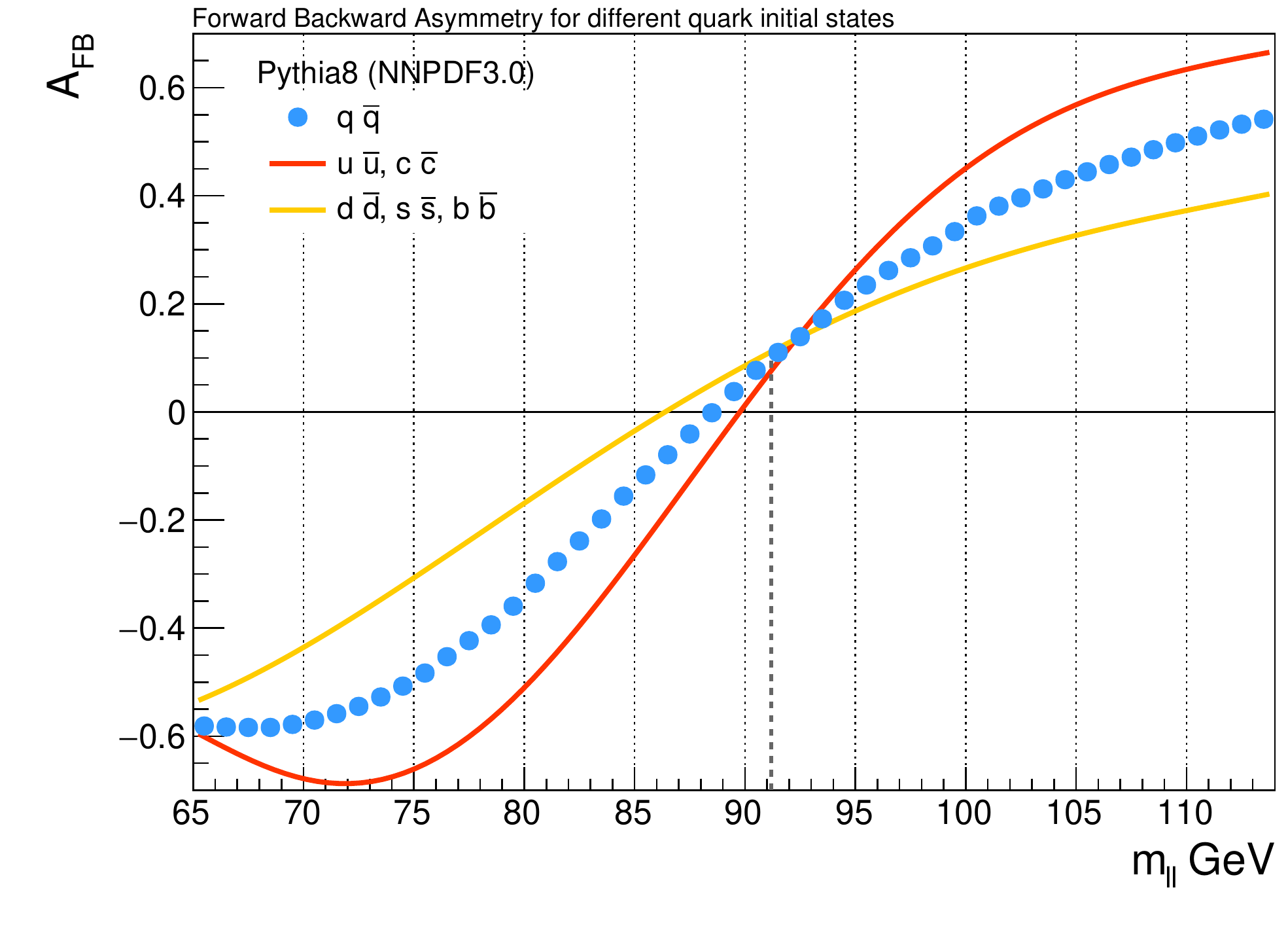}}
\resizebox{0.495\textwidth}{!}{\includegraphics{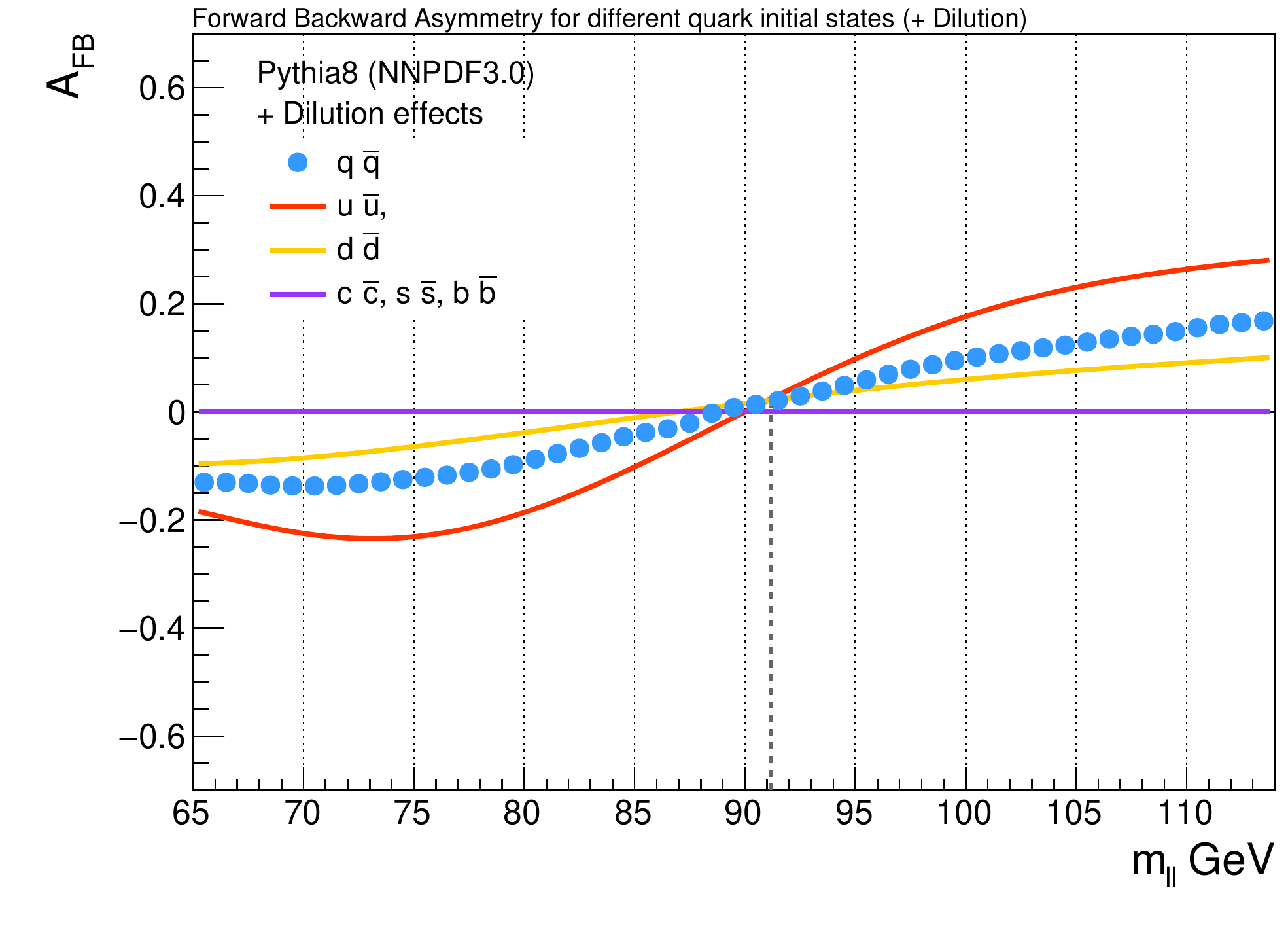}}
\caption{Distributions of $A_{\rm FB}(m_{ll})$ for dilepton events in $pp$ collisions at 13 TeV using \Pythia8 and the NNLOPDF3.0 PDF set with rapidity $|y_{ll}|<2.4$. 
The $A_{\rm FB}(m_{ll})$ distributions for different quark initial states are shown on the left, the observed distributions including dilution effects are shown on the right. 
The incoming quark direction cannot be identified for sea quarks. 
The original version of this plots can be found in Ref.~\cite{CMS:2017zzj}.}
\label{fig:AFBQuarkDependence}
\end{figure}

Since the forward-backward asymmetry in the CS frame is given by a convolution of the $g_A$ and $g_V$ couplings over all the incoming quarks and outgoing leptons, 
it is not only dependent on the value of \sintheta itself, but also on the underlying PDFs. 
A typical dependence of $A_{\rm FB}$ in proton-proton collisions as a function of invariant mass of the decay leptons is shown in Figure~\ref{fig:AFBQuarkDependence}. 
Due to its larger charge the $u$-quark induces a significantly larger variation in $A_{\rm FB}$ than the $d$-quark. 
The offset of $A_{\rm FB}$ at \MZ is therefore the observable which inhibits the highest sensitivity to \sintheta. 
In terms of the angular $A_4$ coefficient, it can be expressed as
\begin{equation}
A_{\rm FB}(m_{ll}) = \frac{\sigma^+(m_{ll}) - \sigma^-(m_{ll})}{\sigma^+(m_{ll}) + \sigma^-(m_{ll})} = \frac{3}{8} A_4 (m_{ll}).
\end{equation}
Hence, the measurement of $A_4$ is equivalent to a measurement of $A_{\rm FB}$ and can be used in principle to determine \sintheta.

In the simplest approach, a purely inclusive $A_{\rm FB}$ is the only basis for the \sinleff extraction. 
Already at this stage, one can distinguish between approaches which perform the fit at the reconstruction level, similar to the kinematic fits of \MW and $\mt$, 
or at the particle level, \ie, where detector effects have been removed through unfolding approaches. 
The latter have the advantage that the actual measurement of \sinleff can be performed independent of the experimental collaborations 
and can be easily updated at later stages with an improved theory modeling. 

$A_{\rm FB}$ is measured at hadron colliders in the electron or muon decay channel of the $Z$ boson as a function of the invariant mass of the dilepton pair. 
The largest sensitivity to \sinleff comes from the measurement directly at the $Z$ pole mass. 
Due to the strong dependence of $A_{\rm FB}$ around \MZ, it is critical to have a precise lepton energy calibration. 
When measuring $A_{\rm FB}$ not only as a function of $m_{ll}$, but also as a function of the pseudo-rapidity of the dilepton system 
and/or as a function of $\theta^*_{CS}$, the extraction of \sinleff can be performed in parallel to constraining proton PDFs, 
leading to reduced model uncertainties as described in Section~\ref{s2wdiscussion}.

MC event generators for Drell-Yan production are used in all experimental approaches for the measurement of \sintheta to generate $A_{\rm FB}$ templates. 
However, common event generators do not include electroweak radiative corrections beyond photon emissions in the initial and final state. 
For example, the \Pythia\ generator uses the same value for the effective couplings \sinleff for all fermions. 
The electroweak corrections, which would need to be implemented in MC event generators are already known since LEP 
and were calculated by~\textsc{Zfitter}~\cite{Arbuzov:2005ma} already in the late 1990s. 
They were used for the corresponding precision measurements at LEP and SLD. 

Since the process $e^+e^- \rightarrow q \bar q$ is analogous to the reaction at hadron colliders $q \bar q \rightarrow e^+e^-$, 
the corresponding electroweak form factors, $\rho_{eq}$, $\kappa_e$, and $\kappa _q$, can be included in the couplings $g_A^f$ and $g_V^f$ 
to improve the Born level expressions of the Drell-Yan process. 
As a refinement, another form factor, $\kappa_{eq}$, multiplying terms proportional to $\sin^4\theta_W$ can be introduced. 
This approach leads to an enhanced Born approximation, which was used for the first time by the CDF collaboration~\cite{Aaltonen:2016nuy}. 
This correction leads to a shift of $+0.00022 \pm 0.00004$ in the extracted value of \sinleff.

\subsubsection{Discussion of recent $\sin^{2} \theta^l_{\rm eff}$ measurements at hadron colliders}
\label{s2wdiscussion}
The latest measurement of \sinleff at the Tevatron was published in 2017 by the \DZero Collaboration~\cite{Abazov:2017gpw} 
using the electron and muon decay channels of the $Z$ boson. 
It is based on a template fit approach, where different $A_{\rm FB}$ distributions have been simulated according to varied input values of \sintheta at the Born level,
using the LO Monte Carlo generator \Pythia~\cite{Sjostrand:2007gs} and the CTEQ6.6 PDF set~\cite{Nadolsky:2008zw}. 
The templates are based on fully simulated events and are obtained by reweighting the invariant mass and the $\cos\theta^*_{CS}$ distributions at the generator level. 
\DZero uses electrons which are reconstructed in the central  and in the forward region of the detector, 
defined by the pseudo-rapidities $|\eta| < 1.1$ and $1.5 < |\eta| < 3.5$, respectively. 
While electron pairs, that are both reconstructed in the forward region are required to have an invariant mass of $81~{\rm GeV} < m_{ee} < 97$~GeV, 
this requirement is relaxed for other electron combinations to $75~{\rm GeV} < m_{ee} < 115$~GeV. 
The measured value, as well as the experimental and model uncertainties, are summarized in Table~\ref{tab:sin2theta}. 
It should be noted that this value was not fully corrected for the underlying assumption of the \Pythia\ event generator of a fixed weak mixing angle for all fermions. 
A partial correction was achieved by comparing the \Pythia\ interpretation with a modified version of \textsc{\Resbos}, 
which uses different values of the effective weak mixing angle for leptons and up- and down-quarks~\cite{Abazov:2014jti}. 
The dominant uncertainty in the final measurements is due to the limited data statistics. 
The leading experimental uncertainty is due to the electron identification. 
The most sensitive measurement is performed for events in which one electron is central and the second electron is reconstructed in the forward region. 

The CDF collaboration published a measurement of \sinleff based on the full Run~II data set of 9.4~fb$^{-1}$,
and is also using both electron and muon decay channels with a similar precision as \DZero~\cite{Abazov:2017gpw}. 
CDF uses muon pairs with a rapidity of $|y_{\mu\mu}| < 1$, and electron pairs with $|y_{ee}| < 1.7$, 
and employs an advanced data-driven event weighting technique~\cite{Bodek2010} for the extraction of $A_{\rm FB}$. 
Here, the asymmetry is evaluated in bins of $|\cos\theta^*_{CS}|$ using,
\begin{equation}
\label{AFB2}
A_{\rm FB} = \frac{ \frac{N^+}{\epsilon^+ A^+} - \frac{N^-}{\epsilon^- A^-} }{ \frac{N^+}{\epsilon^+ A^+} + \frac{N^-}{\epsilon^- A^-}} = \frac{N^+ - N^-}{N^+ + N^-}\ ,
\end{equation}
where $N^\pm$ is the selected number of forward (backward) lepton pairs and $\epsilon^\pm A^\pm$ are the respective detector efficiencies and acceptances.  
No significant differences in the detector performance is expected for high energetic leptons. 
It is therefore assumed that the interchange of particles and anti-particles changes only the sign of $\cos\theta^*_{CS}$. 
Hence, the acceptance and efficiency corrections cancel to first order for each bin in $|\cos\theta^*_{CS}|$, leading to the second equality in Equation~(\ref{AFB2}).
The binned measurements are reformulated into an unbinned, event-by-event weighted expression which is used for the combination of all $\cos\theta^*_{CS}$ bins. 
Resolution and migration effects are unfolded using fully detector simulated MC samples. 
Further details on the employed methodology can be found in~\cite{Bodek2010}. 
The final CDF result on \sintheta including uncertainties is summarized in Table~\ref{tab:sin2theta}. 
Similar to \DZero, the measurement is statistically limited with a significant uncertainty due to PDFs. 
In contrast to \DZero, the electroweak form factor corrections employ the enhanced Born approximation.

\begin{figure*}
\begin{center}
\begin{minipage}{0.48\textwidth}
\resizebox{1.0\textwidth}{!}{\includegraphics{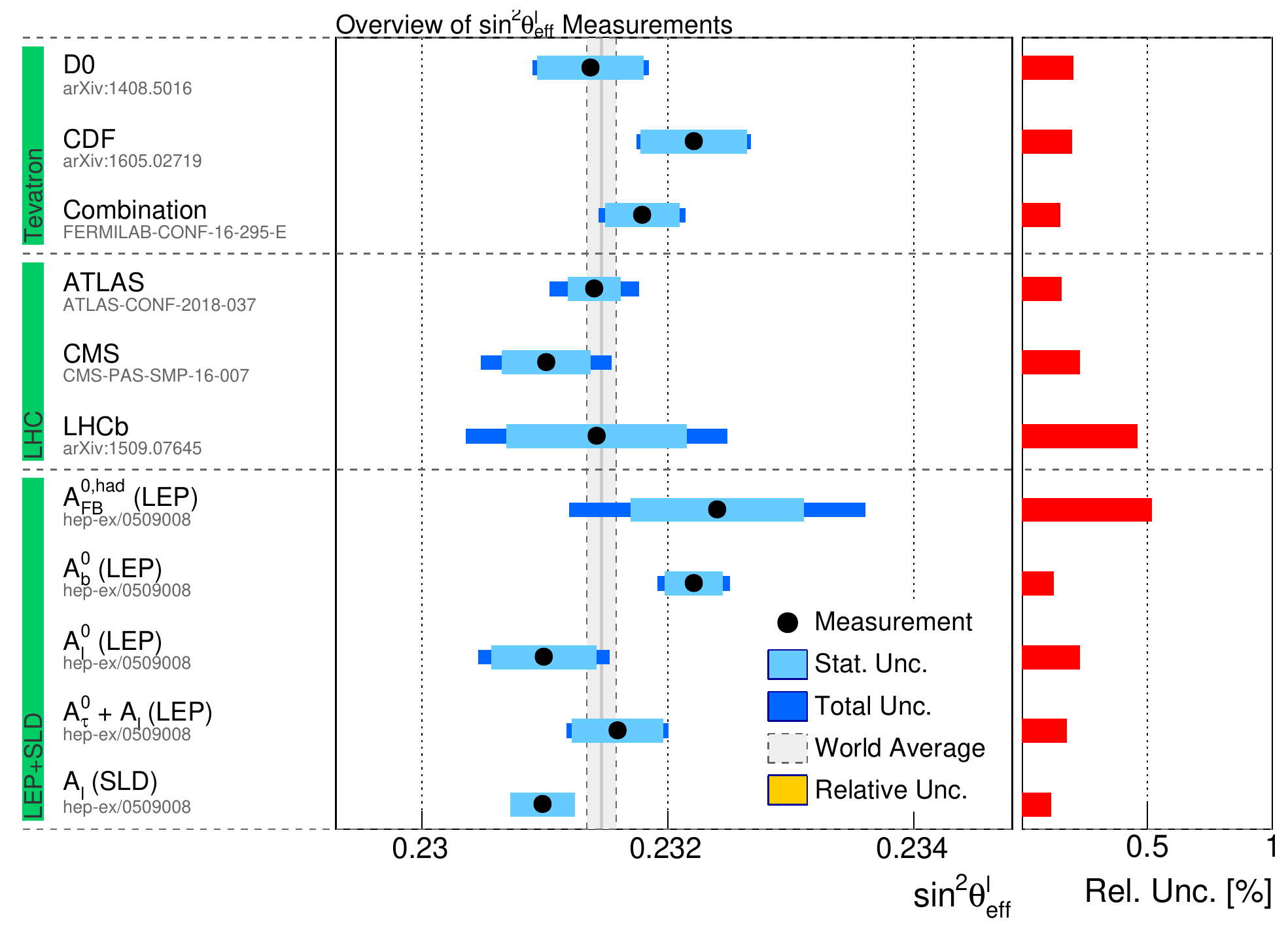}}
\caption{\label{fig:SinThetaAll}Overview of the most precise measurements of \sinleff from the LEP and SLD experiments, 
as well as from the \DZero and CDF collaborations at the Tevatron and from ATLAS, CMS and LHCb at the LHC.}
\end{minipage}
\hspace{0.3cm}
\begin{minipage}{0.48\textwidth}
\resizebox{1.0\textwidth}{!}{\includegraphics{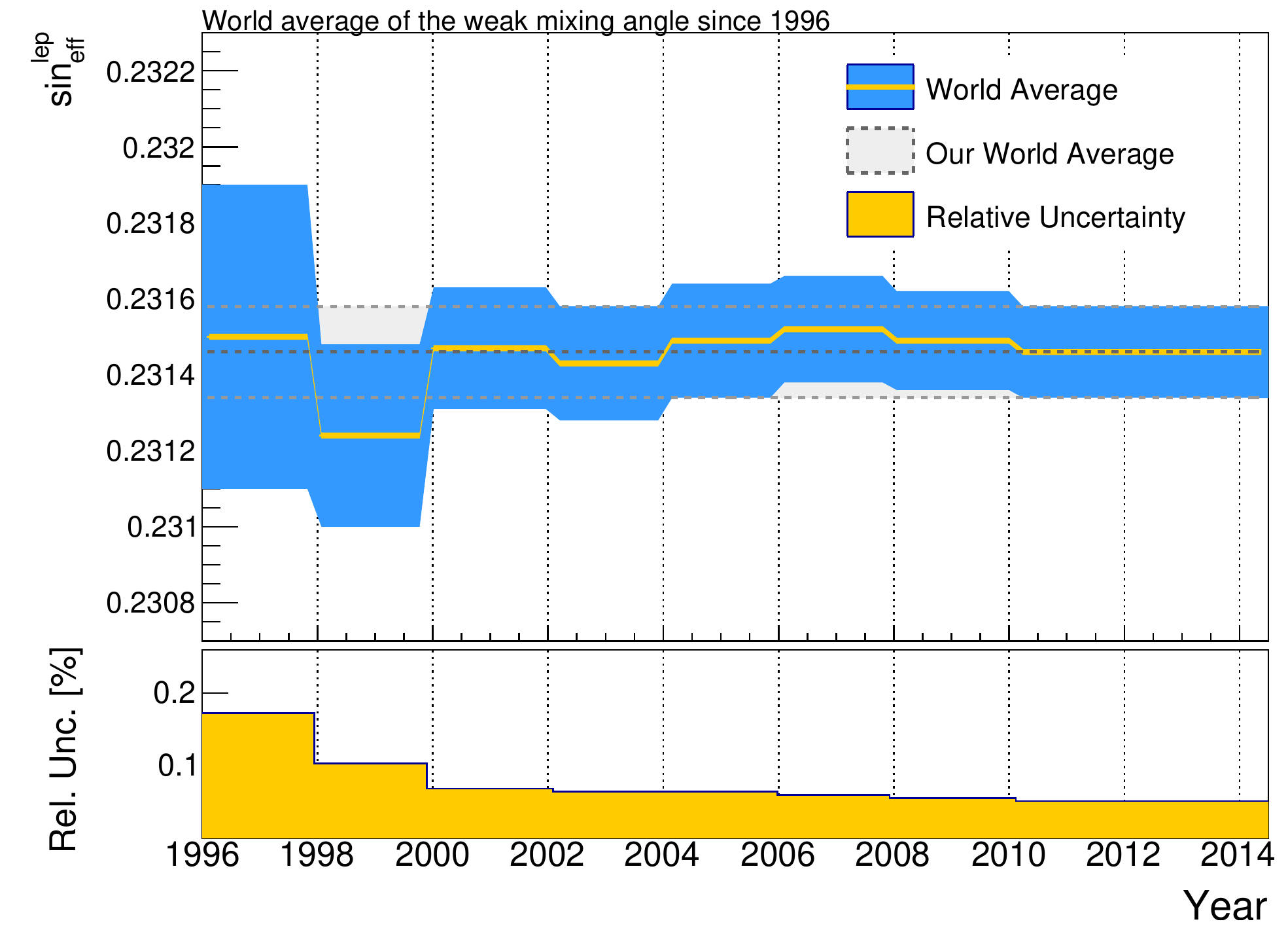}}
\caption{\label{fig:SinThetaEvolution}Evolution of the world average of \sinleff and its uncertainties {\em vs.\/}\ time. 
Values are taken from previous editions of the PDG review~\cite{Olive:2016xmw}.\vspace{0.4cm}}
\end{minipage}
\end{center}
\end{figure*}

The measurements from the CDF and \DZero collaborations have been combined in Ref.~\cite{Aaltonen:2018dxj}. 
The statistical uncertainty is similar for both measurements and dominates the overall precision, where the combination improves the statistical precision from 0.00043 to 0.00027. 
The published value of \DZero was adjusted to account for the difference between the CTEQ6.6 PDF set, 
which was used in the original measurement, and the NNPDF3.0 PDF set, which was used for the CDF measurement. 
In addition, radiative electroweak corrections have been taken into account. 
This leads to an overall change of the central value by $0.00014\pm0.00004$. 
The combined value, $0.23148\pm 0.00033$, is shown in Table~\ref{tab:sin2theta} and illustrated together with all other measurements in Figure~\ref{fig:SinThetaAll},
and falls in between the LEP and SLD measurements which have a similar precision. 

As mentioned, the situation at the LHC is significantly more complicated as the dilution effects from the quark anti-quark assignment are strongly enhanced. 
The latest CMS measurement of the effective weak mixing angle is based on the analysis of Drell-Yan events in the electron and muon decay channels
using nearly the full available data set at 8 TeV~\cite{Sirunyan:2018swq}. 
Similarly to CDF, the event weighting technique is used for the determination of $A_{\rm FB}$, 
which is measured in 12~bins of di-lepton masses in the range between 60~GeV and 120~GeV and 6~equal bins in absolute di-lepton rapidity up to 2.4. 
The weak mixing angle \sinleff is extracted by minimizing the $\chi^2$ value between the data and all template $A_{\rm FB}$ distributions. 
The templates are produced with the \Powheg\,generator~\cite{Frixione:2007vw, Alioli:2010xd} at NLO precision in $\as$, 
using the NNPDF3.0 PDF set~\cite{Ball:2014uwa} interfaced with \Pythia8~\cite{Sjostrand:2007gs} for additional parton showering. 

Since PDF variations can lead to large changes in $A_{\rm FB}$ in the low and high mass regions, these can be used to minimize PDF uncertainties. 
CMS employs a Bayesian $\chi^2$ reweighting technique~\cite{Bodek:2016olg}, 
where PDF variations that better describe the measured $A_{\rm FB}$ distributions receive a larger weight than those which describe the data worse. 
This approach leads to a reduction of the PDF uncertainty by nearly 50\%, while the central value does not change significantly. 
The second largest model uncertainty is due to variations of the factorization and renormalization scales.
Uncertainties caused by unknown higher order electroweak corrections affecting the difference 
between the quark and leptonic effective mixing angles in the MC event generator have been found to be negligible. 

The most precise --- albeit preliminary --- measurement of \sinleff at the LHC has been performed by the ATLAS collaboration 
and is also based on the full 2012 data sample, using the electron and muon decay channels~\cite{ATLAS:2018gqq}. 
Since a high acceptance of $Z$ bosons with large rapidities reduces the dilution of falsely identified quark-directions, ATLAS  also includes forward electrons in their analysis. 
By requiring one electron with $|\eta| < 2.4$ and allowing a second electron within $|\eta| < 4.9$, an acceptance for $Z$ boson events up to a rapidity $|y_{ll}| < 3.6$ is achieved. 
The measurement can therefore be decomposed into two fiducial phase space regions, one where both decay leptons are within a di-lepton rapidity range of $|y_{ll}| < 2.4$ (central) 
and one which extends to higher rapidities of 3.6~using forward electrons. 

The measurement principle relies not on the extraction of \sinleff\ {\em via\/} the $A_{\rm FB}$ distributions, 
but on the measurement of the angular coefficients $A_i$ (in particular $A_4$) in Equation~(\ref{EQN:DECOMP}). 
The extraction is achieved by directly parametrizing $A_4$ in terms of \sinleff. 
The measurement of the $A_i$ coefficients is performed using a profile likelihood ratio method in course bins of the di-lepton mass (3~bins) from 70 to 125~GeV 
and the absolute di-lepton rapidity from 0 to 3.6~(4 bins). 
This binning allows to constrain PDF uncertainties directly in the profiled likelihood fit, where the PDF uncertainties are treated as nuisance parameters. 
This technique which is used with Hessian error PDFs is equivalent to the  Bayesian $\chi^2$ reweighting technique which is used with PDF replicas.
The likelihoods are constructed by comparing templates for each term in Equation~(\ref{EQN:DECOMP}) to the reconstructed angular distributions 
using $8\times8$ bins in ($\cos\theta$, $\phi$) space. 
Further details on the fitting methodology can be found in~\cite{Aad:2016izn}. 
Predictions for the angular coefficients are obtained using \textit{Dyturbo}~\cite{cite:DYTURBO}, which provides fixed-order and re-summed calculations for vector boson production. 
The \textit{Dyturbo} predictions are at leading order in electroweak theory,
while higher orders can be considered using a per-event weighting technique in the improved Born 
approximation~\cite{Bardin:1989tq,Jadach:1999vf} or {\em via\/} the \textit{TauSpinner} approach~\cite{Richter-Was:2018lld}. 
The final result of the ATLAS measurement and its uncertainties is shown in Table~\ref{tab:sin2theta}, separately for the central and forward lepton measurements. 
The dominant uncertainties apart from those related to PDFs are due to limited MC statistics. 
Very similar sized uncertainties are seen, when comparing the latest result of the CMS collaboration to the measurement of ATLAS using central leptons only.

As mentioned, the dilution effect due to the wrong assignment of the quark direction is reduced for $Z$~bosons produced in the forward direction. 
Hence, a measurement using the LHCb detector with its lepton coverage of $2.0 < \eta < 4.5$ offers complementarity. 
LHCb published a measurement of \sintheta using Drell-Yan events in the muon decay channel based on the full available data sets at 7 and 8~TeV center-of-mass energy. 
$A_{\rm FB}$ was measured for invariant di-muon masses between 60 and 160~GeV using a Bayesian unfolding technique. 
These measurements have been compared to predictions of the \Powheg\, event generator for \sinleff values between 0.22 to 0.24~using the NNPDF3.0 PDF set. 
The final value is a combination of the measurements performed at the two center of mass energies (Table~\ref{tab:sin2theta}). 
The statistical uncertainty is still dominant and the associated PDF uncertainties are on a similar level as for the ATLAS and CMS results, 
when no additional PDF constraints are employed. 

Combining all measurements performed at hadron colliders {\em via\/} the \Blue\,method, 
based on the combined value of CDF and \DZero, as well as the measurements by ATLAS, LHCb and CMS, leads to a value of
\begin{equation}
\sinleff = 0.23140 \pm 0.00023,
\end{equation}
with a $\chi^2/{\rm n.d.f.} =0.67/3$. 
Here, no correlation between the experimental systematics, a full correlation of the electroweak correction systematics and a partial correlation of the PDF related 
systematics\footnote{The central value changes by up to 0.00005 when assuming 0 and 100\% correlations of the PDF systematics, \ie. it is well within the given uncertainties.} 
has been assumed. 

\subsubsection{Discussion and prospects for measurements of the weak mixing angle}

The two most precise measurements of \sinleff at the lepton vertex are the LEP~1 measurement of $A^b_{\rm FB}$ 
and the $A_l$ measurement at SLC (Section~\ref{sec:asymLEP}, Table~\ref{tab:sin2theta}), which are in significant tension with each other. 
The combined value of \sinleff from the hadron collider measurements now reaches a similar precision as those individual measurements 
and yield a value in between them, slightly closer to the $A_l$ measurement at the SLC.

Assuming no correlation between the measurements at LEP, SLC and the combined value from the hadron collider measurements, 
a combination with the \Blue\, approach yields the value, 
\begin{equation}
\sinleff = 0.23151 \pm 0.00014,
\end{equation}
with a $\chi^2/{\rm n.d.f.} = 11.5/5$.
The time evolution of the world average of the electroweak mixing angle and its uncertainties is illustrated in Figure~\ref{fig:SinThetaEvolution}, where the measurements 
at the Tevatron and the LHC are indicated separately. 
The measurements at the LHC already reached a precision similar to those at the Tevatron, and they are bound to improve further.
First of all, they are not statistically limited due to the large available data sets at a center of mass energy of 13~TeV. 
Secondly, PDF constraining methods during the determination of \sintheta and improved analysis techniques, 
as already used in the ATLAS and CMS analyses~\cite{Sirunyan:2018swq,ATLAS:2018gqq}, will allow a further reduction of PDF uncertainties. 

\subsection{The top quark mass\label{sec:quarkmasses}}

\subsubsection{On the importance of the top quark mass within the Standard Model}

\begin{figure*}[t]
\begin{center}
\resizebox{0.98\textwidth}{!}{\includegraphics{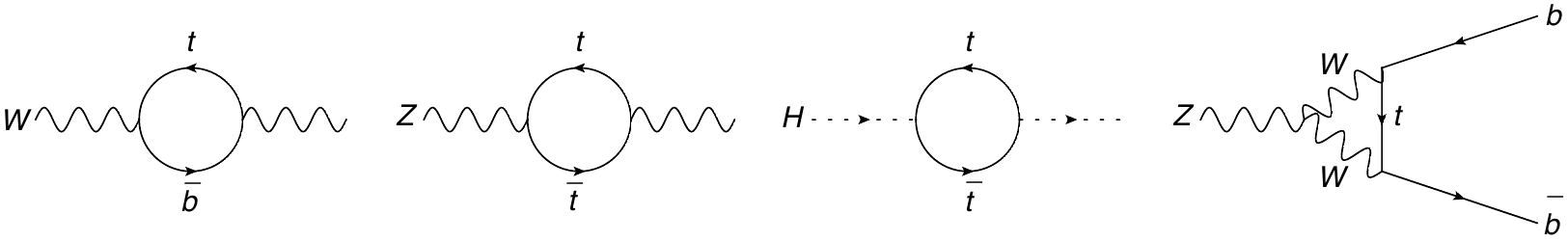}}
\caption{\label{fig:TopDiagrams}Selected Feynman diagrams for loop corrections to precision observables involving the top quark. 
The first two diagrams alter the $W$ and $Z$ boson self-energies, the third diagram is the dominant correction to the self-energy of 
the Higgs boson, while the forth diagram contributes to the effective couplings of bottom quarks to the $Z$ boson.}
\end{center}
\end{figure*}

The top quark plays a special role in electroweak precision physics as it is the heaviest of all elementary particles 
and as such has the largest Yukawa coupling to the Higgs boson. 
It forms no bound states or top-quark flavored hadrons due to its very short lifetime of about $10^{-25}$~s,
and thus can be studied to some extent directly before hadronization. 
Even though several aspects of the top quark properties are of theoretical interest, 
we will focus on its mass as it directly affects the consistency tests of the electroweak sector, as described in Section~\ref{sec:ewpotheory}. 
Examples of loop corrections involving the top quark mass to the propagators and vertices of the electroweak gauge bosons, as well as the Higgs boson, 
are illustrated in Figure~\ref{fig:TopDiagrams}. 
It should be noted that a second reason for the special role of the top quark are its elements of the quark mixing (CKM) matrix, 
which are close to diagonal and trigger decays nearly exclusively to $W$ bosons and $b$ quarks. 
For general review articles on top quark physics, we refer to Refs.~\cite{Cortiana:2015rca,Boos:2015bta}. 

\begin{figure}[t]
\begin{center}
\begin{minipage}{0.55\textwidth}
\resizebox{1.0\textwidth}{!}{\includegraphics{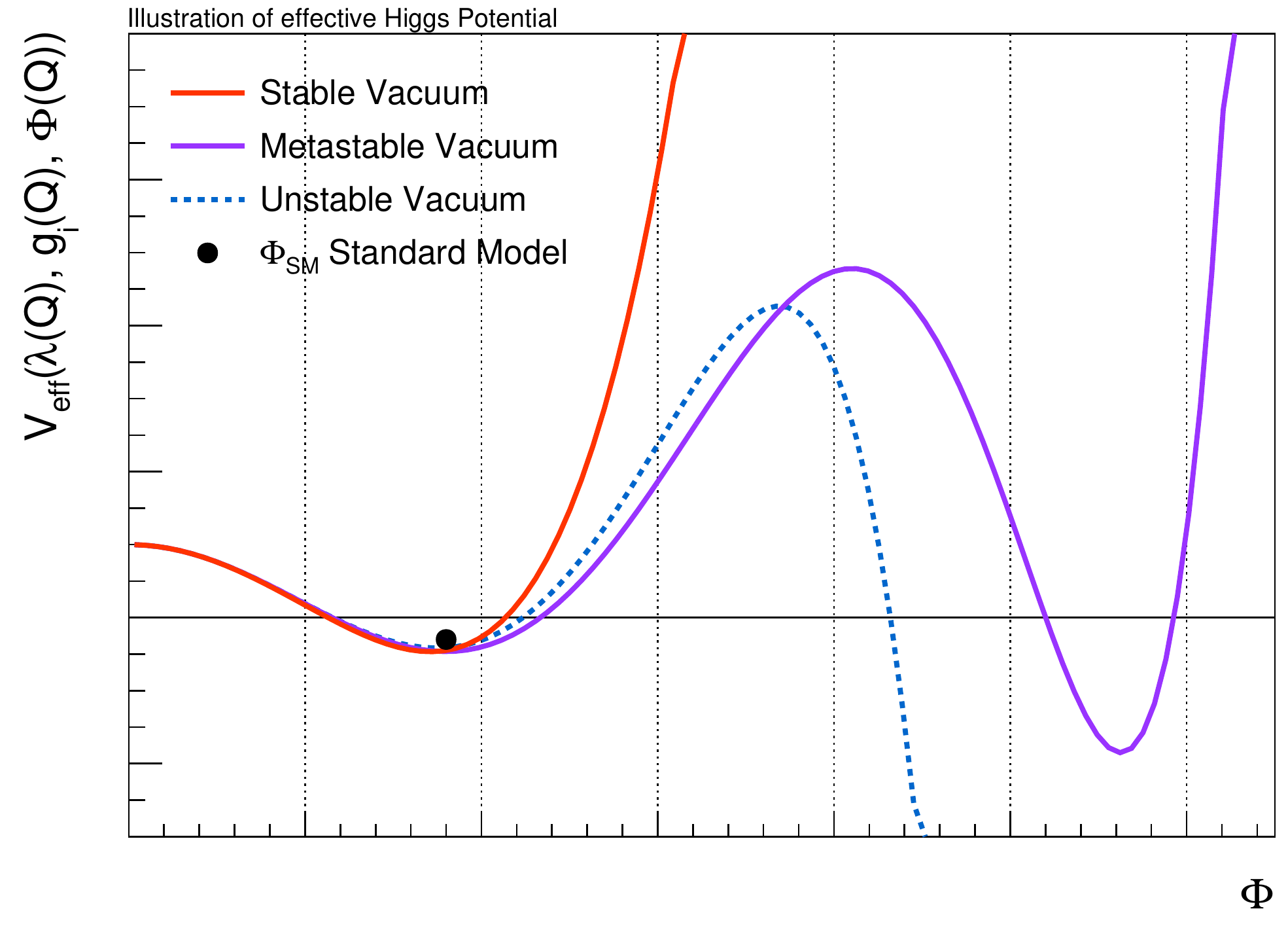}}
\caption{\label{fig:TopVacuum}Illustration of the form of the effective potential for the Higgs field $\phi$ 
for different choices of the Higgs boson self coupling~$\lambda$.}
\end{minipage}
\hspace{0.3cm}
\begin{minipage}{0.40\textwidth}
\resizebox{1.0\textwidth}{!}{\includegraphics{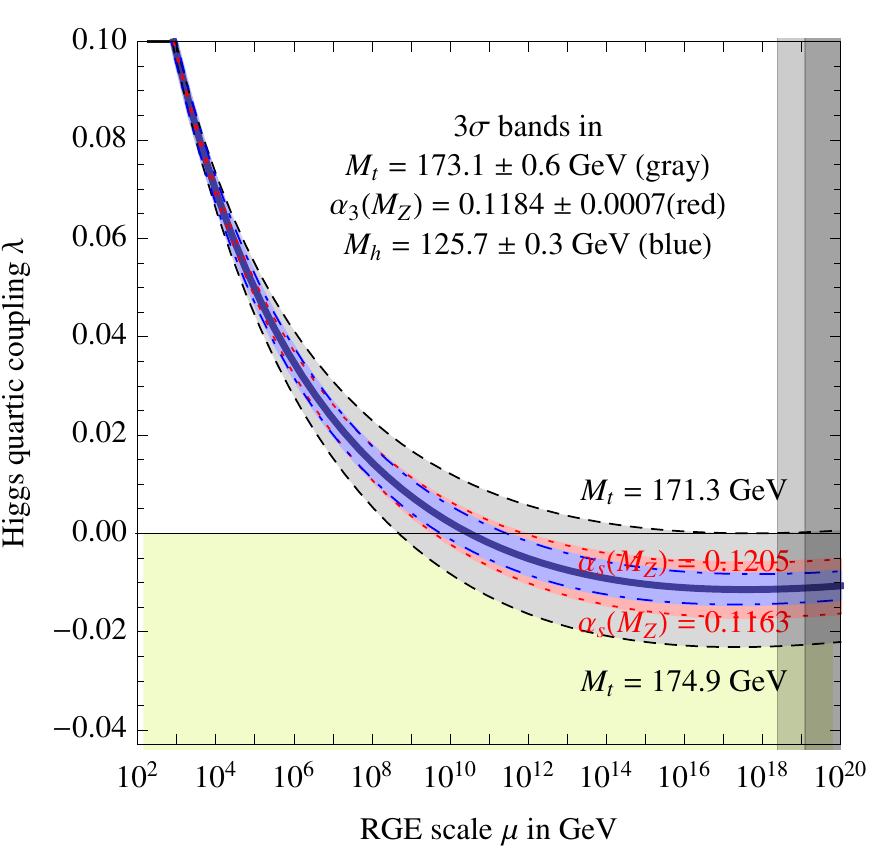}}
\caption{\label{fig:TopInstableVacuumTopMass}Evolution of $\lambda^2$ with energy 
for different values of $\mt$, $\MH$ and $\as$ (from Ref.~\cite{Degrassi:2012ry}).}
\end{minipage}
\end{center}
\end{figure}

In addition to its importance for precision tests of the electroweak sector, the top quark mass also play a decisive role in the extrapolation 
of the Standard Model to high energy scales, far beyond the electroweak scale. 
In the electroweak theory, the ground state of the universe depends on the potential of the Higgs field in Equation~(\ref{eqn:VH}). 
The potential is illustrated for several choices of $\lambda$ in Figure~\ref{fig:TopVacuum}. 
For negative values of $\lambda^2$, the observed minimum of the Higgs potential is only local and a certain tunneling probability through the potential well arises, 
leading to an unstable vacuum~\cite{EliasMiro:2011aa}. 
Due to their large masses the dominant quantum loop corrections to the Higgs boson self-coupling $\lambda$ involve top quarks. 
These corrections can drive $\lambda^2$ to negative values and thus lead to a metastable (long-lived) or to an unstable vacuum. 
However, this argument assumes that no contributions from physics beyond the Standard Model appear up to a very high energy scale, namely the Planck scale. 

It is worthwhile to recall the definition of quark masses in QCD on a pedagogical level, before entering the discussion of the current state of the top quark mass measurements. 
Quark masses enter the QCD Lagrangian as bare parameters and are subject to quantum loop corrections at higher orders. 
Therefore, their values depends on a certain choice of the renormalization scheme. 
A conventional choice in the context of the global electroweak fit is the pole mass, following the standard (and simple) definition of the electron mass in QED. 
This definition of the pole mass is gauge invariant at each order of perturbation theory. 
However, the confinement property of QCD complicates this interpretation, since quarks do not appear as free particles and hence do not generate poles in a complete QCD calculation. 
This ambiguity leads to sizable and irreducible corrections to the pole mass which are on the order of the QCD scale $\Lambda_{\rm QCD}$~\cite{Beneke:1998ui}. 

An alternative way to define the top quark mass is based on the \msbar scheme, 
where the mass is running (scale dependent), analogous to a coupling constant which needs to be specified at a given scale $\mu$ (see Section~\ref{sec:ewpotheory}). 
The pole mass $\mt^{\rm pole}$ and the \msbar mass $\mt^{\msbar}$ are related {\em via\/}
\begin{equation}
\mt^{\rm pole} = \mt^{\msbar} (R,\mu) + \delta\mt(R,\mu),
\end{equation}
where $R$ and $\mu$ are scale parameters \cite{Hoang:2008xm} and the corrections in $\delta \mt$ are known to four loops~\cite{Marquard:2015qpa,Kataev:2015gvt} in QCD. 
The associated uncertainty from converting between both definitions is therefore small but not at all negligible.
Experimentally, the situation is more complicated, since the most precise measurements of $\mt$ rely on template methods, 
similar to those discussed in Section~\ref{sec:wmassdis} for the $W$ boson mass. 
Those are based on a top quark mass parameter incorporated in Monte Carlo event generators, $\mt^{MC}$,
which cannot be simply related to the theoretically better defined $\mt^{\msbar}$ or $\mt^{\rm pole}$. 
The theoretical challenge lies therefore in these relations~\cite{Beneke:2016cbu,Butenschoen:2016lpz,Kataev:2018mob}. 
Details are discussed in the next section.

The experimental uncertainty in $\mt$ is currently about 350~MeV.
This is significantly smaller than the uncertainty in the $\mt$ prediction by the global electroweak fit, which yields an uncertainty of roughly 2~GeV. 
Therefore, an improvement of the current experimental precision will not lead immediately to significant new insights into the internal consistency of the Standard Model. 
However, its precise value can still shed light on the form of the Higgs-potential (Figure~\ref{fig:TopVacuum}). 
The stable and unstable regions of the vacuum in the Standard Model are indicated in Figure~\ref{fig:TopInstableVacuumTopMass} 
together with the experimental values of $\MH$ and $\mt^{\rm pole}$ given in Table~\ref{tab:MTop} with their corresponding uncertainty intervals. 

\subsubsection{Principle and challenges of precision top quark mass measurements}

At the Tevatron and the LHC, top quarks ($t\bar t$) are mainly produced in pairs via gluon fusion and quark-antiquark annihilation processes. 
The branching ratio to one $W$ boson and one $b$ quark is 99.8\% which is due to the corresponding CKM matrix entry of $0.999146$. 
The experimental signature of $t\bar t$ production is therefore defined by the further decay of the $W$ bosons: 
the all hadronic decay channel (46\%) has four light quarks and two $b$ quarks in the final state; 
the semi-lepton decay mode (44\%) involves one hadronic and one leptonic $W$ boson decay and thus has two light quarks, two $b$ quarks, 
one charged lepton and one neutrino in its final state; 
and finally, the full leptonic decay channel (10\%) involves two $b$ quarks, two oppositely charged leptons and two neutrinos. 

After hadronization of the quarks, they are reconstructed as particle jets in the calorimeter systems of the experiments, 
using typically an anti-$k_T$ jet finding algorithm with a typical cone radius of 0.4 or 0.5. 
Based on the properties of the reconstructed secondary vertices and further low energetic lepton reconstruction, 
it is possible to identify particle jets stemming from $b$ quark decays with efficiencies between 0.2 and 0.8, depending on the kinematics, 
the particle detector and the employed tagging approach. 
Charged leptons are reconstructed with a high energy and momentum resolution by the tracking systems of the experiments 
and the electromagnetic calorimeters in the cases of muons and electrons, respectively. 
The neutrinos manifest themselves as missing transverse energy. 

\begin{table}[tb]
\footnotesize
\begin{tabular}{l |c |c |c |c |c |c |c |c |c |c |c |c}

\hline
experiment	&	channel	&	method	& value	& stat.	& syst.	& total	& jet 	& exp. 	& model	& UE + color	&	had.		&	ref.	\\
			&			&			&    		& unc.	& unc.	& unc.	& unc.	& unc.	& unc.	& unc.	&	unc. 		&	\\
\hline
CDF		&	l+jets		&	template		&	172.85	&	0.71	&	0.85		&	1.11	&	0.55	&	0.60	&	0.1	&	0.22	&	0.57	&	\cite{Aaltonen:2012va}	\\
CDF		&	$\nu$+jets 	&	template		&	173.93	&	1.64	&	0.87		&	1.86	&	0.48	&	0.56	&	0.32	&	0.33	&	0.36	&	\cite{Aaltonen:2013aqa}	\\
\DZero	&	l+jets		&	matrix		&	174.98	&	0.58	&	0.49		&	0.76	&	0.29	&	0.32	&	0.19	&	0.12	&	0.26	&	\cite{Abazov:2015spa}	\\
CMS		&	l+jets		&	AMWT		&	172.82	&	0.19	&	1.22		&	1.23	&	0.34	&	0.81	&	0.84	&	0.11	&	0.79	&	\cite{Khachatryan:2015hba}	\\
CMS		&	l+jets		&	ideogram 		&	172.35	&	0.16	&	0.48		&	0.51	&	0.12	&	0.43	&	0.15	&	0.08	&	0.33	&	\cite{Khachatryan:2015hba}	\\
CMS		&	l+jets		&	template 		&	172.22	&	0.18	&	$^{+0.89}_{-0.93}$		&	$^{+0.91}_{-0.95}$	&	0.45	&	0.17	&	0.46	&	0.17	&	0.51 & 	\cite{Sirunyan:2017idq}\\
ATLAS	&	l+jets		&	template 		&	172.33	&	0.75	&	1.03		&	1.27	&	0.64	&	0.62	&	0.48	&	0.19	&	0.18	&	\cite{Aad:2015nba}	\\
\hline
\DZero	&	semi-lep.	&	matrix 		&	173.93	&	1.61	&	0.88		&	1.83	&	0.67	&	0.42	&	0.36	&	0.15	&	0.31	&	\cite{D0:2016ull} 	\\
ATLAS	&	semi-lep.	&	template 		&	172.99	&	0.41	&	0.74		&	0.85	&	0.62	&	0.30	&	0.25	&	0.11	&	0.22	&	\cite{Aaboud:2016igd}	\\
ATLAS	&	semi-lep.	&	template 		&	172.08	&	0.39	&	0.82		&	0.91	&	0.56	&	0.43	&	0.20	&	0.21	&	0.15	&	\cite{Aaboud:2018zbu}	\\
CMS		&	semi-lep.	&	ideogram 		&	172.25	&	0.08	&	0.62		&	0.62	&	0.39	&	0.19	&	0.27	&	0.32	&	0.10	&	\cite{Sirunyan:2018gqx}	\\
\hline
CDF		&	full had. 	&	template 		&	175.07	&	1.19	&	1.55		&	1.95	&	1.12	&	0.98	&	0.28	&	0.32	&	0.29	&	\cite{Aaltonen:2014sea}	\\
ATLAS	&	full had. 	&	template 		&	173.72	&	0.55	&	1.01		&	1.15	&	0.69	&	0.68	&	0.2	&	0.2	&	0.64	&	\cite{Aaboud:2017mae}	\\
CMS 	&	full had.	&	ideogram 		&	172.32	&	0.25	&	0.59		&	0.64	&	0.28	&	0.41	&	0.24	&	0.21	&	0.3	&	\cite{Khachatryan:2015hba}	\\
\hline
\end{tabular}
\centering
\caption{\label{tab:MTop}Overview of kinematic measurements of the top quark mass from the LHC and Tevatron experiments. 
The most precise measurements with a total uncertainty below 2~GeV for each collaboration and for the different decay channels have been selected. 
The entry AMWT refers to an analytical matrix weighting technique.
Experimental uncertainties, which are not associated to particle jets are denoted as 'exp.\ unc.'. 
Model uncertainties summarize uncertainties due to PDFs, scale variation and initial and final state radiation. 
Uncertainties due to color reconnection effects and modeling of underlying event are denoted as 'UE + color'. 
Hadronization uncertainties are summarized under 'had.\ unc.'. 
The relatively large variations in the shown uncertainties are due to different approaches of the collaborations to study systematic effects, 
as well as the interplay between different aspects of the signal modeling, which lead to a different assignment of uncertainties in the various categories. 
All values are given in GeV.}
\end{table}

Thus, a typical experimental selection of top quark pairs involves at least two identified $b$ quark jets 
with a minimal transverse energy of about 30~GeV and further high energetic particles or jets with a similar energy threshold within $|\eta| < 1.0-2.5$, depending on the decay channel. 
In case of at least one hadronically decaying $W$ boson, also a requirement on the invariant mass of the corresponding particle jets is applied. 
Charged leptons in the final state are typically reconstructed within $|\eta| < 2.5$ and are required to have a minimal transverse momentum of $p_T>25$~GeV, 
similar to the minimal $\vec E_T^{\rm miss}$ requirement in case of neutrinos in the final state.

\begin{figure*}[t]
\begin{center}
\begin{minipage}{0.46\textwidth}
\resizebox{1.0\textwidth}{!}{\includegraphics{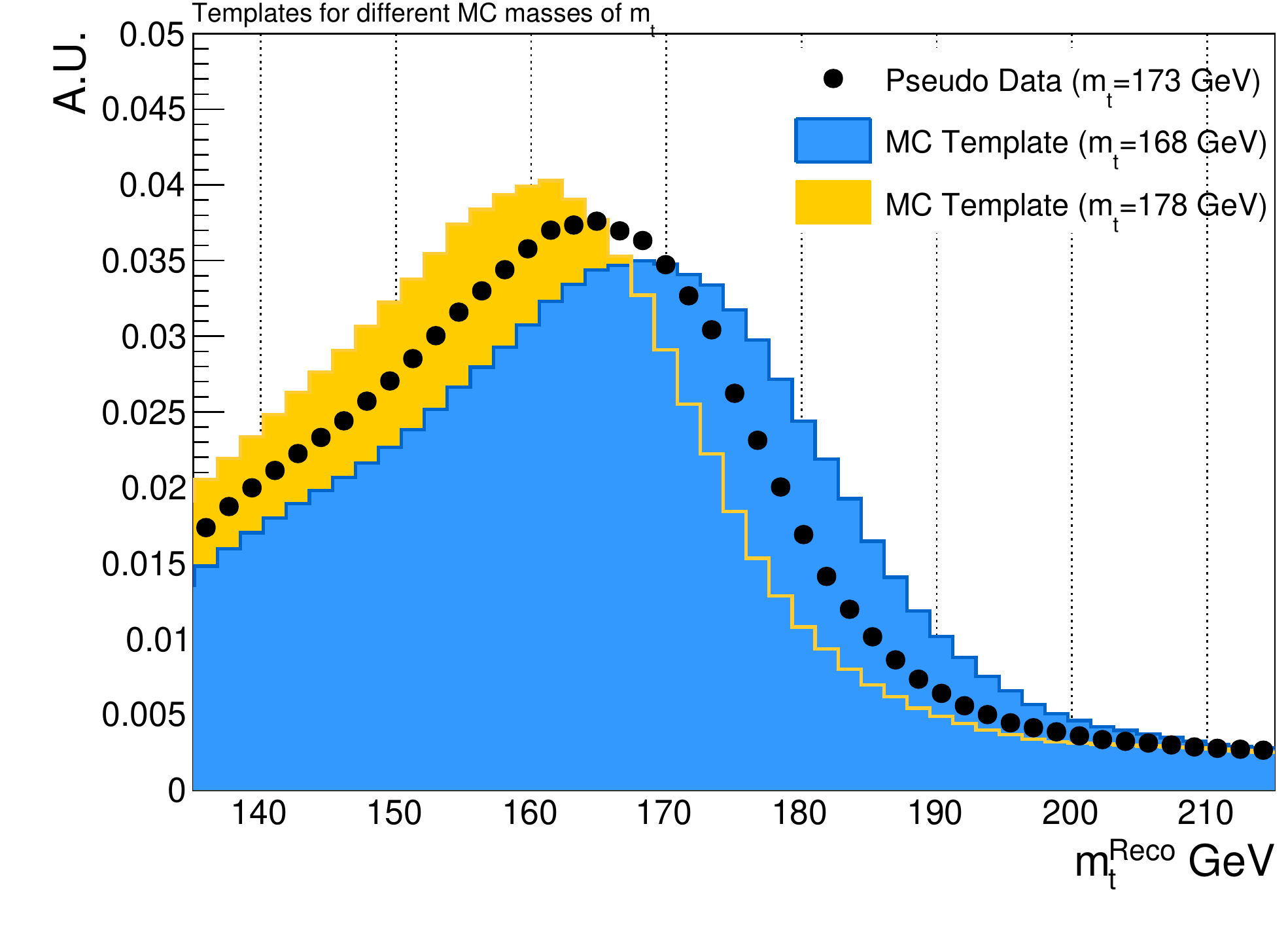}}
\caption{\label{fig:topTemplate}Reconstructed top quark mass for different MC input top quark masses using 
the reconstructed particle jets from a fully hadronically decaying top quark, based on \Pythia8 and the Delphes detector simulation.}
\end{minipage}
\hspace{0.3cm}
\begin{minipage}{0.49\textwidth}
\resizebox{1.0\textwidth}{!}{\includegraphics{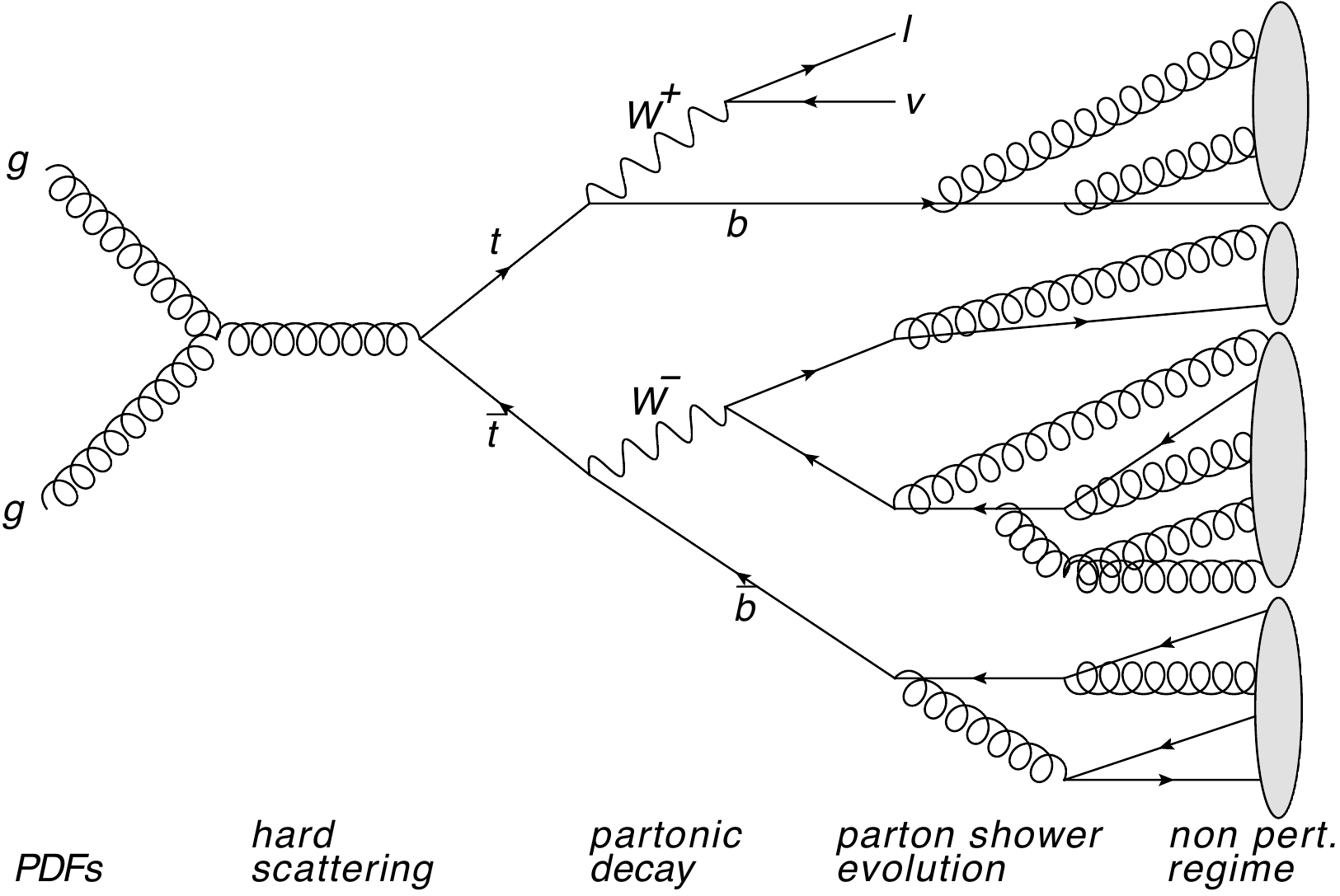}}
\caption{\label{fig:TopMassProblem}Illustration of production and decay of a top quark pair, including the hard scattering processes, 
the partonic decay, the parton shower evolution, as well as the non perturbative regime.}
\end{minipage}
\end{center}
\end{figure*}

Once top quark pair events are selected and background contributions estimated, several approaches to measure $\mt$ can be applied. 
All direct $\mt$ measurements exploit information of the reconstructed kinematics of the measured decay products and their combinations, 
in particular involving reconstructed particle jets and derived observables. 
Three methods have recently been used for the precision measurement of $\mt$:

Similar to the measurement of \MW, the template method relies on distributions which are sensitive to $\mt$, 
such as the invariant mass of two jets stemming from a $W$ boson decay and the associated $b$ quark jet. 
Here, different distributions of kinematic observables corresponding to varying input values of $\mt$ are generated (see Figure~\ref{fig:topTemplate}). 
These distributions involve therefore not only the simulation of the top quark pair decays but also the detector response description. 
The top quark mass is then extracted by comparing these templates to the measured distributions using a $\chi^2$ or a log-likelihood approach. 
In advanced analyses, several input distributions are used in a one-, two- or even multi-dimensional template fit in order to reduce further experimental uncertainties. 

The matrix element method is particular useful for data sets of limited size and was mainly employed by the Tevatron experiments. 
The basic idea of this method is to calculate the probability for observing a given event as a function of the parameters which are to be measured, 
\eg, $\mt$ and quantities that can reduce experimental uncertainties. 
The calculation of such probabilities is based on a (typically leading-order) matrix element, 
incorporating the differential cross sections of the top quark processes relevant to the analysis and the detector resolution~\cite{Abazov:2006bd}. 
The relations between the parton level and the reconstructed four-vectors of final state objects are taken into account using probabilistic transfer functions. 
The maximization of this probability yields the measurement of $\mt$ and further associated parameters. 
The advantage of this method lies in the maximal use of the available statistics of the data samples. 
A dedicated review can be found in Ref.~\cite{Fiedler:2010sg}.

Finally, the ideogram method was partly used by the \DZero and CMS collaborations to measure $\mt$. 
It is based on a heuristic approach for calculating the likelihood of a top quark mass value 
and combines in some sense features of the template method and the matrix element method~\cite{Abazov:2007rk}.
It should be noted that all of these methods determine the top quark mass parameter in the underlying Monte Carlo event generator, $\mt^{MC}$, 
but not directly the pole or the \msbar mass which are actually used in global electroweak fits.

Statistical uncertainties in $\mt^{MC}$ due to the finite size of the data sets are not relevant for the main analyses performed at LHC experiments. 
The associated experimental uncertainties are due to the limited knowledge of the detector response about physics objects used for the event reconstruction. 
While the identification of charged leptons and their energy calibrations are of minor importance, 
the main systematic uncertainty in $\mt^{MC}$ originates from the jet energy scale (JES) of light-quarks ($u, d, c, s$) and gluons,
the scale of $b$-quark-originated jets (bJES), as well as the uncertainties related to the modeling of the $b$-jet identification algorithms. 
The jet energy scale defines the relation between the measured jet energy at reconstruction level to the one at parton level, 
\ie, at generator level before the detector response simulation. 
It corrects for several effects such as irregularities in the calorimeter response, un-instrumented parts of the detector, 
pile-up effects or differences in the electromagnetic and hadronic calorimeters. 
It is typically calibrated with $\gamma$ + jets and $Z$ + jets events, 
where the gauge boson transverse momentum must balance the reconstructed transverse energy of the jets, as well as with studies based on varied MC simulations. 
The uncertainties in the JES vary from 1-3\%, largely depending on the origin of the particle jet. 
It is obvious that this uncertainty will dominate the measurement of $\mt^{MC}$, since jets are a key feature of all $\mt$ sensitive distributions. 
Therefore, most analyses make use of the kinematic constraint of light-quark-flavored jets given by the $W$ boson mass through the $W\rightarrow q \bar q$ decay, 
using a global multiplicative jet energy scale factor in addition to the nominal JES, which is fitted in-situ with the top quark mass measurement. 

In addition to the experimental uncertainties, also uncertainties related to the modeling of top quark pair production and decay have to be considered. 
This modeling depends on the choice of the proton PDFs, the order in $\as$ of the underlying perturbative QCD calculation, 
additional initial and final state radiation and the associated parton shower modeling, as well as the choice of the underlying event and hadronization model. 
A particular problem for top quark decay is the treatment of color reconnection between top and bottom quark color and the color of the other partons in the event. 
Hence, the full event is subject to an interplay between the hard scattering, \ie, a perturbative approach, 
and the modeling of the underlying event, \ie, non-perturbative models (Figure~\ref{fig:TopMassProblem}). 
These non-perturbative models of underlying event and color reconnection can be constrained by studying Drell-Yan events or specific observables in $t\bar t$ events. 
Associated systematic model uncertainties are typically defined by comparing different non-perturbative models. 
Since there is no obvious procedure to decide which non-perturbative models have to be taken into account in such comparisons, 
the final associated uncertainties are ambiguous to a certain extent, leading to different estimates of the different collaborations at the LHC and the Tevatron.

In general, the most precise determinations of $\mt^{MC}$ are achieved in the semi-leptonic decay channel, 
which has a good signal to background ratio and a fully reconstructed event kinematics since the decay neutrino has to match the $W$ boson mass. 
The di-lepton channel typically has the best signal-to-background ratio, but there are two decay neutrinos in the final state, 
leading to an under-constrained system forbidding a complete kinematic event reconstruction. 
The full hadronic decay channel has the worst signal to background ratio. 
Large multi-jet background contributions make sophisticated data-driven techniques for their estimation necessary. 
The advantage of this decay channel lies in the fact that the event kinematics can be fully reconstructed and no neutrinos are involved. 
The uncertainties due to background processes are minor for the leptonic and semi-leptonic decay channels and have only relevance for the fully hadronic decay channel.

During the discussion of the modeling uncertainties in $\mt^{MC}$ several aspects have been mentioned already, 
which become relevant for its interpretation in the context of the global electroweak fit, 
which is based in turn on $\mt^{\rm pole}$ and its relation to a better-defined short-distance definition, such as the \msbar quark mass.
The $\mt^{MC}$ parameter can be interpreted as the mass of the top propagator prior to the top quark decay 
and is not a renormalized field theory parameter~\cite{Jegerlehner:2012kn}. 
The top quark decay process is implemented in all available MC event generators using a parton shower evolution, 
where the splitting probabilities are calculated from perturbative QCD. 
The shower formation stops at a scale below 1~GeV, where one of the available hadronization models takes over, 
as illustrated in Figure~\ref{fig:TopMassProblem}. 
It is important to note that the parton shower, which in some sense describes perturbative QCD corrections, does not take into account any top quark self-energy corrections. 
Therefore, these contributions must be accounted for in $\mt^{MC}$, but only for energy scale above $\sim 1$~GeV, 
since the hadronization model for the event description is employed below that scale. 
As a consequence, $\mt^{MC} \neq \mt^{\rm pole}$, and the numerical difference between these two quantities remains an open question
where values between 0.3 and 1~GeV are under discussion. 
In the following, we assume an additional uncertainty of $320$~MeV in $\mt^{MC}$ within which we identify it with $\mt^{\rm pole}$.
It should be noted that an experimental measurement uncertainty of $\mt^{MC}$ below 300~MeV 
is therefore of limited use in the context of the electroweak fit before its interpretation is clarified. 

\begin{figure*}[t]
\begin{center}
\begin{minipage}{0.48\textwidth}
\resizebox{1.0\textwidth}{!}{\includegraphics{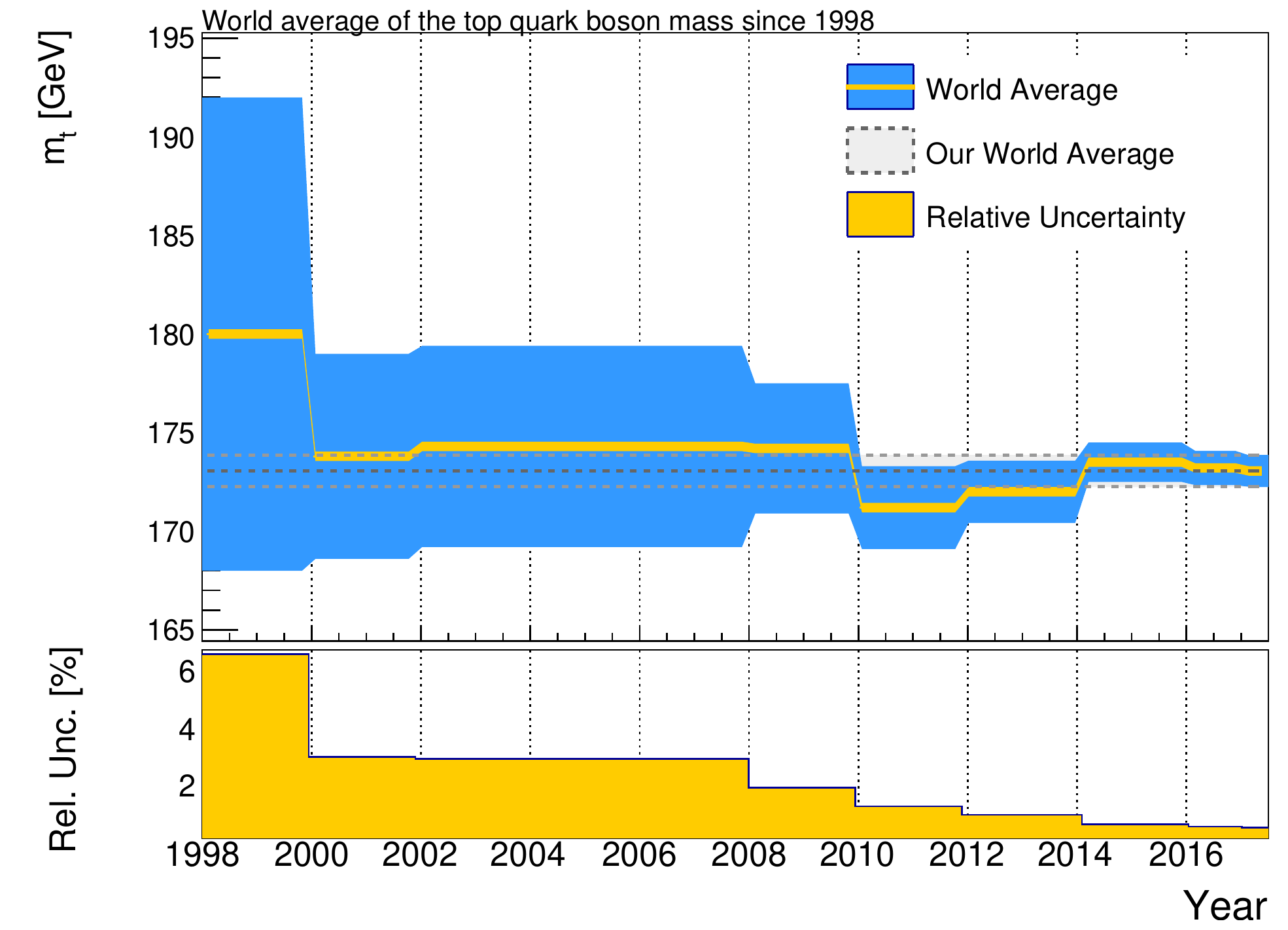}}
\caption{\label{fig:TopMassEvolution}Evolution of the World Average of $\mt$ and its uncertainties {\em vs.\/}\ time. 
Values are taken from previous editions of the PDG review~\cite{Olive:2016xmw} as well as from this article.}
\end{minipage}
\hspace{0.3cm}
\begin{minipage}{0.48\textwidth}
\resizebox{1.0\textwidth}{!}{\includegraphics{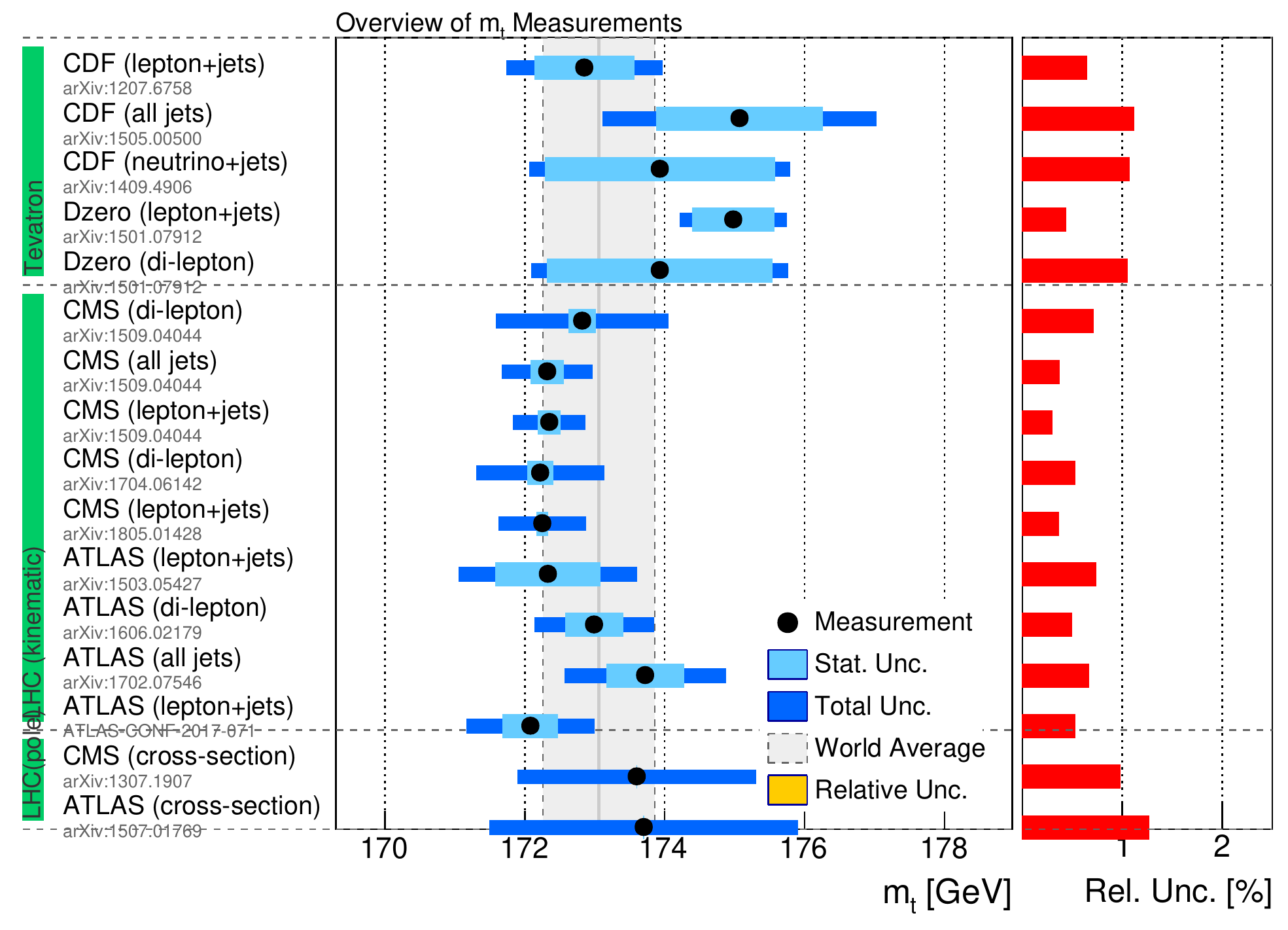}}
\caption{\label{fig:TopSummary}Overview of selected measurements of $\mt$ significantly contributing to the new world average derived here.\vspace{0.3cm}}
\end{minipage}
\end{center}
\end{figure*}

An alternative method to determine $\mt^{\rm pole}$ is based on the mass dependence of the $t\bar t$ production cross section,
$\sigma_{t\bar t}$~\cite{Langenfeld:2009wd, Abazov:2009ae}. 
It is known to next-to-next-to leading order in $\as$ and can be expressed directly in terms of $\mt^{\rm pole}$ or some other convenient mass definition.
Thus, the value of $\mt^{\rm pole}$ can be determined from a precision measurement of $\sigma_{t\bar t}$, 
assuming that the dependence of the measurement has a small (or known) residual dependence on $\mt^{MC}$. 
The ATLAS collaboration used a normalized differential cross section measurement at 8~TeV, in a fiducial region corresponding to the detector acceptance for leptons,
and compared this to NLO fixed-order QCD calculations, exploring the sensitivity of the cross sections to the gluon parton distribution function~\cite{Aaboud:2017ujq}. 
This resulted in $m_t^{\rm pole} = 173.2 \pm 0.9 \pm 0.8 \pm 1.2$~GeV, where the uncertainties are of statistical, experimental and theoretical origin, respectively. 
A similar measurement by CMS yielded $m_t^{\rm pole} = 173.6 \pm 1.7$~GeV, which is illustrated together with the ATLAS result in Figure~\ref{fig:TopSummary} for comparison. 
Once differential NNLO calculations become available, the theoretical uncertainties are expected to shrink significantly, allowing for competitive indirect measurements.

Based on the principle of the cross section determination of $\mt^{\rm pole}$, 
an approach to experimentally estimate the difference between the mass definitions was suggested recently~\cite{Kieseler:2015jzh}. 
A simultaneous determination of $\mt^{MC}$ and of differential or inclusive production cross sections of processes sensitive to $\mt^{\rm pole}$
can constrain their difference to 2~GeV when using current uncertainties. 
In the future, the use of dedicated differential $t\bar t$ distributions could allow to reduce this bound considerably. 

\subsubsection{Discussion of recent top quark mass measurements}

An overview of ten selected precision measurements is given in Table~\ref{tab:MTop}, including a breakdown of the different systematic uncertainties. 
None of the measurements is limited by statistical precision. 
The most precise measurements of $\mt^{MC}$ have been achieved 
in semi-leptonic decay channels using kinematic constrains to reduce jet-energy scale related uncertainties. 
The latter are different among the experiments, even though a similar data set size was used for the calibrations. 
The reason for this are different assumptions about the systematic uncertainties of the underlying calibration methodologies.
It is interesting to note that also the model uncertainties significantly differ between the experiments, even though the physics is similar. 
Again, different assumptions have been made for this evaluation by the experimental collaborations. 

For the combination of $\mt^{MC}$ within this article, we treat the statistical uncertainties as uncorrelated among all measurements. 
Experimental systematic uncertainties are assumed to be uncorrelated among experiments 
but partially correlated among measurements within one experiment. 
The correlations of model uncertainties between experiments cannot be defined unambiguously and we tested several assumptions.

We restrict ourselves to the combined value from the Tevatron~\cite{TevatronElectroweakWorkingGroup:2016lid}, 
as well as to the most precise available combinations of ATLAS~\cite{Aaboud:2018zbu} and CMS~\cite{Khachatryan:2015hba} at Run~I 
and the first precision measurement of CMS at a center-of-mass energy of 13~TeV~\cite{Sirunyan:2018gqx}. 
The ATLAS value of $\mt^{MC} = 172.69 \pm 0.25~(\rm stat.) \pm 0.41~(\rm syst.)$~GeV combines a measurement in the semi-leptonic decay channel 
at $\sqrt{s}=8$~TeV with six previous measurements of ATLAS, where the small overall uncertainty is a result of the careful study of correlations among the measurements. 
The CMS value combines the measurements of $\mt^{MC}$ in the semi-leptonic decay channel at $\sqrt{s} = 7$  and 8~TeV, 
leading to a value of $\mt^{MC} = 172.44 \pm 0.13~(\rm stat.) \pm 0.47~(\rm syst.)$~GeV. 
The latest official combination of the top quark mass by the \DZero and CDF experiments~\cite{TevatronElectroweakWorkingGroup:2016lid} 
yielded $\mt^{MC} = 174.30 \pm 0.35~(\rm stat.) \pm 0.54~(\rm syst.)$~GeV. 
The correlations used for the combination here are estimated using the published values from the LHC and Tevatron working groups~\cite{ATLAS:2014wva}. 
A first combination using the \textsc{Blue}-method yields a value of 
\[\mt^{MC} = 172.90 \pm 0.35~{\rm \GeV}, \]
with a probability of 4.1\%. 
The combined value, the individual measurements, and the published combined values from the experiments 
are illustrated in Figure~\ref{fig:TopSummary}. 
While the results of ATLAS, CMS and CDF are in very good agreement 
with each other\footnote{A combination of ATLAS, CMS and CDF yields a value of $172.59 \pm 0.41$~GeV.}, 
a tension at the $3 \sigma$ level can be observed w.r.t.\ the most precise measurement of \DZero~\cite{Abazov:2015spa}. 

Assuming an additional uncertainty of 320~\MeV to account for the ambiguities in the relations between the various definitions of the top quark 
mass\footnote{As a representative of errors of this type, we take here the size of the forth order term~\cite{Marquard:2015qpa} between $\mt^{\rm pole}$ and $\mt^{\msbar}$.}, 
we find
\[\mt^{\rm pole} = 172.90 \pm 0.47~{\rm \GeV}. \]
The evolution of the world average of the top quark mass and its associated uncertainty from 2000 to 2018 
is illustrated in Figure~\ref{fig:TopMassEvolution}, and is based on values of the PDG and this article. 
Until 2011 the world average value was dominated by the Tevatron experiments. 
The first competitive top quark mass measurements by the LHC collaborations have been published 
in 2012~\cite{Chatrchyan:2012ea} and 2013~\cite{Aad:2015nba}, respectively. 
The aforementioned uncertainty of 320~MeV in $\mt^{MC}$ is already dominant. 
Until the relation between $\mt^{MC}$ and $\mt^{\rm pole}$ has been clarified, 
further precision measurements of $\mt$ using purely kinematic methods will be of limited use in view of the global electroweak fit.
  
Upcoming approaches for direct $\mt^{\rm pole}$ measurements using differential cross section predictions of $\sigma_{t\bar t}$ in higher-order 
perturbation theory might significantly reduce the theory uncertainty in $\mt$. 
Assuming improvements in the measurement methodologies, \eg, using approaches to constrain the difference 
between $\mt^{\rm pole}$ and $\mt^{MC}$, as well as advanced theoretical tools~\cite{Beneke:2016cbu,Butenschoen:2016lpz} in coming years, 
might allow for a total uncertainty in $\mt^{\rm pole}$ well below 500~MeV. 
A significantly higher precision could only be reached at an $e^+ e^-$ collider at sufficiently large energy to perform threshold scans of $t\bar  t$ production.

\subsection{Vacuum polarisation\label{sec:dalpha}}

\begin{figure}[t]
\begin{center}
\begin{minipage}{0.20\textwidth}
\resizebox{1.0\textwidth}{!}{\includegraphics{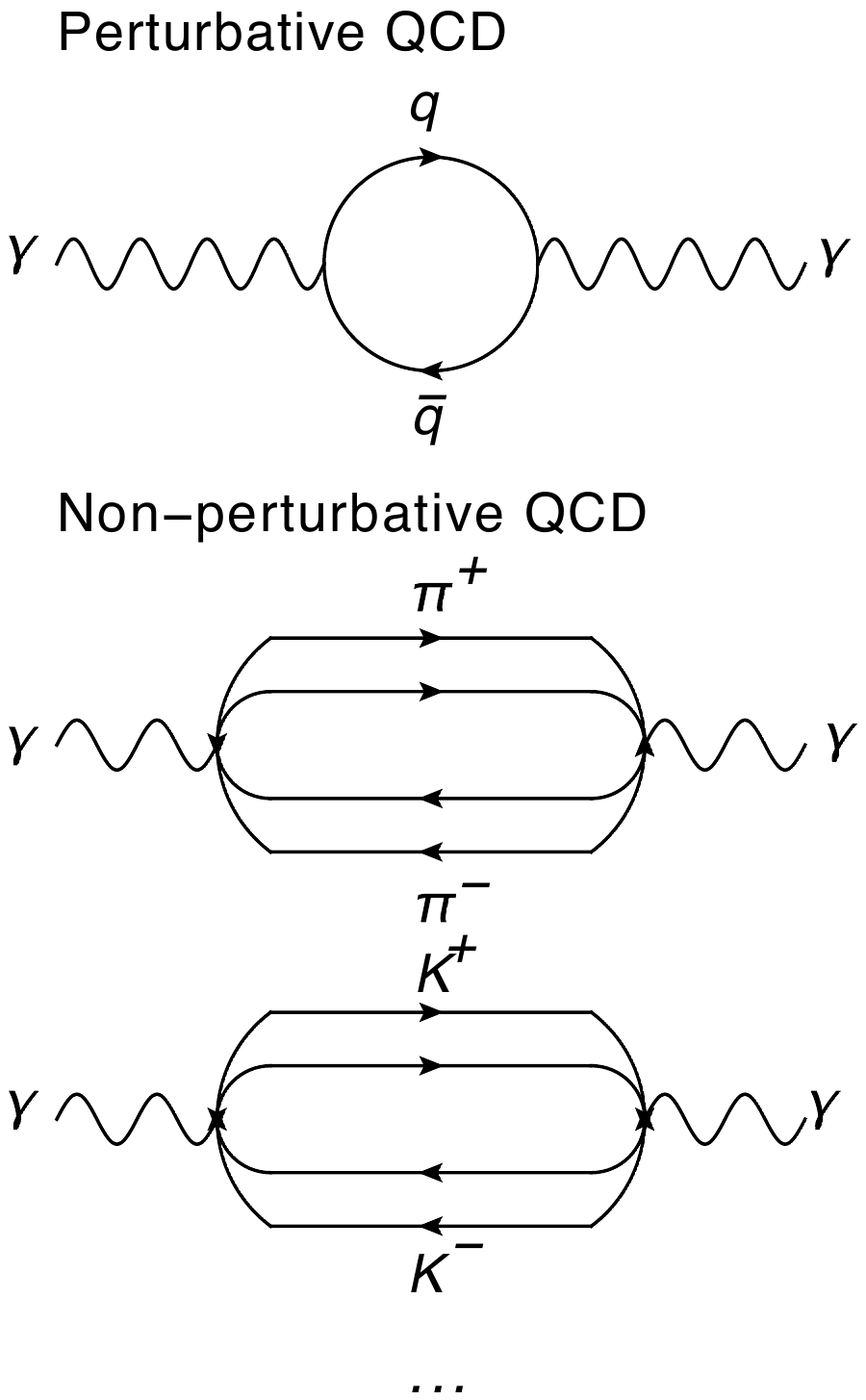}}
\caption{\label{fig:AlphaHad1}Example Feynman diagrams for perturbative and non-perturbative contributions to the running of $\alpha$.
\vspace{0.1cm}}
\end{minipage}
\hspace{0.3cm}
\begin{minipage}{0.22\textwidth}
\resizebox{1.0\textwidth}{!}{\includegraphics{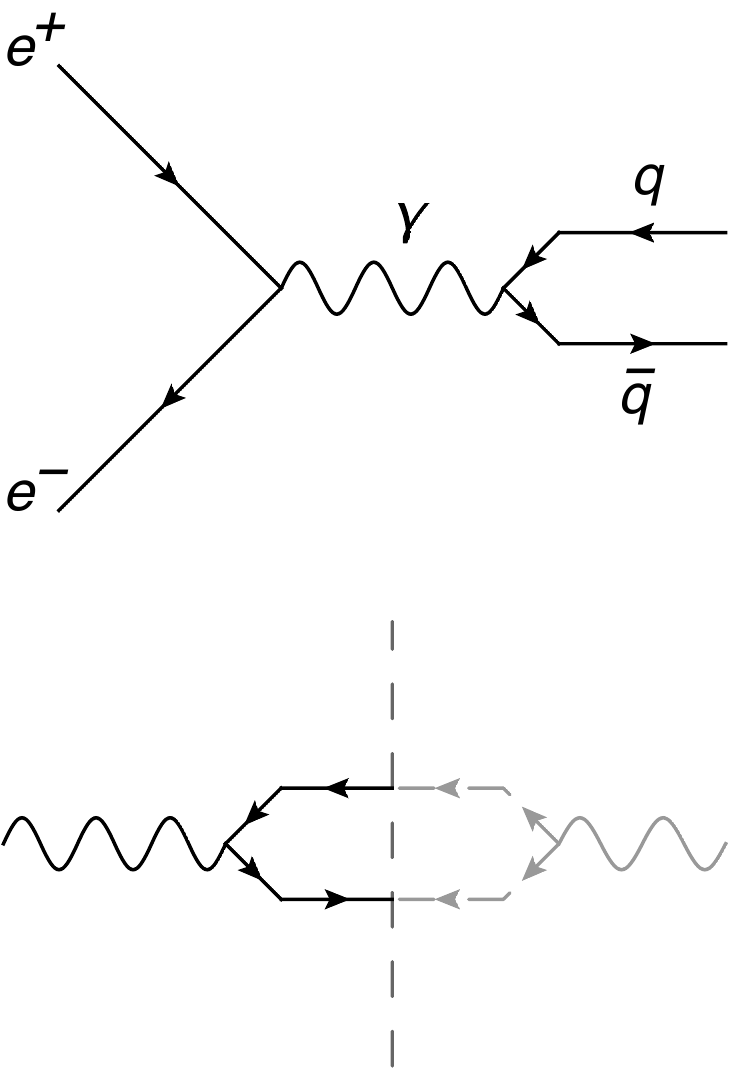}}
\caption{\label{fig:AlphaHad2} Illustration of the usage of inclusive hadron production in $e^+e^-$ for the evaluation 
of the $\Delta \alpha_{\rm had}$ loop contributions.}
\end{minipage}
\hspace{0.3cm}
\begin{minipage}{0.45\textwidth}
\resizebox{1.0\textwidth}{!}{\includegraphics{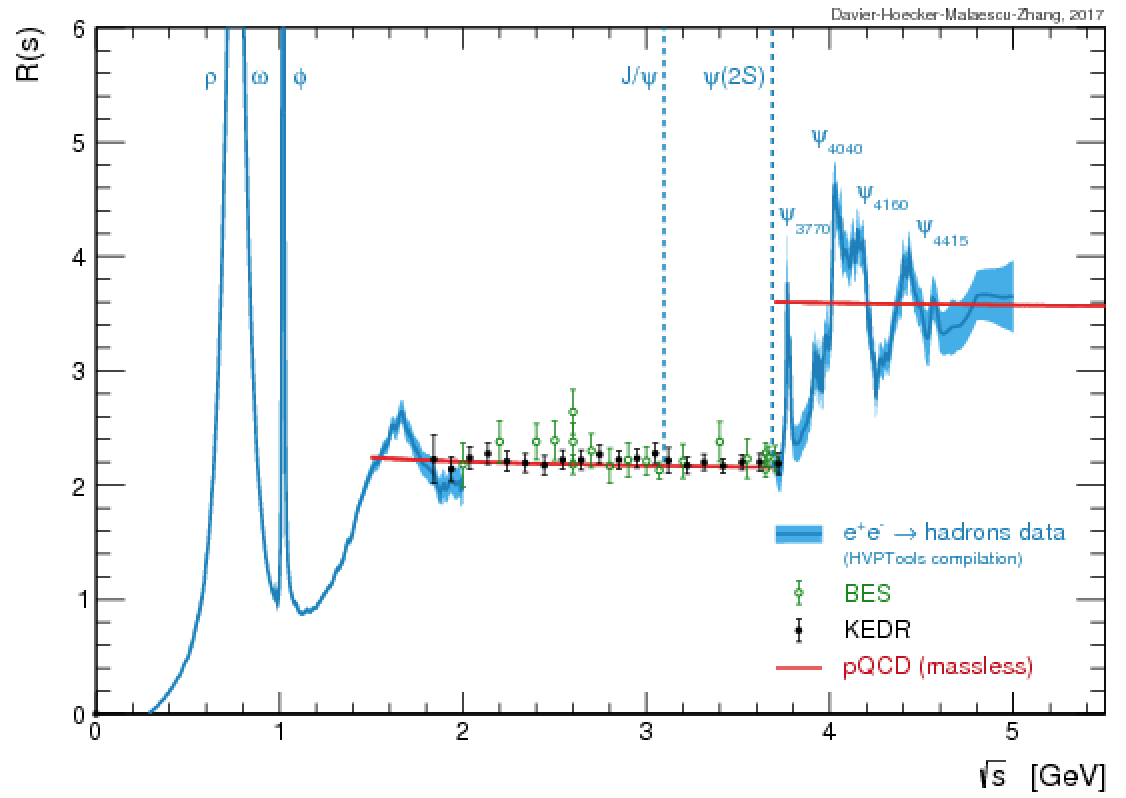}}
\caption{\label{fig:AlphaR} The total hadronic $e^+e^-$ annihilation rate $R$ as a function of $\sqrt{s}$. 
Inclusive measurements from BES and KEDR are shown as data points, while the sum of exclusive channels from the analysis 
in Ref.~\cite{Davier:2017zfy} is given by the narrow blue bands.}
\end{minipage}
\end{center}
\end{figure}

The predictions of the electroweak fit require precise knowledge of the electromagnetic coupling strength at the $Z$ boson mass at the few per mille level or better.
As discussed in Section~\ref{sec:constants}, the dominant uncertainty in the running of $\alpha$ is from the hadronic contribution 
of the five lighter quarks $(u,d,s,c,b)$ to the vacuum polarization, $\Delta \alpha_{\rm had}$. 
While perturbative QCD can be used to calculate $\Delta \alpha_{\rm had}$ for energies above about 2~GeV using the results of calculations up to four-loops 
in $\alpha_s$~\cite{Baikov:2008jh}, experimental data and phenomenological models have to be used to estimate the contributions of the running at energies below that. 
Examples of Feynman diagrams for the perturbative and non-perturbative contributions to the running of $\alpha$ is illustrated in Figure~\ref{fig:AlphaHad1}.

The basic idea of the determination of $\Delta \alpha_{\rm had}$ is the evaluation of energy-squared dispersion integrals ranging from the $\pi^0\gamma$ threshold to
infinity~\cite{Cabibbo:1961sz} using a combination of experimental data from $e^+e^-$ annihilation into hadrons ($\tau$ decay spectral functions can provide additional information). 
By 'mirroring' the underlying Feynman diagrams, information on the corresponding loop contributions can be extracted (Figure~\ref{fig:AlphaHad2}). 
In one of the most recent evaluations~\cite{Davier:2017zfy}, 39~channels of exclusive hadronic cross section measurements were used, 
where the total hadronic $e^+e^-$ annihilation rate $R$ as a function of the center-of-mass energy is shown as an example of parts of this input data in Figure~\ref{fig:AlphaR}.

There are two different approaches how to handle the charm and bottom quark contributions and the perturbative regions.
The more traditional way is to use the experimental electronic widths of the narrow resonances $J/\psi$, $\psi(2S)$, and $\Upsilon(nS)$ for $n=1,2,3$, 
supplemented by the measured rate $R$ in the energy ranges between 3.7 and 5.0~GeV, as well as between 10.6 and 11.2~GeV, in the dispersion integral.  
In the remaining regions $R(s)$ is computed in QCD perturbation theory.
The result are constraints for $\alpha^{(5)}(\MZ)$ (in the on-shell scheme) which can be used as external constraints in electroweak fits (as done, \eg, by ZFITTER and Gfitter).

Alternatively~\cite{Erler:1998sy}, one can gain precision by using the renormalization group equation (RGE) for $\alpha$, \ie, to compute 
the five-flavor definition $\alpha^{(5)}(\MZ)$ (in the \msbar scheme) in terms of the three-flavor quantity $\alpha^{(3)}(2~{\rm \GeV})$. 
The charm and bottom quarks are included by RGE matching conditions at $m_c$ and $m_b$, where the expansion coefficients of 
the anomalous dimension of the photon are updated to include the corresponding quark. 
In this approach $\alpha^{(5)}(\MZ)$ is computed in each call of the fits, which accounts for correlations with $\alpha_s$, $\sin^2\theta_W(0)$, $m_c$, $m_b$, and $a_\mu$ (see below).
This approach is implemented into GAPP and provides the basis of the corresponding results published by the PDG.
Recent values for $\Delta \alpha^{(5)}_{\rm had}$ from different approaches were discussed in Section~\ref{sec:constants}.

The same data constraining $\Delta \alpha_{\rm had}$ are also being used for the prediction of the anomalous magnetic moment of the muon, 
$a_\mu = (g_\mu-2)/2$, where currently a deviation of more than three standard deviations between the experimental value \cite{Bennett:2006fi} 
and the SM prediction~\cite{Jegerlehner:2009ry} is observed. 
In both cases, the errors are dominated by the uncertainties in the experimental data due to the $\pi^+\pi^-$ channel used to calculate the dispersion integral~\cite{Davier:2017zfy}.
However, the kernel functions in the dispersion integrals are different, where lower scales enter with greater weights (roughly by an extra factor of $s^{-1}$) into the calculation of $a_\mu$.

\begin{figure*}[t]
\begin{center}
\begin{minipage}{0.48\textwidth}
\resizebox{1.0\textwidth}{!}{\includegraphics{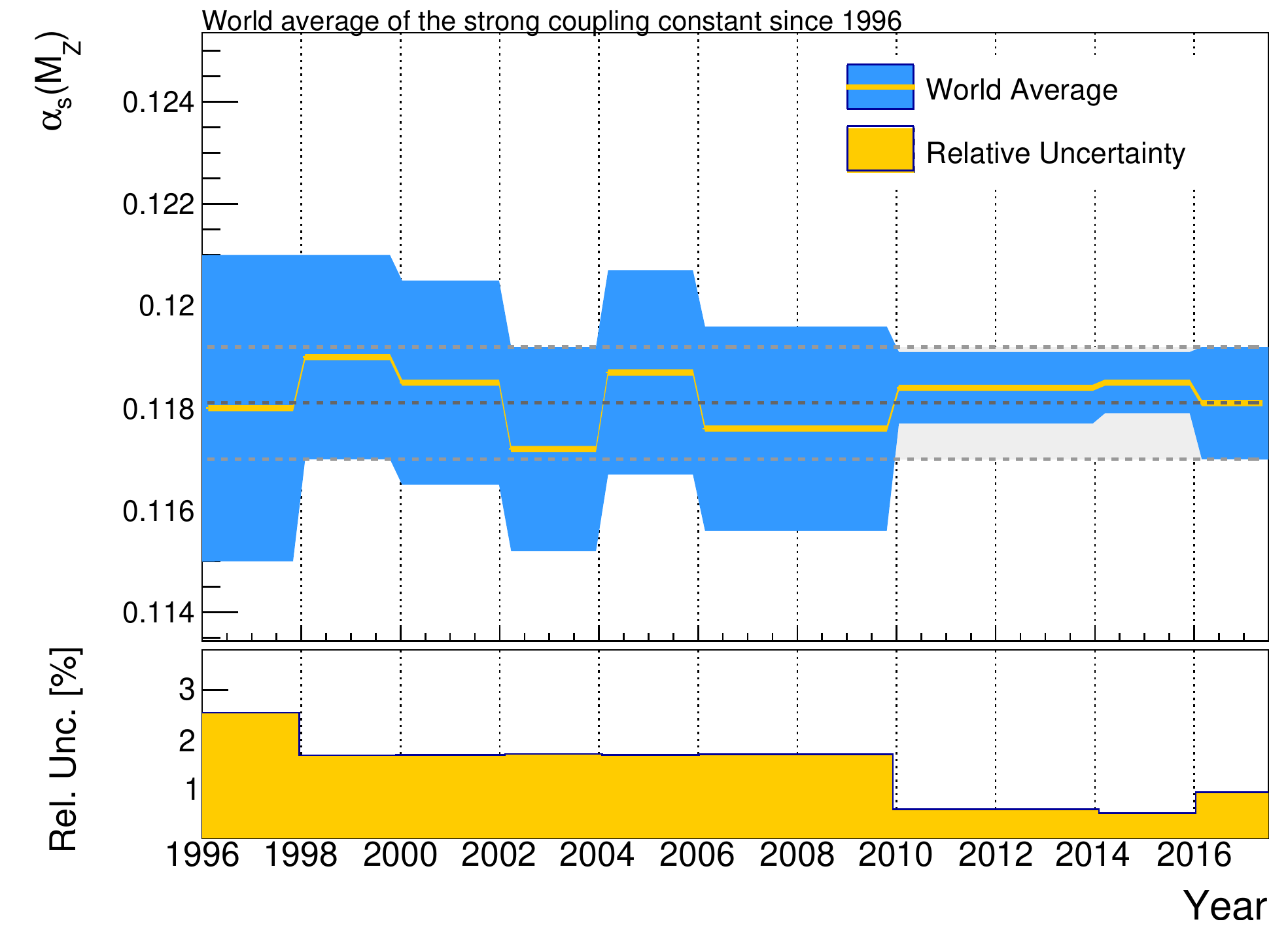}}
\caption{\label{fig:alphaSEvolution}Evolution of the world average of $\as(\MZ)$ and its uncertainties in time. 
Values are taken from previous editions of the PDG review~\cite{Olive:2016xmw}.}
\end{minipage}
\hspace{0.3cm}
\begin{minipage}{0.48\textwidth}
\resizebox{1.0\textwidth}{!}{\includegraphics{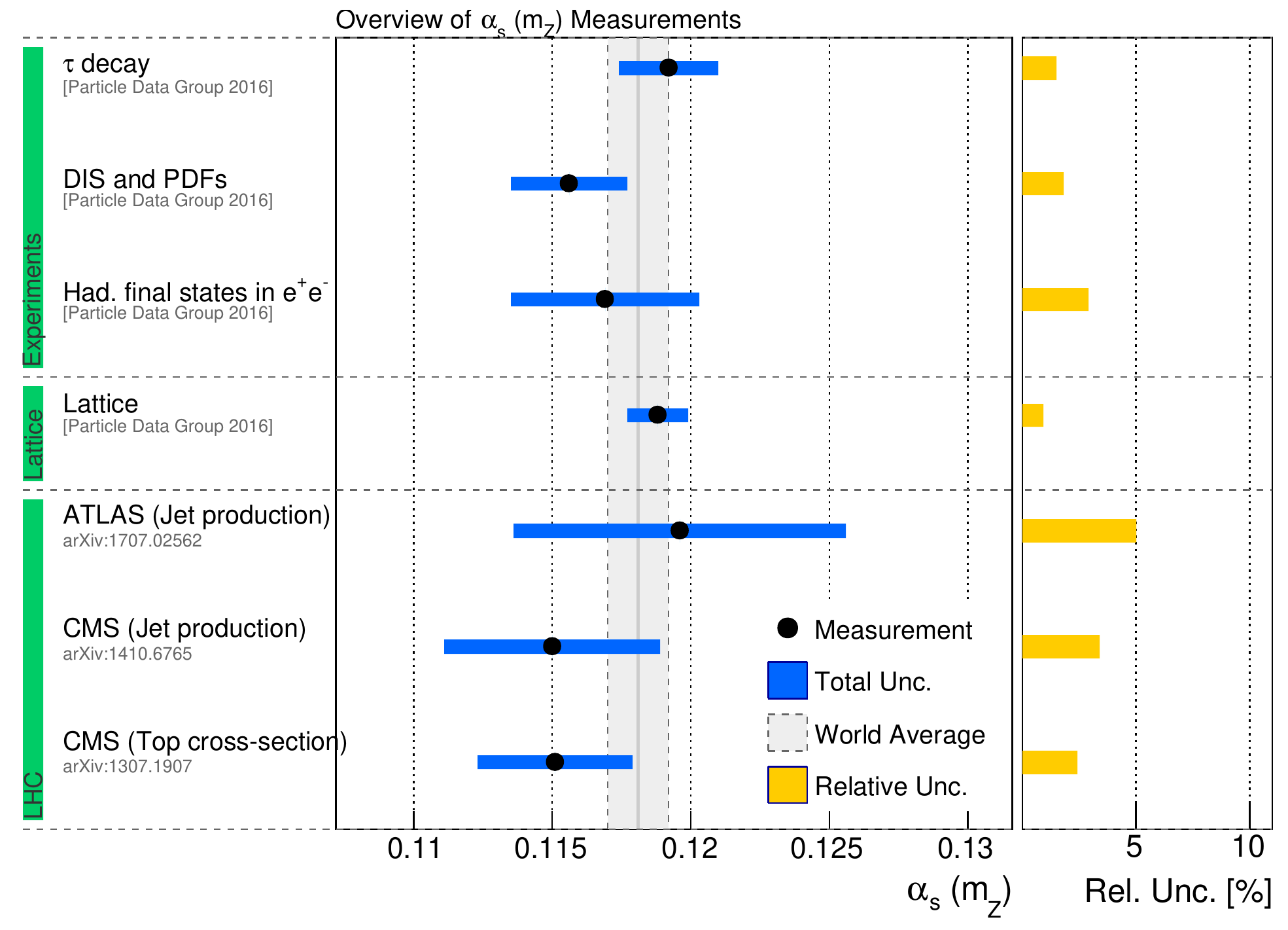}}
\caption{\label{fig:alphaSAverage}Overview of selected measurements of $\as(\MZ)$ in comparison to the world average.\vspace{0.4cm}}
\end{minipage}
\end{center}
\end{figure*}

\subsection{Strong coupling constant\label{sec:alphas}}

All predictions of QCD processes at the LHC rely on perturbative approaches, \ie, they require that the strong coupling constant is sufficiently small. 
While the coupling strength at the $Z$ pole is $\as(\MZ)\approx 0.12$, it rises dramatically at small energy scales near the GeV region. 
As discussed in Section~\ref{sec:constants}, the evolution of $\as(Q)$ is known to four-loop expansion and includes three-loop matching at the quark-flavour thresholds.  
Hence, $\as$ can be determined from several processes at various energy scales and then extrapolated to a reference scale, \eg, $Q = \MZ$, for comparison. 

While the precision of the fine structure constant reaches 32 parts per billion~\cite{Agashe:2014kda}, 
and $\alpha(M_Z)$ is known to a relative precision of $10^{-4}$, $\as(\MZ)$ is only known at the \% level. 
This has not only implications for unification studies of SM gauge couplings at some high energy scale, but also directly impacts the prediction of QCD induced processes at the LHC. 
Determinations of $\as$ have been performed using a wide range of approaches, 
\eg, studying $e^+e^-$ annihilations, deep-inelastic lepton-nucleon scattering, resonance and $\tau$ decays, and hadron collisions. 
For a detailed reviews, we refer to Refs.~\cite{Dissertori:2015tfa,Bethke:2009jm}. 
  
Recent measurements of correlation parameters and angular distributions of particle jets at the LHC can test the evolution of $\as$ to 
the TeV regime~\cite{Aaboud:2017fml, Khachatryan:2014waa}, but reach a relative precision of only 5\% in $\as(\MZ)$ which is due to missing higher order corrections in the predictions. 
Calculations beyond NLO for differential jet cross sections would therefore allow for a significant reduction in the uncertainty of $\as$ also at high energy scales. 

Lattice QCD calculations for observables such as hadron mass splittings currently provide the determinations of $\as$ with the smallest quoted uncertainties.
{\em E.g.}, by comparing data to lattice predictions, the value,
\[\as(\MZ) = 0.1184\pm0.0006,\]
was extracted in Ref.~\cite{Davies:2003ik}, which corresponds to a precision of 0.5\%. 
However, it is currently under discussion, whether the uncertainty is not underestimated~\cite{Dissertori:2015tfa}. 

An alternative approach relies on $Z$ pole observables such as $R_l$, $\Gamma_Z$ and $\sigma_0$,
from where one can determine the strong coupling constant by treating it as a free parameter in electroweak fits.
This is discussed in more detail in Section \ref{sec:ImpactOfFit}. 
An overview of the most precise measurements of $\as$ from different approaches is shown in Figure~\ref{fig:alphaSAverage}, 
and the evolution of the world average in time is displayed in Figure~\ref{fig:alphaSEvolution}. 
It should be noted, that the increase of the uncertainty in the world average of $\as$ in 2016
was due to a change in the combination procedure, where the PDG approach for conflicting measurements was applied.

\subsection{The Fermi constant}

As discussed in Section~\ref{sec:theofermi}, the Fermi constant is extracted accurately from the measurement of the muon lifetime. 
The experimental approach is based on the detection of decay positrons from stopped muons in various targets, 
\ie, stemming from the reaction $\mu^+\rightarrow e^+ \bar \nu_\mu \nu_e$. 
The initial muons, typically provided as a particle beam with a well defined energy, are slowed down in a condensed matter target, 
mainly by ionization and excitation processes. 
In the final stages, the muons exist either in muonium atoms or charged muon ions. 
The count rate of decay positrons {\em vs.\/}\ time can then be used to determine the average muon lifetime.

\begin{figure*}[t]
\begin{center}
\begin{minipage}{0.48\textwidth}
\centering
\resizebox{0.86\textwidth}{!}{\includegraphics{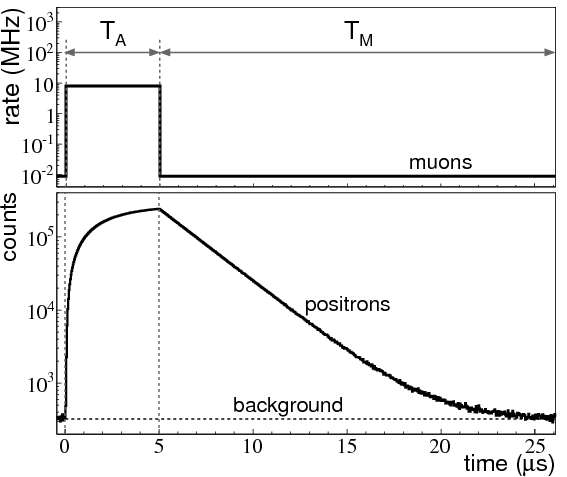}}
\caption{\label{fig:Fermi1}Plot of the muon arrival times (upper panel) and decay positron times (lower panel) that are produced by the pulsed beam technique (from Ref.~\cite{Tishchenko:2012ie}).}
\end{minipage}
\hspace{0.3cm}
\begin{minipage}{0.48\textwidth}
\resizebox{1.0\textwidth}{!}{\includegraphics{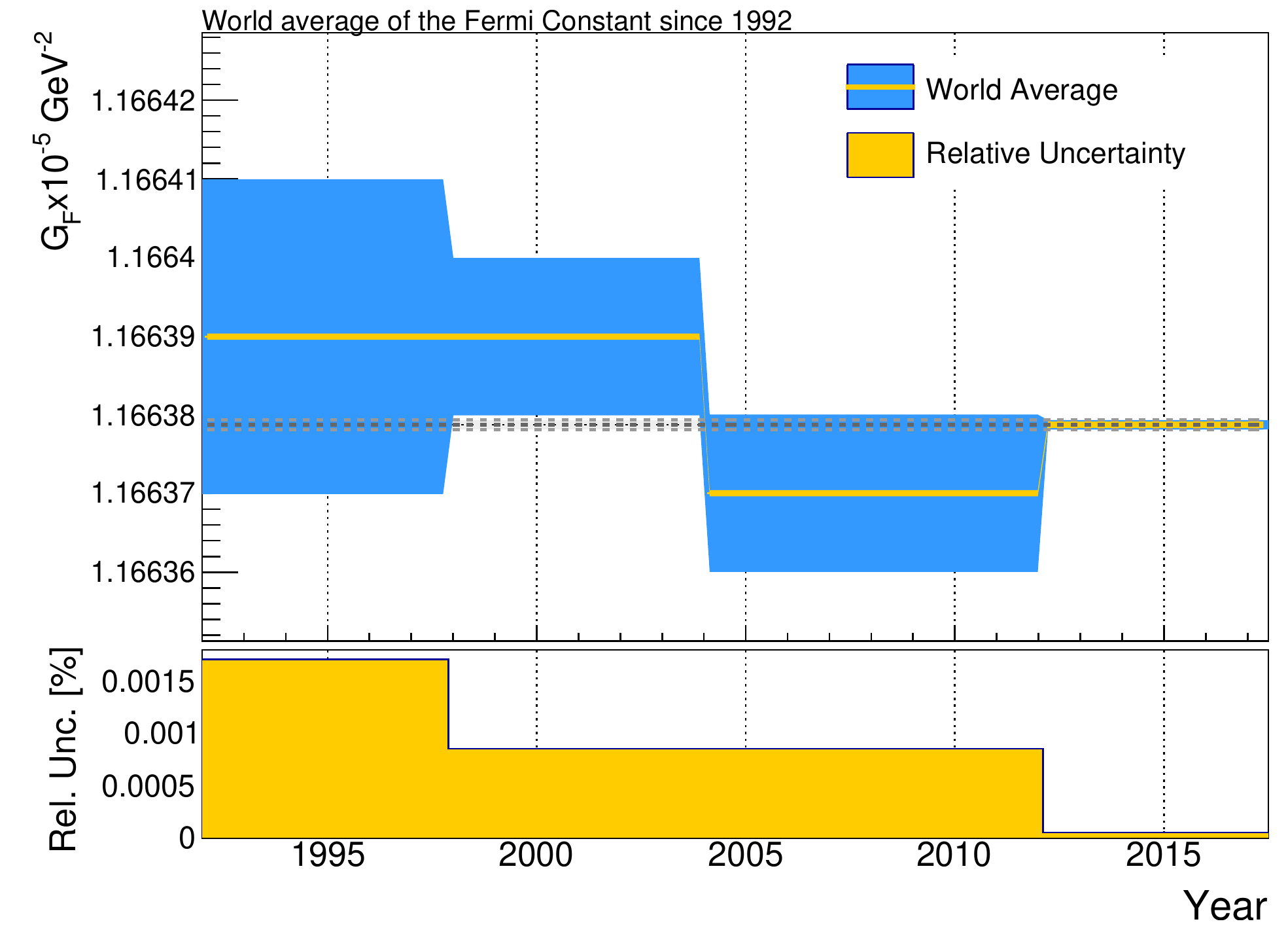}}
\caption{\label{fig:Fermi2}Evolution of the world average of $G_F$ and its uncertainties in time.\vspace{0.4cm}}
\end{minipage}
\label{Labelname}
\end{center}
\end{figure*}

Older experiments~\cite{Balandin:1975fe,Giovanetti:1984yw} used a continuous incoming muon beam 
and therefore had to reduce the rate in order to avoid miscorrelations of decay positrons with their parent muons. 
Hence these measurements were limited by the available data statistics. 
This limitation was overcome by using pulsed beams, where a sufficient number of muons are trapped during the beam period 
and then the decay positrons are counted during a subsequent beam-off period. 
The first measurements, using this approach was performed at Saclay~\cite{Bardin:1984ie}, but also with limited data statistics due to the available duty cycle of the accelerator. 

The most precise measurements of $G_F$ have been performed at the Paul-Scherrer Institute. 
The FAST collaboration~\cite{Barczyk:2007hp} used a finely segmented active target to detect the decay positrons, allowing for a higher beam rate as previous experiment. 
The MuLan collaboration~\cite{Chitwood:2007pa,Webber:2010zf} based their measurement on a dedicated pulsed, 
100\% longitudinally polarized, 29~MeV muon beam, with 5 $\mu$s-long beam-on and subsequent 22 $\mu$s beam-off time segments.
The beam times and the measured count rate for one cycle is illustrated in Figure~\ref{fig:Fermi1}. 
The fast switching between beam-on and -off was achieved by a specially designed electrostatic kicker, which was imposed on the continuous muon beam. 
The positron detector consisted of 170 triangle-shaped plastic scintillators on a sphere geometry around the target. 
Two target materials (Fe-Cr-Co and SiO$_2$) have been used, allowing for the study of several systematic effects due to the initial muon polarization. 
In total, $1.6 \times 10^{12}~\mu^+$ decays have been recorded and analyzed. 
The final result by the MuLan collaboration~\cite{Tishchenko:2012ie} is $\tau_\mu = 2 196 980.3 \pm 2.1(\rm stat.) \pm 0.7(\rm syst.)$~s, which translates to a Fermi constant of
\[G_F=1.1663787(6)\times10^{-5}~\GeV^{-2}~(0.5~{\rm ppm}). \]
This value is compatible with most previous measurements, except for a $2.9\sigma$ discrepancy to the measurement by the FAST Collaboration~\cite{Barczyk:2007hp}. 
An overview of selected $G_F$ measurements, together with the world average, is given in Figure~\ref{fig:Fermi2}.

\section{The Global Electroweak Fit\label{sec:SMfit}}

\subsection{Fitting programs\label{sec:fitprograms}}

As mentioned in the introduction, there are a number of independent computer packages that can be used to perform global electroweak fits,
or are even designed to do so.
The results presented here were based on two of these, namely GAPP and Gfitter.
Indeed, it is important to confirm the results of the fits by using routines that differ in computer language, renormalization scheme, 
implementation, and details regarding the employed data sets. 
In view of these differences which will be briefly described below, it is reassuring that the results from GAPP and Gfitter are generally in very good agreement with each other,
where residual variations are well understood.

Both, GAPP \cite{Erler:1999ug} and Gfitter ~\cite{Flacher:2008zq, Haller:2018nnx}, are generic fitting packages comprising frameworks for the statistical analyses of parameter estimation problems in high energy physics. 
They are specifically designed for involved fitting problems, such as the global SM fit to electroweak precision data. 
Both use the highly efficient minimization program MINUIT~\cite{James:1975dr} to minimize likelihood functions which are
--- at least for the most part --- multivariate Gaussian ($\chi^2$) distributions.
MINUIT returns $1\sigma$ errors as $\Delta \chi^2 = 1$ ranges relative to the global minimum, which are allowed to be asymmetric. 

GAPP is a FORTRAN library written specifically for the calculation of pseudo-observables, \ie, idealized quantities such as the weak mixing angle, the $W$ boson mass, 
branching ratios and asymmetries. 
It uses, with few exceptions, analytical expressions to represent the parameter dependence as faithful as possible.
These are implemented in the $\msbar$ renormalization scheme\footnote{In many cases scheme-conversions of the originally published results were in order.}
due to its convergence properties and the potentially reduced uncertainties from unknown higher order contributions.

\begin{table}[tb]
\footnotesize
\begin{tabular}{l | c | cc | cc}
\hline
parameter	&  measurement	&  \multicolumn{2}{c|}{full EWK fit}	&  \multicolumn{2}{c}{EWK fit excl.\ input in line}\\
	& 	&  without $m_H$	&  with $m_H$	&  without $m_H$	&  with $m_H$\\
\hline
$M_{H}$~[GeV] 		&  $125.09\pm0.15$ 	&  $91\pm19$ 	&  $125.09\pm0.15$ 	&  $91\pm19$ 	&  $91\pm19$ \\
$M_{W}$~[GeV] 		&  $80.380\pm0.013$ 	&  $80.374\pm0.01$ 	&  $80.360\pm0.006$ 	&  $80.364\pm0.017$ 	&  $80.356\pm0.006$ \\
$\Gamma_{W}$~[GeV] 	&  $2.085\pm0.042$ 	&  $2.092\pm0.001$ 	&  $2.091\pm0.001$ 	&  $2.092\pm0.001$ 	&  $2.091\pm0.001$ \\
$m_{t}$~[GeV] 			&  $172.9\pm0.5$ 	&  $172.9\pm0.5$ 	&  $173.1\pm0.5$ 	&  $177.6\pm8$ 	&  $176.5\pm2.1$ \\
$\sin^{2}\theta_{\rm eff}^l$ 	&  $0.2314\pm0.00023$ 	&  $0.2314\pm0.00009$ 	&  $0.23152\pm0.00006$ 	&  $0.2314\pm0.0001$ 	&  $0.23152\pm0.00006$ \\
\hline
$M_{Z}$~[GeV] 		&  $91.188\pm0.002$ 	&  $91.188\pm0.002$ 	&  $91.188\pm0.002$ 	&  $91.185\pm0.024$ 	&  $91.201\pm0.009$ \\
$\sigma^{0}_{\rm had}$  [nb]  &  $41.54\pm0.037$ 	&  $41.482\pm0.015$ 	&  $41.483\pm0.015$ 	&  $41.472\pm0.016$ 	&  $41.474\pm0.016$ \\
$\Gamma_{Z}$~[GeV] 	&  $2.495\pm0.002$ 	&  $2.495\pm0.001$ 	&  $2.495\pm0.001$ 	&  $2.495\pm0.002$ 	&  $2.494\pm0.002$ \\
$A_{c}$ 				&  $0.67\pm0.027$ 	&  $0.6683\pm0.0003$ 	&  $0.6679\pm0.0002$ 	&  $0.6683\pm0.0003$ 	&  $0.6679\pm0.0002$ \\
$A_{b}$ 				&  $0.923\pm0.02$ 	&  $0.9347\pm0.00006$ 	&  $0.93462\pm0.00004$ 	&  $0.9347\pm0.00006$ 	&  $0.93462\pm0.00004$ \\
$A_{l}$ (SLD) 			&  $0.1513\pm0.00207$ 	&  $0.14797\pm0.00073$ 	&  $0.14707\pm0.00044$ 	&  $0.14756\pm0.00079$ 	&  $0.14688\pm0.00045$ \\
$A_{l}$ (LEP) 			&  $0.1465\pm0.0033$ 	&  $0.14797\pm0.00073$ 	&  $0.14707\pm0.00044$ 	&  $0.14756\pm0.00079$ 	&  $0.14688\pm0.00045$ \\
$A_{\rm FB}^{l}$ 			&  $0.0171\pm0.001$ 	&  $0.01642\pm0.00016$ 	&  $0.01622\pm0.0001$ 	&  $0.0164\pm0.00016$ 	&  $0.01621\pm0.0001$ \\
$A_{\rm FB}^{c}$ 		&  $0.0707\pm0.0035$ 	&  $0.0742\pm0.0004$ 	&  $0.0737\pm0.0002$ 	&  $0.0742\pm0.0004$ 	&  $0.0737\pm0.0002$ \\
$A_{\rm FB}^{b}$ 		&  $0.0992\pm0.0016$ 	&  $0.1037\pm0.0005$ 	&  $0.1031\pm0.0003$ 	&  $0.1042\pm0.0006$ 	&  $0.1032\pm0.0003$ \\
$R_{l}^{0}$ 			&  $20.767\pm0.025$ 	&  $20.747\pm0.018$ 	&  $20.744\pm0.018$ 	&  $20.73\pm0.027$ 	&  $20.723\pm0.027$ \\
$R_{c}^{0}$ 			&  $0.1721\pm0.003$ 	&  $0.17226\pm0.00008$ 	&  $0.17225\pm0.00008$ 	&  $0.17226\pm0.00008$ 	&  $0.17225\pm0.00008$ \\
$R_{b}^{0}$ 			&  $0.21629\pm0.00066$ 	&  $0.2158\pm0.00011$ 	&  $0.21581\pm0.00011$ 	&  $0.21579\pm0.00011$ 	&  $0.2158\pm0.00011$ \\
\hline
$\Delta\alpha^{(5)}_{\rm had}$ [$10^{-5}$]	&  $2760\pm9$ 	&  $0.02761\pm9$ 	&  $2757\pm9$ 	&  $2817\pm91$ 	&  $2716\pm36$ \\
\hline
$\alpha_{s}(M_{Z})$ 	&  $0.1181\pm0.0011$ 	&  $0.1198\pm0.003$ 	&  $0.1197\pm0.003$ 	&  $0.1198\pm0.003$ 	&  $0.1196\pm0.003$ \\
\hline
\end{tabular}
\centering
\caption{\label{tab:EWFitOverview}Summary of global electroweak fits for different input parameters, performed with Gfitter~\cite{Haller:2018nnx}. 
The first and second columns summarize the parameter name and its experimental value, discussed in the text. 
The third and forth line show the fit result, using all experimental data, once including and and once excluding the Higgs boson mass. 
The fifth and six column give the fit results without using the corresponding input value of that row, again once with and without using the Higgs boson mass. 
The value of $\sin^{2}\theta_{\rm eff}^l$ corresponds to the average of the hadron collider measurements.}
\end{table}

Gfitter is implemented in object-oriented {\bf\small C\hspace{-1pt}+\hspace{-1pt}+} code and relies on ROOT~\cite{Brun:1997pa} functionality.
Tools for the handling of the data, the fitting, and statistical analyses such as Monte Carlo sampling are provided by a core package, 
where theoretical errors, correlations, and inter-parameter dependencies are consistently dealt with.
Gfitter employs Gaussian nuisance parameters as external constraints to estimate theoretical uncertainties.
Theoretical models are inserted as plugin packages, which may be hierarchically organized (the relevant code of the global electroweak fit is in the Gfitter/\textit{gew} package).
The pseudo-observables are usually expressed in terms of on-shell quantities.

Another difference between the two packages is that the Gfitter/\textit{gew} package assumes lepton universality, while GAPP keeps the flavor and family dependence of the observables.
One of the most important distinctions is the way the QED coupling $\alpha$ is evolved from the Thomson limit to the scale $\MZ$.
While Gfitter uses the hadronic vacuum polarization contribution $\dahadZf$ as an external constraint, GAPP uses $\Delta\alpha_{\rm had}^{(3)}(2~{\rm GeV})$ instead,
and then solves the renormalization group equation to reach the $Z$ scale.
The latter allows to keep the full dependence on $\alpha_s$, as well as the charm and bottom quark masses, all of which are allowed to float in the fits.
It also permits to treat the correlation with the corresponding running of the weak mixing angle and also with the anomalous magnetic moment of the muon $a_\mu$,
whose hadronic vacuum polarization contribution is based on the same data as $\Delta\alpha_{\rm had}^{(3)}(2~{\rm GeV})$ (but weighted differently)
and whose perturbative contribution is re-calculated in each call of the fits ($a_\mu$ is excluded from Gfitter).
The aforementioned $\alpha_s$ dependence of $\dahadZf$ has also been included in Gfitter to linear order.
However, due to the different ways the scale dependence of $\alpha$ is computed, its $\alpha_s$ dependence still differs between the two codes.

GAPP also implemented an update of the two-loop QCD correction to the $b$ quark 
forward-backward asymmetry~\cite{Bernreuther:2016ccf} including the full bottom quark mass dependence.
This reduces the extracted value of the weak mixing angle from this observable by $\approx 1/4~\sigma$.

\begin{figure*}[t!]
\centering
\resizebox{0.48\textwidth}{!}{\includegraphics{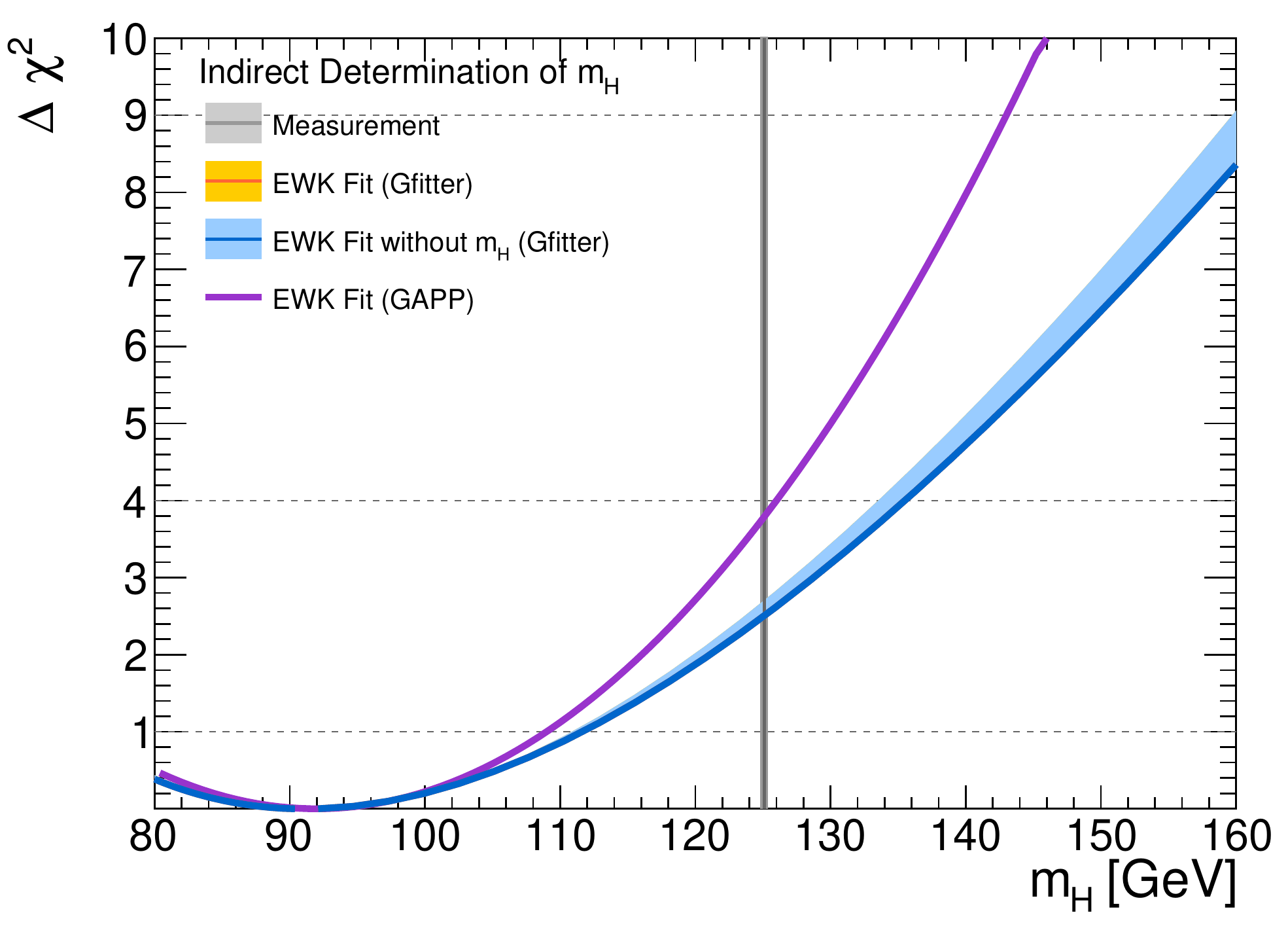}}
\resizebox{0.48\textwidth}{!}{\includegraphics{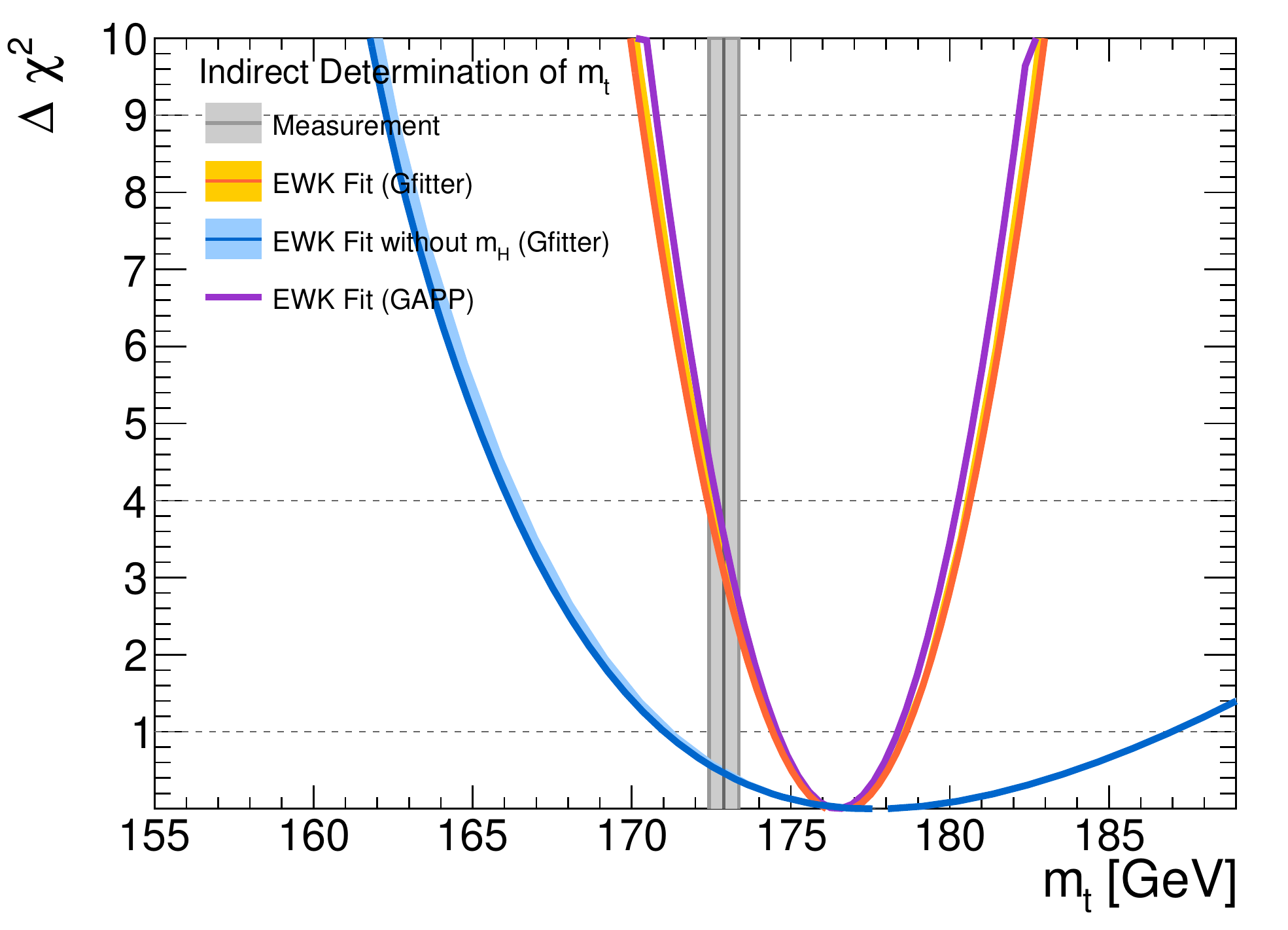}}
\resizebox{0.48\textwidth}{!}{\includegraphics{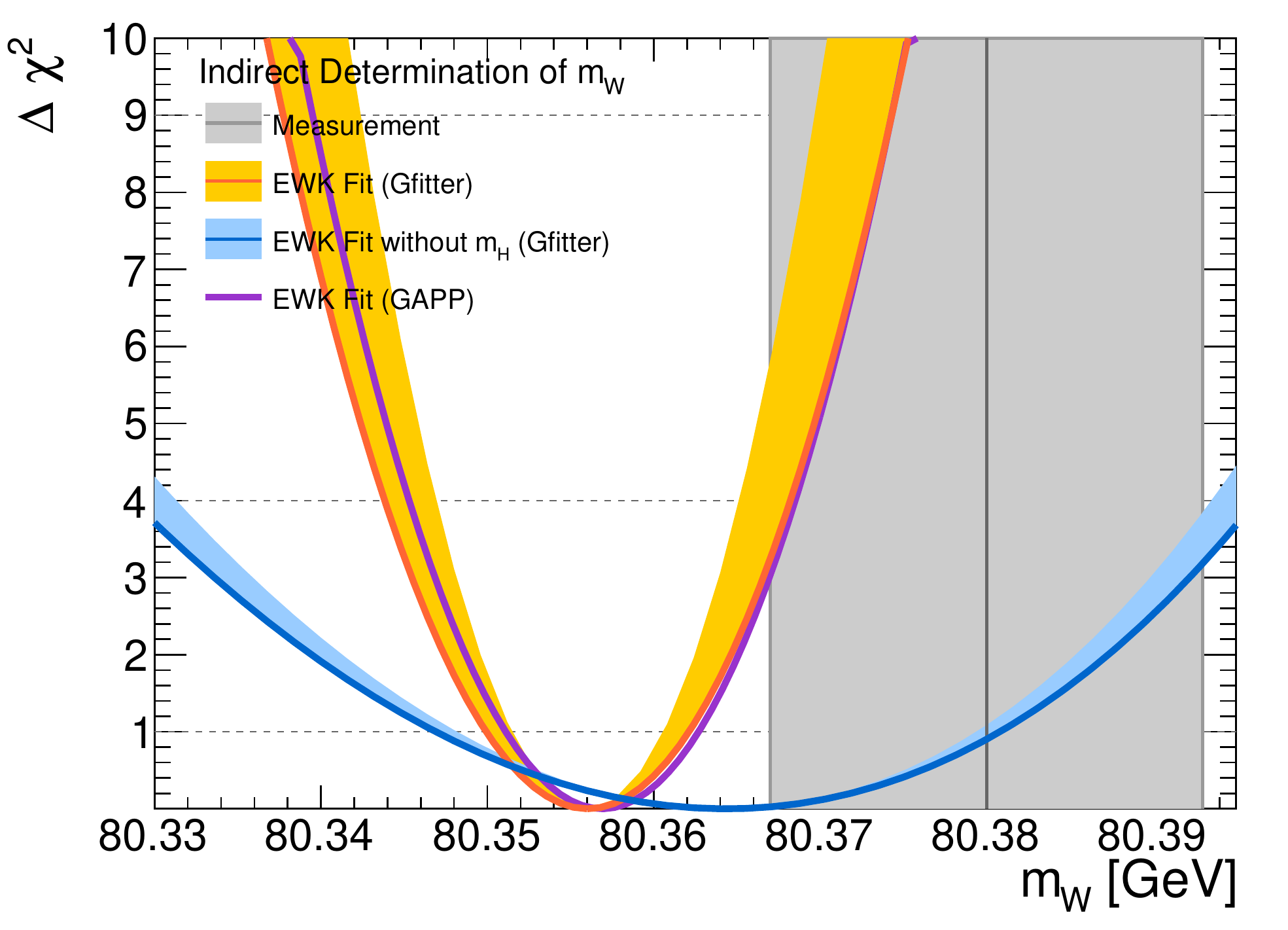}}
\resizebox{0.48\textwidth}{!}{\includegraphics{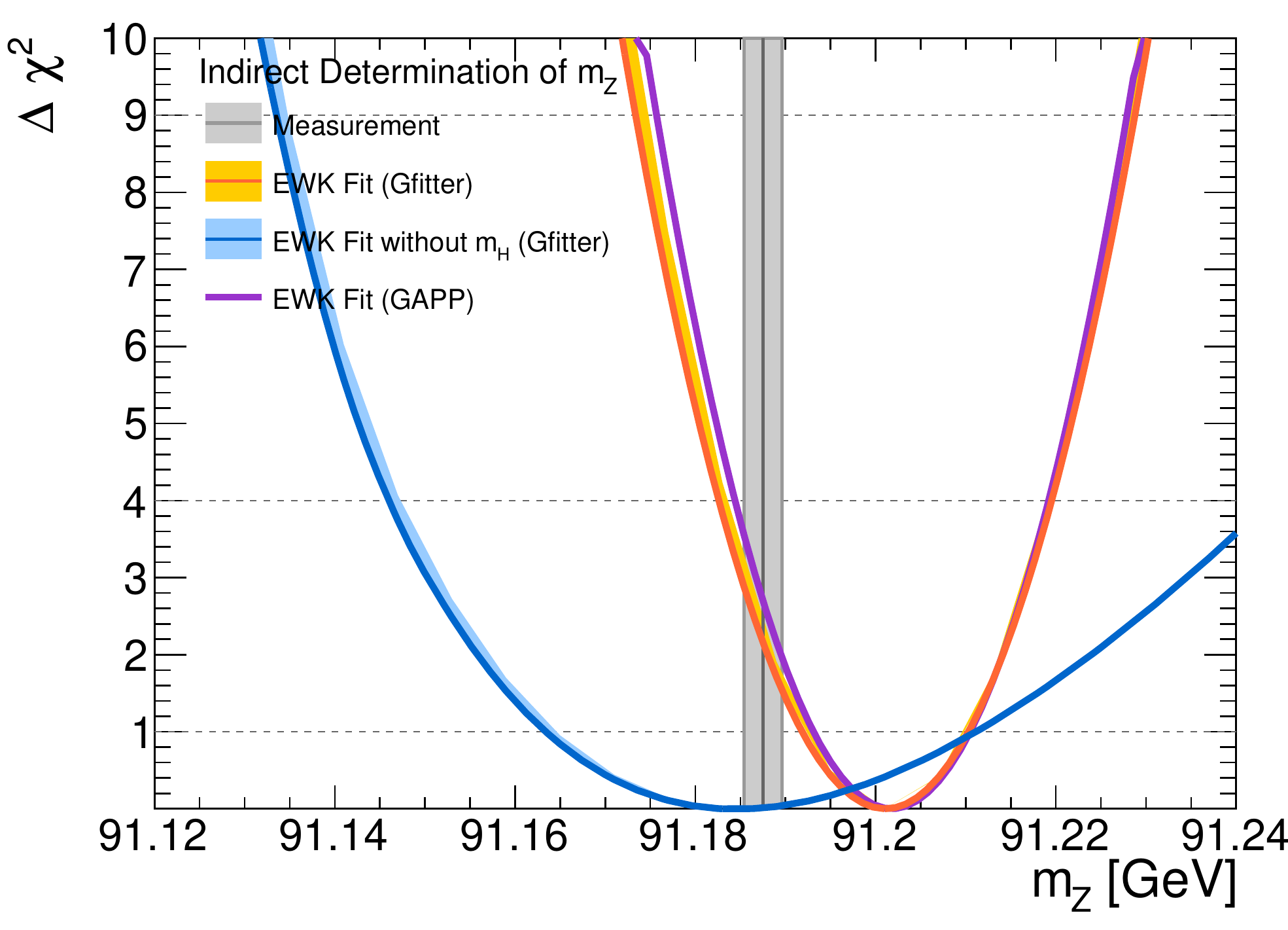}}
\resizebox{0.48\textwidth}{!}{\includegraphics{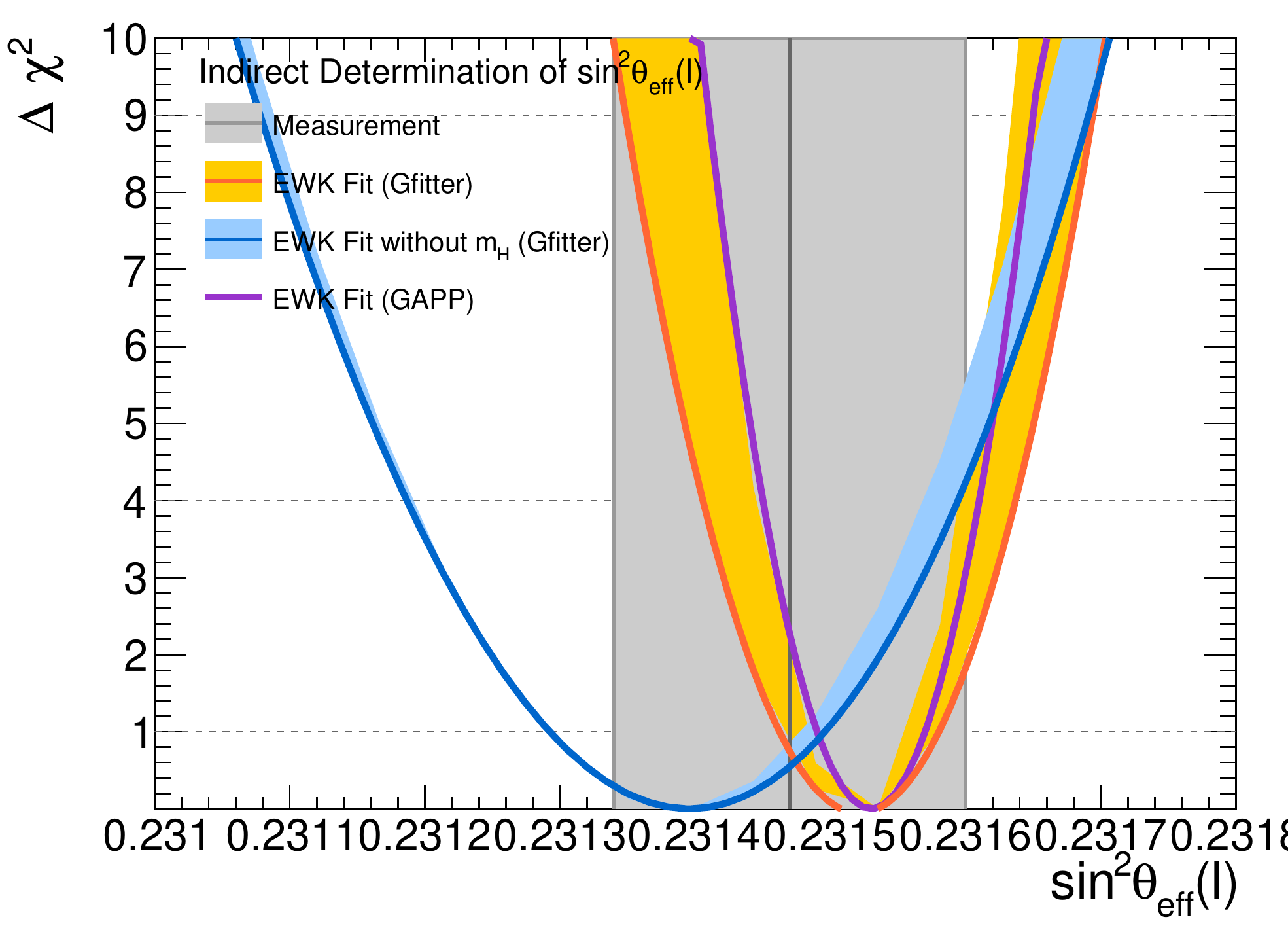}}
\resizebox{0.48\textwidth}{!}{\includegraphics{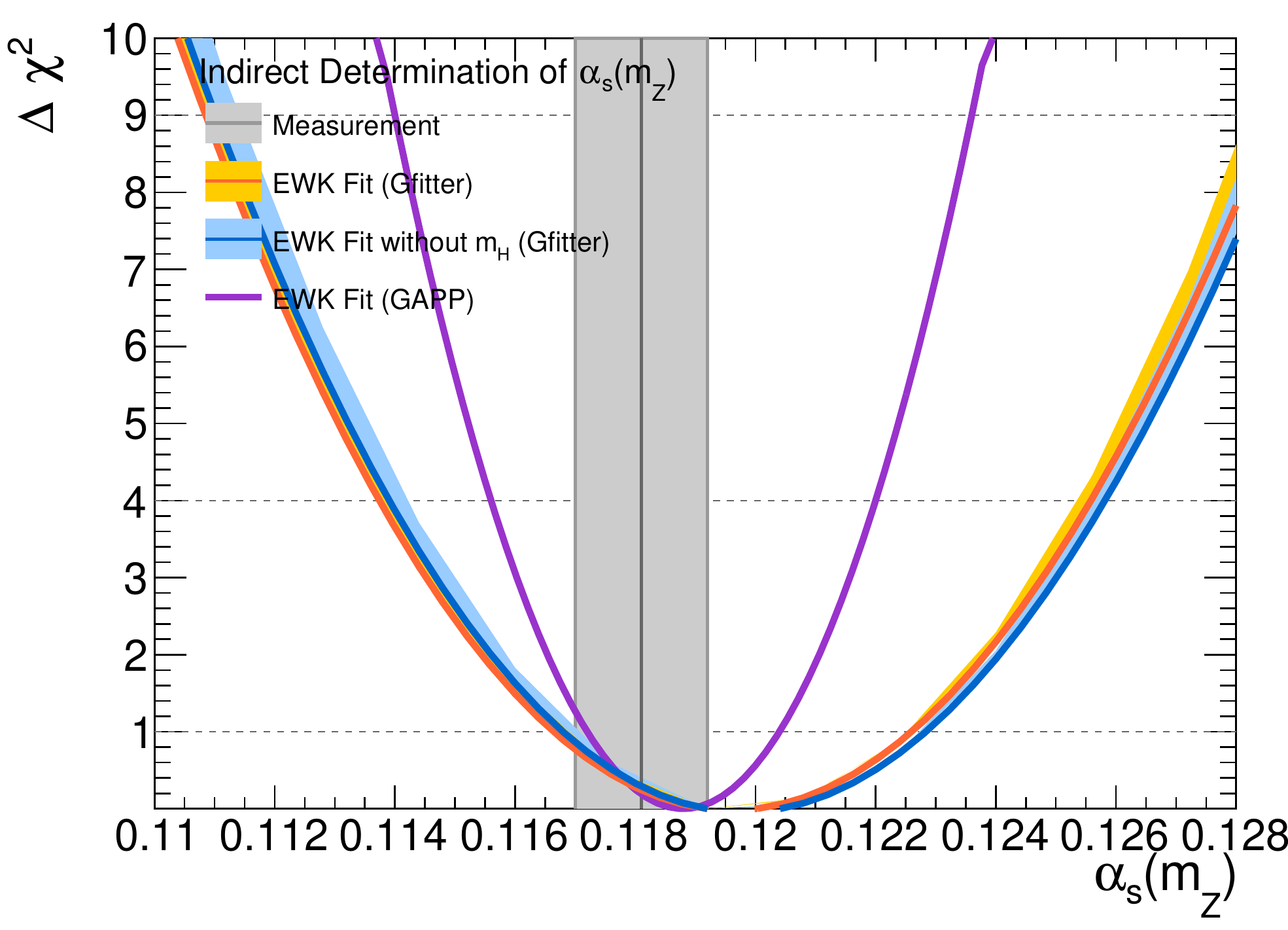}}
\caption{Comparisons of $\chi^2$ distributions for different observables with (blue) and without (orange) including the Higgs boson mass using the Gfitter program \cite{Haller:2018nnx}. 
The theoretical uncertainties are indicated by the filled blue and yellow areas, respectively. 
For comparison, the calculated $\chi^2$ distribution using the GAPP program \cite{Erler:1999ug} is also shown, where a symmetric distribution has been used for simplicity. 
The current world average of the measurements are shown in gray with their $1\sigma$ uncertainties.}
\label{fig:ChiPlot}
\end{figure*}

\subsection{\label{sec:ImpactOfFit}Impact of the discovery of the Higgs boson on the electroweak fit}

Using the input measurements discussed in Section \ref{sec:ewpo}, we performed the global electroweak fit at the $Z$-mass scale 
via the Gfitter package v2.2~\cite{Haller:2018nnx}, in particular its \textit{gew} library, as well as the GAPP code \cite{Erler:1999ug} for comparison. In particular, we focus our discussion on the impact of the Higgs boson mass on the fit, as it was previously studied in Ref.~\cite{Baak:2012kk}, however, based on the measurements available at the time of the Higgs boson discovery.
In both programs, the parameters describing the bosonic sector of the SM are chosen to be $\alpha$, $\MZ$, $G_F$ (\ie, those with the smallest experimental uncertainties), 
$\alpha_s$ and $\MH$. 
The Fermi constant $G_F$ remains fixed during the fit, due to its negligible uncertainty. 
In addition to $\MZ$, $\MH$, $\dahadZf$ (i.e. $\alpha$) and $\alpha_s(\MZ^2)$, also the masses of the heavy quarks, $m_t$, $m_b$ and $m_c$ are left as floating fit parameters. 
No external constraint is used for $\alpha_s(\MZ^2)$, and it is directly determined in the fit. 
GAPP uses the $\tau$ lepton lifetime as an additional input.

\begin{table}[tb]
\footnotesize
\begin{tabular}{l c | | cc | cc}
\hline
parameter& measurement & \multicolumn{2}{c|}{full EWK fit}& \multicolumn{2}{c}{EWK fit excl.\ input in line}\\
& & Gfitter&GAPP& Gfitter& GAPP\\
\hline
$M_{H}$~[GeV] & $125.1\pm0.15$ & $125.09\pm0.15$ & $125.1\pm0.15$ & $91\pm19$ & $91\pm17$ \\
$M_{W}$~[GeV] & $80.380\pm0.013$ & $80.360\pm0.006$ & $80.361\pm0.005$ & $80.356\pm0.006$ & $80.3569\pm0.006$ \\
$\Gamma_{W}$~[GeV] & $2.085\pm0.042$ & $2.091\pm0.001$ & $2.090\pm0.001$ & $2.091\pm0.001$ & $2.0898\pm0.0007$ \\
$m_{t}$~[GeV] & $172.90\pm0.47$ & $173.1\pm0.46$ & $173.10\pm0.46$ & $176.5\pm2.1$ & $176.47\pm1.9$ \\
$\sin^{2}\theta_{\rm eff}^l$ & $0.23140\pm0.00023$ & $0.23152\pm0.00006$ & $0.23153\pm0.00004$ & $0.23152\pm0.00006$ & $0.23153\pm0.00004$ \\
\hline
$M_{Z}$~[GeV] & $91.188\pm0.002$ & $91.188\pm0.002$ & $91.188\pm0.002$ & $91.201\pm0.009$ & $91.2018\pm0.009$ \\
$\sigma^{0}_{\rm had}$~[nb] & $41.540\pm0.037$ & $41.483\pm0.015$ & $41.480\pm0.009$ & $41.474\pm0.016$ & $41.4771\pm0.009$ \\
$\Gamma_{Z}$~[GeV] & $2.4952\pm0.0023$ & $2.495\pm0.001$ & $2.4944\pm0.0008$ & $2.494\pm0.002$ & $2.4942\pm0.0008$ \\
$A_{c}$ & $0.670\pm0.027$ & $0.6679\pm0.0002$ & $0.6677\pm0.0001$ & $0.6679\pm0.0002$ & $0.66773\pm0.0001$ \\
$A_{b}$ & $0.923\pm0.020$ & $0.93462\pm0.00004$ & $0.93471\pm0.00002$ & $0.93462\pm0.00004$ & $0.93471\pm0.00002$ \\
$A_{l}$ (SLD) & $0.15130\pm0.00207$ & $0.14707\pm0.00044$ & $0.14697\pm0.00029$ & $0.14688\pm0.00045$ & $0.14688\pm0.0003$ \\
$A_{l}$ (LEP) & $0.14650\pm0.00330$ & $0.14707\pm0.00044$ & $0.14697\pm0.00029$ & $0.14688\pm0.00045$ & $0.14688\pm0.0003$ \\
$A_{FB}^{l}$ & $0.01710\pm0.00100$ & $0.01622\pm0.00010$ & $0.01620\pm0.00006$ & $0.01621\pm0.00010$ & $0.01619\pm0.00006$ \\
$A_{FB}^{c}$ & $0.07070\pm0.00350$ & $0.0737\pm0.0002$ & $0.07360\pm0.00016$ & $0.0737\pm0.0002$ & $0.0736\pm0.0002$ \\
$A_{FB}^{b}$ & $0.09920\pm0.00160$ & $0.1031\pm0.0003$ & $0.10303\pm0.00020$ & $0.1032\pm0.0003$ & $0.10309\pm0.0002$ \\
$R_{l}^{0}$ & $20.767\pm0.025$ & $20.744\pm0.018$ & $20.738\pm0.010$ & $20.723\pm0.027$ & $20.733\pm0.01$ \\
$R_{c}^{0}$ & $0.17210\pm0.00300$ & $0.17225\pm0.00008$ & $0.17222\pm0.00003$ & $0.17225\pm0.00008$ & $0.17222\pm0.00003$ \\
$R_{b}^{0}$ & $0.21629\pm0.00066$ & $0.21581\pm0.00011$ & $0.21582\pm0.00002$ & $0.21580\pm0.00011$ & $0.21582\pm0.00002$ \\
\hline
\end{tabular}
\centering
\caption{\label{tab:EWFitComparison}Comparison of the results of the global electroweak fit for different parameters, 
performed with Gfitter~\cite{Haller:2018nnx} and GAPP~\cite{Erler:1999ug}. 
The first column indicates the parameter (observable) that is compared and the second column its measurement value. 
The third and fourth line show the fit result, using all experimental data for both fitting codes. 
The fifth and six column give the fit results without using the corresponding input value for both fitting programs.}
\end{table}

When including all experimental data of the second column in Table~\ref{tab:EWFitOverview} except for the Higgs boson mass, 
the fit based on Gfitter converges at a global minimum of $\chi^2_{\rm min} = 16.1$ with 14 degrees of freedom, which corresponds to a p-value of 0.41. 
The individual results of this fit for selected observables are summarized in the third column of Table~\ref{tab:EWFitOverview}, 
where the uncertainties have been estimated using $\chi^2$ profiles of a parameter scan and then symmetrized for simplicity. 
With the discovery of the Higgs boson, the last missing observable of the fit could be included. 
When repeating the fit including the measured value of $\MH$, we find a global minimum of $\chi^2_{\rm min} = 18.4$, obtained for 15 degrees of freedom, \ie, a p-value of 0.24. 
The individual results for the selected observables are shown in the fourth column of Table~\ref{tab:EWFitOverview}. 
The largest impact of the Higgs boson discovery and its mass measurement, is seen, apart from the Higgs mass itself, 
in the central values and uncertainties of $\MW$ and the electroweak mixing angle and related observables. 
The $Z$ boson mass, albeit similar to $\MW$ is not impacted by the inclusion of $\MH$ in the fit, due to its small experimental uncertainty. 

It is instructive to indirectly determine each observable by performing the electroweak fit without using the corresponding measurement value,
similar to the prediction of $\MH$ before its discovery. The results of this indirect determination for each observable, once including $\MH$ and once excluding it, is shown in columns five and six of Table~\ref{tab:EWFitOverview}. 

Technically, the indirect parameter determination is performed by scanning the parameter in a chosen range and calculating the corresponding $\chi^2$ values. 
The value of $\chi^2_{\rm min}$ is not relevant for the uncertainty estimation, but only its difference relative to the global minimum, $\Delta \chi^2 \equiv \chi^2 - \chi^2_{\rm min}$. 
The $\Delta\chi^2=1$ and $\Delta\chi^2=4$ profiles define the $1\sigma$ and $2\sigma$ uncertainties, respectively. 
The $\Delta \chi^2$ distributions of selected observables ($\MH$, $\MW$, $\MZ$, $m_t$, $\sinleff$ and $\alpha_s$) are shown in Figure \ref{fig:ChiPlot}, including and excluding the value of $\MH$ in the fit. These observables are discussed in more detail in the following.

As one of the main results, the Gfitter package predicts,
\begin{equation}
\label{gfitterMH}
\MH = 91.0^{+20}_{-17}~{\rm GeV},
\end{equation}
where the uncertainty is dominated by the uncertainties in $\MW$ and $\sinleff$.
When these observables would be perfectly known, the uncertainty in $\MH$ would reduce to approximately 10 and 12~GeV, respectively.
Currently, we observe a $1.7 \sigma$ tension between the indirectly determined value of $\MH$ and its direct measurement. 
This is mainly driven by the measurements of $\MW$, where a measured value of $80.351$~GeV would result in a predicted Higgs boson mass of 125~GeV. 
It should be noted that after the Higgs boson mass had been measured to GeV precision, the impact of the Higgs boson mass on the global electroweak fit virtually disappeared. 
For example, the minimal $\chi^2$ in the fit varies within 0.005 when changing the experimental uncertainty in $\MH$ from 150~MeV to 1~GeV. 
Hence a further improvement in the measurement of $\MH$ will not alter the SM fit any further. 

Fixing $\MH$ changes the indirectly determined $\MW=80.364\pm 0.017$~GeV to $80.356\pm 0.006$~GeV, 
\ie, leads to a relative reduction in the uncertainty by more than 50\% and a lower central value. 
The tension between the predicted value of the fit and the measurements increases from $1.1\sigma$ to $1.7\sigma$ when $\MH$ is included. 
The uncertainty in the indirectly determined value of $\MW$ is affected by the uncertainty in $m_t$ contributing 2.6~MeV, and $\MZ$ contributing 2.5~MeV. 

\begin{figure}[t]
\centering
\resizebox{0.99\textwidth}{!}{\includegraphics{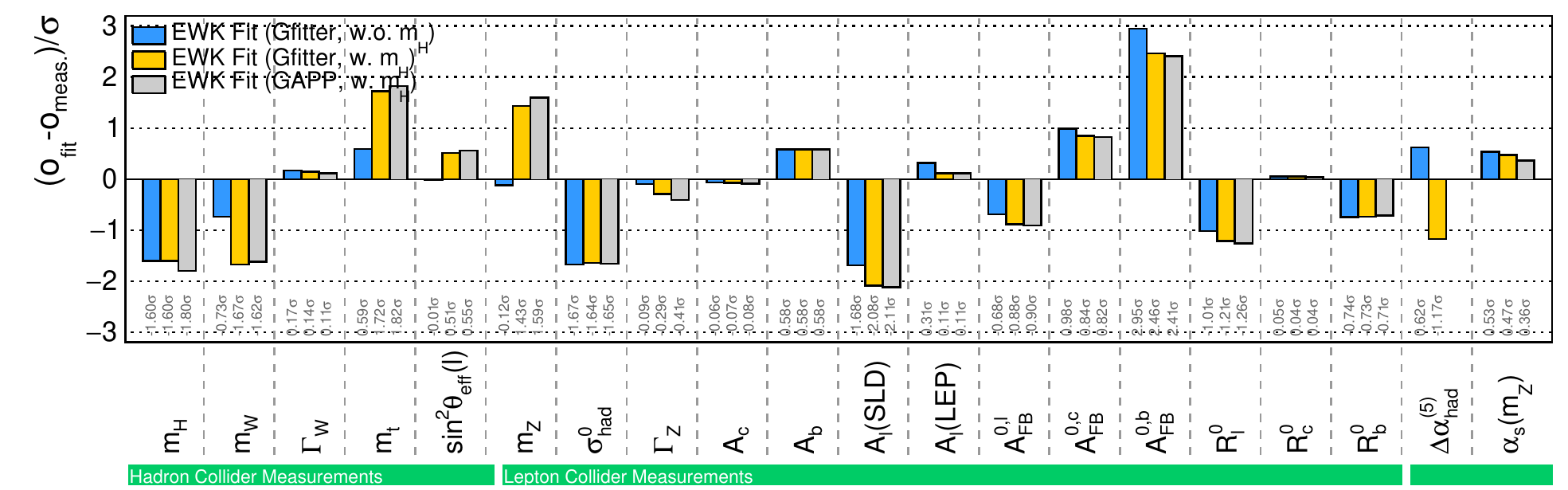}}
\caption{Pull values for different observables, defined as the difference between the measurement values and the indirect determinations using the global electroweak fit, 
normalized to the total errors of the measurements and the indirect determinations of the fit. 
The results by the Gfitter program \cite{Haller:2018nnx} with and without using $\MH$ in the fit are shown in yellow and blue, respectively. 
For comparison the pull distribution using the GAPP program \cite{Erler:1999ug} is shown for the case with the $\MH$ constraint in the fit.}
\label{fig:PollPlot}
\end{figure}

Similarly, the uncertainty in the indirectly determined value of $\MZ$ reduces significantly when fixing $\MH$ in the fit, 
leading to a change of $\MZ=91.185\pm 0.024$~GeV to $91.201\pm0.009$~GeV. 
A perfect knowledge of $\MW$ would reduce this uncertainty from 9~MeV to 4~MeV. 
Given the small experimental uncertainty in $\MZ$ compared to the indirectly determined value, 
an improvement in the measurement precision on the experimental side is not urgently called for. 

The largest impact of the Higgs boson discovery on the electroweak fit is observed for the top quark mass. 
While the central value of its prediction shifts by only 1.3~GeV, its uncertainty reduces significantly, namely from $m_t=177.7^{+9.4}_{-6.6}$~GeV to $m_t=176.5^{+2.1}_{-2.1}$~GeV. 
The remaining uncertainty is dominated by the experimental precision in $\MW$, contributing 1.9~GeV. 
The predicted value of the fit shows a $1.7\sigma$ tension with the measured value. 
However, a more precise experimental measurement will not be able to reduce this tension significantly, as the uncertainty of about 0.5~GeV is already small 
compared to the 2.1~GeV uncertainty in the indirectly determined value, which could be improved further only by a more precise measurement of $\MW$. 

When comparing the indirectly determined values of $\sinleff$ with and without the inclusion of $\MH$ in the fit, 
we find $\sinleff=0.23151\pm 0.00006$ and $\sinleff=0.23139\pm 0.00010$, respectively. 
The remaining uncertainty is dominated by theory uncertainties ($\pm 0.00004$) and uncertainties in \dahadZf ($\pm 0.00004$). 
When comparing the indirectly determined value to the world average of $\sinleff=0.23151\pm 0.00014$ we find perfect agreement. 

As discussed in Section \ref{sec:ewpotheory}, the SM prediction of several observables, in particular $\sigma^0_{\rm had}$, $\Gamma _Z$ and $R^0$, 
depend on the strong coupling constant $\alpha_s(\MZ)$. 
From the electroweak fit using Gfitter without further external constraint we find $\alpha_s(\MZ)=0.1196\pm0.0029$, 
which corresponds to a determination with full two-loop electroweak precision including NNLO plus partial N$^3$LO QCD corrections. 
Since the predictions of $\sigma^0_{\rm had}$, $\Gamma _Z$ and $R^0$ depend only mildly on $\MH$, no significant change is observed when excluding $\MH$ from the fit.

For the most important observables we compared the global electroweak fit performed with Gfitter with the results  from GAPP. 
As mentioned in Section~\ref{sec:fitprograms}, GAPP assumes no lepton universality, expresses the calculations in the $\msbar$ scheme,
and adds several observables such as $a_\mu$ and the $\tau$ lifetime. 
In addition, the estimation of theoretical uncertainties, as well as the treatment of correlated uncertainties is different. 
Using the same input values as listed in Table~\ref{tab:EWFitOverview}, GAPP finds a global minimum with a $\chi^2 = 44.5$ at 41~effective degrees of freedom. 
The $\Delta\chi^2$ distributions for the indirect determinations of observables is illustrated also in Figure~\ref{fig:ChiPlot} as violet lines. 
Here, only Gaussian errors have been derived for simplicity, leading to polynomials of second order as $\chi^2$ distributions. 
A direct comparison between the full fit results for the main observables as well as their indirect determined values using Gfitter and GAPP, respectively, 
is shown in Table~\ref{tab:EWFitComparison}. 
In general a very good agreement in the central values can be observed, while GAPP typically yields smaller uncertainties. 
For example, the indirect determined Higgs boson mass has an uncertainty of $^{+18}_{-16}$~GeV using the GAPP code, while Gfitter yields $^{+21}_{-18}$~GeV. 

\begin{figure}[t]
\centering
\resizebox{0.99\textwidth}{!}{\includegraphics{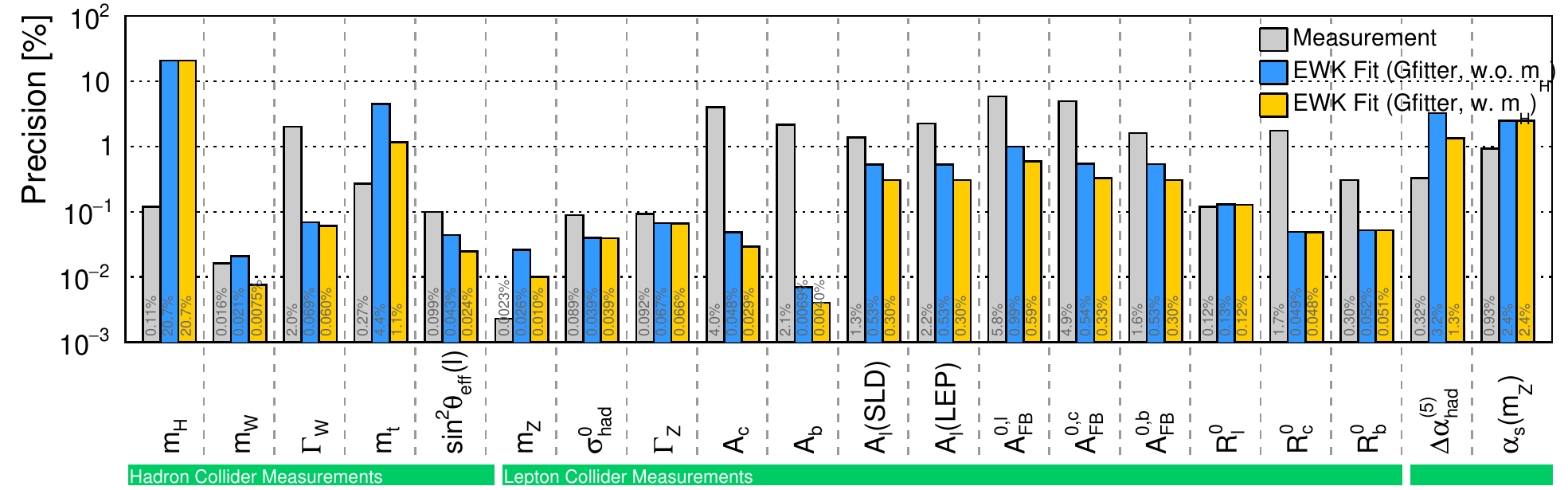}}
\caption{Comparison of experimental and indirectly determined relative precisions for different electroweak precision variables, 
for the cases including and excluding $\MH$ in the fit via the Gfitter code~\cite{Haller:2018nnx}.}
\label{fig:PrecisionPlot}
\end{figure}

A summary of the impact of the Higgs boson mass on the electroweak fit is given in Figures~\ref{fig:PollPlot} and~\ref{fig:PrecisionPlot}. 
Figure~\ref{fig:PollPlot} shows the pull values for different observables, defined as the difference of the measurement value and the indirect determination via the global electroweak fit, 
normalized by the total error of the measurement and the indirect determination, \ie $\sigma_{\rm tot}=\sqrt{\sigma_{\rm exp}^2+\sigma_{\rm ind}^2}$. 
The results by Gfitter~\cite{Haller:2018nnx} with and without using $\MH$ in the fit are shown in blue and yellow. 
For comparison the resulting pull distribution of the GAPP program~\cite{Erler:1999ug} is shown in gray for the case when $\MH$ is included in the fit. 
Figure~\ref{fig:PrecisionPlot} shows the experimental and indirect determined relative precision for the different electroweak precision variables, 
for both cases, including and excluding $\MH$ in the fit, using the Gfitter code. 
Interesting for future measurements are those observables, which have a larger experimental uncertainty 
than their indirect determined value and are accessible with experiments in the next decade. 
These are in particular, the mass $\MW$ and the width $\Gamma_W$ of the $W$ boson, as well as the electroweak mixing angle $\sinleff$.

In the light of the currently perfect agreement of the direct measurement of $\sinleff$ and its indirect determined value, 
one should discuss the potential future measurements of $\sinleff$ at hadron colliders. 
Assuming a future precision of $\sinleff$ on the order of 0.00013, \ie, an improvement by a factor of two compared to the current precision at hadron colliders, 
and a future measurement value within $2\sigma$ of the current world average, the maximal deviation from the SM expectation would be $1.5\sigma$. 
Hence in the midterm future, we cannot expect to observe a significant tension with the SM prediction of this observable at the $Z$ boson mass scale. 
However, measurements probing the running of $\sinleff$ would be highly interesting, as they could shed light on possible deviations at lower energy scales. 

The situation is different for $\MW$ and $\Gamma_W$, where a higher experimental precision could lead to significant deviations from the predicted values by the fit 
and therefore indicate inconsistencies in the Standard Model. 
Similarly important, the reduction of the uncertainty in $\MW$ would reduce the uncertainty in the predicted value of $m_t$,
and hence would then allow to shed light on the currently observed tension of about $1.7\sigma$.

\section{Summary and Outlook\label{sec:summary}}

The long history of the global electroweak fit culminated in the discovery of the Higgs boson in 2012 at the LHC and thus completed the particle spectrum of the Standard Model. 
In this article, we reviewed the status of the global electroweak fit with a special focus on the latest developments of precision measurements at hadron colliders. 
Even though many precision measurements of the $Z$ pole observables date back more than 20 years, 
they still provide the most relevant input for several observables in the global electroweak fit. 
However, the developments on perturbative and non-perturbative aspects of QCD, the better understanding of proton structure, as well as 
the outstanding performance of the hadron collider detectors, allowed for precision measurements of the $W$ boson mass, the $W$ boson width,
the top quark mass and Higgs boson properties. 
We combined all latest measurements and performed a global electroweak fit, yielding a consistency of the Standard Model with a p-value of 0.24. 

While the theoretical uncertainties due to missing higher order corrections in the electroweak precision observables are currently well below the experimental uncertainties, 
no immediate further effort has to be taken to improve these calculations. 
In recent years, however, a different point of view gained traction, in which the SM is viewed as an effective field theory (EFT) with hitherto unknown new particles integrated out. 
The systematic determination of the many new parameters introduced by this formalism has become a vibrant field of research (see, \eg, Ref.~\cite{Falkowski:2017pss}).
Hence, the future of electroweak precision measurements will have to include cross section measurements, 
sensitive to the effect of higher EFT operators, in a new global electroweak EFT fit. 
Hopefully, such an approach will help to guide the way how we should look for signatures for physics beyond the Standard Model.

Looking even further into future, with the advent of possible new high energy lepton colliders it would be possible to increase the precision of 
electroweak fits by an order of magnitude or more~\cite{Fan:2014vta}. 
Under discussion are currently circular colliders that would have the abilities to redo a LEP style physics program with several orders of magnitude greater data statistics,
while at the same time allowing to proceed to higher energies including the $ZH$ production threshold region for high precision Higgs boson studies,
and possibly even reaching the top quark pair production threshold for ultra-precision top quark physics.
Very recently, a conceptual design report appeared for such an $e^+e^-$ accelerator in China, the Circular Electron Positron Collider (CEPC)~\cite{CEPCStudyGroup:2018ghi}.
A similar machine, the Future Circular Collider in $e^+e^-$ mode (FCC-ee), which is an evolution of a concept called TLEP~\cite{Gomez-Ceballos:2013zzn}, is being considered at CERN.
Alternatives are linear options like the International Linear Collider (ILC) which already reached a comparatively mature stage~\cite{Baer:2013cma} 
or the Compact Linear Collider (CLIC)~\cite{Linssen:2012hp},
with the potential to reach the TeV and multi-TeV regions, respectively, and the advantage of high electron (and possibly positron) polarization for further increase in precision.
Should any of these be realized, a world-wide concerted effort to achieve three-loop electroweak precision would then be required on the theory side.
\section*{Acknowledgements}
\noindent
M.S. thanks the Volkswagen Foundation as well as the Deutsche Forschungsgemeinschaft for their support on his research on electroweak precision physics in recent years. Moreover, he thanks Dr. Roman Kogler, who helped during the initial phase of this review article.
J.E. was supported by CONACyT (M\'exico) project 252167--F,
the German--Mexican research collaboration grant SP 778/4--1 (DFG) and 278017 (CONACyT),
and by PASPA (DGAPA--UNAM).
He also gratefully acknowledges the outstanding hospitality and support offered 
by the Mainz Institute for Theoretical Physics (MITP) and the Theoretical High Energy Physics (THEP) group at the Institute of Physics
in Mainz where a significant part of this work has been carried out. 
Both authors benefited greatly from fruitful discussions with Dr. Andreas H{\"o}cker and Prof. Dr. Hubert Spiesberger.
J.E. is indebted to Werner Bernreuther and Long Chen for a dedicated update and a private communication regarding the results of 
Ref.~\cite{Bernreuther:2016ccf} which could be included in the program GAPP.

\newpage

\small 

\section*{References}
\bibliography{EWKReview}

\begin{thebibliography}{100}

\bibitem{Weinberg:1967tq}
S.~Weinberg, {\em Phys. Rev. Lett.} {\bf 19}, 1264  (1967).

\bibitem{Salam:1968rm}
A.~Salam, {\em Conf. Proc.} {\bf C680519}, 367  (1968).

\bibitem{Higgs:1964pj}
P.~W. Higgs, {\em Phys. Rev. Lett.} {\bf 13}, 508  (1964).

\bibitem{Englert:1964et}
F.~Englert and R.~Brout, {\em Phys. Rev. Lett.} {\bf 13}, 321  (1964).

\bibitem{Guralnik:1964eu}
G.~S. Guralnik, C.~R. Hagen and T.~W.~B. Kibble, {\em Phys. Rev. Lett.} {\bf
  13}, 585  (1964).

\bibitem{Glashow:1961tr}
S.~L. Glashow, {\em Nucl. Phys.} {\bf 22}, 579  (1961).

\bibitem{tHooft:1972tcz}
G.~'t~Hooft and M.~J.~G. Veltman, {\em Nucl. Phys.} {\bf B44}, 189  (1972).

\bibitem{Sirlin:1980nh}
A.~Sirlin, {\em Phys. Rev.} {\bf D22}, 971  (1980).

\bibitem{Passarino:1978jh}
G.~Passarino and M.~J.~G. Veltman, {\em Nucl. Phys.} {\bf B160}, 151  (1979).

\bibitem{Aoki:1982ed}
K.~I. Aoki, Z.~Hioki, M.~Konuma, R.~Kawabe and T.~Muta, {\em Prog. Theor. Phys.
  Suppl.} {\bf 73}, 1  (1982).

\bibitem{Hollik:1988ii}
W.~F.~L. Hollik, {\em Fortsch. Phys.} {\bf 38}, 165  (1990).

\bibitem{Kim:1980sa}
J.~E. Kim, P.~Langacker, M.~Levine and H.~H. Williams, {\em Rev. Mod. Phys.}
  {\bf 53},   211  (1981).

\bibitem{Amaldi:1987fu}
U.~Amaldi {\em et~al.}, {\em Phys. Rev.} {\bf D36},   1385  (1987).

\bibitem{Costa:1987qp}
G.~Costa, J.~R. Ellis, G.~L. Fogli, D.~V. Nanopoulos and F.~Zwirner, {\em Nucl.
  Phys.} {\bf B297}, 244  (1988).

\bibitem{Arnison:1983rp}
 UA1 Collaboration (G.~Arnison {\em et~al.}), {\em Phys. Lett.} {\bf B122}, 103
   (1983).

\bibitem{Banner:1983jy}
 UA2 Collaboration (M.~Banner {\em et~al.}), {\em Phys. Lett.} {\bf B122}, 476
  (1983).

\bibitem{Arnison:1983mk}
 UA1 Collaboration (G.~Arnison {\em et~al.}), {\em Phys. Lett.} {\bf B126}, 398
   (1983).

\bibitem{Bagnaia:1983zx}
 UA2 Collaboration (P.~Bagnaia {\em et~al.}), {\em Phys. Lett.} {\bf B129}, 130
   (1983).

\bibitem{ALEPH:2005ab}
 LEP Electroweak Working Group, SLD Electroweak and Heavy Flavour Groups,
  ALEPH, DELPHI, L3, OPAL and SLD Collaborations (S.~Schael {\em et~al.}), {\em
  Phys. Rept.} {\bf 427}, 257  (2006),
  \href{http://arxiv.org/abs/hep-ex/0509008}{{\ttfamily arXiv:hep-ex/0509008
  [hep-ex]}}.

\bibitem{Kennedy:1990ib}
D.~C. Kennedy and P.~Langacker, {\em Phys. Rev. Lett.} {\bf 65}, 2967  (1990),
  [Erratum: {\em ibid.} {\bf 66},395(1991)].

\bibitem{Langacker:1991an}
P.~Langacker and M.~Luo, {\em Phys. Rev.} {\bf D44}, 817  (1991).

\bibitem{Erler:1994fz}
J.~Erler and P.~Langacker, {\em Phys. Rev.} {\bf D52}, 441  (1995),
  \href{http://arxiv.org/abs/hep-ph/9411203}{{\ttfamily arXiv:hep-ph/9411203
  [hep-ph]}}.

\bibitem{Abe:1995hr}
 CDF Collaboration (F.~Abe {\em et~al.}), {\em Phys. Rev. Lett.} {\bf 74}, 2626
   (1995), \href{http://arxiv.org/abs/hep-ex/9503002}{{\ttfamily
  arXiv:hep-ex/9503002 [hep-ex]}}.

\bibitem{D0:1995jca}
 \DZero Collaboration (S.~Abachi {\em et~al.}), {\em Phys. Rev. Lett.} {\bf
  74}, 2632  (1995), \href{http://arxiv.org/abs/hep-ex/9503003}{{\ttfamily
  arXiv:hep-ex/9503003 [hep-ex]}}.

\bibitem{Schael:2013ita}
 LEP Electroweak Working Group, ALEPH, DELPHI, L3 and OPAL Collaborations
  (S.~Schael {\em et~al.}), {\em Phys. Rept.} {\bf 532}, 119  (2013),
  \href{http://arxiv.org/abs/1302.3415}{{\ttfamily arXiv:1302.3415 [hep-ex]}}.

\bibitem{Erler:1998df}
J.~Erler and P.~Langacker, { {Status of the standard model}}, in {\em {Physics
  beyond the standard model. Proceedings, 5th International WEIN Symposium,
  Santa Fe, NM, June 14-19, 1998}\/},  pp. 1--27.
\newblock \href{http://arxiv.org/abs/hep-ph/9809352}{{\ttfamily
  arXiv:hep-ph/9809352 [hep-ph]}}.

\bibitem{Erler:2004cx}
J.~Erler and M.~J. Ramsey-Musolf, {\em Prog. Part. Nucl. Phys.} {\bf 54}, 351
  (2005), \href{http://arxiv.org/abs/hep-ph/0404291}{{\ttfamily
  arXiv:hep-ph/0404291 [hep-ph]}}.

\bibitem{Flacher:2008zq}
H.~Fl{\"a}cher {\em et~al.}, {\em Eur. Phys. J.} {\bf C60}, 543  (2009),
  \href{http://arxiv.org/abs/0811.0009}{{\ttfamily arXiv:0811.0009 [hep-ph]}},
  [Erratum: {\em ibid.} {\bf C71},1718(2011)].

\bibitem{Erler:2010wa}
J.~Erler, {\em Phys. Rev.} {\bf D81},   051301  (2010),
  \href{http://arxiv.org/abs/1002.1320}{{\ttfamily arXiv:1002.1320 [hep-ph]}}.

\bibitem{Erler:2012uu}
J.~Erler, {Weighing in on the Higgs}  (2012),
  \href{http://arxiv.org/abs/1201.0695}{{\ttfamily arXiv:1201.0695 [hep-ph]}}.

\bibitem{Baak:2011ze}
M.~Baak {\em et~al.}, {\em Eur. Phys. J.} {\bf C72},   2003  (2012),
  \href{http://arxiv.org/abs/1107.0975}{{\ttfamily arXiv:1107.0975 [hep-ph]}}.

\bibitem{Aad:2012tfa}
 ATLAS Collaboration (G.~Aad {\em et~al.}), {\em Phys. Lett.} {\bf B716}, 1
  (2012), \href{http://arxiv.org/abs/1207.7214}{{\ttfamily arXiv:1207.7214
  [hep-ex]}}.

\bibitem{Chatrchyan:2012xdj}
 CMS Collaboration (S.~Chatrchyan {\em et~al.}), {\em Phys. Lett.} {\bf B716},
  30  (2012), \href{http://arxiv.org/abs/1207.7235}{{\ttfamily arXiv:1207.7235
  [hep-ex]}}.

\bibitem{Erler:2013xha}
J.~Erler and S.~Su, {\em Prog. Part. Nucl. Phys.} {\bf 71}, 119  (2013),
  \href{http://arxiv.org/abs/1303.5522}{{\ttfamily arXiv:1303.5522 [hep-ph]}}.

\bibitem{Ciuchini:2013pca}
M.~Ciuchini, E.~Franco, S.~Mishima and L.~Silvestrini, {\em JHEP} {\bf 08},
  106  (2013), \href{http://arxiv.org/abs/1306.4644}{{\ttfamily arXiv:1306.4644
  [hep-ph]}}.

\bibitem{Baak:2014ora}
 Gfitter Collaboration (M.~Baak {\em et~al.}), {\em Eur. Phys. J.} {\bf C74},
  3046  (2014), \href{http://arxiv.org/abs/1407.3792}{{\ttfamily
  arXiv:1407.3792 [hep-ph]}}.

\bibitem{Erler:2017ozu}
J.~Erler, { {Theoretical Implications of Precision Measurements}}, in {\em
  {28th International Symposium on Lepton Photon Interactions at High Energies
  (LP17), Guangzhou, China, August 7-12, 2017}\/},
\newblock \href{http://arxiv.org/abs/1710.06503}{{\ttfamily arXiv:1710.06503
  [hep-ph]}}.

\bibitem{Haller:2018nnx}
J.~Haller {\em et~al.}, {\em Eur. Phys. J.} {\bf C78},   675  (2018),
  \href{http://arxiv.org/abs/1803.01853}{{\ttfamily arXiv:1803.01853
  [hep-ph]}}.

\bibitem{Tanabashi:2018}
 Particle Data Collaboration (M.~Tanabashi {\em et~al.}), {\em Phys. Rev.} {\bf
  D98},   010001  (2018).

\bibitem{Bardin:1999yd}
D.~{\relax Yu}. Bardin {\em et~al.}, {\em Comput. Phys. Commun.} {\bf 133}, 229
   (2001), \href{http://arxiv.org/abs/hep-ph/9908433}{{\ttfamily
  arXiv:hep-ph/9908433 [hep-ph]}}.

\bibitem{Arbuzov:2005ma}
A.~B. Arbuzov {\em et~al.}, {\em Comput. Phys. Commun.} {\bf 174}, 728  (2006),
  \href{http://arxiv.org/abs/hep-ph/0507146}{{\ttfamily arXiv:hep-ph/0507146
  [hep-ph]}}.

\bibitem{Akhundov:2013ons}
A.~Akhundov, A.~Arbuzov, S.~Riemann and T.~Riemann, {\em Phys. Part. Nucl.}
  {\bf 45}, 529  (2014), \href{http://arxiv.org/abs/1302.1395}{{\ttfamily
  arXiv:1302.1395 [hep-ph]}}.

\bibitem{Erler:1999ug}
J.~Erler, { {Global fits to electroweak data using GAPP}}, in {\em {QCD and
  weak boson physics in Run II. Proceedings, Batavia, IL, March 4-6, June 3-4,
  November 4-6, 1999}\/},
\newblock \href{http://arxiv.org/abs/hep-ph/0005084}{{\ttfamily
  arXiv:hep-ph/0005084 [hep-ph]}}.

\bibitem{Bardeen:1978yd}
W.~A. Bardeen, A.~J. Buras, D.~W. Duke and T.~Muta, {\em Phys. Rev.} {\bf D18},
    3998  (1978).

\bibitem{Erler:2009jh}
J.~Erler, P.~Langacker, S.~Munir and E.~Rojas, {\em JHEP} {\bf 08},   017
  (2009), \href{http://arxiv.org/abs/0906.2435}{{\ttfamily arXiv:0906.2435
  [hep-ph]}}.

\bibitem{Erler:2010sk}
J.~Erler and P.~Langacker, {\em Phys. Rev. Lett.} {\bf 105},   031801  (2010),
  \href{http://arxiv.org/abs/1003.3211}{{\ttfamily arXiv:1003.3211 [hep-ph]}}.

\bibitem{Langacker:2010zza}
P.~Langacker, {\em {The standard model and beyond}} (Boca Raton, USA: CRC
  Press, 2010).

\bibitem{Kumar:2013yoa}
K.~S. Kumar, S.~Mantry, W.~J. Marciano and P.~A. Souder, {\em Ann. Rev. Nucl.
  Part. Sci.} {\bf 63}, 237  (2013),
  \href{http://arxiv.org/abs/1302.6263}{{\ttfamily arXiv:1302.6263 [hep-ex]}}.

\bibitem{Erler:2014fqa}
J.~Erler, C.~J. Horowitz, S.~Mantry and P.~A. Souder, {\em Ann. Rev. Nucl.
  Part. Sci.} {\bf 64}, 269  (2014),
  \href{http://arxiv.org/abs/1401.6199}{{\ttfamily arXiv:1401.6199 [hep-ph]}}.

\bibitem{Berman:1958ti}
S.~M. Berman, {\em Phys. Rev.} {\bf 112}, 267  (1958).

\bibitem{Kinoshita:1958ru}
T.~Kinoshita and A.~Sirlin, {\em Phys. Rev.} {\bf 113}, 1652  (1959).

\bibitem{vanRitbergen:1998yd}
T.~van Ritbergen and R.~G. Stuart, {\em Phys. Rev. Lett.} {\bf 82}, 488
  (1999), \href{http://arxiv.org/abs/hep-ph/9808283}{{\ttfamily
  arXiv:hep-ph/9808283 [hep-ph]}}.

\bibitem{Ferroglia:1999tg}
A.~Ferroglia, G.~Ossola and A.~Sirlin, {\em Nucl. Phys.} {\bf B560}, 23
  (1999), \href{http://arxiv.org/abs/hep-ph/9905442}{{\ttfamily
  arXiv:hep-ph/9905442 [hep-ph]}}.

\bibitem{Veltman:1977kh}
M.~J.~G. Veltman, {\em Nucl. Phys.} {\bf B123}, 89  (1977).

\bibitem{Marciano:1980pb}
W.~J. Marciano and A.~Sirlin, {\em Phys. Rev.} {\bf D22},   2695  (1980),
  [Erratum: {\em ibid.} {\bf D31},213(1985)].

\bibitem{Sirlin:1983ys}
A.~Sirlin, {\em Phys. Rev.} {\bf D29},  ~89  (1984).

\bibitem{Consoli:1989fg}
M.~Consoli, W.~Hollik and F.~Jegerlehner, {\em Phys. Lett.} {\bf B227}, 167
  (1989).

\bibitem{Degrassi:1990tu}
G.~Degrassi, S.~Fanchiotti and A.~Sirlin, {\em Nucl. Phys.} {\bf B351}, 49
  (1991).

\bibitem{Djouadi:1987gn}
A.~Djouadi and C.~Verzegnassi, {\em Phys. Lett.} {\bf B195}, 265  (1987).

\bibitem{Djouadi:1987di}
A.~Djouadi, {\em Nuovo Cim.} {\bf A100},   357  (1988).

\bibitem{Kniehl:1989yc}
B.~A. Kniehl, {\em Nucl. Phys.} {\bf B347}, 86  (1990).

\bibitem{Halzen:1990je}
F.~Halzen and B.~A. Kniehl, {\em Nucl. Phys.} {\bf B353}, 567  (1991).

\bibitem{Djouadi:1993ss}
A.~Djouadi and P.~Gambino, {\em Phys. Rev.} {\bf D49}, 3499  (1994),
  \href{http://arxiv.org/abs/hep-ph/9309298}{{\ttfamily arXiv:hep-ph/9309298
  [hep-ph]}}, [Erratum: {\em ibid.} {\bf D53},4111(1996)].

\bibitem{vanderBij:1986hy}
J.~J. van~der Bij and F.~Hoogeveen, {\em Nucl. Phys.} {\bf B283}, 477  (1987).

\bibitem{Barbieri:1992nz}
R.~Barbieri, M.~Beccaria, P.~Ciafaloni, G.~Curci and A.~Vicere, {\em Phys.
  Lett.} {\bf B288}, 95  (1992),
  \href{http://arxiv.org/abs/hep-ph/9205238}{{\ttfamily arXiv:hep-ph/9205238
  [hep-ph]}}, [Erratum: {\em ibid.} {\bf B312},511(1993)].

\bibitem{Fleischer:1993ub}
J.~Fleischer, O.~V. Tarasov and F.~Jegerlehner, {\em Phys. Lett.} {\bf B319},
  249  (1993).

\bibitem{Degrassi:1996mg}
G.~Degrassi, P.~Gambino and A.~Vicini, {\em Phys. Lett.} {\bf B383}, 219
  (1996), \href{http://arxiv.org/abs/hep-ph/9603374}{{\ttfamily
  arXiv:hep-ph/9603374 [hep-ph]}}.

\bibitem{Freitas:2000gg}
A.~Freitas, W.~Hollik, W.~Walter and G.~Weiglein, {\em Phys. Lett.} {\bf B495},
  338  (2000), \href{http://arxiv.org/abs/hep-ph/0007091}{{\ttfamily
  arXiv:hep-ph/0007091 [hep-ph]}}, [Erratum: {\em ibid.} {\bf B570},265(2003)].

\bibitem{Freitas:2002ja}
A.~Freitas, W.~Hollik, W.~Walter and G.~Weiglein, {\em Nucl. Phys.} {\bf B632},
  189  (2002), \href{http://arxiv.org/abs/hep-ph/0202131}{{\ttfamily
  arXiv:hep-ph/0202131 [hep-ph]}}, [Erratum: Nucl. Phys.B666,305(2003)].

\bibitem{Awramik:2003ee}
M.~Awramik and M.~Czakon, {\em Phys. Lett.} {\bf B568}, 48  (2003),
  \href{http://arxiv.org/abs/hep-ph/0305248}{{\ttfamily arXiv:hep-ph/0305248
  [hep-ph]}}.

\bibitem{Awramik:2002wn}
M.~Awramik and M.~Czakon, {\em Phys. Rev. Lett.} {\bf 89},   241801  (2002),
  \href{http://arxiv.org/abs/hep-ph/0208113}{{\ttfamily arXiv:hep-ph/0208113
  [hep-ph]}}.

\bibitem{Onishchenko:2002ve}
A.~Onishchenko and O.~Veretin, {\em Phys. Lett.} {\bf B551}, 111  (2003),
  \href{http://arxiv.org/abs/hep-ph/0209010}{{\ttfamily arXiv:hep-ph/0209010
  [hep-ph]}}.

\bibitem{Degrassi:2014sxa}
G.~Degrassi, P.~Gambino and P.~P. Giardino, {\em JHEP} {\bf 05},   154  (2015),
  \href{http://arxiv.org/abs/1411.7040}{{\ttfamily arXiv:1411.7040 [hep-ph]}}.

\bibitem{Martin:2015lxa}
S.~P. Martin, {\em Phys. Rev.} {\bf D91},   114003  (2015),
  \href{http://arxiv.org/abs/1503.03782}{{\ttfamily arXiv:1503.03782
  [hep-ph]}}.

\bibitem{Avdeev:1994db}
L.~Avdeev, J.~Fleischer, S.~Mikhailov and O.~Tarasov, {\em Phys. Lett.} {\bf
  B336}, 560  (1994), \href{http://arxiv.org/abs/hep-ph/9406363}{{\ttfamily
  arXiv:hep-ph/9406363 [hep-ph]}}, [Erratum: {\em ibid.} {\bf B349},597(1995)].

\bibitem{Chetyrkin:1995ix}
K.~G. Chetyrkin, J.~H. K{\"u}hn and M.~Steinhauser, {\em Phys. Lett.} {\bf
  B351}, 331  (1995), \href{http://arxiv.org/abs/hep-ph/9502291}{{\ttfamily
  arXiv:hep-ph/9502291 [hep-ph]}}.

\bibitem{Anselm:1993uq}
A.~Anselm, N.~Dombey and E.~Leader, {\em Phys. Lett.} {\bf B312}, 232  (1993).

\bibitem{Schroder:2005db}
Y.~Schr{\"o}der and M.~Steinhauser, {\em Phys. Lett.} {\bf B622}, 124  (2005),
  \href{http://arxiv.org/abs/hep-ph/0504055}{{\ttfamily arXiv:hep-ph/0504055
  [hep-ph]}}.

\bibitem{Chetyrkin:2006bj}
K.~G. Chetyrkin, M.~Faisst, J.~H. K{\"u}hn, P.~Maierh{\"o}fer and C.~Sturm,
  {\em Phys. Rev. Lett.} {\bf 97},   102003  (2006),
  \href{http://arxiv.org/abs/hep-ph/0605201}{{\ttfamily arXiv:hep-ph/0605201
  [hep-ph]}}.

\bibitem{Boughezal:2006xk}
R.~Boughezal and M.~Czakon, {\em Nucl. Phys.} {\bf B755}, 221  (2006),
  \href{http://arxiv.org/abs/hep-ph/0606232}{{\ttfamily arXiv:hep-ph/0606232
  [hep-ph]}}.

\bibitem{Chetyrkin:1995js}
K.~G. Chetyrkin, J.~H. K{\"u}hn and M.~Steinhauser, {\em Phys. Rev. Lett.} {\bf
  75}, 3394  (1995), \href{http://arxiv.org/abs/hep-ph/9504413}{{\ttfamily
  arXiv:hep-ph/9504413 [hep-ph]}}.

\bibitem{Chetyrkin:1996cf}
K.~G. Chetyrkin, J.~H. K{\"u}hn and M.~Steinhauser, {\em Nucl. Phys.} {\bf
  B482}, 213  (1996), \href{http://arxiv.org/abs/hep-ph/9606230}{{\ttfamily
  arXiv:hep-ph/9606230 [hep-ph]}}.

\bibitem{vanderBij:2000cg}
J.~J. van~der Bij, K.~G. Chetyrkin, M.~Faisst, G.~Jikia and T.~Seidensticker,
  {\em Phys. Lett.} {\bf B498}, 156  (2001),
  \href{http://arxiv.org/abs/hep-ph/0011373}{{\ttfamily arXiv:hep-ph/0011373
  [hep-ph]}}.

\bibitem{Faisst:2003px}
M.~Faisst, J.~H. K{\"u}hn, T.~Seidensticker and O.~Veretin, {\em Nucl. Phys.}
  {\bf B665}, 649  (2003),
  \href{http://arxiv.org/abs/hep-ph/0302275}{{\ttfamily arXiv:hep-ph/0302275
  [hep-ph]}}.

\bibitem{Awramik:2003rn}
M.~Awramik, M.~Czakon, A.~Freitas and G.~Weiglein, {\em Phys. Rev.} {\bf D69},
   053006  (2004), \href{http://arxiv.org/abs/hep-ph/0311148}{{\ttfamily
  arXiv:hep-ph/0311148 [hep-ph]}}.

\bibitem{Awramik:2004ge}
M.~Awramik, M.~Czakon, A.~Freitas and G.~Weiglein, {\em Phys. Rev. Lett.} {\bf
  93},   201805  (2004), \href{http://arxiv.org/abs/hep-ph/0407317}{{\ttfamily
  arXiv:hep-ph/0407317 [hep-ph]}}.

\bibitem{Hollik:2005va}
W.~Hollik, U.~Meier and S.~Uccirati, {\em Nucl. Phys.} {\bf B731}, 213  (2005),
  \href{http://arxiv.org/abs/hep-ph/0507158}{{\ttfamily arXiv:hep-ph/0507158
  [hep-ph]}}.

\bibitem{Awramik:2006ar}
M.~Awramik, M.~Czakon and A.~Freitas, {\em Phys. Lett.} {\bf B642}, 563
  (2006), \href{http://arxiv.org/abs/hep-ph/0605339}{{\ttfamily
  arXiv:hep-ph/0605339 [hep-ph]}}.

\bibitem{Hollik:2006ma}
W.~Hollik, U.~Meier and S.~Uccirati, {\em Nucl. Phys.} {\bf B765}, 154  (2007),
  \href{http://arxiv.org/abs/hep-ph/0610312}{{\ttfamily arXiv:hep-ph/0610312
  [hep-ph]}}.

\bibitem{Degrassi:1996ps}
G.~Degrassi, P.~Gambino and A.~Sirlin, {\em Phys. Lett.} {\bf B394}, 188
  (1997), \href{http://arxiv.org/abs/hep-ph/9611363}{{\ttfamily
  arXiv:hep-ph/9611363 [hep-ph]}}.

\bibitem{Awramik:2006uz}
M.~Awramik, M.~Czakon and A.~Freitas, {\em JHEP} {\bf 11},   048  (2006),
  \href{http://arxiv.org/abs/hep-ph/0608099}{{\ttfamily arXiv:hep-ph/0608099
  [hep-ph]}}.

\bibitem{Czarnecki:1996ei}
A.~Czarnecki and J.~H. K{\"u}hn, {\em Phys. Rev. Lett.} {\bf 77}, 3955  (1996),
  \href{http://arxiv.org/abs/hep-ph/9608366}{{\ttfamily arXiv:hep-ph/9608366
  [hep-ph]}}.

\bibitem{Harlander:1997zb}
R.~Harlander, T.~Seidensticker and M.~Steinhauser, {\em Phys. Lett.} {\bf
  B426}, 125  (1998), \href{http://arxiv.org/abs/hep-ph/9712228}{{\ttfamily
  arXiv:hep-ph/9712228 [hep-ph]}}.

\bibitem{Akhundov:1985fc}
A.~A. Akhundov, D.~{\relax Yu}. Bardin and T.~Riemann, {\em Nucl. Phys.} {\bf
  B276}, 1  (1986).

\bibitem{Bernabeu:1987me}
J.~Bernabeu, A.~Pich and A.~Santamaria, {\em Phys. Lett.} {\bf B200}, 569
  (1988).

\bibitem{Beenakker:1988pv}
W.~Beenakker and W.~Hollik, {\em Z. Phys.} {\bf C40},   141  (1988).

\bibitem{Degrassi:1990ec}
G.~Degrassi and A.~Sirlin, {\em Nucl. Phys.} {\bf B352}, 342  (1991).

\bibitem{Fleischer:1992fq}
J.~Fleischer, O.~V. Tarasov, F.~Jegerlehner and P.~Raczka, {\em Phys. Lett.}
  {\bf B293}, 437  (1992).

\bibitem{Degrassi:1993ij}
G.~Degrassi, {\em Nucl. Phys.} {\bf B407}, 271  (1993),
  \href{http://arxiv.org/abs/hep-ph/9302288}{{\ttfamily arXiv:hep-ph/9302288
  [hep-ph]}}.

\bibitem{Awramik:2008gi}
M.~Awramik, M.~Czakon, A.~Freitas and B.~A. Kniehl, {\em Nucl. Phys.} {\bf
  B813}, 174  (2009), \href{http://arxiv.org/abs/0811.1364}{{\ttfamily
  arXiv:0811.1364 [hep-ph]}}.

\bibitem{Dubovyk:2016aqv}
I.~Dubovyk, A.~Freitas, J.~Gluza, T.~Riemann and J.~Usovitsch, {\em Phys.
  Lett.} {\bf B762}, 184  (2016),
  \href{http://arxiv.org/abs/1607.08375}{{\ttfamily arXiv:1607.08375
  [hep-ph]}}.

\bibitem{Chetyrkin:1994js}
K.~G. Chetyrkin, J.~H. K{\"u}hn and A.~Kwiatkowski, {\em Phys. Rept.} {\bf
  277},   189  (1996), \href{http://arxiv.org/abs/hep-ph/9503396}{{\ttfamily
  arXiv:hep-ph/9503396 [hep-ph]}}.

\bibitem{Appelquist:1973uz}
T.~Appelquist and H.~Georgi, {\em Phys. Rev.} {\bf D8}, 4000  (1973).

\bibitem{Zee:1973sr}
A.~Zee, {\em Phys. Rev.} {\bf D8}, 4038  (1973).

\bibitem{Chetyrkin:1979bj}
K.~G. Chetyrkin, A.~L. Kataev and F.~V. Tkachov, {\em Phys. Lett.} {\bf 85B},
  277  (1979).

\bibitem{Dine:1979qh}
M.~Dine and J.~R. Sapirstein, {\em Phys. Rev. Lett.} {\bf 43},   668  (1979).

\bibitem{Celmaster:1979xr}
W.~Celmaster and R.~J. Gonsalves, {\em Phys. Rev. Lett.} {\bf 44},   560
  (1980).

\bibitem{Surguladze:1990tg}
L.~R. Surguladze and M.~A. Samuel, {\em Phys. Rev. Lett.} {\bf 66}, 560
  (1991), [Erratum: {\em ibid.} {\bf 66},2416(1991)].

\bibitem{Gorishnii:1990vf}
S.~G. Gorishnii, A.~L. Kataev and S.~A. Larin, {\em Phys. Lett.} {\bf B259},
  144  (1991).

\bibitem{Baikov:2008jh}
P.~A. Baikov, K.~G. Chetyrkin and J.~H. K{\"u}hn, {\em Phys. Rev. Lett.} {\bf
  101},   012002  (2008), \href{http://arxiv.org/abs/0801.1821}{{\ttfamily
  arXiv:0801.1821 [hep-ph]}}.

\bibitem{Poggio:1975af}
E.~C. Poggio, H.~R. Quinn and S.~Weinberg, {\em Phys. Rev.} {\bf D13},   1958
  (1976).

\bibitem{Barnett:1980sm}
R.~M. Barnett, M.~Dine and L.~D. McLerran, {\em Phys. Rev.} {\bf D22},   594
  (1980).

\bibitem{Kataev:1992dg}
A.~L. Kataev, {\em Phys. Lett.} {\bf B287}, 209  (1992).

\bibitem{Larin:1994va}
S.~A. Larin, T.~van Ritbergen and J.~A.~M. Vermaseren, {\em Nucl. Phys.} {\bf
  B438}, 278  (1995), \href{http://arxiv.org/abs/hep-ph/9411260}{{\ttfamily
  arXiv:hep-ph/9411260 [hep-ph]}}.

\bibitem{Baikov:2012er}
P.~A. Baikov, K.~G. Chetyrkin, J.~H. K{\"u}hn and J.~Rittinger, {\em Phys. Rev.
  Lett.} {\bf 108},   222003  (2012),
  \href{http://arxiv.org/abs/1201.5804}{{\ttfamily arXiv:1201.5804 [hep-ph]}}.

\bibitem{Kniehl:1989bb}
B.~A. Kniehl and J.~H. K{\"u}hn, {\em Phys. Lett.} {\bf B224}, 229  (1989).

\bibitem{Larin:1993ju}
S.~A. Larin, T.~van Ritbergen and J.~A.~M. Vermaseren, {\em Phys. Lett.} {\bf
  B320}, 159  (1994), \href{http://arxiv.org/abs/hep-ph/9310378}{{\ttfamily
  arXiv:hep-ph/9310378 [hep-ph]}}.

\bibitem{Freitas:2014hra}
A.~Freitas, {\em JHEP} {\bf 04},   070  (2014),
  \href{http://arxiv.org/abs/1401.2447}{{\ttfamily arXiv:1401.2447 [hep-ph]}}.

\bibitem{Freitas:2013dpa}
A.~Freitas, {\em Phys. Lett.} {\bf B730}, 50  (2014),
  \href{http://arxiv.org/abs/1310.2256}{{\ttfamily arXiv:1310.2256 [hep-ph]}}.

\bibitem{Degrassi:1999jd}
G.~Degrassi and P.~Gambino, {\em Nucl. Phys.} {\bf B567}, 3  (2000),
  \href{http://arxiv.org/abs/hep-ph/9905472}{{\ttfamily arXiv:hep-ph/9905472
  [hep-ph]}}.

\bibitem{Denner:1990tx}
A.~Denner and T.~Sack, {\em Z. Phys.} {\bf C46}, 653  (1990).

\bibitem{Bardin:1986fi}
D.~{\relax Yu}. Bardin, S.~Riemann and T.~Riemann, {\em Z. Phys.} {\bf C32},
  121  (1986).

\bibitem{Rosner:1993rj}
J.~L. Rosner, M.~P. Worah and T.~Takeuchi, {\em Phys. Rev.} {\bf D49}, 1363
  (1994), \href{http://arxiv.org/abs/hep-ph/9309307}{{\ttfamily
  arXiv:hep-ph/9309307 [hep-ph]}}.

\bibitem{Kara:2013dua}
D.~Kara, {\em Nucl. Phys.} {\bf B877}, 683  (2013),
  \href{http://arxiv.org/abs/1307.7190}{{\ttfamily arXiv:1307.7190 [hep-ph]}}.

\bibitem{Peskin:1991sw}
M.~E. Peskin and T.~Takeuchi, {\em Phys. Rev.} {\bf D46}, 381  (1992).

\bibitem{Marciano:1990dp}
W.~J. Marciano and J.~L. Rosner, {\em Phys. Rev. Lett.} {\bf 65}, 2963  (1990),
  [Erratum: {\em ibid.} {\bf 68},898(1992)].

\bibitem{Grassi:2000dz}
P.~A. Grassi, B.~A. Kniehl and A.~Sirlin, {\em Phys. Rev. Lett.} {\bf 86}, 389
  (2001), \href{http://arxiv.org/abs/hep-th/0005149}{{\ttfamily
  arXiv:hep-th/0005149 [hep-th]}}.

\bibitem{Fanchiotti:1989wv}
S.~Fanchiotti and A.~Sirlin, {\em Phys. Rev.} {\bf D41},   319  (1990).

\bibitem{Dubovyk:2018rlg}
I.~Dubovyk, A.~Freitas, J.~Gluza, T.~Riemann and J.~Usovitsch, {\em Phys.
  Lett.} {\bf B783}, 86  (2018),
  \href{http://arxiv.org/abs/1804.10236}{{\ttfamily arXiv:1804.10236
  [hep-ph]}}.

\bibitem{Blondel:2018mad}
A.~Blondel {\em et~al.}, { {Standard Model Theory for the FCC-ee:\ The
  Tera-Z}}, in {\em {Mini Workshop on Precision EW and QCD Calculations for the
  FCC Studies: Methods and Techniques, CERN, Geneva, Switzerland, January
  12-13, 2018}\/},
\newblock \href{http://arxiv.org/abs/1809.01830}{{\ttfamily arXiv:1809.01830
  [hep-ph]}}.

\bibitem{Steinhauser:1998rq}
M.~Steinhauser, {\em Phys. Lett.} {\bf B429}, 158  (1998),
  \href{http://arxiv.org/abs/hep-ph/9803313}{{\ttfamily arXiv:hep-ph/9803313
  [hep-ph]}}.

\bibitem{Sturm:2013uka}
C.~Sturm, {\em Nucl. Phys.} {\bf B874}, 698  (2013),
  \href{http://arxiv.org/abs/1305.0581}{{\ttfamily arXiv:1305.0581 [hep-ph]}}.

\bibitem{Davier:2017zfy}
M.~Davier, A.~H{\"o}cker, B.~Malaescu and Z.~Zhang, {\em Eur. Phys. J.} {\bf
  C77},   827  (2017), \href{http://arxiv.org/abs/1706.09436}{{\ttfamily
  arXiv:1706.09436 [hep-ph]}}.

\bibitem{Erler:1998sy}
J.~Erler, {\em Phys. Rev.} {\bf D59},   054008  (1999),
  \href{http://arxiv.org/abs/hep-ph/9803453}{{\ttfamily arXiv:hep-ph/9803453
  [hep-ph]}}.

\bibitem{Fanchiotti:1992tu}
S.~Fanchiotti, B.~A. Kniehl and A.~Sirlin, {\em Phys. Rev.} {\bf D48}, 307
  (1993), \href{http://arxiv.org/abs/hep-ph/9212285}{{\ttfamily
  arXiv:hep-ph/9212285 [hep-ph]}}.

\bibitem{Degrassi:2003rw}
G.~Degrassi and A.~Vicini, {\em Phys. Rev.} {\bf D69},   073007  (2004),
  \href{http://arxiv.org/abs/hep-ph/0307122}{{\ttfamily arXiv:hep-ph/0307122
  [hep-ph]}}.

\bibitem{Keshavarzi:2018mgv}
A.~Keshavarzi, D.~Nomura and T.~Teubner, {\em Phys. Rev.} {\bf D97},   114025
  (2018), \href{http://arxiv.org/abs/1802.02995}{{\ttfamily arXiv:1802.02995
  [hep-ph]}}.

\bibitem{Jegerlehner:2017zsb}
F.~Jegerlehner, { {Variations on Photon Vacuum Polarization}}, in {\em
  {International Workshop on $e^+e^-$ collisions from Phi to Psi, 26-29 June
  2017, Mainz, Germany}\/},
\newblock \href{http://arxiv.org/abs/1711.06089}{{\ttfamily arXiv:1711.06089
  [hep-ph]}}.

\bibitem{Erler:2017knj}
J.~Erler and R.~Ferro-Hern{\'a}ndez, {\em JHEP} {\bf 03},   196  (2018),
  \href{http://arxiv.org/abs/1712.09146}{{\ttfamily arXiv:1712.09146
  [hep-ph]}}.

\bibitem{Gross:1973id}
D.~J. Gross and F.~Wilczek, {\em Phys. Rev. Lett.} {\bf 30}, 1343  (1973).

\bibitem{Politzer:1973fx}
H.~D. Politzer, {\em Phys. Rev. Lett.} {\bf 30}, 1346  (1973).

\bibitem{Caswell:1974gg}
W.~E. Caswell, {\em Phys. Rev. Lett.} {\bf 33},   244  (1974).

\bibitem{Jones:1974mm}
D.~R.~T. Jones, {\em Nucl. Phys.} {\bf B75},   531  (1974).

\bibitem{Tarasov:1980au}
O.~V. Tarasov, A.~A. Vladimirov and A.~{\relax Yu}. Zharkov, {\em Phys. Lett.}
  {\bf 93B}, 429  (1980).

\bibitem{vanRitbergen:1997va}
T.~van Ritbergen, J.~A.~M. Vermaseren and S.~A. Larin, {\em Phys. Lett.} {\bf
  B400}, 379  (1997), \href{http://arxiv.org/abs/hep-ph/9701390}{{\ttfamily
  arXiv:hep-ph/9701390 [hep-ph]}}.

\bibitem{Czakon:2004bu}
M.~Czakon, {\em Nucl. Phys.} {\bf B710}, 485  (2005),
  \href{http://arxiv.org/abs/hep-ph/0411261}{{\ttfamily arXiv:hep-ph/0411261
  [hep-ph]}}.

\bibitem{Baikov:2016tgj}
P.~A. Baikov, K.~G. Chetyrkin and J.~H. K{\"u}hn, {\em Phys. Rev. Lett.} {\bf
  118},   082002  (2017), \href{http://arxiv.org/abs/1606.08659}{{\ttfamily
  arXiv:1606.08659 [hep-ph]}}.

\bibitem{Herzog:2017ohr}
F.~Herzog, B.~Ruijl, T.~Ueda, J.~A.~M. Vermaseren and A.~Vogt, {\em JHEP} {\bf
  02},   090  (2017), \href{http://arxiv.org/abs/1701.01404}{{\ttfamily
  arXiv:1701.01404 [hep-ph]}}.

\bibitem{Schroder:2005hy}
Y.~Schr{\"o}der and M.~Steinhauser, {\em JHEP} {\bf 01},   051  (2006),
  \href{http://arxiv.org/abs/hep-ph/0512058}{{\ttfamily arXiv:hep-ph/0512058
  [hep-ph]}}.

\bibitem{Dissertori:2015tfa}
G.~Dissertori, {\em Adv. Ser. Direct. High Energy Phys.} {\bf 26}, 113  (2016),
  \href{http://arxiv.org/abs/1506.05407}{{\ttfamily arXiv:1506.05407
  [hep-ex]}}.

\bibitem{Baikov:2014qja}
P.~A. Baikov, K.~G. Chetyrkin and J.~H. K{\"u}hn, {\em JHEP} {\bf 10},   076
  (2014), \href{http://arxiv.org/abs/1402.6611}{{\ttfamily arXiv:1402.6611
  [hep-ph]}}.

\bibitem{Liu:2015fxa}
T.~Liu and M.~Steinhauser, {\em Phys. Lett.} {\bf B746}, 330  (2015),
  \href{http://arxiv.org/abs/1502.04719}{{\ttfamily arXiv:1502.04719
  [hep-ph]}}.

\bibitem{Tishchenko:2012ie}
 MuLan Collaboration (V.~Tishchenko {\em et~al.}), {\em Phys. Rev.} {\bf D87},
   052003  (2013), \href{http://arxiv.org/abs/1211.0960}{{\ttfamily
  arXiv:1211.0960 [hep-ex]}}.

\bibitem{vanRitbergen:1999fi}
T.~van Ritbergen and R.~G. Stuart, {\em Nucl. Phys.} {\bf B564}, 343  (2000),
  \href{http://arxiv.org/abs/hep-ph/9904240}{{\ttfamily arXiv:hep-ph/9904240
  [hep-ph]}}.

\bibitem{PhysRevD.16.1519}
B.~W. Lee, C.~Quigg and H.~B. Thacker, {\em Phys. Rev. D} {\bf 16}, 1519 (Sep
  1977).

\bibitem{PhysRevD.40.1725}
W.~Marciano, G.~Valencia and S.~Willenbrock, {\em Phys. Rev. D} {\bf 40}, 1725
  (Sep 1989).

\bibitem{Khachatryan:2016vau}
 ATLAS and CMS Collaborations (G.~Aad {\em et~al.}), {\em JHEP} {\bf 08},   045
   (2016), \href{http://arxiv.org/abs/1606.02266}{{\ttfamily arXiv:1606.02266
  [hep-ex]}}.

\bibitem{Aaboud:2017jvq}
 ATLAS Collaboration (M.~Aaboud {\em et~al.}), {\em Phys. Rev.} {\bf D97},
  072003  (2018), \href{http://arxiv.org/abs/1712.08891}{{\ttfamily
  arXiv:1712.08891 [hep-ex]}}.

\bibitem{Sirunyan:2018hoz}
 CMS Collaboration (A.~M. Sirunyan {\em et~al.}), {\em Phys. Rev. Lett.} {\bf
  120},   231801  (2018), \href{http://arxiv.org/abs/1804.02610}{{\ttfamily
  arXiv:1804.02610 [hep-ex]}}.

\bibitem{Khachatryan:2014jba}
 CMS Collaboration (V.~Khachatryan {\em et~al.}), {\em Eur. Phys. J.} {\bf
  C75},   212  (2015), \href{http://arxiv.org/abs/1412.8662}{{\ttfamily
  arXiv:1412.8662 [hep-ex]}}.

\bibitem{Aad:2015mxa}
 ATLAS Collaboration (G.~Aad {\em et~al.}), {\em Eur. Phys. J.} {\bf C75},
  476  (2015), \href{http://arxiv.org/abs/1506.05669}{{\ttfamily
  arXiv:1506.05669 [hep-ex]}}, [Erratum: {\em ibid.} {\bf C76},152(2016)].

\bibitem{Aaboud:2018urx}
 ATLAS Collaboration (M.~Aaboud {\em et~al.}), {\em Phys. Lett.} {\bf B784},
  173  (2018), \href{http://arxiv.org/abs/1806.00425}{{\ttfamily
  arXiv:1806.00425 [hep-ex]}}.

\bibitem{Aaboud:2018wps}
 ATLAS Collaboration (M.~Aaboud {\em et~al.}), {\em Phys. Lett.} {\bf B784},
  345  (2018), \href{http://arxiv.org/abs/1806.00242}{{\ttfamily
  arXiv:1806.00242 [hep-ex]}}.

\bibitem{Sirunyan:2017exp}
 CMS Collaboration (A.~M. Sirunyan {\em et~al.}), {\em JHEP} {\bf 11},   047
  (2017), \href{http://arxiv.org/abs/1706.09936}{{\ttfamily arXiv:1706.09936
  [hep-ex]}}.

\bibitem{Khachatryan:2014kca}
 CMS Collaboration (V.~Khachatryan {\em et~al.}), {\em Phys. Rev.} {\bf D92},
  012004  (2015), \href{http://arxiv.org/abs/1411.3441}{{\ttfamily
  arXiv:1411.3441 [hep-ex]}}.

\bibitem{Aad:2015zhl}
 ATLAS and CMS Collaborations (G.~Aad {\em et~al.}), {\em Phys. Rev. Lett.}
  {\bf 114},   191803  (2015),
  \href{http://arxiv.org/abs/1503.07589}{{\ttfamily arXiv:1503.07589
  [hep-ex]}}.

\bibitem{Olive:2016xmw}
 Particle Data Collaboration (C.~Patrignani {\em et~al.}), {\em Chin. Phys.}
  {\bf C40},   100001  (2016).

\bibitem{Lyons:1988rp}
L.~Lyons, D.~Gibaut and P.~Clifford, {\em Nucl. Instrum. Meth.} {\bf A270},
  110  (1988).

\bibitem{Nisius:2014wua}
R.~Nisius, {\em Eur. Phys. J.} {\bf C74},   3004  (2014),
  \href{http://arxiv.org/abs/1402.4016}{{\ttfamily arXiv:1402.4016
  [physics.data-an]}}.

\bibitem{Aaltonen:2012bp}
 CDF Collaboration (T.~Aaltonen {\em et~al.}), {\em Phys. Rev. Lett.} {\bf
  108},   151803  (2012), \href{http://arxiv.org/abs/1203.0275}{{\ttfamily
  arXiv:1203.0275 [hep-ex]}}.

\bibitem{Abazov:2012bv}
 \DZero Collaboration (V.~M. Abazov {\em et~al.}), {\em Phys. Rev. Lett.} {\bf
  108},   151804  (2012), \href{http://arxiv.org/abs/1203.0293}{{\ttfamily
  arXiv:1203.0293 [hep-ex]}}.

\bibitem{Aaboud:2017svj}
 ATLAS Collaboration (M.~Aaboud {\em et~al.}), {\em Eur. Phys. J.} {\bf C78},
  110  (2018), \href{http://arxiv.org/abs/1701.07240}{{\ttfamily
  arXiv:1701.07240 [hep-ex]}}, [Erratum: {\em ibid.} {\bf C78},898(2018)].

\bibitem{Sjostrand:2007gs}
T.~Sj{\"o}strand, S.~Mrenna and P.~Skands, {\em Comput. Phys. Commun.} {\bf
  178}, 852  (2008), \href{http://arxiv.org/abs/0710.3820}{{\ttfamily
  arXiv:0710.3820 [hep-ph]}}.

\bibitem{deFavereau:2013fsa}
 DELPHES 3 Collaboration (J.~de~Favereau {\em et~al.}), {\em JHEP} {\bf 02},
  057  (2014), \href{http://arxiv.org/abs/1307.6346}{{\ttfamily arXiv:1307.6346
  [hep-ex]}}.

\bibitem{Calame_2006}
C.~M.~C. Calame, G.~Montagna, O.~Nicrosini and A.~Vicini, {\em Journal of High
  Energy Physics} {\bf 2006}, 016 (Dec 2006).

\bibitem{Ladinsky:1993zn}
G.~Ladinsky and C.~Yuan, {\em Phys. Rev.} {\bf D50},   4239  (1994),
  \href{http://arxiv.org/abs/hep-ph/9311341}{{\ttfamily arXiv:hep-ph/9311341
  [hep-ph]}}.

\bibitem{Balazs:1997xd}
C.~Balazs and C.~Yuan, {\em Phys. Rev.} {\bf D56}, 5558  (1997),
  \href{http://arxiv.org/abs/hep-ph/9704258}{{\ttfamily arXiv:hep-ph/9704258
  [hep-ph]}}.

\bibitem{Besson:2008zs}
 ATLAS Collaboration (N.~Besson {\em et~al.}), {\em Eur. Phys. J.} {\bf C57},
  627  (2008), \href{http://arxiv.org/abs/0805.2093}{{\ttfamily arXiv:0805.2093
  [hep-ex]}}.

\bibitem{Chang:1981qq}
T.~H. Chang, K.~J.~F. Gaemers and W.~L. van Neerven, {\em Nucl. Phys.} {\bf
  B202}, 407  (1982).

\bibitem{Renton:2008ub}
P.~Renton, {Updated SM calculations of $\sigma_W/\sigma_Z$ at the Tevatron and
  the $W$ boson width}  (2008),
  \href{http://arxiv.org/abs/0804.4779}{{\ttfamily arXiv:0804.4779 [hep-ph]}}.

\bibitem{Aaltonen:2007ai}
 CDF Collaboration (T.~Aaltonen {\em et~al.}), {\em Phys. Rev. Lett.} {\bf
  100},   071801  (2008), \href{http://arxiv.org/abs/0710.4112}{{\ttfamily
  arXiv:0710.4112 [hep-ex]}}.

\bibitem{Abazov:2009vs}
 \DZero Collaboration (V.~M. Abazov {\em et~al.}), {\em Phys. Rev. Lett.} {\bf
  103},   231802  (2009), \href{http://arxiv.org/abs/0909.4814}{{\ttfamily
  arXiv:0909.4814 [hep-ex]}}.

\bibitem{Aaltonen:2013iut}
 CDF and \DZero Collaborations (T.~Aaltonen {\em et~al.}), {\em Phys. Rev.}
  {\bf D88},   052018  (2013), \href{http://arxiv.org/abs/1307.7627}{{\ttfamily
  arXiv:1307.7627 [hep-ex]}}.

\bibitem{Catani:2009sm}
S.~Catani, L.~Cieri, G.~Ferrera, D.~de~Florian and M.~Grazzini, {\em Phys. Rev.
  Lett.} {\bf 103},   082001  (2009),
  \href{http://arxiv.org/abs/0903.2120}{{\ttfamily arXiv:0903.2120 [hep-ph]}}.

\bibitem{CDFWZ}
 CDF Collaboration (A.~Abulencia {\em et~al.}), {\em J. Phys. G} {\bf 34}, 2457
   (2007), \href{http://arxiv.org/abs/hep-ex/0508029}{{\ttfamily
  arXiv:hep-ex/0508029 [hep-ex]}}.

\bibitem{D0WZ}
 \DZero Collaboration (B.~Abbott {\em et~al.}), {\em Phys. Rev. D} {\bf 61},
  072001  (2000), \href{http://arxiv.org/abs/hep-ex/9906025}{{\ttfamily
  arXiv:hep-ex/9906025 [hep-ex]}}.

\bibitem{CMSWZ2}
 CMS Collaboration (S.~Chatrchyan {\em et~al.}), {\em Phys. Rev. Lett.} {\bf
  112},   191802  (2014), \href{http://arxiv.org/abs/1402.0923}{{\ttfamily
  arXiv:1402.0923 [hep-ex]}}.

\bibitem{Camarda:2016twt}
S.~Camarda, J.~Cuth and M.~Schott, {\em Eur. Phys. J.} {\bf C76},   613
  (2016), \href{http://arxiv.org/abs/1607.05084}{{\ttfamily arXiv:1607.05084
  [hep-ex]}}.

\bibitem{Bardin:1990fu}
D.~{\relax Yu}. Bardin {\em et~al.}, {\em Nucl. Phys.} {\bf B351}, 1  (1991),
  \href{http://arxiv.org/abs/hep-ph/9801208}{{\ttfamily arXiv:hep-ph/9801208
  [hep-ph]}}.

\bibitem{Arnaudon:1994zq}
L.~Arnaudon {\em et~al.}, {\em Z. Phys.} {\bf C66}, 45  (1995).

\bibitem{Barate:1999ce}
 ALEPH Collaboration (R.~Barate {\em et~al.}), {\em Eur. Phys. J.} {\bf C14}, 1
   (2000).

\bibitem{Akers:1993is}
 OPAL Collaboration (R.~Akers {\em et~al.}), {\em Z. Phys.} {\bf C61}, 19
  (1994).

\bibitem{Bouchiat:1974zz}
M.~A. Bouchiat and C.~Bouchiat, {\em J. Phys. ({\rm France)}} {\bf 35}, 899
  (1974).

\bibitem{Wood:1997zq}
C.~S. Wood {\em et~al.}, {\em Science} {\bf 275}, 1759  (1997).

\bibitem{Guena:2004sq}
J.~Gu{\'e}na, M.~Lintz and M.~A. Bouchiat, {\em Phys. Rev.} {\bf A71},   042108
   (2005), \href{http://arxiv.org/abs/physics/0412017}{{\ttfamily
  arXiv:physics/0412017 [physics.atom-ph]}}.

\bibitem{Ginges:2003qt}
J.~S.~M. Ginges and V.~V. Flambaum, {\em Phys. Rept.} {\bf 397}, 63  (2004),
  \href{http://arxiv.org/abs/physics/0309054}{{\ttfamily arXiv:physics/0309054
  [physics]}}.

\bibitem{Budker:2018}
D.~Antipas {\em et~al.}, {\em Nat. Phys.} {\bf 14}  (2018),
  \href{http://arxiv.org/abs/1804.05747}{{\ttfamily arXiv:1804.05747
  [physics.atom-ph]}}.

\bibitem{Beise:2004py}
E.~J. Beise, M.~L. Pitt and D.~T. Spayde, {\em Prog. Part. Nucl. Phys.} {\bf
  54}, 289  (2005), \href{http://arxiv.org/abs/nucl-ex/0412054}{{\ttfamily
  arXiv:nucl-ex/0412054 [nucl-ex]}}.

\bibitem{Prescott:1979dh}
C.~Y. Prescott {\em et~al.}, {\em Phys. Lett.} {\bf 84B}, 524  (1979).

\bibitem{Wang:2014bba}
 PVDIS Collaboration (D.~Wang {\em et~al.}), {\em Nature} {\bf 506}, 67
  (2014).

\bibitem{Souder:2016xcn}
P.~A. Souder, {\em Int. J. Mod. Phys. Conf. Ser.} {\bf 40},   1660077  (2016).

\bibitem{Androic:2018kni}
 Qweak Collaboration (D.~Androi{\'c} {\em et~al.}), {\em Nature} {\bf 557}, 207
   (2018).

\bibitem{Anthony:2005pm}
 SLAC--E158 Collaboration (P.~L. Anthony {\em et~al.}), {\em Phys. Rev. Lett.}
  {\bf 95},   081601  (2005),
  \href{http://arxiv.org/abs/hep-ex/0504049}{{\ttfamily arXiv:hep-ex/0504049
  [hep-ex]}}.

\bibitem{Becker:2018ggl}
D.~Becker {\em et~al.}, {\em Eur. Phys. J.} {\bf A54},   208  (2018),
  \href{http://arxiv.org/abs/1802.04759}{{\ttfamily arXiv:1802.04759
  [nucl-ex]}}.

\bibitem{Benesch:2014bas}
MOLLER Collaboration, J.~Benesch {\em et~al.}, {The MOLLER Experiment: An
  Ultra-Precise Measurement of the Weak Mixing Angle Using M{\o}ller
  Scattering}  (2014), \href{http://arxiv.org/abs/1411.4088}{{\ttfamily
  arXiv:1411.4088 [nucl-ex]}}.

\bibitem{Erler:2004in}
J.~Erler and M.~J. Ramsey-Musolf, {\em Phys. Rev.} {\bf D72},   073003  (2005),
  \href{http://arxiv.org/abs/hep-ph/0409169}{{\ttfamily arXiv:hep-ph/0409169
  [hep-ph]}}.

\bibitem{Zeller:2001hh}
 NuTeV Collaboration (G.~P. Zeller {\em et~al.}), {\em Phys. Rev. Lett.} {\bf
  88},   091802  (2002), \href{http://arxiv.org/abs/hep-ex/0110059}{{\ttfamily
  arXiv:hep-ex/0110059 [hep-ex]}}, [Erratum: {\em ibid.} {\bf
  90},239902(2003)].

\bibitem{Gambino:1993dd}
P.~Gambino and A.~Sirlin, {\em Phys. Rev.} {\bf D49}, 1160  (1994),
  \href{http://arxiv.org/abs/hep-ph/9309326}{{\ttfamily arXiv:hep-ph/9309326
  [hep-ph]}}.

\bibitem{Blondel:1989ev}
A.~Blondel {\em et~al.}, {\em Z. Phys.} {\bf C45}, 361  (1990).

\bibitem{Allaby:1987vr}
 CHARM Collaboration (J.~V. Allaby {\em et~al.}), {\em Z. Phys.} {\bf C36},
  611  (1987).

\bibitem{McFarland:1997wx}
 CCFR Collaboration (K.~S. McFarland {\em et~al.}), {\em Eur. Phys. J.} {\bf
  C1}, 509  (1998), \href{http://arxiv.org/abs/hep-ex/9701010}{{\ttfamily
  arXiv:hep-ex/9701010 [hep-ex]}}.

\bibitem{Paschos:1972kj}
E.~A. Paschos and L.~Wolfenstein, {\em Phys. Rev.} {\bf D7}, 91  (1973).

\bibitem{Agashe:2014kda}
 Particle Data Collaboration (K.~Olive {\em et~al.}), {\em Chin. Phys.} {\bf
  C38},   090001  (2014).

\bibitem{Abe:1996nj}
 SLD Collaboration (K.~Abe {\em et~al.}), {\em Phys. Rev. Lett.} {\bf 78}, 2075
   (1997), \href{http://arxiv.org/abs/hep-ex/9611011}{{\ttfamily
  arXiv:hep-ex/9611011 [hep-ex]}}.

\bibitem{Abe:2005nqa}
 SLD Collaboration (K.~Abe {\em et~al.}), {\em Phys. Rev.} {\bf D71},   112004
  (2005), \href{http://arxiv.org/abs/hep-ex/0503005}{{\ttfamily
  arXiv:hep-ex/0503005 [hep-ex]}}.

\bibitem{Bernreuther:2016ccf}
W.~Bernreuther, L.~Chen, O.~Dekkers, T.~Gehrmann and D.~Heisler, {\em JHEP}
  {\bf 01},   053  (2017), \href{http://arxiv.org/abs/1611.07942}{{\ttfamily
  arXiv:1611.07942 [hep-ph]}}.

\bibitem{Altarelli:1992fs}
G.~Altarelli and B.~Lampe, {\em Nucl. Phys.} {\bf B391}, 3  (1993).

\bibitem{Ravindran:1998jw}
V.~Ravindran and W.~L. van Neerven, {\em Phys. Lett.} {\bf B445}, 214  (1998),
  \href{http://arxiv.org/abs/hep-ph/9809411}{{\ttfamily arXiv:hep-ph/9809411
  [hep-ph]}}.

\bibitem{Catani:1999nf}
S.~Catani and M.~H. Seymour, {\em JHEP} {\bf 07},   023  (1999),
  \href{http://arxiv.org/abs/hep-ph/9905424}{{\ttfamily arXiv:hep-ph/9905424
  [hep-ph]}}.

\bibitem{Weinzierl:2006yt}
S.~Weinzierl, {\em Phys. Lett.} {\bf B644}, 331  (2007),
  \href{http://arxiv.org/abs/hep-ph/0609021}{{\ttfamily arXiv:hep-ph/0609021
  [hep-ph]}}.

\bibitem{Jersak:1979uv}
J.~Jersak, E.~Laermann and P.~M. Zerwas, {\em Phys. Lett.} {\bf 98B}, 363
  (1981).

\bibitem{Djouadi:1989uk}
A.~Djouadi, J.~H. K{\"u}hn and P.~M. Zerwas, {\em Z. Phys.} {\bf C46}, 411
  (1990).

\bibitem{Abazov:2017gpw}
 \DZero Collaboration (V.~M. Abazov {\em et~al.}), {\em Phys. Rev. Lett.} {\bf
  120},   241802  (2018), \href{http://arxiv.org/abs/1710.03951}{{\ttfamily
  arXiv:1710.03951 [hep-ex]}}.

\bibitem{Aaltonen:2016nuy}
 CDF Collaboration (T.~Aaltonen {\em et~al.}), {\em Phys. Rev.} {\bf D93},
  112016  (2016), \href{http://arxiv.org/abs/1605.02719}{{\ttfamily
  arXiv:1605.02719 [hep-ex]}}, [Addendum: {\em ibid.} {\bf 95},119901(2017)].

\bibitem{Aaltonen:2018dxj}
 CDF and \DZero Collaborations (T.~Aaltonen {\em et~al.}), {\em Phys. Rev.}
  {\bf D97},   112007  (2018),
  \href{http://arxiv.org/abs/1801.06283}{{\ttfamily arXiv:1801.06283
  [hep-ex]}}.

\bibitem{Sirunyan:2018swq}
 CMS Collaboration (A.~M. Sirunyan {\em et~al.}), {\em Eur. Phys. J.} {\bf
  C78},   701  (2018), \href{http://arxiv.org/abs/1806.00863}{{\ttfamily
  arXiv:1806.00863 [hep-ex]}}.

\bibitem{ATLAS:2018gqq}
ATLAS Collaboration, {\em {Measurement of the effective leptonic weak mixing
  angle using electron and muon pairs from $Z$-boson decay in the ATLAS
  experiment at $\sqrt s = 8$ TeV}}, Tech. Rep. ATLAS-CONF-2018-037, CERN
  (2018).

\bibitem{Aaij:2015lka}
 LHCb Collaboration (R.~Aaij {\em et~al.}), {\em JHEP} {\bf 11},   190  (2015),
  \href{http://arxiv.org/abs/1509.07645}{{\ttfamily arXiv:1509.07645
  [hep-ex]}}.

\bibitem{Collins:1977iv}
J.~Collins and D.~Soper, {\em Phys.Rev.} {\bf D16},   2219  (1977).

\bibitem{Mirkes:1994eb}
E.~Mirkes and J.~Ohnemus, {\em Phys.Rev.} {\bf D50}, 5692  (1994),
  \href{http://arxiv.org/abs/hep-ph/9406381}{{\ttfamily arXiv:hep-ph/9406381
  [hep-ph]}}.

\bibitem{Mirkes:1994dp}
E.~Mirkes and J.~Ohnemus, {\em Phys.Rev.} {\bf D51}, 4891  (1995),
  \href{http://arxiv.org/abs/hep-ph/9412289}{{\ttfamily arXiv:hep-ph/9412289
  [hep-ph]}}.

\bibitem{Schott:2014sea}
M.~Schott and M.~Dunford, {\em Eur. Phys. J.} {\bf C74},   2916  (2014),
  \href{http://arxiv.org/abs/1405.1160}{{\ttfamily arXiv:1405.1160 [hep-ex]}}.

\bibitem{Aaltonen:2011nr}
 CDF Collaboration (T.~Aaltonen {\em et~al.}), {\em Phys. Rev. Lett.} {\bf
  106},   241801  (2011), \href{http://arxiv.org/abs/1103.5699}{{\ttfamily
  arXiv:1103.5699 [hep-ex]}}.

\bibitem{Khachatryan:2015paa}
 CMS Collaboration (V.~Khachatryan {\em et~al.}), {\em Phys. Lett.} {\bf B750},
  154  (2015), \href{http://arxiv.org/abs/1504.03512}{{\ttfamily
  arXiv:1504.03512 [hep-ex]}}.

\bibitem{Aad:2016izn}
 ATLAS Collaboration (G.~Aad {\em et~al.}), {\em JHEP} {\bf 08},   159  (2016),
  \href{http://arxiv.org/abs/1606.00689}{{\ttfamily arXiv:1606.00689
  [hep-ex]}}.

\bibitem{CMS:2017zzj}
CMS Collaboration, {\em {Measurement of the weak mixing angle with the
  forward-backward asymmetry of Drell-Yan events at 8 TeV}}, Tech. Rep.
  CMS-PAS-SMP-16-007, CERN  (2017).

\bibitem{Nadolsky:2008zw}
P.~Nadolsky {\em et~al.}, {\em Phys.Rev.} {\bf D78},   013004  (2008),
  \href{http://arxiv.org/abs/0802.0007}{{\ttfamily arXiv:0802.0007 [hep-ph]}}.

\bibitem{Abazov:2014jti}
 \DZero Collaboration (V.~M. Abazov {\em et~al.}), {\em Phys. Rev. Lett.} {\bf
  115},   041801  (2015), \href{http://arxiv.org/abs/1408.5016}{{\ttfamily
  arXiv:1408.5016 [hep-ex]}}.

\bibitem{Bodek2010}
A.~Bodek, {\em Eur. Phys. J.} {\bf C67}, 321  (2010),
  \href{http://arxiv.org/abs/0911.2850}{{\ttfamily arXiv:0911.2850 [hep-ex]}}.

\bibitem{Frixione:2007vw}
S.~Frixione, P.~Nason and C.~Oleari, {\em JHEP} {\bf 0711},   070  (2007),
  \href{http://arxiv.org/abs/0709.2092}{{\ttfamily arXiv:0709.2092 [hep-ph]}}.

\bibitem{Alioli:2010xd}
S.~Alioli, P.~Nason, C.~Oleari and E.~Re, {\em JHEP} {\bf 1006},   043  (2010),
  \href{http://arxiv.org/abs/1002.2581}{{\ttfamily arXiv:1002.2581 [hep-ph]}}.

\bibitem{Ball:2014uwa}
 NNPDF Collaboration (R.~D. Ball {\em et~al.}), {\em JHEP} {\bf 04},   040
  (2015), \href{http://arxiv.org/abs/1410.8849}{{\ttfamily arXiv:1410.8849
  [hep-ph]}}.

\bibitem{Bodek:2016olg}
A.~Bodek, J.~Han, A.~Khukhunaishvili and W.~Sakumoto, {\em Eur. Phys. J.} {\bf
  C76},   115  (2016), \href{http://arxiv.org/abs/1507.02470}{{\ttfamily
  arXiv:1507.02470 [hep-ex]}}.

\bibitem{cite:DYTURBO}
S.~Camarda, {DYTurbo:\ Fast predictions for Drell Yan processes,} {talk
  presented at the {\em LHC EW Precision sub-group meeting, CERN, November
  13-15, 2018}}.

\bibitem{Bardin:1989tq}
D.~{\relax Yu}. Bardin, M.~S. Bilenky, T.~Riemann, M.~Sachwitz and H.~Vogt,
  {\em Comput. Phys. Commun.} {\bf 59}, 303  (1990).

\bibitem{Jadach:1999vf}
S.~Jadach, B.~F.~L. Ward and Z.~Was, {\em Comput. Phys. Commun.} {\bf 130}, 260
   (2000), \href{http://arxiv.org/abs/hep-ph/9912214}{{\ttfamily
  arXiv:hep-ph/9912214 [hep-ph]}}.

\bibitem{Richter-Was:2018lld}
E.~Richter-Was and Z.~Was, {The TauSpinner approach for electroweak corrections
  in LHC $Z \to ll$ observables}  (2018),
  \href{http://arxiv.org/abs/1808.08616}{{\ttfamily arXiv:1808.08616
  [hep-ph]}}.

\bibitem{Cortiana:2015rca}
G.~Cortiana, {\em Rev. Phys.} {\bf 1}, 60  (2016),
  \href{http://arxiv.org/abs/1510.04483}{{\ttfamily arXiv:1510.04483
  [hep-ex]}}.

\bibitem{Boos:2015bta}
E.~Boos, O.~Brandt, D.~Denisov, S.~Denisov and P.~Grannis, {\em Phys. Usp.}
  {\bf 58}, 1133  (2015), \href{http://arxiv.org/abs/1509.03325}{{\ttfamily
  arXiv:1509.03325 [hep-ex]}}.

\bibitem{EliasMiro:2011aa}
J.~Elias-Miro {\em et~al.}, {\em Phys. Lett.} {\bf B709}, 222  (2012),
  \href{http://arxiv.org/abs/1112.3022}{{\ttfamily arXiv:1112.3022 [hep-ph]}}.

\bibitem{Beneke:1998ui}
M.~Beneke, {\em Phys. Rept.} {\bf 317}, 1  (1999),
  \href{http://arxiv.org/abs/hep-ph/9807443}{{\ttfamily arXiv:hep-ph/9807443
  [hep-ph]}}.

\bibitem{Hoang:2008xm}
A.~H. Hoang and I.~W. Stewart, {\em Nucl. Phys. Proc. Suppl.} {\bf 185}, 220
  (2008), \href{http://arxiv.org/abs/0808.0222}{{\ttfamily arXiv:0808.0222
  [hep-ph]}}.

\bibitem{Marquard:2015qpa}
P.~Marquard, A.~V. Smirnov, V.~A. Smirnov and M.~Steinhauser, {\em Phys. Rev.
  Lett.} {\bf 114},   142002  (2015),
  \href{http://arxiv.org/abs/1502.01030}{{\ttfamily arXiv:1502.01030
  [hep-ph]}}.

\bibitem{Kataev:2015gvt}
A.~L. Kataev and V.~S. Molokoedov, {\em Eur. Phys. J. Plus} {\bf 131},   271
  (2016), \href{http://arxiv.org/abs/1511.06898}{{\ttfamily arXiv:1511.06898
  [hep-ph]}}.

\bibitem{Beneke:2016cbu}
M.~Beneke, P.~Marquard, P.~Nason and M.~Steinhauser, {\em Phys. Lett.} {\bf
  B775}, 63  (2017), \href{http://arxiv.org/abs/1605.03609}{{\ttfamily
  arXiv:1605.03609 [hep-ph]}}.

\bibitem{Butenschoen:2016lpz}
M.~Butenschoen {\em et~al.}, {\em Phys. Rev. Lett.} {\bf 117},   232001
  (2016), \href{http://arxiv.org/abs/1608.01318}{{\ttfamily arXiv:1608.01318
  [hep-ph]}}.

\bibitem{Kataev:2018mob}
A.~L. Kataev and V.~S. Molokoedov, {Dependence of five and six-loop estimated
  QCD corrections to the relation between pole and running masses of heavy
  quarks on the number of light flavours}  (2018),
  \href{http://arxiv.org/abs/1811.02867}{{\ttfamily arXiv:1811.02867
  [hep-ph]}}.

\bibitem{Degrassi:2012ry}
G.~Degrassi {\em et~al.}, {\em JHEP} {\bf 08},   098  (2012),
  \href{http://arxiv.org/abs/1205.6497}{{\ttfamily arXiv:1205.6497 [hep-ph]}}.

\bibitem{Aaltonen:2012va}
 CDF Collaboration (T.~Aaltonen {\em et~al.}), {\em Phys. Rev. Lett.} {\bf
  109},   152003  (2012), \href{http://arxiv.org/abs/1207.6758}{{\ttfamily
  arXiv:1207.6758 [hep-ex]}}.

\bibitem{Aaltonen:2013aqa}
 CDF Collaboration (T.~Aaltonen {\em et~al.}), {\em Phys. Rev.} {\bf D88},
  011101  (2013), \href{http://arxiv.org/abs/1305.3339}{{\ttfamily
  arXiv:1305.3339 [hep-ex]}}.

\bibitem{Abazov:2015spa}
 \DZero Collaboration (V.~M. Abazov {\em et~al.}), {\em Phys. Rev.} {\bf D91},
   112003  (2015), \href{http://arxiv.org/abs/1501.07912}{{\ttfamily
  arXiv:1501.07912 [hep-ex]}}.

\bibitem{Khachatryan:2015hba}
 CMS Collaboration (V.~Khachatryan {\em et~al.}), {\em Phys. Rev.} {\bf D93},
  072004  (2016), \href{http://arxiv.org/abs/1509.04044}{{\ttfamily
  arXiv:1509.04044 [hep-ex]}}.

\bibitem{Sirunyan:2017idq}
 CMS Collaboration (A.~M. Sirunyan {\em et~al.}), {\em Phys. Rev.} {\bf D96},
  032002  (2017), \href{http://arxiv.org/abs/1704.06142}{{\ttfamily
  arXiv:1704.06142 [hep-ex]}}.

\bibitem{Aad:2015nba}
 ATLAS Collaboration (G.~Aad {\em et~al.}), {\em Eur. Phys. J.} {\bf C75},
  330  (2015), \href{http://arxiv.org/abs/1503.05427}{{\ttfamily
  arXiv:1503.05427 [hep-ex]}}.

\bibitem{D0:2016ull}
 \DZero Collaboration (V.~M. Abazov {\em et~al.}), {\em Phys. Rev.} {\bf D94},
   032004  (2016), \href{http://arxiv.org/abs/1606.02814}{{\ttfamily
  arXiv:1606.02814 [hep-ex]}}.

\bibitem{Aaboud:2016igd}
 ATLAS Collaboration (M.~Aaboud {\em et~al.}), {\em Phys. Lett.} {\bf B761},
  350  (2016), \href{http://arxiv.org/abs/1606.02179}{{\ttfamily
  arXiv:1606.02179 [hep-ex]}}.

\bibitem{Aaboud:2018zbu}
 ATLAS Collaboration (M.~Aaboud {\em et~al.}), {\em {\rm submitted to} Eur.
  Phys. J.} {\bf C}  (2018), \href{http://arxiv.org/abs/1810.01772}{{\ttfamily
  arXiv:1810.01772 [hep-ex]}}.

\bibitem{Sirunyan:2018gqx}
 CMS Collaboration (A.~M. Sirunyan {\em et~al.}), {\em Eur. Phys. J.} {\bf
  C78},   891  (2018), \href{http://arxiv.org/abs/1805.01428}{{\ttfamily
  arXiv:1805.01428 [hep-ex]}}.

\bibitem{Aaltonen:2014sea}
 CDF Collaboration (T.~Aaltonen {\em et~al.}), {\em Phys. Rev.} {\bf D90},
  091101  (2014), \href{http://arxiv.org/abs/1409.4906}{{\ttfamily
  arXiv:1409.4906 [hep-ex]}}.

\bibitem{Aaboud:2017mae}
 ATLAS Collaboration (M.~Aaboud {\em et~al.}), {\em JHEP} {\bf 09},   118
  (2017), \href{http://arxiv.org/abs/1702.07546}{{\ttfamily arXiv:1702.07546
  [hep-ex]}}.

\bibitem{Abazov:2006bd}
 \DZero Collaboration (V.~M. Abazov {\em et~al.}), {\em Phys. Rev.} {\bf D74},
   092005  (2006), \href{http://arxiv.org/abs/hep-ex/0609053}{{\ttfamily
  arXiv:hep-ex/0609053 [hep-ex]}}.

\bibitem{Fiedler:2010sg}
F.~Fiedler, A.~Grohsjean, P.~Haefner and P.~Schieferdecker, {\em Nucl. Instrum.
  Meth.} {\bf A624}, 203  (2010),
  \href{http://arxiv.org/abs/1003.1316}{{\ttfamily arXiv:1003.1316 [hep-ex]}}.

\bibitem{Abazov:2007rk}
 \DZero Collaboration (V.~M. Abazov {\em et~al.}), {\em Phys. Rev.} {\bf D75},
   092001  (2007), \href{http://arxiv.org/abs/hep-ex/0702018}{{\ttfamily
  arXiv:hep-ex/0702018 [hep-ex]}}.

\bibitem{Jegerlehner:2012kn}
F.~Jegerlehner, M.~{\relax Yu}. Kalmykov and B.~A. Kniehl, {\em Phys. Lett.}
  {\bf B722}, 123  (2013), \href{http://arxiv.org/abs/1212.4319}{{\ttfamily
  arXiv:1212.4319 [hep-ph]}}.

\bibitem{Langenfeld:2009wd}
U.~Langenfeld, S.~Moch and P.~Uwer, {\em Phys. Rev.} {\bf D80},   054009
  (2009), \href{http://arxiv.org/abs/0906.5273}{{\ttfamily arXiv:0906.5273
  [hep-ph]}}.

\bibitem{Abazov:2009ae}
 \DZero Collaboration (V.~M. Abazov {\em et~al.}), {\em Phys. Rev.} {\bf D80},
   071102  (2009), \href{http://arxiv.org/abs/0903.5525}{{\ttfamily
  arXiv:0903.5525 [hep-ex]}}.

\bibitem{Aaboud:2017ujq}
 ATLAS Collaboration (M.~Aaboud {\em et~al.}), {\em Eur. Phys. J.} {\bf C77},
  804  (2017), \href{http://arxiv.org/abs/1709.09407}{{\ttfamily
  arXiv:1709.09407 [hep-ex]}}.

\bibitem{Kieseler:2015jzh}
J.~Kieseler, K.~Lipka and S.~Moch, {\em Phys. Rev. Lett.} {\bf 116},   162001
  (2016), \href{http://arxiv.org/abs/1511.00841}{{\ttfamily arXiv:1511.00841
  [hep-ph]}}.

\bibitem{TevatronElectroweakWorkingGroup:2016lid}
CDF and \DZero Collaborations, T.~Aaltonen {\em et~al.}, {Combination of CDF
  and \DZero results on the mass of the top quark using up 9.7~fb$^{-1}$ at the
  Tevatron}  (2016), \href{http://arxiv.org/abs/1608.01881}{{\ttfamily
  arXiv:1608.01881 [hep-ex]}}.

\bibitem{ATLAS:2014wva}
ATLAS, CDF, CMS and \DZero Collaborations, {First combination of Tevatron and
  LHC measurements of the top-quark mass}  (2014),
  \href{http://arxiv.org/abs/1403.4427}{{\ttfamily arXiv:1403.4427 [hep-ex]}}.

\bibitem{Chatrchyan:2012ea}
 CMS Collaboration (S.~Chatrchyan {\em et~al.}), {\em Eur. Phys. J.} {\bf C72},
    2202  (2012), \href{http://arxiv.org/abs/1209.2393}{{\ttfamily
  arXiv:1209.2393 [hep-ex]}}.

\bibitem{Cabibbo:1961sz}
N.~Cabibbo and R.~Gatto, {\em Phys. Rev.} {\bf 124}, 1577  (1961).

\bibitem{Bennett:2006fi}
 Muon g-2 Collaboration (G.~W. Bennett {\em et~al.}), {\em Phys. Rev.} {\bf
  D73},   072003  (2006), \href{http://arxiv.org/abs/hep-ex/0602035}{{\ttfamily
  arXiv:hep-ex/0602035 [hep-ex]}}.

\bibitem{Jegerlehner:2009ry}
F.~Jegerlehner and A.~Nyffeler, {\em Phys. Rept.} {\bf 477}, 1  (2009),
  \href{http://arxiv.org/abs/0902.3360}{{\ttfamily arXiv:0902.3360 [hep-ph]}}.

\bibitem{Bethke:2009jm}
S.~Bethke, {\em Eur. Phys. J.} {\bf C64}, 689  (2009),
  \href{http://arxiv.org/abs/0908.1135}{{\ttfamily arXiv:0908.1135 [hep-ph]}}.

\bibitem{Aaboud:2017fml}
 ATLAS Collaboration (M.~Aaboud {\em et~al.}), {\em Eur. Phys. J.} {\bf C77},
  872  (2017), \href{http://arxiv.org/abs/1707.02562}{{\ttfamily
  arXiv:1707.02562 [hep-ex]}}.

\bibitem{Khachatryan:2014waa}
 CMS Collaboration (V.~Khachatryan {\em et~al.}), {\em Eur. Phys. J.} {\bf
  C75},   288  (2015), \href{http://arxiv.org/abs/1410.6765}{{\ttfamily
  arXiv:1410.6765 [hep-ex]}}.

\bibitem{Davies:2003ik}
 Fermilab Lattice, HPQCD, UKQCD and MILC Collaborations (C.~T.~H. Davies {\em
  et~al.}), {\em Phys. Rev. Lett.} {\bf 92},   022001  (2004),
  \href{http://arxiv.org/abs/hep-lat/0304004}{{\ttfamily arXiv:hep-lat/0304004
  [hep-lat]}}.

\bibitem{Balandin:1975fe}
M.~P. Balandin, V.~M. Grebenyuk, V.~G. Zinov, A.~D. Konin and A.~N. Ponomarev,
  {\em Sov. Phys. JETP} {\bf 40}, 811  (1975).

\bibitem{Giovanetti:1984yw}
K.~L. Giovanetti {\em et~al.}, {\em Phys. Rev.} {\bf D29}, 343  (1984).

\bibitem{Bardin:1984ie}
G.~Bardin {\em et~al.}, {\em Phys. Lett.} {\bf 137B}, 135  (1984).

\bibitem{Barczyk:2007hp}
 FAST Collaboration (A.~Barczyk {\em et~al.}), {\em Phys. Lett.} {\bf B663},
  172  (2008), \href{http://arxiv.org/abs/0707.3904}{{\ttfamily arXiv:0707.3904
  [hep-ex]}}.

\bibitem{Chitwood:2007pa}
 MuLan Collaboration (D.~B. Chitwood {\em et~al.}), {\em Phys. Rev. Lett.} {\bf
  99},   032001  (2007), \href{http://arxiv.org/abs/0704.1981}{{\ttfamily
  arXiv:0704.1981 [hep-ex]}}.

\bibitem{Webber:2010zf}
 MuLan Collaboration (D.~M. Webber {\em et~al.}), {\em Phys. Rev. Lett.} {\bf
  106},   041803  (2011), \href{http://arxiv.org/abs/1010.0991}{{\ttfamily
  arXiv:1010.0991 [hep-ex]}}.

\bibitem{James:1975dr}
F.~James and M.~Roos, {\em Comput. Phys. Commun.} {\bf 10}, 343  (1975).

\bibitem{Brun:1997pa}
R.~Brun and F.~Rademakers, {\em Nucl. Instrum. Meth.} {\bf A389}, 81  (1997).

\bibitem{Baak:2012kk}
M.~Baak {\em et~al.}, {\em Eur. Phys. J.} {\bf C72},   2205  (2012),
  \href{http://arxiv.org/abs/1209.2716}{{\ttfamily arXiv:1209.2716 [hep-ph]}}.

\bibitem{Falkowski:2017pss}
A.~Falkowski, M.~Gonz{\'a}lez-Alonso and K.~Mimouni, {\em JHEP} {\bf 08},   123
   (2017), \href{http://arxiv.org/abs/1706.03783}{{\ttfamily arXiv:1706.03783
  [hep-ph]}}.

\bibitem{Fan:2014vta}
J.~Fan, M.~Reece and L.~Wang, {\em JHEP} {\bf 09},   196  (2015),
  \href{http://arxiv.org/abs/1411.1054}{{\ttfamily arXiv:1411.1054 [hep-ph]}}.

\bibitem{CEPCStudyGroup:2018ghi}
CEPC Collaboration, M.~Abbrescia {\em et~al.}, {CEPC Conceptual Design Report:
  Volume 2 - Physics \& Detector}  (2018),
  \href{http://arxiv.org/abs/1811.10545}{{\ttfamily arXiv:1811.10545
  [hep-ex]}}.

\bibitem{Gomez-Ceballos:2013zzn}
 TLEP Design Collaboration (M.~Bicer {\em et~al.}), {\em JHEP} {\bf 01},   164
  (2014), \href{http://arxiv.org/abs/1308.6176}{{\ttfamily arXiv:1308.6176
  [hep-ex]}}.

\bibitem{Baer:2013cma}
H.~Baer {\em et~al.}, {The International Linear Collider Technical Design
  Report - Volume 2: Physics}  (2013),
  \href{http://arxiv.org/abs/1306.6352}{{\ttfamily arXiv:1306.6352 [hep-ph]}}.

\bibitem{Linssen:2012hp}
L.~Linssen {\em et~al.}, {Physics and Detectors at CLIC: CLIC Conceptual Design
  Report}  (2012), \href{http://arxiv.org/abs/1202.5940}{{\ttfamily
  arXiv:1202.5940 [physics.ins-det]}}.

\end{thebibliography}
\bibliographystyle{ws-ijmpa}

\end{document}